\documentclass[final,times]{elsarticle} 
\usepackage{amssymb}
\journal{Physics Reports}
\pdfoutput=1 

\newcommand{\be}{\begin{equation}}
\newcommand{\ee}{\end{equation}}
\newcommand{\simless}{\lower.5ex\hbox{$\; \buildrel < \over \sim\;$}}
\newcommand{\simgreat}{\lower.5ex\hbox{$\; \buildrel > \over \sim\;$}} 
\newcommand{\starmass}{ M_\star } 
\newcommand{\mpro}{m_{\rm p}}
\newcommand{\mbar}{\langle m \rangle} 
\newcommand{\mdm}{ m_{\rm dm} }
\newcommand{\mion}{ m_{\rm ion} }
\newcommand{\mpion}{ m_\pi }
\newcommand{\bcon}{f_{\rm g}} 
\newcommand{\prob}{ {\cal P}} 
\newcommand{\dimnum}{\mathfrak{D}} 
\newcommand{\rhov}{\rho_{\scriptscriptstyle\Lambda} } 
\newcommand{\rhom}{\rho_{\rm M}}
\newcommand{\rhob}{\rho_{\rm b}}
\newcommand{\rhor}{\rho_{\rm R}} 
\newcommand{\rhod}{\rho_{\rm dm}} 
\newcommand{\omegav}{\Omega_{\scriptscriptstyle\Lambda} } 
\newcommand{\omegam}{\Omega_{\rm M}}
\newcommand{\omegab}{\Omega_{\rm b}}
\newcommand{\omegar}{\Omega_{\rm R}} 
\newcommand{\omegad}{\Omega_{\rm dm}} 
\newcommand{\mplanck}{M_{\rm pl}}
\newcommand{\tplanck}{t_{\rm pl}}
\newcommand{\lplanck}{\ell_{\rm pl}}
\newcommand{\timebbn}{t_{\scriptscriptstyle{BBN}} } 
\newcommand{\umass}{ m_{\rm u}}
\newcommand{\dmass}{ m_{\rm d}}
\newcommand{\tmass}{ m_{\rm t}}
\newcommand{\mhiggs}{m_{\scriptscriptstyle H}} 
\newcommand{\dmzero}{ m_{\rm d0}}
\newcommand{\emass}{ m_{\rm e}}
\newcommand{\delem}{ \Delta_{\rm em}}
\newcommand{\bind}{ B }
\newcommand{\tempeq}{ T_{\rm eq} }
\newcommand{\timeeq}{ t_{\rm eq} }
\newcommand{\timecol}{ t_{\rm col} }
\newcommand{\rhoeq}{\rho_{\rm eq}} 
\newcommand{\thetacen}{\Theta_{\rm c}}
\newcommand{\tcent}{T_{\rm c}}
\newcommand{\mzero}{\mu_\star} 
\newcommand{\betacon}{\ell_\star} 
\newcommand{\conlum}{{\mathcal{C}_{\scriptscriptstyle{\kern-0.1em\star}}}} 
\newcommand{\effish}{{\cal E}} 
\newcommand{\sigdark}{\langle\sigma{v}\rangle_{\rm dm}}
\newcommand{\sigone}{\sigma_{\rm dm}}
\newcommand{\sigmasb}{\sigma_{\rm sb}} 
\newcommand{\ck}{c_n} 
\newcommand{\intwo}{\hskip 18pt} 
\newcommand{\metal}{\raisebox{0.2ex}{--}\kern-0.55em Z}

\begin{document}

\begin{frontmatter}

\title{{\bf The Degree of Fine-Tuning in our Universe -- and Others}} 

\author{Fred C. Adams$^{1,2}$}

\address{$^1$Physics Department, University of Michigan, 
Ann Arbor, MI 48109, USA} 
\address{$^2$Astronomy Department, University of Michigan, 
Ann Arbor, MI 48109, USA} 

\begin{abstract}

Both the fundamental constants that describe the laws of physics and
the cosmological parameters that determine the properties of our
universe must fall within a range of values in order for the cosmos to
develop astrophysical structures and ultimately support life. This
paper reviews the current constraints on these quantities. The
discussion starts with an assessment of the parameters that are
allowed to vary. The standard model of particle physics contains both
coupling constants $(\alpha,\alpha_{\rm s},\alpha_{\rm w})$ and
particle masses $(\umass,\dmass,\emass)$, and the allowed ranges of
these parameters are discussed first. We then consider cosmological
parameters, including the total energy density of the universe
$(\Omega)$, the contribution from vacuum energy $(\rhov)$, the
baryon-to-photon ratio $(\eta)$, the dark matter contribution
$(\delta)$, and the amplitude of primordial density fluctuations
$(Q)$.  These quantities are constrained by the requirements that the
universe lives for a sufficiently long time, emerges from the epoch of
Big Bang Nucleosynthesis with an acceptable chemical composition, and
can successfully produce large scale structures such as galaxies. On
smaller scales, stars and planets must be able to form and
function. The stars must be sufficiently long-lived, have high enough
surface temperatures, and have smaller masses than their host
galaxies. The planets must be massive enough to hold onto an
atmosphere, yet small enough to remain non-degenerate, and contain
enough particles to support a biosphere of sufficient
complexity. These requirements place constraints on the gravitational
structure constant $(\alpha_G)$, the fine structure constant
$(\alpha)$, and composite parameters $(\conlum)$ that specify nuclear
reaction rates. We then consider specific instances of possible
fine-tuning in stellar nucleosynthesis, including the triple alpha
reaction that produces carbon, the case of unstable deuterium, and the
possibility of stable diprotons. For all of the issues outlined above,
viable universes exist over a range of parameter space, which is
delineated herein. Finally, for universes with significantly different
parameters, new types of astrophysical processes can generate energy
and thereby support habitability.

\end{abstract}


\end{frontmatter}

\newpage
\noindent
{\large \bf Table of Contents} 
\bigskip
\medskip 

\noindent{\bf 1. Introduction} \dotfill 5

\smallskip{\it 1.1. Types of Fine-Tuning Arguments} 

\smallskip{\it 1.2. Scales of the Universe} 

\smallskip{\it 1.3. Formulation of the Fine-Tuning Problem} 

\smallskip{\it 1.4. Scope of this Review} 

\bigskip\noindent {\bf 2. Particle Physics Parameters} \dotfill 13

\smallskip{\it 2.1. The Standard Model of Particle Physics} 

\smallskip{\it 2.2. Constraints on Light Quark Masses} 

\smallskip{\it \intwo 2.2.1. Stability of Quarks within Hadrons} 

\smallskip{\it \intwo 2.2.2. Stability of Protons and Neutrons within Nuclei} 

\smallskip{\it \intwo 2.2.3. Stability of Free Protons and Hydrogen}

\smallskip{\it \intwo 2.2.4. Unbound Deuterium and Bound Diprotons}

\smallskip{\it \intwo 2.2.5. Plane of Allowed Quark Masses}

\smallskip{\it \intwo 2.2.6. Summary of Quark Constraints}

\smallskip{\it \intwo 2.2.7. Mass Difference between the Neutron and Proton}

\smallskip{\it \intwo 2.2.8. Constraints on the Higgs Parameters}

\smallskip{\it 2.3. Constraints on the $\alpha$-$\beta$ Plane} 

\smallskip{\it 2.4. Constraints on the Strong Coupling Constant} 

\smallskip{\it 2.5. Additional Considerations} 

\smallskip{\it \intwo 2.5.1. Charge Quantization}

\smallskip{\it \intwo 2.5.2. Constraint from Proton Decay}

\bigskip\noindent {\bf 3. Cosmological Parameters and the Cosmic Inventory} 
\dotfill 33

\smallskip{\it 3.1. Review of Parameters} 

\smallskip{\it 3.2. Constraints on the Cosmic Inventory} 

\smallskip{\it 3.3. The Flatness Problem} 

\smallskip{\it 3.4. Quantum Fluctuations and Inflationary Dynamics}

\smallskip{\it 3.5. Eternal Inflation} 

\bigskip\noindent {\bf 4. The Cosmological Constant and/or Dark Energy} 
\dotfill 49

\smallskip{\it 4.1. The Cosmological Constant Problem} 

\smallskip{\it 4.2. Bounds on the Vacuum Energy Density from Structure Formation} 

\newpage 

\bigskip\noindent {\bf 5. Big Bang Nucleosynthesis} \dotfill 55

\smallskip{\it 5.1. BBN Parameters and Processes} 

\smallskip{\it 5.2. BBN Abundances with Parameter Variations} 

\smallskip{\it \intwo 5.2.1. Variations in the Baryon to Photon Ratio}

\smallskip{\it \intwo 5.2.2. Variations in the Gravitational Constant}

\smallskip{\it \intwo 5.2.3. Variations in the Neutron Lifetime}

\smallskip{\it \intwo 5.2.4. Variations in the Fine Structure Constant}

\smallskip{\it \intwo 5.2.5. Variations in both $G$ and $\eta$}

\smallskip{\it 5.3. BBN without the Weak Interaction} 

\bigskip\noindent {\bf 6. Galaxy Formation and Large Scale Structure}
\dotfill 68

\smallskip{\it 6.1. Mass and Density Scales of Galaxy Formation} 

\smallskip{\it 6.2. Structure of Dark Matter Halos} 

\smallskip{\it 6.3. Bounds on the Amplitude of Primordial Fluctuations} 
{\it from Planet Scattering} 

\smallskip{\it 6.4. Constraints from the Galactic Background Radiation} 

\smallskip{\it 6.5. Variations in the Abundances of Dark Matter and Baryons} 

\smallskip{\it 6.6. Gravitational Potential of Galaxies} 

\smallskip{\it 6.7. Cooling Considerations} 

\bigskip\noindent {\bf 7. Stars and Stellar Evolution} \dotfill 83 

\smallskip{\it 7.1. Analytic Model for Stellar Structure} 

\smallskip{\it 7.2. Minimum Stellar Temperatures} 

\smallskip{\it 7.3. Stellar Lifetime Constraints} 

\smallskip{\it 7.4. The Triple-Alpha Reaction for Carbon Production} 

\smallskip{\it 7.5. Effects of Unstable Deuterium and Bound Diprotons on Stars}

\smallskip{\it \intwo 7.5.1. Universes with Stable Diprotons}

\smallskip{\it \intwo 7.5.2. Universes with Unstable Deuterium}

\smallskip{\it 7.6. Stellar Constraints on Nuclear Forces} 

\smallskip{\it \intwo 7.6.1. Stellar Evolution without the Weak Interaction} 

\smallskip{\it \intwo 7.6.2. Stellar Constraint on the Weak Interaction} 

\smallskip{\it \intwo 7.6.3. Supernova Constraint on the Weak Interaction}

\smallskip{\it \intwo 7.6.3. Supernova Constraint on the Nucleon Potential}

\bigskip\noindent {\bf 8. Planets} \dotfill 110 

\smallskip{\it 8.1. Mass Scale for Non-Degenerate Planets} 

\smallskip{\it 8.2. Mass Scale for Atmospheric Retention} 

\smallskip{\it 8.3. Allowed Range of Parameter Space for Planets} 

\smallskip{\it 8.4. Planet Formation} 

\smallskip{\it 8.5. Planets and Stellar Convection} 

\newpage  

\bigskip\noindent {\bf 9. Exotic Astrophysical Scenarios} \dotfill 118 

\smallskip{\it 9.1. Dark Matter Halos as Astrophysical Objects} 

\smallskip{\it \intwo 9.1.1. Power from Dark Matter Annihilation}

\smallskip{\it \intwo 9.1.2. Time Evolution of Dark Matter Halos} 

\smallskip{\it 9.2. Dark Matter Capture and Annihilation in White Dwarfs} 

\smallskip{\it 9.3. Black Holes as Stellar Power Sources} 

\smallskip{\it 9.4. Degenerate Dark Matter Stars} 

\smallskip{\it 9.5. Nuclear-Free Universe} 

\bigskip\noindent {\bf 10. Conclusion} \dotfill 134

\smallskip{\it 10.1. Summary of Fine-Tuning Constraints} 

\smallskip{\it 10.2. General Trends} 

\smallskip{\it 10.3. Anthropic Arguments} 

\smallskip{\it 10.4. Is our Universe Maximally Habitable? } 

\smallskip{\it 10.5. Open Issues} 

\smallskip{\it 10.6. Insights and Perspective} 

\bigskip\noindent 
{\bf Appendix A. Mass Scales in terms of Fundamental Constants} \dotfill 150 

\bigskip\noindent 
{\bf Appendix B. Number of Space-Time Dimensions} \dotfill 155

\smallskip{\it B.1. Stability of Classical Orbits} 

\smallskip{\it B.2. Stability of Atoms: Bound Quantum States} 

\bigskip\noindent 
{\bf Appendix C. Chemistry and Biological Molecules} \dotfill 162

\bigskip\noindent 
{\bf Appendix D. Global Bounds on the Structure Constants} \dotfill 166

\bigskip\noindent 
{\bf Appendix E. Probability Considerations} \dotfill 169 

\bigskip\noindent 
{\bf Appendix F: Nuclei and the Semi-Empirical Mass Formula}
\dotfill 173 

\bigskip\noindent {\bf References} \dotfill 178 

\newpage 

\vskip1.0truein

\bigskip 
\section{Introduction}
\label{sec:intro}

The laws of physics in our universe support the development and
operations of biology --- and hence observers --- which in turn
require the existence of a range of astrophysical structures. The
cosmos synthesizes light nuclei during its early history and later
produces a wide variety of stars, which forge the remaining entries of
the periodic table. On larger scales, galaxies condense out of the
expanding universe and provide deep gravitational potential wells that
collect and organize the necessary ingredients. On smaller scales,
planets form alongside their host stars and provide suitable
environments for the genesis and maintenance of life. Within our
universe, the laws of physics have the proper form to support all of
these building blocks that are needed for observers to arise.
However, a large and growing body of research has argued that
relatively small changes in the laws of physics could render the
universe incapable of supporting life. In other words, the universe
could be fine-tuned for the development of complexity. The overarching
goal of this contribution is to review the current arguments
concerning the possible fine-tuning of the universe and make a
quantitative assessment of its severity.

\begin{figure}[tbp]
\centering 
\includegraphics[width=1.0\textwidth,trim=0 150 0 150]{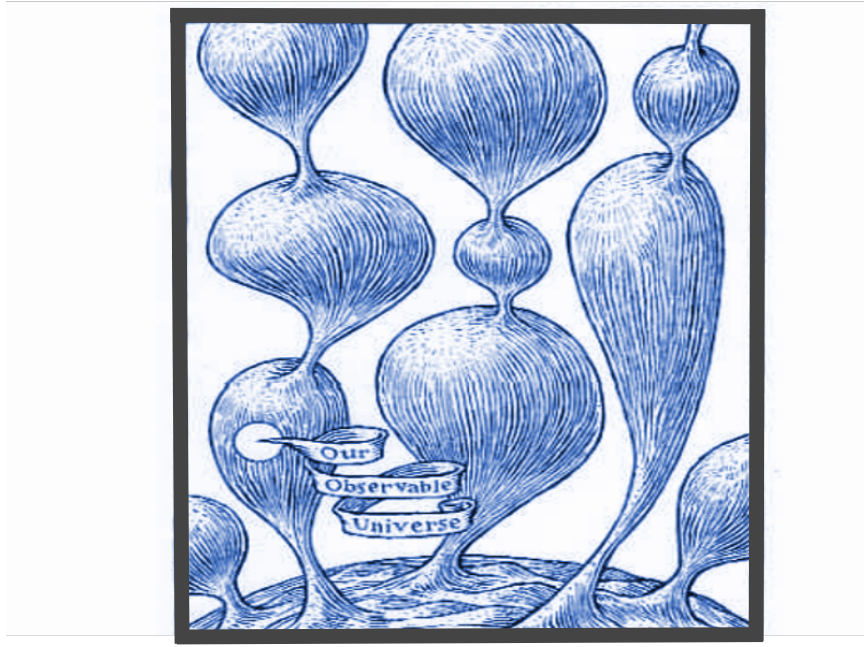} 
\vskip-48pt
\caption{Schematic representation of a small portion of the multiverse 
(adapted from \cite{book2}). Each individual universe within the larger 
ensemble is represented here as a separate expanding bubble. Every
such region could in principle have a different realization of the
laws of physics and/or different values for the cosmological parameters. 
The observable portion of our universe, depicted here as a white disk,
is a small fraction of our entire universe -- the region of space-time
that has the same version of the laws of physics.  The manner in which
the various components of the multiverse are connected is not known,
so that this depiction is heuristic. The number of theoretically
expected universes is vastly larger than the number shown here. }
\label{fig:multiverse} 
\end{figure}

\begin{figure}[tbp]
\centering 
\includegraphics[width=1.0\textwidth,trim=0 150 0 150]{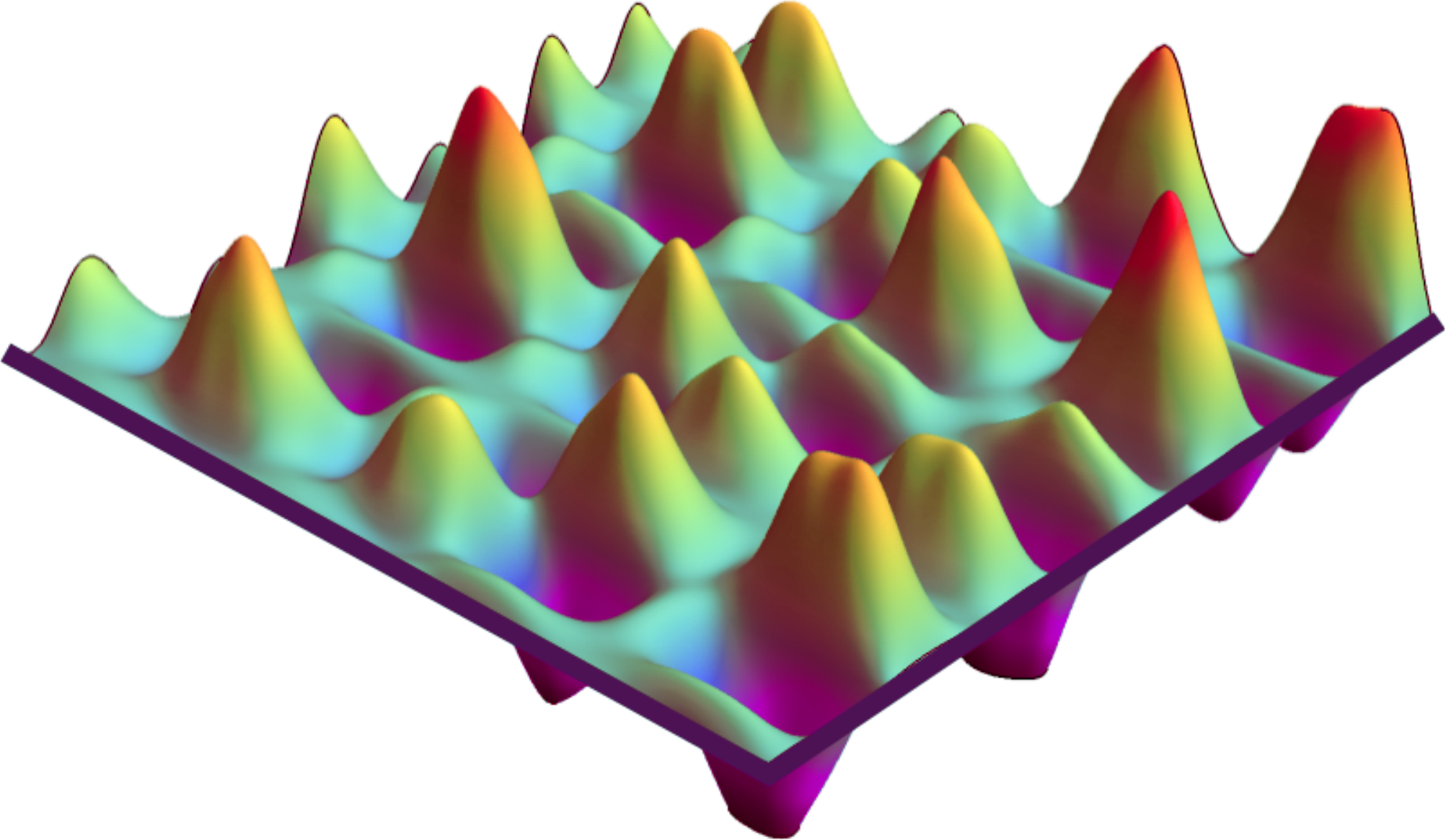}
\vskip 4.0truecm 
\caption{Schematic representation of the landscape of possible vacuum 
states of the universe (highly simplified). Each local minimum of the 
function shown here depicts a possible vacuum state for the universe, 
and hence a different possible realization of the fundamental parameters 
that determine the laws of physics. The number of vacua is expected to 
be enormously large \cite{denefdouglas,douglas2003,grana,lindevanchurin,
schellekens2008}, with typical estimates of order 
$N_{\scriptscriptstyle V}\sim10^{500}$, or larger \cite{halverson}, 
so that only a small portion of the entire landscape is shown here. } 
\label{fig:landscape} 
\end{figure}

Current cosmological theories argue that our universe may be only one
component of a vast collection of universes that make up a much larger
region of space-time, often called the ``multiverse'' or the 
``megaverse'' \cite{carrellis,davies2004,deutsch,donoghuethree,
ellis2004,garriga2008,hallnomura,lindemultiverse,reesbefore}. 
This ensemble is depicted schematically in Figure 
\ref{fig:multiverse}. Parallel developments in string theory
and its generalizations indicate that the vacuum structure of the
universe could be sampled from an enormous number of possible states
\cite{boussopolchinski,halverson,hogan2006,kachru,schellekens,susskind}. 
The potential energy function for this configuration space is depicted
schematically in Figure \ref{fig:landscape}, where each minimum
corresponds to a different low-energy universe.  If each individual
universe within the multiverse (represented by a particular bubble in
Figure \ref{fig:multiverse}) samples the underlying distribution of
possible vacuum states (by choosing a particular local minimum
represented in Figure \ref{fig:landscape}), the laws of physics could
vary from region to region within the ensemble. In this scenario, our
universe represents one small subdomain of the entire space-time with
one particular implementation of the possible versions of physical
law. Other domains could have elementary particles with different
properties and/or different cosmological parameters.  A fundamental
question thus arises: What versions of the laws of physics are
necessary for the development of astrophysical structures, which are
in turn necessary for the development of life?

\newpage 

\subsection{Types of Fine-Tuning Arguments} 
\label{sec:overview} 

Fine-tuning arguments have a long history 
\cite{dicke,dirac1937,dirac1938,gamow}.  Although many
previous treatments have concluded that the universe is fine-tuned 
for the development of life
\cite{barnes2012,barrow2002,bartip,boussoetal,carr,carter1974,carter1983,
davies2006,dirac1974,donoghue,hogan,lewbarn,reessix,schellekens,uzan,uzantwo},
it should be emphasized that different authors make this claim with
widely varying degrees of conviction (see also \cite{bradford2011,
carroll2006,davies2004,gleiser,hogan2006,linde,liviorees2018,rees2003}). 
We also note that this topic has been addressed through the lens of 
philosophy (see \cite{craigcarroll,ellisphil,friederich,leslie,smeenk} 
and references therein), although this present discussion will focus
on results from physics and astronomy.  In any case, the concept of
fine-tuning is not precisely defined. Here we start the discussion by
making the distinction between two types of tuning issues: 

The usual meaning of ``fine-tuning'' is that small changes in the
value of a parameter can lead to significant changes in the system as
a whole.  For example, if the strong nuclear force were somewhat
weaker, then the deuterium nucleus would no longer be bound. If the
strong force were somewhat stronger, then diprotons would be bound.
In both of these examples, relatively small changes (here, several
percent variations in the strong force) lead to different nuclear
inventories.  A second type of tuning arises when a parameter of
interest has a vastly different value from that expected (usually on
theoretical grounds). The cosmological constant provides an example of
this issue: The observed value of the cosmological constant is smaller
than some expectations of its value by $\sim120$ orders of magnitude.
This second type of tuning is thus hierarchical. In the first example,
the strong nuclear force can apparently vary by only $\sim10$ percent 
without rendering deuterium unstable or diprotons stable. 
Nuclear structure thus represents a possible instance of 
{\sl Sensitive Fine-Tuning}. In the second example, the value of the
cosmological constant could be a million times smaller or larger (if
the fluctuation amplitude $Q$ is also allowed to vary) and nothing
catastrophic would happen, but the values would still be much smaller
than the Planck scale (by $\sim100$ orders of magnitude or more). 
The cosmological constant is thus an example of 
{\sl Hierarchical Fine-Tuning}. 

In addition, when an unexpected hierarchy arises due to some quantity
being much smaller than its natural scale, one way to get such an
ordering is for two large numbers to almost-but-not-quite cancel. This
near cancellation of large quantities can be extremely sensitive to
their exact values and could thus require some type of tuning. This
state of affairs arises, for example, in the cosmological constant
problem \cite{caldwellkam,weinberg89} (see Section \ref{sec:rhovac}). 
This general concept is known as {\sl Naturalness}. Although many
definitions exist in the literature, the basic idea is that a quantity
in particle physics is considered unnatural if the quantum corrections
are larger than the observed value (for recent discussions of this
issue, see \cite{dine2015,wellstune} and references therein; for a
more critical point of view, see \cite{hossenfelder}). In such a
situation, the quantum corrections must (mostly) cancel in order to
allow for the observed small value to emerge.  This cancellation is
not automatic, so that it requires some measure of fine-tuning. One
way to codify this concept, due to 't Hooft, is to state a Principle
of Naturalness: A physical quantity should be small if and only if the
underlying theory becomes more symmetric in the limit where that
quantity approaches zero \cite{thooft}.

\begin{figure}[tbp]
\centering 
\includegraphics[width=1.0\textwidth,trim=0 150 0 150]{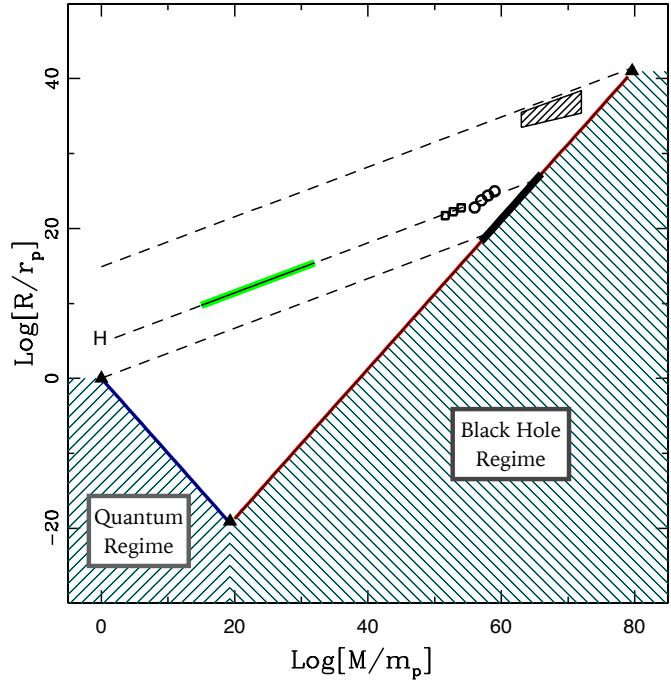}
\caption{Length and mass scales of the universe, given in units of the 
proton mass and size (adapted from \cite{carr,liviorees2018}).  The
left end point of the blue line corresponds to the mass and radius of
the proton (marked by the triangle at the origin), and its right end
point corresponds to a Planck mass black hole (lower triangle).  The
red line extends from the location of the Planck black hole to the
size/mass of the entire observable universe (upper right triangle).
Below the red line, objects are smaller than their Schwarzschild radii
and are black holes.  Below the blue line, objects are smaller than
their Compton wavelength and lie in the quantum regime. The three
dashed lines show contours of constant density, for nuclear density
(lower), atomic density (middle), and cosmic density (upper). The left
end point of the middle dashed line corresponds to the Hydrogen atom.
The range of life forms is depicted as the heavy green line segment. 
The range of known black holes is depicted as the heavy black line
segment. The circles show the locations of stars, with masses $M_\ast$
= 0.1, 1, 10, and 100 $M_\odot$ (left to right). The squares show the
locations of planets, analogs of Earth, Neptune, and Jupiter (left to
right). Galaxies lie in the shaded region in the upper right portion
of the diagram. }
\label{fig:astroscales} 
\end{figure}  

\subsection{Scales of the Universe} 
\label{sec:scales} 

The physical constituents of our universe display a hierarchy of
scales that allows it to function \cite{carr,liviorees2018,rees1980}. 
Before considering the details of fine-tuning, it is useful to assess
the scope of our particular universe. Figure \ref{fig:astroscales}
depicts the range of length scales and mass scales that allow our
universe to operate. The masses and lengths are given in units of the
proton mass and the proton size, respectively. The triangular symbol at
the origin $(0,0)$ thus marks the location of the proton. At the other
end of the diagram, the mass and size of the observable universe is
marked by the triangle at $(80,41)$.  Objects that are smaller than
their event horizons $(r<2GM/c^2$) fall below the red line, and lie in
the black hole regime.  Objects that are smaller than their Compton
wavelengths ($r<h/c\mpro$) fall below the blue line and lie in the
quantum regime.  These two regions meet at the location of a Planck
mass black hole, marked by the lower triangle at $(19,-19)$. Contours
of constant density are shown by the dashed lines in the figure.  A
number of macroscopic bodies lie near the line of atomic density (middle
dashed curve), which extends from the Hydrogen atom on the left to the
black hole boundary on the right. In between, the green line segment
shows the regime of known life forms, ranging from bacteria to whales.
Planets are depicted by the square symbols and stars are depicted by
the circles. The range of known black holes in shown by the heavy
black line segment. Note that this segment is much shorter than the
total possible range of black holes, which could span the entire red
line. Finally, the region sampled by galactic structures is shown as
the shaded region in the upper right portion of the diagram.

Figure \ref{fig:astroscales} illustrates both the challenges and
limitations posed by the scales of the universe. The full mass range
spans approximately 80 decades. The range in radial scale, while
large, is more constrained. The lower dashed curve shows the contour
of nuclear density. At large mass scales, where gravity can crush
material to higher density, objects become black holes. For lower
masses, the nuclear forces dominate, so that our universe does not
generally produce entities with sizes below the line of nuclear
density. The upper dashed curve corresponds to the density of the
universe as a whole. Objects above this curve would have densities
lower than that of background space, and would be subject to tidal
destruction.  As result, our universe does not generally produce
entities that fall above this line. The range of possible sizes for a
given mass thus spans `only' about 15 decades (with a smaller range at
high masses because of the black hole limit). The universe, with its
myriad structure, supports a parameter space that is about
$80\times15$ decades in extent. This large range of length and mass
scales is enabled by the large hierarchy between the strength of
gravity and the electromagnetic force.  As emphasized previously
\cite{carr,liviorees2018}, if gravity were stronger, this range of 
scales would be correspondingly smaller: The red line would move
upward in Figure \ref{fig:astroscales} and the real estate available
for astrophysical structures would shrink accordingly.

Another feature of the universe illustrated by Figure 
\ref{fig:astroscales} is that the regions occupied by particular 
types of terrestrial and astrophysical objects are relatively small.
The diagram shows the locations in parameter space populated by life
forms, planets, stars, black holes, and galaxies. Moreover, the
regions populated by atoms are tightly clustered near the point shown
for the Hydrogen atom. Similarly, nuclei are clustered near the
location of the proton. All of these regions are small compared to the
total available parameter space and are widely separated from each
other.

\subsection{Formulation of the Fine-Tuning Problem} 
\label{sec:formulation} 

The overarching question under review is whether the parameters of
physics in our universe are fine-tuned for the development of life.
This question, which can be stated simply, is fraught with
complications. This section outlines the basic components of the 
fine-tuning problem. 

The first step is to specify what parameters of physics and
astrophysics are allowed to vary from universe to universe. It is well
known that the Standard Model of Particle Physics has at least 26
parameters, but the theory must be extended to account for additional
physics, including gravity, neutrino oscillations, dark matter, and
dark energy (Section \ref{sec:particlephys}).  Specification of such
extensions requires additional parameters. On the other hand, not all
of the parameters are necessarily vital to the functioning of the
low-energy universe (which does not depend on the exact masses of the
heavy quarks). One hope --- not yet realized --- is that a more
fundamental theory would have fewer parameters, and that the large
number of Standard Model parameters could be derived or calculated
from the smaller set. As a result, the number of parameters could be
larger or smaller than the well-known 26. In addition to the
parameters of particle physics, the Standard Cosmological Model has
its own set of quantities that are necessary to specify the properties
of the universe (Section \ref{sec:cosmology}). These parameters
include the baryon-to-photon ratio $\eta$, the analogous ratio for
dark matter $\delta$, the amplitude $Q$ of primordial density
fluctuations, the energy density $\rhov$ of background space, and so
on.  In principle, some or all of these quantities could be calculated
from a fundamental theory, but this program cannot be carried out at
present. Even if the cosmological parameters are calcuable, their
values could depend on the expansion history of the particular
universe in question, so that these values depend on the initial
conditions (presumably set at the Planck epoch).

Once the adjustable parameters of physics and cosmology have been
identified, a full description of the problem must consider their
probability distributions. In the case of a single parameter, we need
to know the underlying probability distribution for a universe to
realize a given value of that parameter. For example, if the
underlying probability distribution is a delta function, which would
be centered on the value measured in our universe, then all universes
must be the same in this regard. In the more general case of interest
for fine-tuning arguments, the probability distributions are assumed
to be sufficiently wide that large departures from our universe are
possible. In particular, the range of possible parameters values (the
minima and maxima of the distributions) must be specified.  A full
assessment of fine-tuning requires knowledge of these fundamental
probability distributions, one for each parameter of interest
(although they are not necessarily independent). Unfortunately, these
probability distributions are not available at the present time.

The probability distributions described above are {\it priors}, i.e.,
theoretically predicted distributions that apply to a random point in
space-time at the end of the inflationary epoch (or more generally
whatever epoch of the ultra-early universe sets up its initial
conditions). As emphasized by Ref. \cite{tegmark}, one must also
consider selection effects in order to test the theoretical
predictions through experiment. For example, if a parameter affects
the formation of planets, then the probability distribution for that
parameter will be different when evaluated at a random point in
space-time or at a random planet.

The crucial next step is to determine what range of the parameters
allow for observers to develop. The question of what constitutes an
observer represents yet another complication. For the sake of
definiteness, this review considers a universe to be successful
(equivalently, viable or habitable) if it can support the full range
of astrophysical structures necessary for life or some type of
complexity to arise. We then implicitly assume that observers will
arise if the requisite structures are in place, and we won't worry 
whether the resulting observers are mice or dolphins or androids. The
list of required structures includes complex nuclei, planets, stars,
galaxies, and the universe itself. In addition to their existence,
these structures must have the right properties to support
observers. Stable nuclei must populate an adequate fraction of the
periodic table. Stars must be sufficiently hot and live for a long
time. The galaxies must have gravitational potential wells that are
deep enough to retain heavy elements produced by stars and not overly
dense so that planets can remain in orbit. The universe itself must
allow galaxies to form and live long enough for complexity to arise.
And so on. The bulk of this review describes the constraints on the
parameters of physics and astrophysics enforced by these requirements.

To summarize this discussion: In order to make a full assessment of
the degree of fine-tuning of the universe, one must address the 
following components of the problem:

\medskip\noindent[$\mathbb{I}$]
Specification of the {\sl relevant parameters} of physics and
astrophysics that can vary from universe to universe.

\medskip\noindent[$\mathbb{II}$] 
Determination of the {\sl allowed ranges of parameters} that allow for
the development of complexity and hence observers.

\medskip\noindent[$\mathbb{III}$] 
Identification of the {\sl underlying probability distributions} 
from which the fundamental parameters are drawn, including the full
possible range that the parameters can take.

\medskip\noindent[$\mathbb{IV}$] 
Consideration of {\sl selection effects} that allow the interpretation
of observed properties in the context of the {\it a priori} probability
distributions. 

\medskip\noindent[$\mathbb{V}$] 
{\sl Synthesis} of the preceding ingredients to determine the overall 
likelihood for universes to become habitable. 

\medskip\noindent
This treatment focuses primarily on first two of these considerations.
For both the Standard Model of Particle Physics and the current
Consensus Model of Cosmology, we review the full set of parameters and
identify those that have the most influence in determining the
potential habitability of the universe. Most of the manuscript then
reviews the constraints enforced on the allowed ranges of the relevant
parameters by requiring that the universe can produce and maintain
complex structures. Unfortunately, the underlying probability
distributions are not known for either the fundamental parameters of
physics or the cosmological parameters. As a result, these
distributions and how they influence selection effects are considered
only briefly. Similarly, selection effects depend on the probability
distributions for the fundamental parameters and cannot be properly 
addressed at this time. 

\subsection{Scope of this Review}  
\label{sec:scope} 

The consideration of possible alternate universes, here with different
incarnations of the laws of physics, is by definition a counterfactual
enterprise. This review considers the ranges of physical parameters
that allow such a universe to be viable. Since alternate universes are
not observable, this endeavor necessarily lies near the boundary of
science \cite{carrellis,ellis2004}. Nonetheless, this discussion is
useful on several fronts: First, one can take the existence of the
multiverse seriously, so that other universes are considered to
actually exist, and the question of their possible habitability is
relevant \cite{reesbefore}.  Moreover, if multiverse theory becomes
sufficiently developed, then one could in principle predict the
probability for a universe to have a particular realization of the
laws of physics, and hence estimate the probability of a universe
becoming habitable. Second, anthropic arguments \cite{carr,bartip} 
are currently being used as an explanation for why the universe has
its observed version of the laws of physics. In order to understand
both of these issues, the first step is to determine the ranges of
parameters that allow a universe to develop structure and complexity. 
Finally, and perhaps most importantly, studying the degree of tuning
necessary for the universe to operate provides us with a greater
understanding of how it works.

In this review, the term {\sl multiverse} refers to the ensemble of
other possible universes represented schematically in Figure
\ref{fig:multiverse} --- other regions of space-time that are far
away and largely disconnected from our own universe. For completeness,
we note that the Many Worlds Interpretation of quantum mechanics 
\cite{dewitt,everett} describes physical reality as bifurcating into 
multiple copies of itself and this collection of possibilities is
sometimes also called a multiverse \cite{deutsch}.  Here we consider
the multiverse only in the first, cosmological sense. The philosophy
of quantum mechanics, and hence the second type of multiverse, is
beyond the scope of this present treatment.

This review is organized as follows: We first consider the Standard
Model of Particle Physics in Section \ref{sec:particlephys}. After
discussing the full range of parameters, we focus on the subset of
quantities that have the greatest influence in determining the
properties of complex structures and then discuss constraints on those
parameters resulting primarily from considerations of particle
physics.  Additional constraints resulting from astrophysical
requirements are discussed in subsequent sections. The standard model
of cosmology is presented in Section \ref{sec:cosmology}. The full
range of cosmological parameters is reviewed, along with an assessment
of the most important quantities for producing structure and some
basic constraints on the cosmic inventory. The case of the
cosmological constant (dark energy) is of particular interest and is
considered separately in Section \ref{sec:rhovac}. The epoch of Big
Bang Nucleosynthesis (BBN) is also considered separately in Section 
\ref{sec:bbn}, which assesses how the abundances of the light
elements change with varying values for the input cosmological
parameters. Galaxy formation and galactic structure are considered in
Section \ref{sec:galaxies}, which provides constraints on both
fundamental and cosmological parameters due to required galactic
properties. Section \ref{sec:stars} considers the constraints due to
the necessity of working stars, which are required to have stable
nuclear burning configurations, sufficiently long lifetimes, and hot
photospheres.  This section also revisits the classic issues of the
triple alpha resonance for carbon production, the effects of unstable
deuterium, and the effects of stable diprotons. The required
properties of planets are considered in Section \ref{sec:planets},
where the parameter constraints are found to be similar to --- but
less limiting than -- those from stellar considerations. More exotic
scenarios are introduced in Section \ref{sec:exotica}, including
alternate sources of energy generation such as dark matter
annihilation and black hole radiation. The paper concludes in
Section \ref{sec:conclude} with a summary of the fine-tuning
constraints and a discussion of their implications. A series of
Appendices provides more in-depth discussion, and presents some
ancillary issues, including a summary of astrophysical mass scales in
terms of the fundamental constants (\ref{sec:massscales}), the number
of space-time dimensions (\ref{sec:dimensions}), molecular
bio-chemistry (\ref{sec:chemistry}), global bounds on the structure
constants (\ref{sec:globalbounds}), a brief discussion of the
underlying probability distributions for the tunable parameters
(\ref{sec:probability}), and the range of possible nuclei
(\ref{sec:semfappend}).

\medskip
{\sl A note on notation:} The particle physics literature generally
uses natural units where $\hbar=1$, $c=1$, and $G=\mplanck^{-2}$. Most
of our discussion of particle physics topics follows this convention.
On the other hand most of the astrophysical literature uses cgs units,
so the discussion of stars and planets includes the relevant factors
of $\hbar$ and $c$.

\bigskip 
\section{Particle Physics Parameters} 
\label{sec:particlephys}

A full assessment of the parameters of particle physics --- along with
an analysis of their degree of possible fine-tuning --- is complicated
by the current state of development of the field. On one hand, the
Standard Model of Particle Physics provides a remarkably successful
description of most experimental results to date. In addition to its
myriad successes, the theory is elegant and well motivated.  On the
other hand, this theory is incomplete. We already know that extensions
to the minimal version of the Standard Model are required to include 
neutrino oscillations, non-baryonic dark matter, dark energy, and
quantum gravity. Additional extensions are likely to be necessary to
account for cosmic inflation, or whatever alternate construction
explains the relevant cosmological problems, along with baryon number
violating processes that lead to the observed cosmic asymmetry.
Against this background, this section reviews the parameters of
particle physics that are known to be relevant, along with the
sensitivity of the universe to their possible variations. Allowed
ranges of parameter space are discussed for the fine structure
constant, light quark masses, the electron to proton mass ratio, and
the strong coupling constant. We also briefly consider constraints
arising from physics beyond the Standard Model, including charge
quantization and nucleon decay.

\subsection{The Standard Model of Particle Physics} 
\label{sec:standardmodel} 

Specification of the the Standard Model itself requires a large number
of parameters \cite{gaillard,kanebook}.  In the absence of neutrino
masses, the Lagrangian of the minimal Standard Model contains 19
parameters \cite{hogan}, and the inclusion of neutrinos raises the
number to 26 \cite{tegmark}.  Fortunately, however, only a subset of
these parameters appear to require critical values for the successful
functioning of the universe.  Here we first review the full set of
parameters (see \cite{particlegroup} for current values) and then
discuss the minimal subset necessary to consider variations in
alternate universes.

In this treatment, we assume that the entire set of parameters can
vary independently from universe to universe. Keep in mind, however,
that if the current Standard Model of particle physics is the
low-energy manifestation of a more fundamental theory, then the number
of parameters could be smaller --- or larger --- than that considered
here. Moreover, their variations could be correlated or even fully
coupled. In any case, following previous treatments \cite{hogan,tegmark}, 
the Standard Model parameters can be organized and enumerated as follows: 

\medskip\noindent$\bullet$
The masses of the six quarks and three leptons are specified by Yukawa
coupling coefficients. Although the coefficients appear in the
Standard Model Lagrangian, the masses appear in most phenomenological
discussions of fine-tuning. In either case, this subset of parameters
can be denoted as ${\cal S}_{\rm mass}$ = 
$\left\{u,d,s,c,t,b,e,\mu,\tau\right\}$. Here we denote the coupling  
coefficients as $G_k$, where the subscript $k=u,d,s,\dots$ refers to 
the type of particle. The corresponding particle masses are given by 
$m_k={\cal V}G_k/\sqrt{2}$, where ${\cal V}$ is the Higgs vacuum 
expectation value. 

\medskip\noindent$\bullet$ 
The Higgs mechanism allows for non-zero particle masses. The Higgs
parameters can be taken to be the Higgs mass and vacuum
expectation value, 
${\cal S}_{\rm higgs} = \left\{ \mhiggs, {\cal V} \right\}$, 
or, equivalently, the quadratic and quartic coefficients of the 
Higgs potential 
${\cal S}_{\rm higgs} = \left\{ \mu^2, \lambda \right\}$. 
The two choices of parameters are related by $\mhiggs$ =
$(-\mu^2/2)^{1/2}$ and ${\cal V}=(-\mu^2/\lambda)^{1/2}$. One should
keep in mind that more complicated versions of the Higgs potential are
possible \cite{branco_higgs}.  

\medskip\noindent$\bullet$
The quark mixing matrix (generally called the CKM matrix
\cite{cabibbo,kobayashi}) specifies the strength of flavor-changing
interactions among the quarks. The matrix is determined by three
angles and one phase, and thus requires the specification of four
parameters, which can be written in the form 
${\cal S}_{\rm ckm}$ =
$\left\{\sin\theta_{12},\sin\theta_{23},\sin\theta_{31},\delta_q\right\}$.

\medskip\noindent$\bullet$
The remaining parameters include a phase angle for the QCD vacuum and
coupling constants for the gauge group
$U(1)\times \mathit{SU}(2)\times \mathit{SU}(3)$.
The latter three parameters are often specified by the strong and weak
coupling constants (evaluated at a particular energy scale) and the
Weinberg angle, so that the remaining subset of parameters can be
written in the form ${\cal S}_{\rm g}$ = 
$\left\{ \theta_{\rm qcd}, g_{\rm s}, g_{\rm w}, \theta_W \right\}$.
The latter three parameters, in conjunction with the Higgs expectation
value, define more familiar entities. The mass of the $W^{\pm}$
particles are given by $m_W={\cal V}g_{\rm w}/2$ and the mass of the
$Z$ particle is $m_Z={\cal V}g_{\rm w}/(2\cos\theta_W)$. The
electromagnetic coupling constant, evaluated at $m_Z$, is given by
$e=g_{\rm w}\sin\theta_W$.  The corresponding electromagnetic
interaction strength then becomes $\alpha(m_Z)=e^2/4\pi$ = 
$g_{\rm w}^2\sin^2\theta_W/4\pi\approx1/128$; when scaled to zero
energy one obtains $\alpha\approx1/137$.  The weak interaction
strength can be written in the form $\alpha_{\rm w}$ = 
$g_{\rm w}^2/4\pi$. The strong interaction strength is given by
$\alpha_{\rm s}=g_{\rm s}^2/4\pi$.  Finally, we have the Fermi
constant $G_F=1/(\sqrt{2}{\cal V}^2)\approx(293{\rm GeV})^{-2}$.

\medskip\noindent$\bullet$ 
The neutrino sector includes Yukawa coupling constants to specify the 
mass of each neutrino, three mixing angles for the neutrino matrix, 
and an additional phase. Neutrino physics can thus be characterized 
by seven parameters, which can be written in the form ${\cal S}_\nu$ = 
$\left\{\nu_e,\nu_\mu,\nu_\tau,\sin\theta_{\nu12},
\sin\theta_{\nu23},\sin\theta_{\nu31},\delta_\nu\right\}$. 

\medskip\noindent$\bullet$
Although gravity is not part of the Standard Model of particle
physics, a full accounting of the forces of nature requires a
specification of its strength. Most approaches either use the
gravitational constant $G$ or, equivalently, the gravitational
fine-structure constant 
\be
\alpha_G = {G\mpro^2\over\hbar c} \approx 5.9\times10^{-39} \,.
\label{alphag} 
\ee
The incredibly small value of this dimensionless parameter is 
the source of many instances of Hierarchical Fine-Tuning. 

\medskip 

In addition to the parameters of particle physics, a number of
cosmological parameters are required to specify the properties of the
universe (see Section \ref{sec:cosmology}). These properties include
the inventory of baryons and dark matter in the universe 
\cite{feng,jungman}, as well as the amplitude of the primordial spectrum 
of density fluctuations.  With a more comprehensive theory, these
abundances --- or perhaps their distribution of allowed values ---
could in principle be calculated from the parameters of particle
physics. In the absence of such an overarching theory, however,
current approaches consider the particle physics parameters and the
cosmological parameters as separate and allow them to vary
independently (e.g., see the discussions in
\cite{bartip,hogan,tegmark}, as well as references therein).

The successful operation of a universe does not depend on the specific
values for all of the particle physics parameters found in the
Standard Model. For example, the mass of the top quark plays little
role in everyday life or in any astrophysical processes operating at
the present epoch. As a result, when considering the possible
fine-tuning of the universe, we can substantially reduce the set of 26
Standard Model parameters. Although not all existing treatments of
this issue are identical (compare \cite{cahn},
\cite{hogan}, \cite{reessix}, \cite{tegmark}, and others), the 
following reduction of parameters is representative: 

The Higgs parameters and the Yukawa coupling constants determine the
masses of quarks and leptons.  Since only the first generation
survives to form astrophysical structures (including nuclei), the
reduced set of parameters must include masses for the up quark, the
down quark, and the electron.  All of the neutrino sector can be
ignored, provided that the neutrino masses are small enough to not be
cosmologically interesting \cite{lesgourgues,pogosian,tegmarkneutrino}.  
In practice, this constraint requires 
\be
\sum_{k} m_{\nu k} \simless 1 {\rm eV} \,.
\ee
The four parameters of the CKM matrix determine how rapidly the
heavier quarks decay into the lighter ones.  As long as the decay
mechanisms operate, so that we only need to consider the first
generation of particles, the particular values of the mixing matrix
need not be fine-tuned. The decay width $\Gamma_q$ for a heavy quark 
of mass $m_q$ can be written in the general form 
\be
\Gamma_q \sim C V_{qp}^2 G_F^2 m_q^5 \,,
\ee
where $C$ is a dimensionless factor and $V_{qp}$ is the matrix element
corresponding to the decay $q\to p$. The CKM matrix represents the
inverse of a fine-tuning problem. Unless the matrix elements were
exactly zero, the heavier quarks would decay into lighter ones. We 
are also implicitly assuming that the masses of the heavy quarks are
large compared to $\umass$ and $\dmass$. With these reductions, the
minimal set of parameters can be written in the form 
\be
{\cal S}_{\rm min} = \left\{ \emass, \umass, \dmass, 
\alpha, \alpha_{\rm w}, \alpha_{\rm s}, \alpha_G \right\} \,. 
\label{minparameter} 
\ee
The value of the gravitational coupling constant is given by equation
(\ref{alphag}). The remaining coupling constants depend on energy. One 
common reference scale is the mass of the $Z$ particle, where current 
experimental measurements provide the values
\be
\left\{\alpha(m_Z), \alpha_{\rm w}(m_Z), \alpha_{\rm s}(m_Z)\right\}
\approx \left\{ 0.007818, 0.03383, 0.1186 \right\}\,.
\ee 
On the other hand, the coupling constants are sometimes given by their
effective values at zero energy. In this limit, the fine structure
constant approaches it usual value, $\alpha\rightsquigarrow1/137$. 
For the strong and weak forces, particle interactions in the low
energy limit can be described by potential energy functions of the
forms  
\be
U_{\rm s} = - {g_N^2 \over 4\pi r} \exp\left[ -m_\pi r \right] 
\qquad {\rm and} \qquad 
U_{\rm w} = {g_F^2 \over 4\pi r} \exp\left[ -m_{\rm w} r \right] \,.
\label{yukawa} 
\ee
In this treatment, the pion mass $m_\pi$ and the $W$ or $Z$ masses
(represented here as a single value $m_{\rm w}$) determine the effective 
range of the forces. The coefficient for the weak force is related to 
the Fermi constant according to $g_F^2=4\pi G_F \mpro^2$ so that the 
weak coupling constant in this limit is given by $\alpha_{\rm w}$ = 
$G_F\mpro^2\approx10^{-5}$. Similarly, the constant $g_N$ is the 
effective charge of the nucleon-nucleon interaction, and the corresponding 
coupling constant $\alpha_{\rm s}\approx15$. As a result, the values for 
the coupling constants are sometimes quoted in the form 
$\{\alpha,\alpha_{\rm w},\alpha_{\rm s}\}\approx\{10^{-2},10^{-5},10\}$.

Another derived parameter that plays a role in many fine-tuning 
discussions is the ratio of the electron mass to the proton mass. 
This quantity, 
\be
\beta \equiv {\emass \over \mpro} \approx {1 \over 1836}\,,
\label{definebeta} 
\ee
is a function of the more fundamental parameters given in equation
(\ref{minparameter}). Note that some authors define $\beta$ as the
inverse of that given in equation (\ref{definebeta}) and the ratio 
is sometimes denoted by the symbol $\mu$. 

\subsection{Constraints on Light Quark Masses}  
\label{sec:quarklimit} 

A large body of previous work has placed constraints on the allowed
range of particle masses, including quark masses
\cite{barrkhan,bedaque,berengut,damour,donoghue,hogan,jaffe}, the Higgs 
mass \cite{donoghuetwo}, the proton mass \cite{page}, and the Standard
Model in general \cite{hallnomura,hallnomura2010}.  This section
reviews and reconsiders the conventional arguments for the allowed
range of light quark masses. Constraints are imposed by the
requirements that protons and neutrons do not decay within nuclei, 
and that both free protons and hydrogen atoms are stable. Previous
work often invokes the additional requirements that deuterium nuclei
are bound, and that diprotons must remain unstable 
\cite{bartip,hogan,reessix}.  However, recent studies of stellar 
evolution in other universes indicate that stars continue to operate
with both stable diprotons \cite{barnes2015} and unstable deuterium 
\cite{agdeuterium}, so that the corresponding constraints on quark 
masses should be removed (see Section \ref{sec:stars}). With this
generalized treatment, the allowed region in parameter space for the
light quark masses is larger and thus exhibits less evidence for
fine-tuning.

As discussed above, the Standard Model does not specify the values of
the quark masses (or, equivalently, the values of the coupling
constants that determine the quark masses). Moreover, the distribution
of possible quark masses is also unknown. As a starting point, only
the masses of the lightest two quarks (up and down) are allowed to
vary in the discussion below. 

\begin{figure}[tbp]
\centering 
\includegraphics[width=1.0\textwidth,trim=0 150 0 150,clip]{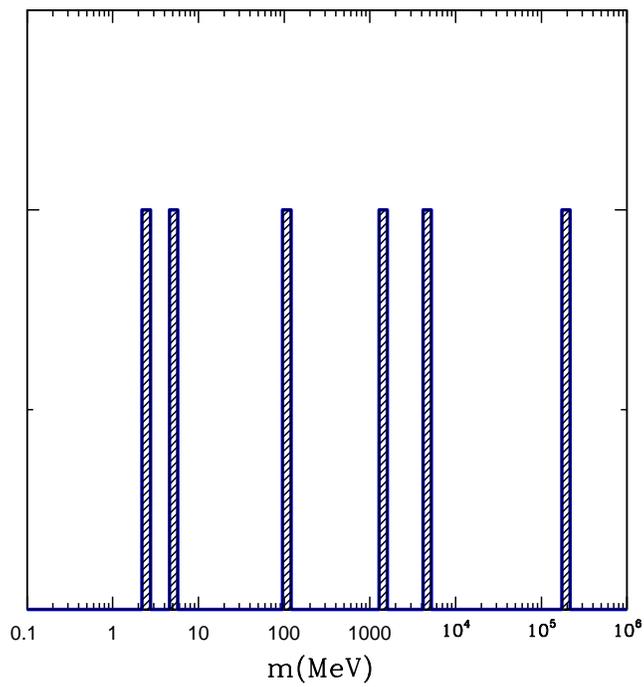}
\caption{Experimentally measured quark masses. The spikes correspond to 
the six known quarks $(u,d,s,c,b,t)$, from left to right. The quark
masses are relatively evenly spaced on a logarithmic scale (see 
\cite{donoghuemass,jaffe}). }
\label{fig:quarkhist} 
\end{figure}

Although we do not need to specify the possible distribution of quark
masses to determine the range of possible values, it is useful to plot
the allowed parameter space in logarithmic units. If the allowed quark
masses were distributed in a log-random manner, then the allowed areas
in such diagrams would reflect the probability of successful
realizations of the parameters. The only direct input we have on this
issue is the experimentally determined masses for the six known
quarks.  The distribution of these masses is shown in Figure 
\ref{fig:quarkhist}, which indicates that the logarithmic quantities 
$\log{m_q}$ are relatively evenly spaced. This apparent trend holds up
under more rigorous statistical tests \cite{donoghuemass}. As a result,
as stated in \cite{jaffe}, ``there is reason to assume that the
logarithms of the quark masses are smoothly distributed over a range
of masses small compared to the Planck scale'' (see their Figure 2).
The measured lepton masses $(e,\mu,\tau)$ are also distributed in a
manner that is consistent with log-uniform \cite{donoghuemass}, but
definitive conclusions are difficult with only three values.  Notice
also that the masses of all the quarks and leptons are small compared 
to the Planck scale (by 17 to 22 orders of magnitude), so that some 
degree of hierarchical fine-tuning is present. 

\subsubsection{Stability of Quarks within Hadrons}
\label{sec:hadrons} 

If the mass difference between up quarks and down quarks is too large,
then the heavier quark can decay into the lighter one within a hadron
(such as a proton or neutron). In order to prevent such decays, and
allow for long-lived particles of interest, there exists an upper
limit to the mass difference between the quarks, as outlined
below \cite{hogan,barrkhan}.

Down quarks can beta decay into up quarks inside of hadrons 
so that protons and neutrons could not exist. In this limit, 
the only stable particles would consist of only up quarks, 
so the universe would be composed of $\Delta^{++}$ = $(uuu)$.
This condition places an upper limit on the down quark mass, 
which can be written in the from 
\be
\dmass < \umass + \emass + E_3 \,, 
\ee
where $E_3$ is the energy required to produce an anti-symmetric 
state of three quarks. In our universe, we have $E_3\approx300$ MeV. 

In the opposite limit where the mass of the up quark is much larger
than that of the down quark, the opposite decay can happen. This 
condition thus places an analogous upper limit on the up quark mass, 
\be
\umass < \dmass + \emass + E_3\,.
\ee

\subsubsection{Stability of Protons and Neutrons within Nuclei} 
\label{sec:nuclei} 

Another constraint arises by requiring that both protons and neutrons
are stable within atomic nuclei. If the mass differences between the
up and down quark are too large, then beta decay can take place within 
nuclei. 

First consider the usual beta decay of a neutron inside a nucleus: 
\be
(A,Z) \to (A-1,Z) + p + e^{+} + {\bar \nu_e}\,.
\ee
The requirement that this decay is not allowed on energetic grounds 
can be written in the form 
\be
\dmass < \umass + \emass + \delem + \bind \,,
\label{uppernuke} 
\ee
where $\delem$ is the contribution to the mass difference between 
the proton and neutron due to the electromagnetic force, and where 
$\bind$ is the binding energy of the proton within the nucleus. 
In our universe $\delem\approx1.7$ MeV. The binding energy varies 
with the nuclear species in question, but has a typical value of 
$B\approx10$ MeV. 

Similarly, protons should also be stable within atomic nuclei, 
so that the process
\be
(A,Z) \to (A-1,Z-1) + n + e^{+} + \nu_e
\ee
should be suppressed. This requirement, in turn, places an upper 
limit on the mass of the up quark, 
\be
\umass < \dmass + \emass - \delem + B \,. 
\label{lowernuke} 
\ee

\subsubsection{Stability of Free Protons and Hydrogen} 
\label{sec:protons} 

In order for Hydrogen to exist, protons cannot spontaneously 
decay into neutrons via $p \to n + e^{+} + \nu_e$.
Preventing this reaction from occurring implies the limit 
\be
\dmass > \umass - \emass + \delem \,.
\label{lowerproton} 
\ee
We get a similar but slightly stronger constraint by requiring that 
hydrogen atoms cannot convert themselves into neutrons through 
the reaction $p + e^{-} \to n + \nu_e$. This requirement implies
the constraint 
\be
\dmass > \umass + \emass + \delem \,. 
\label{lowerhydro} 
\ee

\subsubsection{Unbound Deuterium and Bound Diprotons} 
\label{sec:twonuke} 

Many previous treatments consider a universe to be uninhabitable if
deuterium becomes unbound or if the diproton becomes bound. Although 
these constraints are not necessary for a universe to be habitable, 
it is nonetheless instructive to consider the conditions required for 
unbound deuterium or bound diprotons.  

The customary argument for the first case is that deuterium is a
necessary stepping stone for nuclear reactions. The universe starts
with only protons and neutrons, although the latter decay through the
weak interaction. The reaction $p + p \to d + e^+ + \nu$ is the first
step of the reaction chain in the Sun, whereas $p+n\to d+\gamma$ is the
first step in BBN. If deuterium is unstable, then -- the argument goes
-- no complex nuclei can be made.  However, recent work shows that
stars can continue to make complex nuclei even if deuterium is
unstable (see Section \ref{sec:stars} and
Refs. \cite{agdeuterium,barnes2017}). Nonetheless, it is instructive
to review the constraints that would be met if deuterium is required
to be stable (see \cite{beane} for a more detailed treatment).  The
stability of deuterium to beta decay is essentially the same as
equation (\ref{lowernuke}) for the case where the binding energy is
that of deuterium in our universe, so that $B$ = $B_D\approx2.2$ MeV
and the constraint becomes

\be
\umass < \dmass + \emass - \delem + B_D \,. 
\label{deutbeta} 
\ee
A weaker but more convincing constraint arises from the requirement 
that deuterium nuclei are stable to decay from the strong interaction
where $d \to p + n$. This constraint requires that the binding energy 
of deuterium is positive. One model \cite{barrkhan} writes the 
modified binding energy ${\widetilde B}_D$ in the form 
\be
{\widetilde B}_D = B_D - b 
\left( {\umass+\dmass\over(\umass+\dmass)_0} - 1 \right) \,, 
\ee
where $B_D=2.2$ MeV is the binding energy for deuterium in our 
universe. The parameter $b$ is not well-determined, but lies in 
the range $b = 1.3-5.5$ MeV. The constraint thus has the form 
\be
{\umass+\dmass\over(\umass+\dmass)_0} < 1 + {B_D \over b} \,, 
\ee
and requires the sum of the quark masses to be less about about 
twice their measured values. 

Going in the other direction, if the quark masses are lighter, then 
the pion mass is smaller, and the strong force has a greater range. 
For sufficiently small quark masses, diprotons are stable, so one 
obtains the constraint 
\be
{\umass+\dmass\over(\umass+\dmass)_0} \simgreat 0.4\,.
\label{quarkdiproton} 
\ee
The value appearing on the right hand side of this inequality 
varies with the author (compare \cite{barrkhan} and \cite{hogan}). 
As discussed below (Section \ref{sec:stars}), stable diprotons are 
not problematic for habitability, so the constraint of equation 
(\ref{quarkdiproton}) is not required to be enforced (Section 
\ref{sec:stars}).  

\subsubsection{Constraints on Quark Masses} 
\label{sec:quarkplane} 

\begin{figure}[tbp]
\centering 
\includegraphics[width=1.0\textwidth,trim=0 150 0 150,clip]{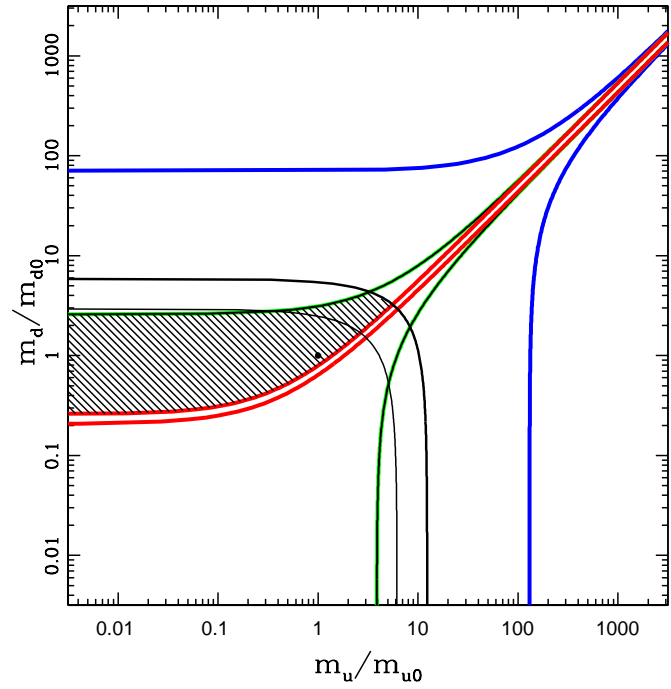}
\caption{Allowed range of quark masses, where the masses are scaled to 
the values in our universe. The blue curves show the constraints resulting 
from the requirement that quarks cannot decay within hadrons. The green 
curves show the constraints resulting from the requirement that nuclei 
are stable. The red curves result from the requirement that free protons 
are stable (lower curve) and that protons within hydrogen atoms are stable
(upper curve). The black dot marks the location of our universe in the 
diagram. The black curves are the contours where the sum of the quark
masses are two times larger (lower curve) and four times larger (upper
curve) than the observed value in our universe. The shaded region
depicts the most likely habitable region in the diagram (see text). }
\label{fig:updown} 
\end{figure}

The treatment thus far allows for a three dimensional parameter 
space $(\umass,\dmass,\emass)$. However, symmetry considerations 
\cite{barrkhan} suggest that the electron mass could be a fixed
fraction of the mass of the down quark, so that the ratio
$f=\emass/\dmass$ is constant under variations of the quark
masses. Under this assumption we can evaluate the above constraints.
Using the value $f\approx0.107$ appropriate for our universe, the
resulting parameter space is shown in Figure \ref{fig:updown}. The
black dot marks the location of our universe in the diagram. (Keep 
in mind that other choices for $f=\emass/\dmass$ are possible, and 
would lead to corresponding changes in the diagram.) 

In the figure, the blue curves delimit the region for which quarks
cannot decay within hadrons, where protons and neutrons are of primary
interest (Section \ref{sec:hadrons}). The allowed region falls between
the two curves. These constraints are not as confining as the others
under consideration here due to the large value of the energy $E_3$
required to produce a bound state of three quarks. The green curves in
the figure show the region for which nuclei are stable (Section
\ref{sec:nuclei}).  The allowed region again falls between the two
curves. In the region above the upper curve, neutrons are unstable
within nuclei, whereas in the region below the lower curve, protons
are unstable. The most stringent constraints result from the
requirement that protons cannot decay (Section \ref{sec:protons}). In
the region below the lower red curve, free protons can decay into
neutrons and positrons. In the region below the upper red curves,
protons in hydrogen atoms can combine with the bound electron to form
neutrons. This latter constraint is the most confining. Significantly,
our universe lies close to this limit. If the down quark (and hence
the neutron) were lighter by $\sim1$ MeV, hydrogen atoms would decay
via this channel.

Note that the two most important constraints are the upper limit on
the down quark mass necessary to keep neutrons from decaying within
nuclei (equation [\ref{uppernuke}]) and the lower limit necessary to
keep atomic hydrogen from combining into a neutron (equation
[\ref{lowerhydro}]). We can thus write a combined constraint on 
the down quark mass
\be
{\umass + \delem \over (1-f) \dmzero} \le {\dmass \over \dmzero} 
\le {\umass + \delem + B \over (1-f) \dmzero} \,. 
\label{comconstraint} 
\ee
In the limit of small mass for the up quark (left side of Figure 
\ref{fig:updown}), the allowed range for the down quark mass can 
be written in the form 
\be
{\delem \over (1-f) \dmzero} \le {\dmass \over \dmzero} \le
{\delem + B \over (1-f) \dmzero} \qquad \Rightarrow \qquad 
0.40 \simless {\dmass \over \dmzero} \simless 2.7 \,. 
\label{smallup} 
\ee
In the opposite limit of large up quark mass, we obtain 
\be
{\umass \over (1-f) \dmzero} \le {\dmass \over \dmzero} \le
{\umass \over (1-f) \dmzero} \qquad \Rightarrow \qquad 
{\dmass \over \dmzero} \approx 0.54 {\umass \over m_{\rm u0}} \,. 
\label{largeup} 
\ee
These asymptotic forms show that in the limit of small $\umass$, 
the mass $\dmass$ of the down quark can vary by a factor
$1+B/\delem\sim6.9$. In the limit of large $\umass$, the allowed range
of values narrows to (essentially) a line in the plane of parameters. 
Significantly, the up quark mass can vary (to lower values) by several
orders of magnitude while the down quark mass has a range of $\sim7$.
The allowed parameter space is not overly restrictive.

Notice also that the two most restrictive bounds (from equations
[\ref{comconstraint}--\ref{largeup}]) provide a bound on the composite
parameter $(1-f)\dmass/m_{\rm d0}$. In this treatment, we specified
the electron mass to be a fixed ratio ($f$) of the down quark mass,
where $f=m_{\rm e0}/m_{\rm d0}$.  For other choices of the ratio $f$,
the ranges of allowed quark masses are similar, with the allowed
region in the plane of Figure \ref{fig:updown} moving up or down
accordingly. 

\subsubsection{Summary of Quark Constraints} 
\label{sec:quarksummary} 

The allowed ranges for the light quark masses, shown in Figure
\ref{fig:updown}, are significantly larger than reported in some 
earlier assessments. The region of allowed quark masses spans a factor
of $\sim7$ for the down quark over a range of several orders of
magnitude for the up quark. One reason for this expanded range,
compared with previous treatments, is that this work removes the
unnecessary restrictions that deuterium must be stable and that
diprotons must be unstable.  Although these two constraints would
reduce the allowed range of parameter space \cite{hogan,barrkhan},
recent work shows that stars -- and hence universes -- can operate
with either stable diprotons \cite{barnes2015} or unstable
deuterium \cite{agdeuterium} (see also \cite{barnes2017}).

Although stars can operate with unstable deuterium, which requires
$\umass+\dmass$ to increase by a factor of $\sim2$ (e.g., see Figure
11 of \cite{epelbaum2003}), the sum of the quark masses cannot be made
arbitrarily large. The quark masses determine the pion mass, which in
turn sets the range of the strong force. If the quarks become too
heavy, then the range of the strong force could become short enough to
render all nuclei unstable. Although the required increase in quark
masses has not been unambiguously determined, Figure \ref{fig:updown}
shows the contours where the sum of the light quark masses increases
by factors of 2 and 4 (given by the lower and upper black curves in the
diagram). This additional constraint cuts off only the tail of
parameter space at large quark masses, and leaves most of the range
viable.

The discussion thus far has considered only the two lightest quarks.
For the case of three light quarks, the range of viable universes is
even greater \cite{jaffe}. The band of congeniality found in that work
is about 29 MeV wide in terms of the mass difference between the
lightest two quarks (see \cite{jaffe} for further discussion; see also 
\cite{ali2013} for a less optimistic viewpoint). Note that an even wider 
range of possible universes may be viable if one considers more light
quark masses, but such models have not been worked out.

Finally, notice that most of the constraints summarized in Figure 
\ref{fig:updown} result from some type of beta decay, where neutrons 
and protons are transformed into each other. Universes can remain
viable in the absence of the weak force \cite{weakless}, and such
universes would not be subject to beta decay. As a result, for
scenarios that are somewhat removed from our expectations, these
constraints on the light quark masses could be significantly weaker 
(see also \cite{grohsweakless}). 

\subsubsection{Mass Difference between the Neutron and Proton} 
\label{sec:dneutpro} 

Recent work has provided an ab initio calculation of the mass
difference $\Delta m$ between the neutron and proton using lattice QCD
and QED calculations \cite{borsanyi}. Historically, the calculation of
$\Delta m$ has been notoriously difficult. Even this state-of-the-art
treatment provides a mass splitting estimate of 1.5 MeV, which is
somewhat larger than the measured value of $\Delta m$ = 1.29 MeV. In
addition to the successful calculation of this quantity, these results
provide estimates for the separate contributions to the mass differences 
from QCD effects (setting $\alpha=0$ and $\dmass-\umass\ne0$) and
electromagnetic effects (setting $\alpha\ne0$ and $\dmass-\umass=0$). 
The result is that $(\Delta m)_{qcd} \approx -2.5 (\Delta m)_{em}$.

\begin{figure}[tbp]
\centering 
\includegraphics[width=1.0\textwidth,trim=0 150 0 150,clip]{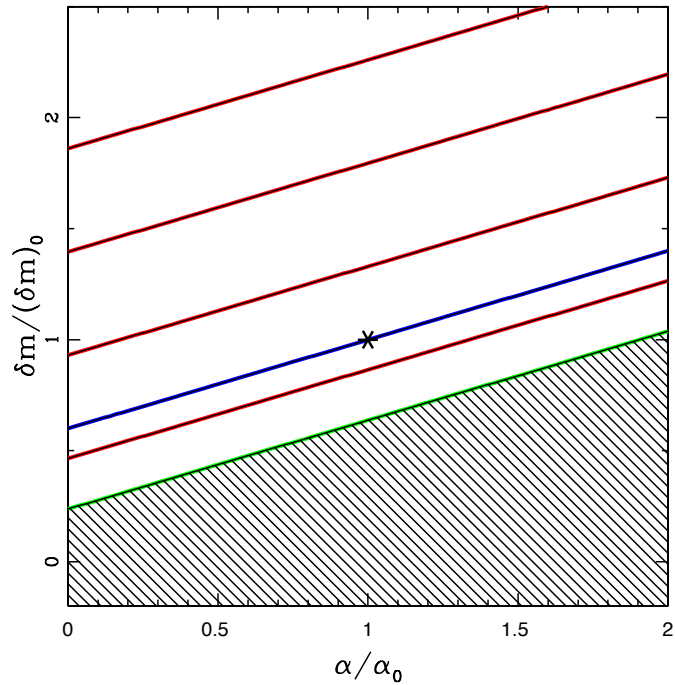}
\caption{Mass splitting between the neutron and proton as a function 
of quark mass difference $\delta m = \dmass-\umass$ and fine structure
constant $\alpha$ (see Figure 3 of \cite{borsanyi}). Both quantities
are scaled to the values in our universe, so that its location lies in
the center of the diagram marked by the star symbol.  The red curves
show the contours of constant mass splitting for $\Delta m$ = 
$m_{\rm n}-\mpro$ = 1 -- 4 MeV, from bottom to top. The blue curve
shows the contour for the mass difference $\Delta m$ = 1.29 MeV
measured in our universe. If the mass difference is too small, as
delimited by the shaded region under the green curve, then the neutron
is susceptible to inverse $\beta$ decay, and the corresponding universe
is not viable. The contour slope is estimated from lattice
quantum-chromodynamics and quantum-electrodynamics calculations
\cite{borsanyi}. }
\label{fig:massalpha} 
\end{figure}

One can use the results outlined above to determine how the mass
difference between the neutron and proton depend on the mass
difference between their constituent quarks (i.e., $\delta{m}$ =
$\dmass-\umass$) and the electromagnetic coupling ($\alpha$). The 
result is shown in Figure \ref{fig:massalpha} (analogous to Figure 3
of Ref. \cite{borsanyi}). The figure shows the contours of constant
neutron-proton mass difference in the plane of parameters. For the
value of $\alpha$ in our universe, the quark mass difference can only
vary downward by a factor of $\sim2$. Greater changes lead to inverse
$\beta$ decay. Larger values of the quark mass differences are
unconstrained in this diagram, although other considerations come into
play (see Figure \ref{fig:updown}). If the quark masses are held
constant, then the fine structure constant can only become larger by a
factor of $\sim2$, but has no lower limit in this diagram.

\subsubsection{Constraints on the Higgs Parameters} 
\label{sec:higgs} 

Instead of variations in the masses of the light quarks (and/or the
electron), one can also consider possible changes in the parameters of
the Higgs potential, e.g., the expectation value ${\cal V}$ =
$\sqrt{-\mu^2/\lambda}$.  As the value of ${\cal V}$ increases, the
mass difference between neutrons and protons increases, so that
neutrons are more likely to decay within nuclei. Larger values of
${\cal V}$ also increase the pion mass, which decreases the effective
range of the strong force.  Both of these effects lead to nuclei that
are more unstable.  The maximum allowed increase in the expectation
value is estimated to be ${\cal V}\simless5{\cal V}_0$ 
\cite{agrawalprl,agrawal}, where the subscript corresponds to the 
value in our universe. For larger ${\cal V}$, the mass difference
$\Delta{m}=m_{\rm n}-\mpro$ is larger than the typical binding energy
of a nucleon within an atomic nucleus. As a result, neutrons can decay
into protons within bound nuclei, thereby leaving hydrogen as the only
truly stable nucleus. Somewhat tighter bounds are derived in Ref. 
\cite{damour} based on considerations of nuclear stability. 

Although the range of the vacuum expectation values ${\cal V}$ is not
overly restrictive, the observed value ${\cal V}_0\approx246$ GeV and
the maximum allowed value ${\cal V}_{\rm max}\sim1200$ GeV remain
small compared to the Planck scale (at $\mplanck\sim10^{19}$ GeV)
and/or the GUT scale (at $M_{GUT}\sim10^{16}$ GeV). A problematic
issue arises: In simple grand unified models, the Standard Model
parameter $\mu^2$, which determines ${\cal V}$, has a naturalness
problem. The quantum corrections are expected to be ${\cal
O}(M_{GUT})$, so that the relevant terms must cancel to high accuracy
in order to produce the observed value (see the discussion
of \cite{agrawal} and references therein). Such models are fine-tuned
in the sense that small changes in the other model parameters would
presumably alter this precise cancellation and lead to typical values
of $\mu^2$ and ${\cal V}$ that are much larger than those observed in
our universe.

Additional constraints on the Higgs parameters arise from stability
considerations. Sufficiently large changes to the Higgs potential
could result in vacuum instability \cite{chigusa,coleman,sher1989},
which would have important consequences for the habitability of the
universe. For example, the Higgs potential generally has more than one
minimum. If the Higgs field resides in a higher energy minimum (a
false vacuum state), then the field can tunnel into a lower (true)
vacuum state sometime in the future. In order for the universe to
remain viable, however, the vacuum must be either stable or
sufficiently long-lived (if the false vacuum is metastable). The
quantum tunneling rate depends on the shape of the Higgs Potential,
which in turn depends on the input parameters. As one example, for the
case where the Higgs potential has a quartic form, the highest order
term $\lambda\phi^4$ must be positive to ensure vacuum stability. The
coefficient $\lambda$ in the classical potential depends on the Higgs
mass, but quantum corrections can modify its value and even render
$\lambda<0$ \cite{alekhin,branchina,buttazzo}.  These corrections
depend on the Yukawa couplings, where that of the top quark makes the
largest contribution. As a result, the shape of the Higgs potential
and the fate of the cosmic vacuum state depend on the Higgs mass
$\mhiggs\sim125$ GeV and the top quark mass $\tmass\sim173$ GeV. 
The resulting constraints are determined by the form of the Higgs
potential, which is not fully specified (and could have alternate
forms in other universes). Recent work \cite{alekhin,branchina,buttazzo} 
indicates that vacuum stability requires the ratio $\mhiggs/\tmass$ to
be sufficiently large, where the measured values in our universe are 
close to the limit. 

\subsection{Constraints on the $\alpha$-$\beta$ Plane} 
\label{sec:alphabeta} 

The Standard Model of Particle Physics describes interactions at the
fundamental level of quarks and leptons. At lower energies, however,
the basic properties of atoms and molecules, and hence chemistry, are
determined by the values of the fine structure constant $\alpha$ and
the mass ratio $\beta=\emass/\mpro$. Since the neutron and proton have
similar masses, the neutron does not introduce a third parameter in
this context. In this section, we review basic constraints on the
constants $\alpha$ and $\beta$, and find the allowed region in the
plane of parameter space.

Many authors (e.g., \cite{bartip,tegmarktoe}) have argued that both
$\alpha\ll1$ and $\beta\ll1$ in order for chemistry to operate (in a
manner roughly similar to chemistry in our universe). Several
arguments imply that the $\alpha$ must be small. Since the kinetic
energy of electrons in atoms scales as $K\propto\alpha^2 \emass$, the
constant $\alpha$ must be smaller than unity in order for electrons 
to remain non-relativistic. In addition, as discussed in Section 
\ref{sec:stars} (see also \cite{adams}), the fine structure constant
must be smaller than unity in order for stars to function as nuclear
burning objects. If the stars are required to have sufficiently high
surface temperatures, the constraint on $\alpha$ is somewhat tighter
\cite{adamsnew}. Of course, if $\alpha$ becomes too large relative to
the strong nuclear force, then large nuclei would cease to exist
(\ref{sec:semfappend}).  Finally, for completeness, we note that the
fine structure constant $\alpha$ must be less than unity in order for
bulk matter to remain stable \cite{liebyau,liebyaualt}.  All of these
considerations restrict $\alpha\ll1$. For purposes of this discussion, 
we thus adopt the particular bound $\alpha \simless 1/3$.

Small values of the mass ratio $\beta$ are required for the existence
of stable ordered structures, such as a solid or a living cell
\cite{tegmarktoe}.  For the structure to be well ordered, the
fluctuation amplitude of a nucleus must be much smaller than the
distance between the atoms. This constraint requires that
$\beta^{1/4}\ll1$. Following \cite{tegmarktoe}, we enforce the
constraint $\beta^{1/4}<1/3$ so that $\beta<1/81$. For completeness,
note that the localization requires a large mass ratio, but that one
could in principle have the electron heavier than the proton. As a
result, a second window of allowed parameter space opens up for
large mass ratios $\beta\simgreat81$. 

The constants $(\alpha,\beta)$ also appear in the equations of stellar
structure \cite{chandra,clayton,hansen,kippenhahn,phil} and are thus
constrained by stellar considerations.  Although stellar masses in our
universe can vary by a factor of $\sim1000$, if $\alpha$ is too large,
or $\beta$ is too small, then the minimum mass of a star would exceed
the maximum stellar mass, thereby preventing the existence of working
stellar entities. The minimum and maximum stellar masses are given in
\ref{sec:massscales}. Combining equations (\ref{massminstar}) and 
(\ref{massmaxstar}), this constraint can be written in the form 
\be
\beta^{-3/4} \alpha^{3/2} \simless 100 
\qquad {\rm or} \qquad \beta \simgreat 10^{-8/3} \alpha^2 \,.
\label{abstarlimit} 
\ee
Stable nuclear burning stars can fail to exist for another reason: If
the fine structure constant is too small, then the electrical barrier
for quantum mechanical tunneling becomes too small and stars would
burn all of their nuclear fuel at once \cite{adams}. The constraint
required to avoid this circumstance can be written in the form
\be
\alpha^2 \beta \simgreat constant \approx 2.62 \times 10^{-12} 
\approx 9.03 \times 10^{-5} \alpha_0^2 \, \beta_0 \,, 
\ee
where the numerical constant can be evaluated from the equations 
of stellar structure (see the Appendix of Ref. \cite{adamsnew}). 

The constants $(\alpha,\beta)$ determine, in part, how the gas in 
a forming galaxy can dissipate energy, cool, and collapse. This 
requirement places a limit/estimate for the mass scale of galaxies
\cite{reesost,tegmarkrees}, as outlined in \ref{sec:massscales}.
Since the mass of the galaxy must be larger than the minimum mass 
of a star, we can combine equations (\ref{massminstar}) and 
(\ref{massgalcool}) to derive a constraint of the form 
\be 
\alpha_G^{-1/2} \alpha^5 \beta^{-1/2} \simgreat {1\over2} 
\beta^{-3/4} \alpha^{3/2} \qquad {\rm or} \qquad 
\beta \simgreat {1 \over 16} \alpha_G^2 \alpha^{-14}\,.
\label{abgalaxylimit} 
\ee 
Note that this bound depends on the gravitational fine structure
constant $\alpha_G$ in addition to $(\alpha,\beta)$. For the sake of
definiteness in the following analysis, we fix $\alpha_G$ to be its
value in our universe. 

\begin{figure}[tbp]
\centering 
\includegraphics[width=1.0\textwidth,trim=0 150 0 150,clip]{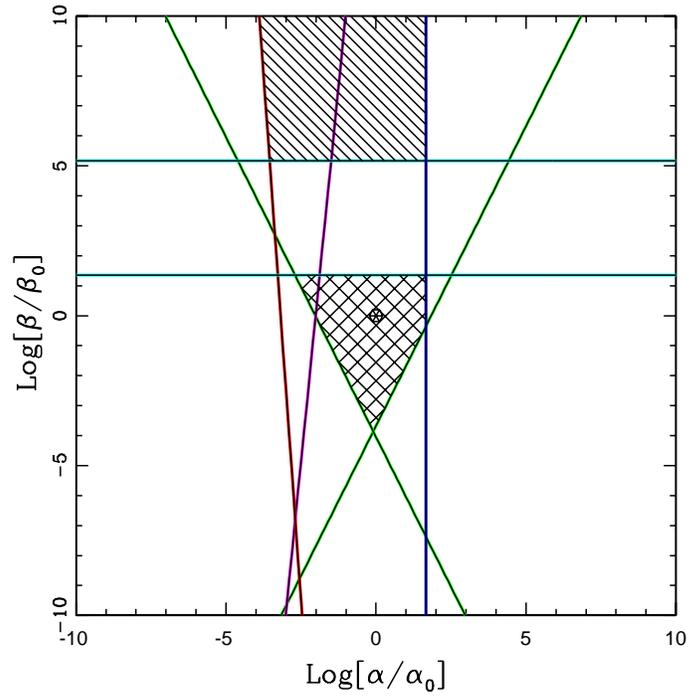}
\caption{Allowed region of parameter space in the $\alpha$-$\beta$ plane,
where $\alpha$ is the fine structure constant and $\beta$ is the
electron to proton mass ratio. The allowed region is hatched and the
location of our universe is marked by the star symbol in the center of
the diagram.  The region is bounded from the requirements that
$\alpha\ll1$ (blue line) and $\beta\ll1$ (lower cyan line). Additional 
constraints arise from the requirement that stars exist (green curves) 
and that galaxies are larger in mass than the smallest stars (red curve). 
For completeness, the purple curve shows the locus of points where the 
galactic mass scale is equal to the typical stellar mass. } 
\label{fig:abplane} 
\end{figure}

Figure \ref{fig:abplane} shows the allowed parameter space in the
$\alpha$-$\beta$ plane subject to the constraints outlined above.  The
requirement that both $\alpha\ll1$ and $\beta\ll1$ limit the parameter
space to the lower left quadrant of the figure, as delimited by the
cyan and blue lines. The requirement that the minimum stellar mass is
less than the maximum stellar mass requires $\beta$ to lie above the
green curve with positive slope. In order for stars to exist as
long-lived, stable, nuclear-burning entities, the mass ratio $\beta$
must lie above the green curve with negative slope.  For completeness,
note that the minimum point of the two green curves would be slightly
rounded off if one uses results from a full stellar structure
calculation.  The requirement that galaxies are larger than the
minimum stellar mass requires $\beta$ to lie above the red curve. This
latter curve is so steep that it enforces an effective lower bound on
$\alpha$, although the nuclear burning constraint is more restrictive
for small values of $\beta$.  For completeness, the figure also shows
the limit where the galactic mass scale is equal to the typical
stellar mass scale (marked by the purple curve).

In Figure \ref{fig:abplane}, the location of our universe is marked by
the star symbol. The allowed region of parameter space surrounding
that point has a nearly triangular shape, where the base (range of
$\alpha$) and altitude (range of $\beta$) span about 4 orders of
magnitude.  Notice that Figure \ref{fig:abplane} includes a second
allowed region of parameter space in the upper central part of the
diagram. This regime corresponds to the case where the electron is
much heavier than the proton $\emass\gg\mpro$. Universes with
parameters in this region are likely to be quite different from our
own, but the constraints enforced here do not rule them out as viable.

The constraints depicted in Figure \ref{fig:abplane} are based on the
existence of known structures, including galaxies, stars, and atoms.
However, another type of constraint can be placed on the fine
structure constant based on purely theoretical considerations. The
three gauge coupling constants of the Standard Model are energy
dependent. If one enforces the requirement of Grand Unification ---
that the three constants have the same value at some large energy
scale --- then the value of $\alpha$ measured at low energy is highly
constrained (see the recent review of \cite{donoghuethree}). These
limits also assume that proton decay occurs at the GUT scale, with a
new heavy $X$-boson, but that protons are stable on stellar
timescales. The constraints on the fine structure constant obtained
through this argument are more more restrictive than those presented
in Figure \ref{fig:abplane} and are centered around the observed
value. Previous estimates for the allowed range include 
$120\simless\alpha^{-1}\simless170$ \cite{ellisnano} and
$85\simless\alpha^{-1}\simless180$ \cite{bartip}.  At the present
time, however, no experimental evidence exists for Grand Unified
theories \cite{donoghuethree} and the Standard Model in its current
form does not allow for unification (one needs to invoke new physics
such as supersymmetry). As a result, the status of these tighter
bounds on $\alpha$ remains undetermined.

\subsection{Constraints on the Strong Coupling Constant} 
\label{sec:strongcon} 

\begin{figure}[tbp]
\centering 
\includegraphics[width=1.0\textwidth,trim=0 150 0 150,clip]{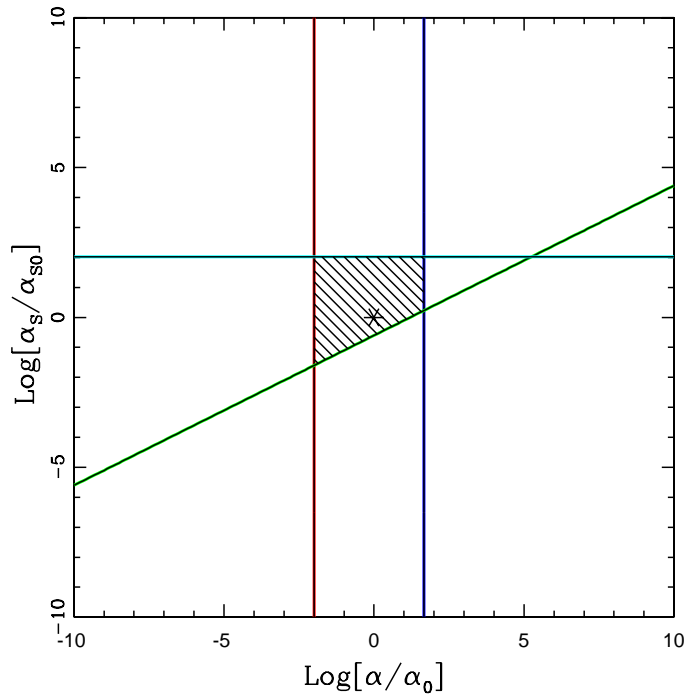}
\caption{Allowed region of parameter space in the $\alpha$-$\alpha_{\rm s}$ 
plane, where $\alpha$ is the fine structure constant and $\alpha_{\rm s}$ 
is the analogous structure constant for the strong force. The shaded
region corresponds to the allowed range of parameters, where the
location of our universe is marked by the star symbol. The allowed 
region is bounded by the constraint that $\alpha\ll1$ (blue line)
and the requirement that $\alpha$ is large enough for stars to
function (red line). In addition, the strong force must be strong
enough that nuclei are stable against fission (green curve) and weak
enough that binding energies per nucleon correspond to
non-relativistic energies (cyan line). }
\label{fig:asplane} 
\end{figure} 

This section considers limits on the magnitude of the strong coupling
constant $\alpha_{\rm s}$. One well-known constraint arises from the
requirement that nuclei are stable against fission. This constraint is
generally derived by using the Semi-Empirical Mass Function as a model
for atomic nuclei \cite{semf} and then requiring that the binding
energy of a nucleus is larger than the binding energy of two separated
nuclei with half the particles \cite{bartip,tegmarktoe}. This 
consideration results in a limit on the strong force coupling constant
as a function of the fine structure (electromagnetic) constant such
that 
\be
{\alpha_{\rm s} \over \alpha_{{\rm s}0}} \simgreat {1\over4} 
\left( {\alpha \over \alpha_0} \right)^{1/2}\,, 
\label{nofission} 
\ee
where the subscripts denote the values in our universe. The numerical
coefficient depends on the largest nucleus that is required to have a
bound state, where equation (\ref{nofission}) uses the value 
corresponding to carbon-12. 

Additional constraints on the strong coupling constant 
$\alpha_{\rm s}$ arise from the required ordering of atomic size and
energy scales. In order for bulk matter to have its observed form in
our universe, the size scale of atoms, given by the extent of
electronic orbits, must be larger than atomic nuclei. Electron orbits
have radii $r_e=1/(\alpha\emass)$, whereas nuclei have radii given 
approximately by $r_N\approx1/(\alpha_{\rm s}\mpro)$. The ordering of
size scales thus implies the constraint
\be
\alpha \beta \ll \alpha_{\rm s} \,,
\label{atomnukesize} 
\ee
where we have used $\beta=\emass/\mpro$. Similarly, the energy scales
for chemical reactions are much lower than those of nuclear reactions.
If the opposite were true, then chemical reactions, which provide the
basis for life, would instigate nuclear reactions and thereby change
the elements that make up life forms during the course of biological
processes. The required ordering of energy scales leads to the 
analogous constraint 
\be
\alpha^2 \beta \ll \alpha_{\rm s}^2 \,. 
\label{atomnukeenergy} 
\ee
If equations (\ref{atomnukesize}) and (\ref{atomnukeenergy}) did not
hold, it is possible that a universe could remain habitable, but it
would be much different than our own. However, the previous section 
shows that $\beta\ll1$ in viable universes, so that these constraints 
are less restrictive than that of equation (\ref{nofission}). 

Going in the other direction, the strong force cannot be too much
greater than that realized in our universe without changing the manner
in which nuclear processes occur. The most tightly bound nucleus is
iron-56, which has a binding energy per nucleon of $E_{56}\approx8.8$
MeV.  If $\alpha_{\rm s}$ is too large, however, then the binding
energy per nucleon would become larger than the nucleon mass, and
energy levels of the nucleus would become relativistic.  Although
nuclear reactions can still take place under relativistic conditions,
the way in which they occur in stars (and BBN) would be vastly
different than in our universe. This consideration thus places an
upper limit on the strength of the strong force. Here we invoke this
constraint in the conservative form 
\be
\alpha_{\rm s} \simless {\mpro \over E_{56}} \alpha_{{\rm s}0} 
\approx 100\,\alpha_{{\rm s}0}\,.
\label{nonrelnuke} 
\ee

The constraints on the strong coupling constant outlined above are
depicted in Figure \ref{fig:asplane}, which shows the allowed region
in the plane of parameters $(\alpha,\alpha_{\rm s})$. The range of the
fine structure constant $\alpha$ is limited by the same constraints
used in the previous section. First we require $\alpha\ll1$ (see
Section \ref{sec:alphabeta}), so that the allowed region falls to 
the left of the blue line in the diagram. On the other hand, $\alpha$
must be large enough that stellar structure solutions exist
\cite{adams,adamsnew}. Working stars thus limit the allowed region 
to the right of the red line. Next we require that the binding energy
per particle is small enough that the constituent particles in nuclei 
remain non-relativistic, so that the allowed region falls below the
cyan line. Finally, the strong force must effectively compete with the
electromagnetic force to prevent nuclear fission (see equation
[\ref{nofission}]), as marked by the green curve. The resulting 
region of parameter space spans a factor of $\sim1000$ in both 
$\alpha$ and $\alpha_{\rm s}$. 

For completeness, we note that many authors invoke tighter limits on
the strong coupling constant through considerations of nuclei with
mass number $A=2$ (see \cite{dyson1971,bartip,reessix,tegmarktoe} and 
many others; see also Section \ref{sec:twonuke}). If the strong force
were somewhat stronger, diprotons would be bound, and the cross
sections for nuclear reactions in stars would enhanced by an enormous
factor.  In spite of many claims of disaster, this enhancement would
lead to only a modest decrease in the operating temperatures of
stellar cores (from $\sim15 \times10^6$ K down to $\sim10^6$ K) and a
modest decrease in stellar lifetimes (see Section \ref{sec:stars}
and \cite{barnes2015}). If the strong force were somewhat weaker,
then deuterium would no longer have a bound state, and the usual
pathways for nucleosynthesis would be altered. Nonetheless, this
scenario also allows stars to provide both nuclear processing and
long-lived supplies of energy (see Section \ref{sec:stars}
and \cite{agdeuterium,barnes2017}). 

Estimates for the changes to the strong coupling constant required to
make diprotons bound or deuterons unstable depends on the model of the
nucleus. In the square well approximation for the nuclear potential,
6\% increases in $\alpha_{\rm s}$ lead to stable diprotons whereas 4\%
decreases lead to unbound deuterium \cite{davies1972}.  For nuclear
potentials of Yukawa form \cite{hulthen,pochet}, the required increase
(decrease) in $\alpha_{\rm s}$ becomes 17\% (6\%). Other authors find
similar requirements \cite{agdeuterium,reessix}. Bound states of the
$A=2$ nuclei can also be altered with corresponding changes in quark
masses, which result in different ranges for the strong force. If the
sum of the light quark masses $(\umass+\dmass)$ is decreased by 25\%,
then diprotons become bound, whereas mass increases of 40\% lead to
unbound deuterium \cite{barrkhan} (see also Section \ref{sec:quarklimit}). 

\subsection{Additional Considerations} 
\label{sec:moreconstraints} 

For completeness, this section considers additional constraints on the
parameters of particle physics. Specifically, the issue of charge
quantization is discussed in Section \ref{sec:charge}. We then present
a constraint on the energy scale of Grand Unified Theories from the
requirement that nucleons have sufficiently long lifetimes
(Section \ref{sec:pdecay}).

\subsubsection{Charge Quantization} 
\label{sec:charge} 

Our understanding of the laws of physics remains incomplete. One
important unresolved issue that could affect the habitability of the
universe is the specification of electromagnetic charges on the
fundamental particles.  In our universe, charge is observed to be
quantized. All free particles (notably protons, neutrons, and
electrons) have charges that are integer multiples of the electron
charge ($Q=ne$ for some $n\in\mathbb{Z}$). More generally, the charges
for all Standard Model particles are integer multiples of the charge
of the down quark $Q_d=e/3$.

Charge quantization is important for the operation of the universe, as
it allows for the existence of atoms that are electrically neutral. In
turn, neutral atoms allow for the construction of working stars and
other bulk matter. However, in conventional quantum electrodynamics
--- including the Standard Model --- electric charges are not
specified by fundamental considerations, but rather are input
parameters. On the other hand, as outlined below, both Grand Unified
Theories and the removal of anomalies imply constraints on the charges
of fundamental particles and can thus provide mechanisms for charge
quantization.

Many types of Grand Unified Theories have been put forward
\cite{particlegroup}. One general feature of unification models 
is that the quark and leptons are incorporated into a larger symmetry
group, so that their properties are related due to constraints on the
theory. As one example, in the case of $\mathit{SU}(5)$ unification,
the electric charge operator is the sum of diagonal $\mathit{SU}(2)$
and $U(1)$ generators (e.g., see \cite{kanebook} for a textbook
treatment). Since the generators must be traceless, the sum of the
eigenvalues of electric charge must vanish, which leads to a charge
quantization condition of the form 
\be
Q(e^{-}) + Q(\nu_e) + 3 Q({\bar d}) = 0 \,. 
\label{gutcharge} 
\ee
The neutrino has no charge in our universe, so that $Q({\bar d})$ =
$-Q(e^{-})/3$ = $e/3$. This model not only implies charge
quantization, but also provides the fractional charges measured for
quarks. More generally, charge quantization must arise in any unified
theory \cite{georgiguts}. Although charge quantization is a highly
desirable feature, Grand Unified Theories have not been experimentally
verified, so it is not known if they provide the explanation for 
charge quantization in our universe. However, any universe described 
by a unified theory of this class will have its charges quantized. 

Another way to achieve charge quantization is through the requirement
that anomalies vanish in the theory. A full discussion of this topic
requires a rather lengthy formalism and is beyond the scope of this
review (for further detail, see Chapter 22 of \cite{weinbergbook}).
Briefly, the conditions for anomaly cancellation lead to constraints
on the sum of the particle charges, roughly analogous to that of
equation (\ref{gutcharge}), and such conditions imply charge
quantization (see also \cite{foot,npanomaly}). 

Particle physics theories thus support two different classes of
constraints --- those arising from Grand Unification and those due to
anomaly cancellation. Both considerations enforce charge quantization
and thereby lead to viable universes. With the current state of the
field, all anomalies must cancel to avoid the prediction of infinite
quantities, whereas Grand Unified Theories are not yet experimentally
necessary.

In addition to quantization of charge, our universe displays the
related properties of charge conservation and charge neutrality.
Conservation of charge follows from the symmetries of the Lagrangian
of the underlying theory \cite{noether}, so that the class of theories
with this property is large and well-defined.  The relative numbers of
positive and negative charges in the universe are determined early in
cosmic history through a number of processes, including baryogenesis 
and leptogenesis. At the present time, these mechanisms remain under 
study \cite{dolgov,steigman}, but observations indicate that the
universe as a whole is close to neutral \cite{oritoyoshi}. The excess
charge density per baryon is bounded from above. One such estimate
\cite{caprini} takes the form $Q_0/n_{\rm b}\simless10^{-26}e$, 
where $n_{\rm b}$ is the number density of baryons, and where this
limit holds for uniformly distributed excess charge. In principle, a
universe could have a net electric charge and still obey charge
conservation and charge quantization. Moreover, such a universe could
remain viable provided that the excess charge is not too large (see
also \cite{lyttleton}). Although the range of $Q_0/n$ that allows for
habitability requires further specification, it includes the value
$Q_0/n=0$, which could be considered special --- and perhaps even
likely --- in the space of all possible universes \cite{barnes2012}.

\subsubsection{Constraint from Proton Decay}
\label{sec:pdecay}  

In any Grand Unified Theory, conservation of baryon number is
necessarily violated \cite{gellmann}, which allows for the possibility
of nucleon decay (see also \cite{langacker81,nath2007,particlegroup}
and references therein). The requirement that nucleons are
sufficiently long-lived thus places constraints on the theory. Since
the number of possible theories --- and operators that violate baryon
number --- is large, we consider only a representative example.  For
the simplest class of interactions that drive proton decay, the time
scale can be written in the form 
\be
\tau_{\rm p} = C_{\rm p} 
{M_{\rm gut}^4 \over \alpha_{\rm gut}^2 \mpro^5}\,,
\label{prodecay} 
\ee
where $C_{\rm p}$ is a dimensionless constant of order unity, 
$M_{\rm gut}$ is the GUT scale ($\sim10^{16}$ GeV), and 
$\alpha_{\rm gut}$ is the coupling constant evaluated at that
scale. Current measurements  \cite{superkamio} indicate that 
the proton lifetime in our universe has a lower limit of 
$\tau_{\rm p}\simgreat1.6\times10^{33}$ yr for the decay channel 
$p\to e+\pi^0$ and $\tau_{\rm p}\simgreat7.7\times10^{33}$ yr 
for the channel $p\to \mu+\pi^0$. 
 
If we require that protons (nucleons) live long enough for life to
evolve, then we must enforce the limit $\tau_{\rm p}>N_At_A$, where
$t_A$ is the atomic time scale and $N_A$ is the number of atomic time
scales required for successful biological evolution. Here we take
$N_A\sim10^{33}$, corresponding to an time scale of $\sim1$ Gyr
(see \cite{lunine,knoll,scharf} and Section \ref{sec:lifetime} for
further discussion). This constraint can be written in the form 
\be
N_A \alpha_{\rm gut}^2 \mpro^4 < \alpha^2 \beta M_{\rm gut}^4 
\qquad {\rm or} \qquad 
M_{\rm gut} > N_A^{1/4} \left({\alpha_{\rm gut}\over\alpha}\right)^{1/2} 
\beta^{-1/4} \mpro \,,
\label{gutlimit} 
\ee
where $\alpha$ is the fine structure constant in the low energy
limit. In our universe, the quantity 
$(\alpha_{\rm gut}/\alpha)^{1/2}\beta^{-1/4}\approx0.4$.  This
quantity is close to unity and the constraint of interest is not
overly sensitive to its value.  Moreover, the appropriate value of 
$N_A$ is not precisely known.  As a result, we obtain the approximate
bound $M_{\rm gut}>N_A^{1/4}\mpro\sim10^8\mpro$. Any viable universe
must have a significant hierarchy between the GUT scale and the mass
of the proton in order to keep nucleons stable long enough for life to
evolve. However, this minimum hierarchy (a factor of $\sim10^8$) is
much smaller than that realized in our universe (where 
$M_{\rm gut}\sim10^{16}\mpro$). A similar situation arises for 
proton decay in supersymmetric theories: the anthropically preferred
time scale is much shorter than the observed proton lifetime
(see \cite{banksdinegorb,susskindproton} for further discussion).  
In addition, a number of scenarios have been put forth to allow for 
proton stability \cite{nath2007} (e.g., in theories with large extra
dimensions), so that the bound implied by equation (\ref{gutlimit}) 
is not ironclad.

\bigskip 
\section{Cosmological Parameters and the Cosmic Inventory} 
\label{sec:cosmology}

This section outlines the cosmological parameters that are required to
describe a universe as a member of the multiverse.  We start with a
review of the cosmological parameters that are necessary to specify
the current state of our own universe. However, some of these
parameters have relatively little effect on structure formation and
are not necessary for an arbitrary universe to be habitable. We thus
define the subset of parameters that are relevant for considerations
of fine-tuning across the multiverse. The section then outlines the
flatness problem and related cosmological issues, and briefly
describes how an early inflationary epoch can drive a universe to
become spatially flat. We also discuss how inflationary models can
provide cosmological perturbations and elucidate the relationship
between the parameters of the inflaton potential and the amplitude of
primordial fluctuations. 

\subsection{Review of Parameters} 
\label{sec:cosreview} 

The current state of the universe can be characterized by a relatively
small collection of parameters. The expansion of the universe is 
governed by the Friedmann equation 
\be
\left({{\dot a}\over a}\right)^2 = 
{8\pi G\over3} \rho - {k \over a^2} \,, 
\label{adotovera} 
\ee
where $a(t)$ is the scale factor, the curvature constant $k=0,\pm1$,
and the energy density $\rho$ includes contributions from all sources.
If we are concerned only with the expansion and evolution of the
universe as a whole --- and not the formation of structure within it
--- then the current state of the universe can be specified by
measuring all of the contributions to the energy density $\rho$ 
and the Hubble constant 
\be
H_0 = \left( {{\dot a}\over a}\right)_0\,,
\label{hubble} 
\ee
where all of these quantities are evaluated at the present epoch.
Note that once $H_0$ and $\rho$ are specified, then the equation of
motion (\ref{adotovera}) determines the curvature constant $k$. 
Following cosmological convention, we take $a=1$ at the present epoch.
The total energy density $\rho$ contains a number of components,
including matter $\rhom$, radiation $\rhor$, and vacuum energy
$\rhov$. The matter density $\rhom$ includes at least two
contributions, from baryons ($\rhob$) and from dark matter ($\rhod$).
Notice also that the dark matter could have contributions from
different types of particles, including neutrinos and some type of
cold dark matter. Many candidates have been put forward, including 
the lightest super-symmetric partner and axions (e.g., see 
\cite{baer,feng,jungman,steffen} and references therein). 

The various components of the energy density evolve differently in the
presence of cosmic expansion. The matter components, both dark matter
and baryons, vary with the scale factor according to $\rhom\propto
a^{-3}$, whereas the radiation component varies according to
$\rhor\propto a^{-4}$. Unfortunately, the behavior of the dark energy
$\rhov$ remains unknown \cite{frieman}. Current observations indicate
that the energy density of the vacuum energy evolves slowly over
cosmic time, so that it acts like a cosmological constant. For
simplicity, we assume here that $\rhov\approx$ {\sl constant}.  One
should keep in mind, however, that more complicated behavior is
possible and could be realized in other universes even if $\rhov$ is
essentially constant in our own.

Instead of working in terms of energy densities themselves, 
one can also define a critical density, 
\be
\rho_{\rm crit} \equiv {3 H_0^2 \over 8\pi G} \,, 
\label{rhocrit} 
\ee
and write the energy densities as ratios 
\be
\Omega_j \equiv {\rho_j \over \rho_{\rm crit}} \,, 
\ee
where the subject identifies the component of the universe (dark
matter, radiation, etc). The set of parameters ${\cal S}_{\rm exp}$
necessary to determine the expansion properties of the universe thus
becomes
\be
{\cal S}_{\rm exp} = 
\bigl\{ H_0, \omegab, \omegad, \omegar, \omegav \bigr\} \,. 
\ee
With these parameters specified, note 
that the curvature constant $k$ is given by 
\be
k = H_0^2 \left[1-\omegab-\omegad-\omegar-\omegav\right]\,. 
\ee
Notice also that the total mass density is determined
\be
\omegam = \omegab + \omegad \, . 
\ee

The discussion thus far only accounts for the expansion of a universe,
and implicitly assumes that space-time is homogeneous and isotropic.
In order for structure to form, the universe in question must contain
deviations from homogeneity. In our universe, the starting amplitude
of these fluctuations is extremely small. Moreover, as discussed
below, considerations of structure formation indicate that such
fluctuations should be small in any successful universe. As a result, 
the expansion of the universe proceeds largely independently of the 
formation of structure on smaller scales. 

The primordial fluctuations can be described in a number of ways. 
In our universe, these seeds of structure formation are found to be
Gaussian-distributed adiabatic fluctuations to a high degree of
approximation \cite{planck2014,planck2016,wmap} (see also 
\cite{copi,muir}).  Moreover, the fluctuations have a nearly
scale-invariant spectrum. The theory of inflation (see below) 
tends to produce such a spectrum, but the scale-invariant hypothesis
was proposed as a working model of the fluctuations much earlier 
\cite{harrison1970,zeldovich}. In any case, the spectrum of 
perturbations can be written in the form 
\be
{\cal P}_S(k) = A_S \left({k\over k_0}\right)^{n_S-1}\,, 
\label{spower} 
\ee
where $k$ is the wavenumber (equivalently, spatial scale) of the
fluctuation. In our universe the spectrum is nearly independent of
spatial scale so that $|n_S-1|\ll1$. 

In general, the universe must also contain a contribution to the
fluctuations due to gravitational waves, often known as tensor modes.
The dimensionless spectrum of tensor modes can be written in a form 
similar to that considered previously, i.e.,  
\be
{\cal P}_T(k) = A_T \left({k\over k_0}\right)^{n_T}\,. 
\label{tpower} 
\ee
In our universe, tensor modes have not (yet) been observed, but we
expect that $n_T\sim0$ and the amplitude $A_T \ll A_S$.

The set of parameters ${\cal S}_{\rm per}$ required to specify 
the departures of the universe from homogeneity thus involves 
at least four parameters and can be written 
\be
{\cal S}_{\rm per} = 
\bigl\{ A_S, n_S, A_T, n_T, \dots \bigr\} \,. 
\ee
The number of parameters required for a full specification could be
larger if the fluctuations are non-gaussian. 

The experiments that determine the cosmological parameters in our
universe rely heavily on observations of the cosmic microwave
background \cite{planck2014,cobe,wmap}.  These measurements depend on
another cosmological parameter $\tau$, which is the scattering optical
depth of the universe due to reionization. The optical depth $\tau$ =
0.09 in our universe and must be determined in order to make precise
estimates for the other cosmological parameters of interest. In the
present context, however, the scattering optical depth $\tau$ does
not play an important role in structure formation. Reionization 
occurs only because structure --- first galaxies and then massive 
stars --- is able to form. In any case, we will not include the 
scattering optical depth as a relevant variable for purposes of 
studying fine-tuning.  

\bigskip 
\begin{table} 
\centering 
{\bf Observed Cosmological Parameters in our Universe} 
\medskip
\def\arraystretch{1.1} 
\begin{tabular}{ l c c }
\\
\hline 
\hline 
quantity & symbol & observed value \\
\hline 
Baryon density & $\omegab$ & 0.044 \\
Dark matter density & $\omegad$ & 0.24 \\
Radiation density & $\omegar$ & $10^{-4}$\\ 
Vacuum energy density & $\omegav$ & 0.72\\ 
Hubble constant (scaled) & $h$ & 0.7 \\
Fluctuation amplitude & $Q$ & $10^{-5}$ \\ 
Power spectrum index & $n_S$ & 0.96\\ 
Scattering optical depth & $\tau$ & 0.09\\
\hline 
\hline 
\end{tabular}
\caption{Table of the observed cosmological parameters in
our universe. These quantities have been measured by a number 
of cosmological experiments 
\cite{des,perlmutter,planck2014,riess,riess2,cobe,wmap}. 
The values are listed with few enough significant digits 
that they are consistent with all current measurements. } 
\label{table:cosparam} 
\end{table}  

The most important cosmological parameters are summarized in Table 
\ref{table:cosparam}. This list contains quantities that define the 
cosmological inventory, the current expansion rate, and the
characteristics of the primordial density fluctuations (see also 
\cite{liddle2004,lahavliddle}. Note that the inventory is not complete, 
as one can consider the various types of stellar objects and gaseous
phases that make up the baryonic component, as well as the radiation
fields produced by a wide range of astrophysical processes 
\cite{fukugita}.  On the other hand, not all of the parameters listed 
in Table \ref{table:cosparam} are important for discussions of
fine-tuning. As discussed below, we can reduce the number of
cosmological parameters to a minimal set.  

We first note that the Hubble constant $H_0$, while vital to
understanding the current state of our universe, essentially defines
the current cosmological epoch. In considerations of other universes,
however, we only need to consider whether or not structure formation
occurs at any epoch. As a result, the Hubble constant, which varies
with time, does not need to take on a specific value. 

Next, our universe is observed to be nearly spatially flat. We can
also argue that successful universes must be close to flat: Some
solution to the flatness problem, either by inflation \cite{guth1981} 
or some other mechanism, is assumed to be operational in any viable
universe (see Section \ref{sec:flatness} for further discussion). 
We can also assume that the horizon problem and monopole problem
(unwanted relics) are not impediments to a successful universe.  
As a result, we can take $k=0$ and hence enforce the constraint 
\be
\omegab+\omegad+\omegar+\omegav = 1\,.
\ee
In other words, only three of the density contributions are
independent. The inventory of a universe is thus specified by three
quantities. The values of $\Omega_j$, however, vary with time or
equivalently scale factor. As outlined below, the early universe must
emerge from the epoch of Big Bang Nucleosynthesis with an acceptable
chemical composition, which in turn depends on the baryon to photon
ratio $\eta$, which is (nearly) constant. We can thus use $\eta$ to
specify the baryonic component. We can then use the ratio
$\omegad/\omegab$ to specify the dark matter content. Note that even
though the $\Omega_j$ vary with time, the ratio of any two matter
components does not.  Alternately, we can define a dark matter
parameter $\delta\equiv\eta\omegad/\omegab$ so that we have symmetric
definitions for the baryonic component $\eta$ and the dark matter
component $\delta$.  Finally, the dark energy density is assumed to be
constant and can be specified through its value $\rhov$. The inventory
of the universe is thus specified by the reduced set of parameters 
\be
{\cal S}_{\rm inv} \to 
\Bigl\{ \eta, {\omegad \over \omegab}, \rhov \Bigr\} = 
\Bigl\{ \eta, \delta, \rhov \Bigr\} \,. 
\ee

The number of parameters necessary to specify the spectrum of density
fluctuations can also be significantly reduced. Given that the tensor
modes are subdominant in our universe, and that the spectrum of
perturbations for both contributions is relatively flat, we can
characterize the primordial fluctuations with a single parameter
$Q\sim\sqrt{A_S}$, where $Q\approx10^{-5}$ in our universe. The set 
of parameters for structure formation thus collapses to the form 
${\cal S}_{\rm per}\to\left\{Q\right\}$. This simplification assumes
that the index $n_S$ remains close to unity. For much larger (smaller)
values of $n_S$, the spectrum of perturbations will have significantly
more power on smaller (larger) spatial scales, and will lead to
corresponding changes in structure formation. Unfortunately, a
comprehensive assessment of the allowed range of the index $n_S$ has
not been carried out. Nonetheless, before the value of $n_S$ was
well-determined observationally, explorations of structure formation
with a range of indices $n_S\ne1$ (e.g., \cite{bardeenbig,blumenthal}) 
did not find that the universe becomes uninhabitable. 

With our present level of understanding of physics and cosmology, the
parameters ${\cal S}_{\rm cos}=\left\{\eta,\delta,Q,\rhov\right\}$
represent the most important dials that can be adjusted to specify the
properties of a given universe. In a complete theory, the values of
these parameters --- or more likely the distributions of the parameter
values --- could be calculable from physics beyond the Standard
Model. In the meantime, for this discussion, the values of
$\left\{\eta,\delta,Q,\rhov\right\}$ are left as free parameters.

The baryon to photon ratio $\eta$ is nonzero because the universe
experienced an epoch of baryogenesis that broke the symmetry between
matter and antimatter (unless $\eta\ne0$ results from highly unusual
initial conditions). Baryogenesis, in turn, requires three essential
ingredients \cite{sakharov}: The first requirement is that baryon
number is not conserved. The second is that both C and CP conservation
must be violated, where `P' is the discrete symmetry of parity and `C'
is that of charge conjugation. Finally, the universe must depart from
thermal equilibrium during the epoch(s) when non-conservation of the
aforementioned quantities takes place. Although these three features
are known to be required for successful baryogenesis, an accepted
theory of this process is not yet available (see \cite{kolbturner} for
additional detail and \cite{dine2003,steigman} for more recent
reviews). Grand Unified Theories \cite{langacker81}, theories of
quantum gravity \cite{hawkingpagepope}, and other approaches allow for
the violation of baryon number conservation, so that new physics
should eventually predict the expected distribution of the baryon to
photon ratio $\eta$.

The abundance of dark matter $\delta$ is also determined by processes
taking place in the early universe. The simplest scenario occurs when
the universe has a single dark matter species that is produced in
thermal equilibrium.  In that scenario, the abundance of dark matter
is determined when the weak interactions become too slow to maintain
statistical equilibrium, typically at cosmic times $t\simless1$ sec
and temperature $T\simgreat1$ MeV \cite{kolbturner}. The dark matter
abundance depends on the particle properties (mass, interaction cross
section, etc.), which are not known at the present time 
\cite{jungman}.  Again, a description of these properties requires 
extensions of the Standard Model, so that new physics may eventually
predict the possible abundances of dark matter.

The value of the amplitude $Q$ of the primordial density fluctuations
also cannot be predicted using known physics. In a large class of
inflationary theories \cite{lindeinflate}, quantum fluctuations in the
inflaton field produce density perturbations, so that the amplitude
$Q$ could be calculated in principle. In this case, a large number of
inflationary scenarios are possible, so that the possible distribution
of $Q$ is similarly enormous. In addition, even for a given set of
inflaton fields, the spectrum of density fluctuations can depend on
the initial conditions, i.e., the manner in which the universe enters
into its inflationary epoch. Although the details are both complicated
and unknown, the value of $Q$ is unlikely to have the same value in
all universes, so that the density fluctuations must be described by a
distribution of values across the multiverse. Moreover, this
discussion assumes a scale-invariant spectrum of fluctuations within a
given universe. In addition to the overall amplitude $Q$ of the
spectrum having a distribution of values, the form of the fluctuation
spectrum itself (see equations [\ref{spower},\ref{tpower}]) could
also vary from universe to universe.

Finally, the value of the vacuum energy density $\rhov$ is not
understood in the context of known physics. This issue is essentially
the cosmological constant problem \cite{caldwellkam,weinberg89}, which
has a long history and no accepted resolution (see Section 
\ref{sec:rhovac}).

\subsection{Constraints on the Cosmic Inventory} 
\label{sec:cosconstraint} 

\begin{figure}[tbp]
\centering 
\includegraphics[width=1.0\textwidth,trim=0 150 0 150,clip]{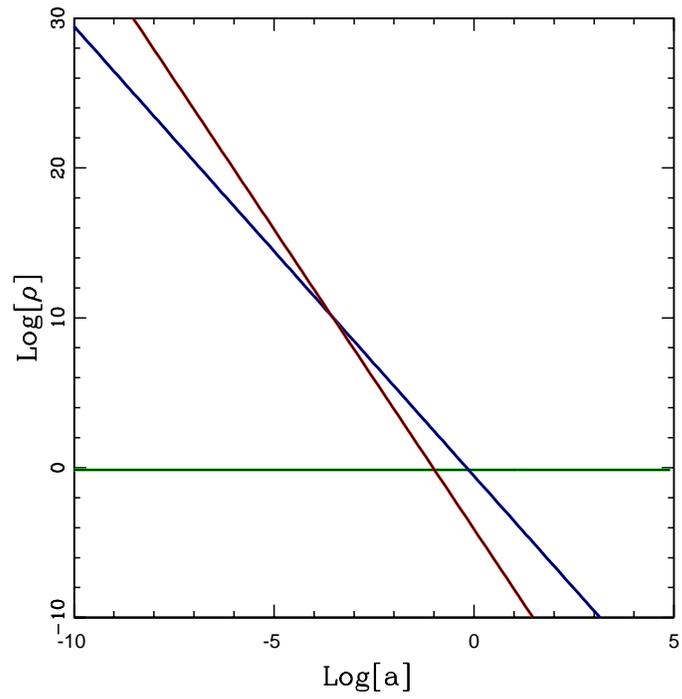}
\caption{Evolution of density components with the expansion of the 
universe. Energy density is plotted versus scale factor $a$ for 
radiation (red), matter (blue), and vacuum energy (green). The scale 
factor is taken to be unity at the present epoch; the densities are 
scaled by the current value of the critical density (see equation 
[\ref{rhocrit}]). } 
\label{fig:scalefact} 
\end{figure}

Figure \ref{fig:scalefact} shows the evolution of the various density
components of the universe. This figure is scaled to the observed
values in our universe, where $a=1$ corresponds to the current
cosmological epoch. The universe is radiation dominated at early
times, transitions into a matter dominated era at intermediate times,
and has just recently become dominated by its vacuum energy.  When
viewed across the relatively large span of cosmic time shown here, two
things are evident: First, the fact that the universe is dominated by
the vacuum energy at the present epoch is hard to discern -- this near
equality of matter and vacuum energy at the present epoch is a
manifestation of the well-known cosmological constant problem
\cite{weinberg89,padmanabhan}. Second, the duration of the matter 
dominated era is relatively short. For some values of the cosmological
parameters, the universe could move directly from its radiation era
into a vacuum dominated era. Such a universe would have no period of
matter domination and hence no structure formation. This scenario ---
without a matter dominated era --- is an extreme version of cases
where the vacuum energy density is too large relative to the
primordial fluctuation amplitude to allow for structure formation 
(see Section \ref{sec:rhovac} and Refs. 
\cite{adamsrhovac,efstathiou,garriga1999,garriga2000,garriga2006,
graesser,martel,mersini,piran,weinberg87}). 

We can derive a constraint that depends only on the cosmic inventory
--- independent of the fluctuation amplitude $Q$ --- by requiring that
the universe have a matter dominated era. Equivalently, the energy
density of the vacuum must be smaller than the energy density of the
universe at the epoch of equality. If we write the vacuum energy
density in terms of an energy scale $\lambda$, i.e., 
\be
\rhov \equiv \lambda^4 \,, 
\label{rhovenergy} 
\ee
then the constraint takes the form 
\be
\lambda < \left(2 a_R\right)^{1/2} \eta \mpro {\omegam\over\omegab}\,,
\label{needmatter} 
\ee
where $\eta$ is the baryon to photon ratio and where $a_R=\pi^2/15$ is
the radiation constant.  In our universe, the energy scale of the
vacuum $\lambda\sim3\times10^{-3}$ eV. For comparison, the right hand
side of equation (\ref{needmatter}) is $\sim4$ eV. The universe is
thus safe by a factor of $\sim1000$ for the energy scale $\lambda$ 
(and hence a factor of $\sim10^{12}$ for energy density $\rhov$).  

The constraint of equation (\ref{needmatter}) is necessary but not
sufficient. Even if the universe has a matter dominated era, structure
formation can be suppressed if the vacuum energy density is too large
relative to the amplitude of the primordial density fluctuations.
This issue is taken up in Section \ref{sec:rhovac} and provides
stronger constraints on the energy density of the vacuum. On the other
hand, the baryon to photon ratio $\eta$ can be larger in other
universes (Section \ref{sec:bbn}), which would allow for even larger
values of the vacuum energy scale $\lambda$. If the baryon to photon 
ratio becomes too large, however, the epoch of matter domination 
will take place before Big Bang Nucleosynthesis. If we approximate 
the energy scale of BBN as $E_{\rm bbn}\sim\Delta{m}\sim\emass$ 
\cite{bartip,carr}, this constraint can be written in the form 
\be
\eta \le \left(2 a_R\right)^{-1/2} {\omegab\over\omegam} 
{\emass \over \mpro} \approx 10^{-4} \,, 
\label{needbbn} 
\ee
where the numerical value corresponds to the parameters of our 
universe. Note that a universe in which matter domination occurs 
before the epoch of BBN would not necessarily be sterile, but it 
would represent a significant departure from the usual ordering of 
cosmological time scales. 

\subsection{The Flatness Problem} 
\label{sec:flatness} 

One of the classic fine-tuning problems in cosmology is sometimes
known as the flatness problem. This issue can be illustrated by
writing the equation of motion (\ref{adotovera}) for the scale 
factor in the form  
\be
\Omega = {1 \over 1 - 3k/8\pi{G}\rho{a^2}} 
= {1 \over 1 - \chi(t)} \,,
\ee 
where the second equality defines the parameter $\chi(t)$. Note that,
in general, the density parameter $\Omega$ is not constant in
time. More specifically, for a radiation dominated universe,
$\rho\propto a^{-4}$, so that $\chi\propto a^2$. Similarly, for a
matter dominated universe $\rho\propto a^{-3}$ and $\chi\propto a$.
The parameter $\chi$ thus increases as the universe expands for the
case of both matter and radiation. On the other hand, if the universe
is dominated by vacuum energy, then $\chi \propto a^{-2}$ and hence 
decreases with time. 

In order for the universe to remain flat, equivalently for $\Omega$ to
remain close to unity, the quantity $\chi$ must remain small. But if 
$\chi$ is small at a given epoch, then it must have been even smaller
at an earlier epoch (for universes dominated by either radiation or 
matter). To leading order, for $\chi\ll1$, 
\be
\big| \Omega - 1 \big| = {\cal O} (\chi) \,.
\ee
In order for $\chi$ to be small at the present cosmological epoch, 
as observed, this quantity must have been extremely small at earlier
times. When the cosmic age was $\timebbn\sim1$ sec, near the beginning
of Big Bang Nucleosynthesis, the parameter $\chi(\timebbn)\sim10^{-16}$. 
If we go all the way back to the Planck era, with age
$\tplanck\sim10^{-43}$ sec, the parameter
$\chi(\tplanck)\sim10^{-60}$.

\begin{figure}[tbp]
\centering 
\includegraphics[width=1.0\textwidth,trim=0 150 0 150,clip]{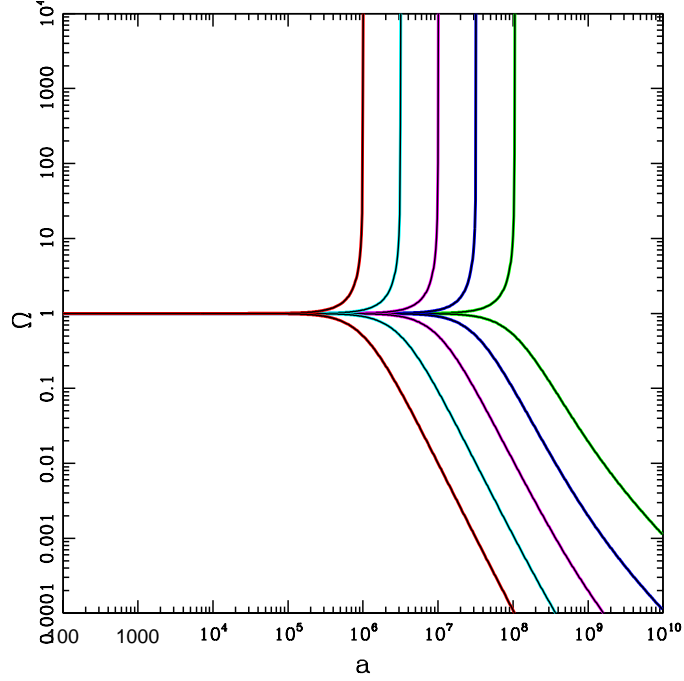}
\caption{Evolution of density parameter $\Omega$ with increasing 
scale factor $a$. Curves are plotted for five initial values of 
the parameter $\chi=10^{-16}-10^{-11}$, and for both positively 
curved universes (top curves) and negatively curved universes 
(bottom curves). } 
\label{fig:omegaevolve} 
\end{figure}  

Figure \ref{fig:omegaevolve} shows the evolution of the density
parameter $\Omega$ as the universe expands. The scale factor is taken
to be unity at the start of the evolution and five initial values of
the parameter $\chi$ are used, specifically, $\log_{10}\chi(a=1)1$ =
$-16,-15,\dots,-11$. The evolution in $\Omega$ is shown for universes
with both positive (upper curves) and negative (lower curves)
curvature. The density parameter $\Omega$ remains close to unity until
the parameter $\chi$ becomes significant, and then $\Omega$ evolves
rapidly. For positive curvature, the value of $\Omega$ becomes
increasingly large and universe eventually recollapses. For negative
curvature, $\Omega$ steadily decreases and the expansion becomes that
corresponding to an empty universe.

The paradigm of inflation \cite{albrechtstein,guth2000,linde1983} was
developed to alleviate this issue of the sensitive fine-tuning of the
density parameter $\Omega$ (although other motivations, such as the
monopole problem, were also important). Since this topic has been
discussed extensively elsewhere \cite{guth2000,guth2007,lindeinflate}, 
the present treatment will be brief. As outlined above, our universe
requires the parameter $\chi\simless10^{-60}$ at the Planck epoch in
order to evolve into its present state. In contrast, the value of
$\chi$ is expected to be of order unity at this time. If the universe
experiences an epoch of rapid expansion due to vacuum domination, and
$\chi\propto{a}^{-2}$ during that epoch, then successful inflation
requires the scale factor $a$ to grow by a factor of $\sim10^{30}$.
This growth factor is generally expressed in terms of the number $N$
of e-foldings so that the requirement becomes 
\be
{\rm e}^{2N} \simgreat 10^{60} \qquad \Rightarrow \qquad 
N \simgreat 30 \ln(10) \approx 70 \,. 
\label{enumber} 
\ee
The exact number of required e-foldings of the scale factor depends on
the energy scale of the inflaton field, where lower energy scales lead
to somewhat less stringent requirements. However, the number of
e-foldings $N$ appears in the exponential, so that the required minimum
number $N_{\rm min}$ generally falls in the range $N_{\rm min}$
$\approx60-70$. Moreover, most inflation models tend to produce far
more e-foldings so that $N \gg N_{\rm min}$. 

Another problem facing our observed universe is the so-called horizon
problem: Observations of the Cosmic Microwave Background show that the
universe is isotropic to high precision. Without the aforementioned
inflationary epoch, regions that are now observed in opposite sides of
the microwave sky would have never been in causal contact. But they have
the same temperature within about one part in $10^5$, where we can 
now measure the spectrum of these small deviations to high precision  
\cite{planck2014,planck2016,cobe,wmap}.  Since one effect of the 
inflationary epoch is to accelerate the expansion and effectively move
regions out of the horizon, such regions could have been in contact at
earlier epochs provided that the duration of the inflationary epoch is
long enough.  The number of e-foldings of the scale factor required to
alleviate the flatness problem (from equation [\ref{enumber}]) is
nearly the same as that required to alleviate the horizon problem 
\cite{kolbturner,baumann}. As a result, universes that emerge from an
early inflationary epoch with sufficiently small $\chi$, which can
survive to old ages, will also generally be close to isotropic. Our
universe could in principle be habitable without the extreme level of
isotropy that is observed, but smoothness and flatness tend to arise
together in the simplest versions of inflation.

The paradigm of inflation alleviates the flatness problem --- and
hence the apparent fine-tuning of the density parameter $\Omega$ ---
as long as the universe can accelerate for the required number of
e-foldings (equation [\ref{enumber}]) and then subsequently evolve
according to the standard, radiation dominated hot Big Bang model. In
order for inflation to be successful \cite{kolbturner,steinturner}, a
number of constraints must be met:

The universe must first be able to enter into an inflationary state,
where the energy density is dominated by the vacuum and the scale
factor accelerates (grows faster than linear with time). This
superluminal phase must last long enough to solve the flatness,
horizon, and monopole problems in our universe. Although other
universes are not required to be be as isotropic as our own, so that
the horizon problem is not as severe, all of these issues are
addressed by the same large growth factors. 

The likelihood that the universe can enter into an inflationary state
remains under intense debate (for further discussion, see
\cite{brandenberger,carroll2014,carroll2010,corichi,gibbons,
hawkingpage,hollandswald,schiffrin,steinhardt2011,turok2002}).  
The required fine-tuning for achieving successful inflation can be
described in terms of the space of possible trajectories for the
expansion history of the universe.  Some authors argue that if the
universe starts at the Planck epoch with reasonable assumptions, then
successful inflation becomes ``exponentially unlikely'' \cite{ijjas}.
In other words, given that the universe is required to have desirable
late-term properties, only a small fraction of the possible starting
conditions lead to acceptable cosmological histories. Other authors
conclude that the paradigm of cosmic inflation remains on a strong
footing \cite{guth2013}. The key question is whether or not the
conditions required for successful inflation are more constraining
than the cosmological problems that the paradigm seeks to alleviate.
These conditions include both the necessary parameter values of the
theory (e.g., the properties of the inflaton potential) and the
requisite initial conditions.  This issue remains unresolved. On a
related note, the fraction of cosmological trajectories that lead to
smooth universes at late times is dominated by those that are not
smooth at early epochs \cite{carroll2014}, which changes the 
constraints on cosmological initial conditions.

During the inflationary expansion phase, quantum fluctuations in the
inflaton field produce density perturbations in the background
universe \cite{bardeen,guthpi,mukhanov}. These perturbations must be
sufficiently small in amplitude in order for the inflationary epoch to
begin, and this condition is expected to hold in only a small fraction
of realistic cosmologies \cite{vachaspati}. Provided that inflation is
successful, in the late universe these density fluctuations grow into
galaxies, clusters, and other large scale structures.  Moreover, these
fluctuations must have an amplitude $Q$ that falls within the
approximate range $10^{-6}\simless{Q}\simless10^{-2}$ in order for the
universe to produce galaxies with acceptable densities and hence be
habitable \cite{coppess,tegmarkrees,tegmark} (see Section 
\ref{sec:galaxies}).  The relation between the amplitude $Q$ and the 
parameters that appear in the inflaton potential is described in
Section \ref{sec:fluctuate}.

After the universe has expanded by the required factor, it must leave
its accelerating state and begin to expand in the usual subluminal
manner. In order for the universe to become potentially habitable,
first by producing heavy nuclei and later by forming galaxies and
stars, it must become radiation dominated after inflation ends. The
rapid, accelerated expansion of the inflationary phase leaves the
universe with an extremely low temperature, of order
$T\sim10^{-60}T_{\rm pl}\sim10^{-28}$ K, so that all of the energy is
locked up in the vacuum. This vacuum energy must be converted into
radiation and particles through a process known as reheating. The
conversion must be efficient enough to reheat the universe to a
sufficiently high temperature such that baryogenesis can take place.
This minimum temperature is often taken to be the scale of the
electroweak phase transition $\sim100$ GeV. Successful reheating of
the universe to this high temperature is by no means automatic, so
this requirement represents another hurdle that a successful universe
must negotiate.

\subsection{Quantum Fluctuations and Inflationary Dynamics} 
\label{sec:fluctuate} 

As outlined in the previous subsection, an early epoch of inflation
can potentially alleviate a number of cosmological problems, although
achieving successful inflation is not without its own issues.
Although alternate explanations exist, it is useful to illustrate
inflationary dynamics in greater detail. Toward that end, this section
describes the semi-classical dynamics for the evolution of the scalar
field that drives inflation, generally called the inflaton field. Next
we elucidate the relationship between the parameters that appear in
the inflaton potential and the spectrum of fluctuations produced
during the inflationary epoch.

The equation of motion for the evolution of the inflaton field $\phi$ 
form is generally written in the from 
\be
{\ddot \phi} + 3H {\dot \phi} = - {\partial V \over \partial \phi} \,,
\label{inflatephi} 
\ee
where $H$ is the Hubble parameter and $V$ is the potential. 
During the inflationary epoch, the energy density of the universe
is dominated by the potential of the inflaton, so that the Hubble
parameter is given by 
\be
H^2 = {8\pi\over3\mplanck^2} \left[ V(\phi) + 
{1\over2} {\dot\phi}^2 \right] \,. 
\label{inflatehub} 
\ee
Under a wide range of conditions, the first term in equation
(\ref{inflatephi}) and the last term in equation (\ref{inflatehub})
can be neglected, and evolution takes place during what are called
slow-roll conditions. The number of e-foldings of the inflationary
epoch is then given by 
\be
N = \int H dt = {8\pi\over\mplanck^2} \int V 
\left(- {\partial V \over \partial \phi}\right)^{-1} d\phi \,, 
\ee
where the integrals are taken over the time interval (values of
$\phi$) corresponding to the inflationary epoch.

To fix ideas, we can illustrate the scalar field dynamics of 
inflation with a simple power-law form for the potential
\cite{linde1983}, 
\be
V(\phi) = q_4 \phi^4 \,,
\label{vquartic} 
\ee
where $q_4$ is a dimensionless coefficient. With this choice, 
the number of e-foldings takes the form 
\be
N = {\pi\over\mplanck^2} \left[\phi_0^2 - \phi_f^2\right] 
\approx \pi \left({\phi_0\over\mplanck}\right)^2  \,,
\label{nefold} 
\ee
where $\phi_0$ ($\phi_f$) is the initial (final) value of the inflaton
field. The requirement of sufficient inflation $N\simgreat60$ thus
implies that the starting value of the inflaton field satisfies the 
constraint $\phi_0\simgreat(60/\pi)^{1/2}\approx4.4\mplanck$. The
requirement of sufficient inflation thus requires that the value of
the scalar field $\phi$ is comparable to --- and somewhat larger than
--- the Planck scale. As summarized below, however, constraints on the
perturbation spectrum require the energy density to be well below the
Planck scale.

In addition to allowing the universe to become connected and flat, the
inflationary epoch also imprints density fluctuations on the otherwise
smooth background of the universe \cite{guthpi}. To leading order, the
spectra of cosmological perturbations produced by inflation can be
written in the forms 
\be
{\cal P}_S = {H^4 \over (2\pi{\dot\phi})^2} 
\qquad {\rm and} \qquad 
{\cal P}_T = {2\over\pi^2} \left({H\over\mplanck}\right)^2 \,, 
\ee
for scalar and tensor contributions, respectively \cite{baumann}. 
These quantities are evaluated at the epoch when a perturbation with a
particular length scale crosses outside the horizon \cite{kolbturner}.
Using the equation of motion (\ref{inflatephi}) for the scalar field
and the definition (\ref{inflatehub}) of the Hubble parameter (in the 
slow-roll approximation), the quantities $H$ and ${\dot\phi}$ can be
written in terms of the inflationary potential. For the particular
potential given by equation (\ref{vquartic}), the expressions for the
perturbation spectra take the forms 
\be
{\cal P}_S = {8\pi\over3} q_4 \left({\phi \over \mplanck}\right)^6 
\qquad {\rm and} \qquad 
{\cal P}_T = {16\over3\pi} q_4 \left({\phi \over \mplanck}\right)^4\,.
\label{inflatepower} 
\ee
The ratio of scalar to tensor perturbations is thus given by 
\be
{{\cal P}_S \over {\cal P}_T} = {\pi^2 \over 2} 
\left({\phi\over\mplanck}\right)^2 \approx {\pi\over2}N \gg 1\,,
\ee
where $N$ is the number of e-foldings from equation (\ref{nefold}).
Since the perturbations that are constrained by measurements of the
cosmic background radiation are those that left the horizon about
$N\sim60$ e-foldings before the end of inflation \cite{kolbturner}, 
the value of $\phi$ used to evaluate equation (\ref{inflatepower}) is
of order $\phi\sim\phi_0\sim4\mplanck$. Using this result to evaluate
the scalar perturbation ${\cal P}_S$ allows us to specify the
amplitude $Q$ in terms of inflationary parameters, 
\be
Q \approx \left[{\cal P}_S\right]^{1/2} =
\left({8\pi q_4\over3}\right)^{1/2} 
\left({\phi \over \mplanck}\right)^3 \approx 200 \sqrt{q_4} \,.
\label{qinflate} 
\ee
As outlined above (see Table \ref{table:cosparam}), the primordial
fluctuations in our universe have amplitude $Q\sim10^{-5}$, so that
the required value for the dimensionless constant $q_4$ must be
extremely small, roughly $q_4\ll10^{-14}$.

In the absence of special circumstances, however, the dimensionless
parameter $q_4$ is expected to be of order unity. The requirement that
its value must be incredibly small thus leads to a fine-tuning problem.  
More specifically, the problem is one of naturalness (see \cite{thooft} 
and Section \ref{sec:overview}). The value of the parameter is
expected to become of order unity due to quantum corrections unless
the small required value is protected by a special symmetry. Moreover,
one can show that the constraint $Q\sim\sqrt{q_4}$ and hence $q_4\ll1$
holds more generally, and that any successful inflation model in this
class must have a small parameter \cite{afguth}.

Given the tension between the requirement of a small value of the
dimensionless constant ($q_4\sim10^{-14}$) and its much larger
expected value ($q_4\sim1$), one would expect other universes to have
larger values of $q_4$. As a result, it is natural (in both the
technical and colloquial sense) for the amplitude $Q$ of cosmological
fluctuations to be larger in other universes.  Larger values of $Q$
lead to earlier structure formation and denser galaxies. Of course,
the amplitude $Q$ could sometimes be smaller, so that the universe
produces more rarefied galaxies.  The consequences of these changes,
and accompanying constraints, are discussed in Section
\ref{sec:galaxies}.

Note that for viable universes the value of $Q$ is bounded from above:
If the amplitude $Q$ becomes of order unity, fluctuations are close to
non-linearity --- and hence ready for collapse --- as soon as they
enter the horizon after inflation. In this case, a large fraction of
the energy within the horizon could become locked up within black
holes in the early universe. The analysis of \cite{tegmarkrees}
indicates that the resulting density of black holes could dominate the
density of dark matter and baryons if $Q\simgreat0.1$. This bound
assumes that the spectrum of density perturbations remains relatively
flat, with index $n_S\sim1$, down to the mass scales characteristic of
the horizon just after inflation (see \cite{carrkuhnsand,greenliddle} 
for further detail). 

Finally, we note that although the tensor perturbations are often
subdominant, they provide an important constraint on the energy scale
of inflation (see also \cite{liddle1994}). The requirement that 
${\cal P}_T\simless Q^2$ implies that the inflaton potential must 
obey the bound
\be
V(\phi) \simless Q^2 \mplanck^4 \,. 
\ee
If we characterize the energy density of the potential by defining 
an energy scale $\mu \equiv V^{1/4}$, then the constraint becomes 
$\mu \simless \sqrt{Q} \mplanck \sim 0.003 \mplanck$ for our universe. 
In other words, for applications to our universe, the energy scale of
inflation is bounded from above by the GUT scale and must be
substantially below the Planck scale. In other universes with larger
fluctuation amplitudes $Q$, the required hierarchy between the Planck
scale and the inflation scale could be less pronounced. 

\begin{figure}[tbp]
\centering 
\includegraphics[width=1.0\textwidth,trim=0 150 0 150]{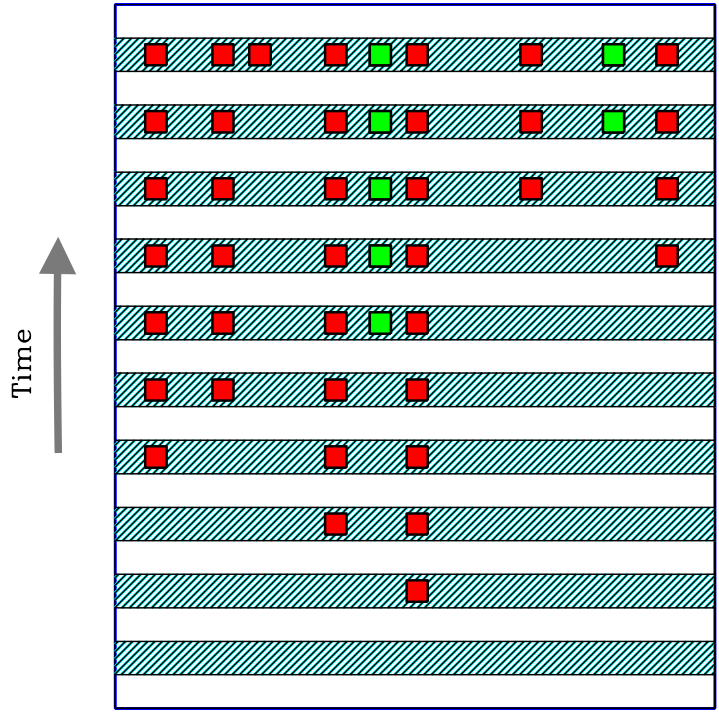}
\caption{Schematic representation of the eternal inflation scenario, 
which provides a possible mechanism for the production of multiple
universes.  In this diagram, each horizontal band represents a
snapshot of the multiverse at a particular epoch, where time increases
along the vertical axis. In each shaded band, the background
represents the portion of the multiverse that is dominated by its
vacuum energy and experiences superluminal expansion. The squares
represent regions of space-time that have emerged from their
inflationary epochs and are dominated by either radiation or matter. 
These regions appear at random time intervals, so that their number
increases with time (some of these regions could disappear as they
recollapse, but this behavior is not shown).  Although the background
is expanding exponentially and the universes (squares) are expanding
as power-laws in time, this expansion is suppressed in the diagram.
Some fraction of the universes will have the right parameters to
support habitability, as depicted by the green squares. The spatial
location of the individual universes (horizontal direction) and their
times of nucleation (vertical direction) are randomly distributed. }
\label{fig:eternal} 
\end{figure}  

\subsection{Eternal Inflation} 
\label{sec:eternal} 

An important generalization of the inflationary universe paradigm is
that (in many cases) most of the volume of the entire space-time is in
a state of superluminal expansion, so that the inflationary epoch can
be eternal \cite{guth2007,linde1986}. Sub-regions of space-time detach
from the background and form separate `pocket universes', which can
evolve to become similar to our own.  Moreover, this feature can be
generic for some classes of inflationary theories \cite{vilenkin83}.
The scenario of eternal inflation provides one specific mechanism for
generating multiple universes and is thus of interest for the problem
of fine-tuning and the multiverse.

To illustrate the manner in which inflation can be eternal, we
consider the simple potential of equation (\ref{vquartic}). The scalar
field obeys the equation of motion (\ref{inflatephi}), so that its
classical trajectory would be to slowly evolve to smaller $\phi$ and
hence lower potential energy $V(\phi)$. The scalar field is said to
`slowly roll downhill'. As the scalar field evolves down the
potential, however, quantum fluctuations are superimposed on the
classical motion. If these fluctuations are large enough, then parts
of the space-time can remain in an inflationary state for an
indefinitely long span of time. The conditions required for this
scenario to operate are illustrated below.

Following the discussion of \cite{guth2000,guth2007}, consider the
time interval corresponding to one local Hubble time, i.e., $\Delta t$
= $H^{-1}$. At the start of that time interval, the scalar field will
have a value $\phi_0$ corresponding to its average over a Hubble
volume $V_H \sim H^{-3}$. During the Hubble time $\Delta t$, the scale
factor grows by one factor of $e$ and the Hubble volume grows by a
factor of $e^3$. At the same time, the scalar field will evolve. The
change in the scalar field due to its classical trajectory can be
denoted as $(\Delta\phi)_{\rm cl}$, whereas the change due to quantum
fluctuations is $(\Delta\phi)_{\rm qm}$. The total change in the
scalar field during one Hubble time is thus given by
\be
\Delta\phi = (\Delta\phi)_{\rm cl} + (\Delta\phi)_{\rm qm} \,. 
\ee
To leading order, the quantum fluctuations will have gaussian
probability distribution, with width given approximately by $H/(2\pi)$
\cite{starobinsky}. For small fluctuations, the scalar field will
evolve to lower values. For sufficiently large fluctuations, however,
the scalar field can move up the potential to higher values over the
course of one Hubble time. In order for eternal inflation to operate,
the probability of moving to larger values of $\phi$ must be high
enough that at least one of $e^3\sim20$ new Hubble volumes will have 
this property.  This condition implies the constraint 
\be
(\Delta\phi)_{\rm qm} \approx {H \over 2\pi} \simgreat 
0.6 (\Delta\phi)_{\rm cl} \approx 0.6 \big|{\dot\phi}_{\rm cl}\big| 
H^{-1} \,. 
\ee
Using the equation of motion (\ref{inflatephi}) and the definition 
(\ref{inflatehub}) of the Hubble parameter, this constraint takes 
the form 
\be
\phi \simgreat {1 \over 5} q_4^{-1/6} \mplanck \,. 
\ee
Although this required value of the field is larger than the 
Planck scale, the energy density is given by 
\be
V (\phi) \approx {\,q_4^{1/3} \over 625} \,\,\mplanck^{\,4} \,. 
\ee
As a result, the energy density of the scalar field required for
eternal inflation is much smaller than that given by the Planck
scale. If the universe starts out at high temperatures close to the
Planck scale, then it will be born with enough energy density to
enter and maintain eternal inflation. 

In this paradigm, at any given time, the energy density in most of the
volume of the multiverse is dominated by the potential energy of the
scalar field $\phi$ and is inflating.  Some regions will evolve far
enough down the potential so that quantum fluctuations do not push
them back up to higher vacuum energy densities. These regions can
experience inflation as described in the previous section, with
decreasing $\phi$ following its classical trajectory to lower values
of $V(\phi)$.  Some subset of these regions will successfully convert
vacuum energy into particles and radiation, and eventually evolve
according to classical cosmological theory like our own. In this
manner, the background space-time of the universe continually gives
rise to new universes.

This evolutionary picture is illustrated in Figure \ref{fig:eternal},
which shows a sequence of snapshots of the multiverse over time. The
shaded regions depict the background energy density of the vacuum,
which causes the universe to inflate. The volume of the multiverse in
this rapidly expanding state grows exponentially with time, so that
most of the space-time resides in this state dominated by vacuum energy.
Some regions can evolve to lower-energy vacuum states and eventually
become radiation dominated.  These regions thus become universes that
evolve according to some realization of cosmology, and are depicted as
the square symbols in Figure \ref{fig:eternal}. Although the
background space expands exponentially, and the universes expand as
well, this growth is suppressed in the diagram. The different
universes can in principle have different realizations of the laws of
physics and/or different values of the cosmological parameters. Only
some fraction of these regions have suitable choices of the parameters
for the universe to be habitable, as illustrated by the green regions
in the figure.
(As an aside: Many versions of eternal inflation result in the
production of an infinity of universes, and could result in an
infinite number of both red and green squares in Figure
\ref{fig:eternal}. Any assessment of the fraction of habitable 
universes depends on how the counting is carried out.) 

\bigskip 
\section{The Cosmological Constant and/or Dark Energy} 
\label{sec:rhovac} 

Although our universe can be specified by relatively few cosmological
parameters (see Section \ref{sec:cosmology}), one of the necessary
ingredients is a substantial energy density of the vacuum --- often
called dark energy. This quantity acts like a cosmological constant
and is currently the dominant component of the cosmic inventory (see
Table \ref{table:cosparam}). The dark energy is driving the currently
observed acceleration of the universe and will have enormous
consequences in the future \cite{busha2005,busha2007,nagamine}.
Moreover, the existence and nature of this counter-intuitive component
poses an interesting and important problem for fundamental physics.
On the other hand, astrophysical processes in our universe --- for
example, the formation of galaxies and other large scale structure ---
have been influenced more by dark matter than by dark energy thus far
in cosmic history (see the discussion of \cite{liviorees2018}).

Even though the theory of general relativity allows empty space to
have a nonzero energy density, its existence poses (at least) two
coupled problems:

\medskip \noindent
[A] If the vacuum energy density plays a significant role at the
present cosmological epoch, its value must be exceedingly small
relative to theoretically expected values. This extreme ordering of
energy scales is one manifestation of the cosmological constant
problem, and is an example of a Hierarchical Fine-Tuning problem
(Section \ref{sec:cosconprob}).

\medskip \noindent 
[B] If the vacuum energy density is large enough to affect the
cosmological expansion, it acts to suppress the formation of dark
matter halos and other cosmic structures. In a universe with too much
energy density in the vacuum, cosmological structure is unable to
separate itself from the expanding background and form gravitationally
bound entities. This consideration constrains the allowed energy
density of the vacuum, where the severity of the bound depends on the
other cosmological parameters (Section \ref{sec:qvbounds}). 

\medskip
In spite of being at the center of the cosmological constant problem,
the vacuum energy density of the universe is not necessarily constant
in time. In some models, this dark energy evolves smoothly, usually as
a decreasing function of time, through the evolution of scalar fields 
\cite{sahni,quintessence}. In other models, the vacuum energy density 
can evolve suddenly, e.g., through cosmological phase transitions 
\cite{coleman,voloshin,sakharov1984,zeldovich2}. Although the dark 
energy could display a wide range of temporal evolution in other
universes, for the sake of simplicity this treatment focuses on the
case of constant dark energy.

\subsection{The Cosmological Constant Problem} 
\label{sec:cosconprob} 

We start with a short review of the cosmological constant problem 
(for a more comprehensive discussion, see \cite{weinberg89} and
\cite{boussovac}). In General Relativity, the field equations 
take the form   
\be
R_{\mu\nu} - {1\over2} R g_{\mu\nu} + \Lambda g_{\mu\nu} = 
8 \pi G T_{\mu\nu} \,, 
\label{grfield} 
\ee
where the symbols have their usual meanings. In particular, the
parameter $\Lambda$ is the cosmological constant, which is
non-vanishing in general. Classically, the cosmological constant could
have any value. The only other constant appearing in the field
equation is the gravitational constant $G=\mplanck^{-2}$. On dimensional
grounds, one expects the cosmological constant to be determined by
$G$. In natural units, the quantity $\Lambda/G=\rhov$ has units of 
energy density (the quantity measured in cosmology). We can thus write 
\be
\Lambda/G = \rhov = {\cal O} \left( \mplanck^4 \right) \sim 10^{112} 
\, {\rm eV}^4 \,.
\ee
This expected value stands in sharp contrast to the value inferred
from cosmological experiments, which indicate that $\rhov\sim10^{-10}$
eV$^4$. The mismatch between these two values of the vacuum energy
density is one manifestation of the cosmological constant problem.

Another way to present the problem is to start from the 
definitions of the basic units of time and length from the
gravitational constant, i.e., the Planck time 
\be
\tplanck = \left({\hbar G \over c^5}\right)^{1/2} 
\approx 5.4 \times 10^{-44} \,{\rm sec}\,,
\label{tplanck} 
\ee
and the Planck length
\be
\lplanck = \left({\hbar G \over c^3}\right)^{1/2} 
\approx 1.6 \times 10^{-33} \,{\rm cm}\,.
\label{lplanck} 
\ee
The universe is observed to be large compared to the Planck length,
with size scale $R_{\rm univ} = c/H_0\approx 10^{28}$ cm
$\sim10^{61}\,\lplanck$. Similarly, the universe is inferred to be 
old compared to the Planck time, where the cosmic age 
$t_{\rm univ}\approx$ 13.7 Gyr $\sim10^{61}\,\tplanck$. 

Note that the magnitude of the cosmological constant problem depends
on what quantity is under consideration. The observed vacuum energy
density is smaller than the expected value by 122 orders of
magnitude. But the age and size of the universe are larger than their
expected values by `only' 61 orders of magnitude. If we were to define
an energy scale $\lambda$ for the vacuum, so that $\rhov=\lambda^4$
(see equation [\ref{rhovenergy}]), then the observed value of the
scale $\lambda\approx3\times10^{-3}$ eV is smaller than the expected
value ($\lambda\sim\mplanck$) by 30 orders of magnitude. Of course, these
discrepancies represent the same underlying hierarchy. On a related note, 
one could consider any function of the parameter (e.g., $\log\rhov$ or 
even $\log[\log\rhov]$), which would change the numerical value of the 
hierarchy, but not the underlying issue \cite{norton}. This choice of 
variable also affects the probability distributions of the parameters 
(see \ref{sec:probability}).  

Although classical general relativity is silent on the preferred value
of the cosmological constant $\Lambda$, aside from the above
dimensional considerations, quantum field theory indicates that the
issue must be taken seriously. In general, the vacuum (although
somewhat poorly named) has a non-zero energy density $\rho_V$, 
which can be written in the form 
\be
\langle T_{\mu\nu} \rangle_V = - \rho_V g_{\mu\nu}\,.
\label{qftrhov} 
\ee
The vacuum contribution to the energy-momentum tensor (left side of
the equation) is proportional to the metric (indicated by the right
side of equation) due to Lorentz invariance. The constant of
proportionality is the energy density of the vacuum, and the sign
convention is chosen so that if $\Lambda\equiv8\pi\rhov$, then we 
can identify $\rho_V\sim\rhov$. 

The key issue is that quantum field theory predicts that 
$\langle T_{\mu\nu}\rangle\ne0$. In a manner roughly analogous to the
zero-point energy of a quantum harmonic oscillator, all of the modes
for all of the free fields produce a zero-point energy that
contributes to the energy density of the vacuum. In the standard 
treatment of this effect (e.g., \cite{carrollpress,padmanabhan}), 
the size of each component has the from 
\be
\big|\langle T_{\mu\nu} \rangle\big| \propto \big| \rho_V \big| 
\sim \int_0^{M_{\rm max}} {4\pi k^2 dk \over (2\pi)^3} {1\over2}
\sqrt{k^2+m^2} \propto M_{\rm max}^4 \,, 
\label{zeropoint} 
\ee
where $M_{\rm max}$ is the maximum energy scale or cutoff scale. 
The mass $m$ of the field is assumed to be small such that 
$m\ll M_{\rm max}$. For completeness, we note that the arguments 
leading to equation (\ref{zeropoint}) do not obey Lorentz invariance
and do not produce the well-known equation of state $\langle{p}\rangle$
= $-\langle\rho\rangle$ for the vacuum. As a result, a more detailed
treatment is necessary. For example, an argument using dimensional
regularization \cite{martin} gives qualitatively similar results
(specifically, a quartic dependence on a large mass scale, but with a
logarithmic correction).

For quantum fluctuations in gravity, one expects $M_{\rm max}\sim\mplanck$
so that $|\langle T_{\mu\nu}\rangle|\sim\mplanck^4$. Even if one makes the
argument that our current understanding of quantum gravity is too
primitive to accept this result, the known particles of the Standard
Model are expected to make vacuum contributions. In this case, the
cutoff energies are comparable to scales appearing in the Higgs
potential, $M_{\rm max}\sim100$ GeV, so that the vacuum energy density
is expected to be $\rho_V \sim 10^8$ GeV$^4$, which is larger than the
observed value $\rhov$ by a factor of $\sim10^{54}$.  Many
calculations of this type assume that supersymmetry holds so that the
cutoff scale is the supersymmetric scale $M_{\rm ss}$ (which must be
larger than $\sim1$ TeV). For example, one class of potentials in
M-theory implies that the vacuum energy density has the form
$\rhov\sim\mplanck^2 M_{\rm ss}^2$ \cite{acharya}.  In any case,
quantum field theory predicts many contributions to the vacuum energy
density, where the known/expected contributions are larger than the
observed vacuum energy density by at least 54 orders of magnitude, and
perhaps much more.

For completeness, we note that vacuum energy density can arise on both
sides of the field equation (\ref{grfield}): If the original parameter
from General Relativity $\Lambda\ne0$, then the contribution
originates on the left side of the equation and is usually called the
cosmological constant. On the other hand, if the energy-momentum
tensor has a contribution from quantum fluctuations (as in equation
[\ref{qftrhov}]), then the term arises on the right side of equation
(\ref{grfield}) and is usually called a vacuum energy density. Since
the universe is (essentially) homogeneous and isotropic, the field
equation (\ref{grfield}) reduces to a single equation of motion for
the scale factor (the Friedmann equation [\ref{adotovera}]), and both
contributions provide a single term to the cosmic energy inventory.
Current experiments indicate that $\rhov\ne0$, but make no distinction 
between contributions from `pure $\Lambda$' and $\langle T_{\mu\nu}\rangle$. 

Although many ideas have been put forth, the cosmological constant
problem does not have a definitive solution at the present time
\cite{weinberg89}. Possible explanations include dynamical approaches 
using evolving scalar fields \cite{picon,sahni}, extra space-time
dimensions \cite{rubakov}, vacuum structure produced by parallel branes 
\cite{tyewassarman}, and many others \cite{sola}. Before experiments 
discovered that our universe is accelerating \cite{riess}, which 
indicates a nonzero vacuum energy density, one hope was that some
physical mechanism or symmetry principle would show that the
cosmological constant must be exactly zero. Given the currently
measured $\rhov\ne0$, however, the cosmological constant problem now
has two components \cite{garriga2001}: We need an explanation for why
its value so much smaller than the Planck scale {\it and} why it has
the particular small value $\rhov\sim(0.003{\rm eV})^4$.

Given the large number of possible mechanisms by which the universe
could determine the magnitude of its vacuum energy density, it is
likely that other regions of space-time (other universes in the
multiverse) would have different values of the cosmological constant
\cite{weinberg87}. The value of $\rhov$ (equivalently $\Lambda$ or 
$\lambda$) is thus expected to vary from universe to universe. In
addition, unless the size of the cosmological constant is determined
by as-yet-unknown physics, we expect many other universes to have much
larger values of $\rhov$ than that observed locally. As a result, the
most common universes, with large $\rhov$, will not be able to produce
structure (see the following section) and will quickly evolve to
become almost empty de Sitter space \cite{carrollbook}. In this
context, our universe is unusual in its profligate complexity.

\subsection{Bounds on the Vacuum Energy Density from Structure Formation} 
\label{sec:qvbounds} 

Although the energy density $\rhov$ of the vacuum is not predicted by
any known theory, its value in our universe is nonetheless constrained
by cosmological considerations. A relatively weak upper bound follows
from the requirement that the universe must have a matter dominated
era (see equation [\ref{needmatter}]). A stronger bound results from
the requirement that galaxy formation occurs before the the expansion
of the universe starts to accelerate due to the vacuum energy. This
section reviews bounds of this type.

The original constraints of this type
\cite{banks,daviesunwin,sakharov,weinberg87} require that
quasars form by redshift $z=4.5$ and can be written in the form  
\be
\rhov < {500 \over 729} \rhoeq Q^3 \,,
\label{weinbound} 
\ee
where $\rhoeq$ is the density at the epoch of matter/radiation
equality and where the numerical coefficient is derived using a simple
spherical infall model for galaxy formation \cite{peebles67}.  For the
the amplitude of density fluctuations observed in our universe,
$Q=10^{-5}$, one obtains the upper limit $\rhov\simless200\rhov(obs)$.

This bound on $\rhov$ was subsequently generalized to estimate the
likely values of the cosmological constant \cite{martel}. This updated 
treatment finds the probability of an observer measuring a given value
of $\rhov$, conditioned by the requirement that a given universe can
produce observers.  In this context, the formation of collapsed
cosmological structures --- essentially dark matter halos --- serves
as a proxy for the production of observers. Because values of $\rhov$
much larger than the observed value (in our universe) tend to shut
down structure formation, observers in any universe are likely to
measure values of $\rhov$ relatively close to the observed value. 
This conclusion continues to hold with alternate assumptions, 
for example using the amount of material in stars as a proxy for 
observers and using more detailed models of structure formation 
\cite{peacock,sudoh}. 

The conclusion that typical observers will measure a value of $\rhov$
roughly comparable to our own follows from assuming that the other
cosmological parameters are held constant. Bounds of this type become
substantially weaker if the amplitude $Q$ of primordial density
fluctuations is allowed to vary \cite{aguirretegmark,liviorees,mersini}.
Specifically, the bound of equation (\ref{weinbound}) scales as
$Q^3$. For each factor of ten increase in $Q$, the limit becomes
weaker by 3 orders of magnitude. Since $Q\sim10^{-5}$ in our universe,
larger values are not only possible, but likely \cite{garriga2006}
(see also Section \ref{sec:fluctuate}).  Constraints on the maximum
allowed values of the amplitude $Q$ follow from requiring structure
formation produce galaxies with favorable properties (see
Section \ref{sec:galaxies}).  Universes with larger $Q$ tend to form
denser galaxies, which can disrupt habitability by scattering
planets \cite{tegmarkrees,tegmark} and by producing strong background
radiation fields \cite{coppess}. Nonetheless, the amplitude $Q$ can
become as large as $Q\sim10^{-2}$ without compromising habitability
altogether. This increase in $Q$ allows the bound on $\rhov$ to become
a billion times less constraining. One thus obtains a limit of the
form $\rhov\simless10^{11}\rhov(obs)$.

\begin{figure}[tbp]
\centering 
\includegraphics[width=.95\textwidth,trim=0 150 0 150,clip]{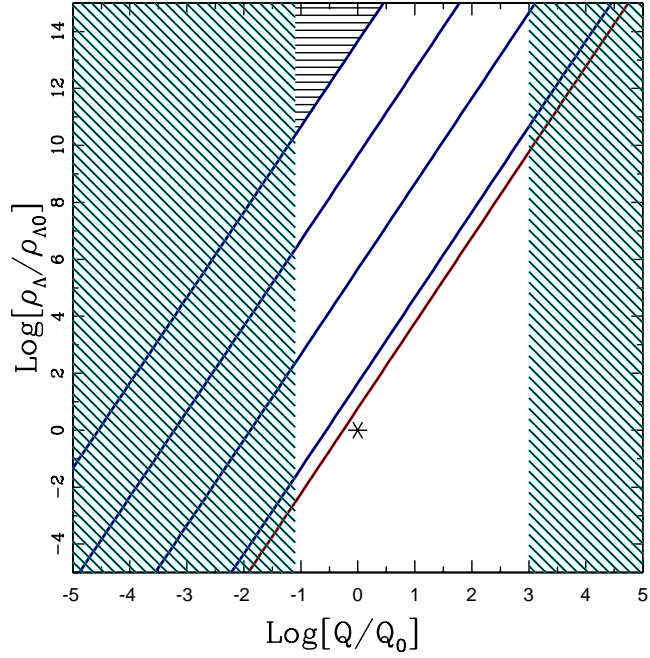}
\caption{Constraints on the allowed value of the vacuum energy density 
$\rhov$, also known as dark energy, as a function of amplitude $Q$ of
the primordial fluctuations.  The allowed region lies between the
shaded regions and below the lines (shown here for varying values of
$\eta$).  The red curve depicts the constraint for universes with the
same baryon to photon ratio as our universe, $\eta=6\times10^{-10}$.
The blue lines show weaker constraints corresponding to larger values
$\eta$ = $10^{-9}$, $10^{-8}$, $10^{-7}$, and $10^{-6}$, from bottom
to top. The parameters $Q$ and $\rhov$ are scaled to their observed
values, so that our universe lies at the origin as marked by the star
symbol.  Large values of $Q$ are ruled out because structure formation
would overproduce black holes \cite{bhrees}, whereas small values of
$Q$ are ruled out because gas within galaxies would be too rarefied to
cool and condense into stars \cite{reessix,tegmarkrees,tegmark}. Large
values of $\rhov$ are ruled out because structure formation is
compromised by the accelerating expansion
\cite{adamsrhovac,liviorees,martel,mersini,weinberg87}. } 
\label{fig:qvlambda} 
\end{figure}  

The discussion thus far assumes that the transition between the
radiation dominated and matter dominated eras occurs at the same epoch
as in our universe. In general, however, the baryon to photon ratio
$\eta$ can vary from universe to universe, although its value is
constrained by the requirement that the universe emerge from the epoch
of Big Bang Nucleosynthesis with an acceptable chemical composition
(Section \ref{sec:bbn}).  Larger values of $\eta$ lead to earlier
matter domination, and thereby give density fluctuations a longer time
to grow. This generalization thus allows a universe to have even
larger vacuum energy densities without compromising structure
formation. Including this effect, the bound on $\rhov$ takes the form
\be
\rhov < (2 A^3 a_R) Q^3 \eta^4 \mpro^4 
\left({\Omega_M \over \Omega_b}\right)^4 \,,
\label{genweinbound} 
\ee
where $A$ is a dimensionless constant of order unity and $a_R$ is 
the radiation constant. Since the baryon to photon ratio $\eta$ can
increase by (at least) three orders of magnitude without rendering 
the universe inhospitable (see Section \ref{sec:bbn} and Figure 
\ref{fig:bbnplane}), the bound on $\rhov$ becomes weaker by an
additional factor of one trillion \cite{adamsrhovac}.

The bounds discussed in this section are summarized in Figure 
\ref{fig:qvlambda}, which plots the maximum allowed energy density 
$\rhov$ of the vacuum as a function of the amplitude $Q$ of the
density fluctuations. The allowed values of $Q$ (see Section 
\ref{sec:galaxies}) correspond to the area between the shaded
regions on either side of the diagram and below the curves. 
Larger values of the fluctuation amplitude (right side) allow the
universe to produce denser galaxies \cite{peebles67}, which lead to
solar system disruption \cite{tegmarkrees,tegmark,coppess} and even
catastrophic black hole formation \cite{bhrees}. Smaller values of the
fluctuation amplitude (left side) lead to rarefied galaxies, with gas
that is too diffuse to cool, condense, and make structures
\cite{reesost,tegmarkrees}. The red curve shows the bound
corresponding to universes with the same matter content
$(\eta,\omegad/\omegab)$ as our universe. The star symbol shows the
location of our universe in the diagram. With the measured value of
$Q$, $\eta$, and other properties of our universe, the vacuum energy
density cannot be much larger than observed and still allow for
structure formation to occur. Much larger values are possible for
larger fluctuation amplitudes $Q$, where $\rhov$ can be larger than
the observed value by a factor of $\sim10^{10}$ and still allow for
structure formation.

If the baryon to photon ratio $\eta$ is increased, the bound on
$\rhov$ becomes substantially weaker, as shown by the blue curves 
in Figure \ref{fig:qvlambda}. Considerations of BBN (Section  
\ref{sec:bbn}) show that $\eta$ can become as large as 
$\eta=10^{-6}$ without rendering the universe uninhabitable 
\cite{adamsrhovac}. In the extreme case with the largest allowed 
values of $(Q,\eta)$, the energy density of the vacuum would be larger
than the value in our universe by a factor of $\sim4\times10^{22}$
(which lies off the top of the figure). Although such universes might
avoid catastrophic black hole formation (with $Q<0.01$), the resulting
galaxies would be much denser than those in our universe, and only a
fraction of the galactic environment would allow for habitable planets
(see Section \ref{sec:galaxies}).  The upper bound on $\rhov$ thus
depends on the minimum number of habitable planets that are necessary
for a habitable universe. As a result, a conservative (optimistic)
approach allows $\rhov$ to be larger than its observed value by a
factor of $10^{10}$ ($10^{22}$). The corresponding allowed increase 
in the energy scale $\lambda$ (defined in equation [\ref{rhovenergy}])
is approximately a factor of 300 ($3\times10^5$). 

Notice also that as $\rhov$ (equivalently $\lambda$) increases, the
total mass contained within the late-time cosmological horizon
decreases (see \ref{sec:massscales}). The number of stellar mass units
inside this horizon is given by equation (\ref{massuniverse}) and
approaches unity when $\lambda\to\mpro$, i.e., when the energy scale
of the vacuum is larger than the value in our universe by a factor of
$\sim4\times10^{11}$. For the approximate limits on $\lambda$ given
above, the number of stellar mass units within the horizon decreases
from $N_\star\sim10^{23}$ in our universe down to only about
$N_\star\sim10^{18}$ ($10^{12}$).  As a result, for the largest
possible contribution of dark energy consistent with structure
formation, the horizon mass decreases to that of a large galaxy.

The hierarchical fine-tuning of the cosmological constant arises
because the observed energy scale of the vacuum is small compared to
the Planck scale. One way to make this problem less severe is for the
universe to have a smaller Planck mass \cite{adamsrhovac},
corresponding to stronger gravity. Considerations of BBN (Section 
\ref{sec:bbn}) and stellar structure (Section \ref{sec:stars}) allow 
for the gravitational constant to be larger by a factor of $\sim10^6$,
so that the Planck mass could be one thousand times smaller. Another 
resolution of the hierarchical fine-tuning issue would result if the 
possible energy scales (or energy densities) of the vacuum are 
distributed in a log-uniform manner (see \ref{sec:probability}). 

The discussion thus far considers only positive values $\rhov>0$.
As discussed above, however, contributions to this term can come from
both geometric and matter sources. As a result, a negative
contribution from the geometric side could be large enough to overcome
the matter contribution, and thereby produce a cosmological constant
term such that $\rhov<0$. In this scenario, the vacuum energy acts
to slow down the cosmic expansion and does not impede the formation of
structure. On the other hand, the entire universe could collapse to a
singularity in a finite time. For example, the total lifetime 
$t_{\rm s}$ for a flat ($k=0$) universe containing matter and a
negative cosmological constant is given by
\be
t_{\rm s} = {2\pi \sqrt{3}\over3} 
\left( 8\pi G |\rhov|\right)^{-1/2} 
\sim \left( G |\rhov| \right)^{-1/2} \, .
\ee
In order for the universe to be habitable, it must live for a
sufficiently long time \cite{bartip,weinberg87}. This lower limit on
the cosmic lifetime $t_{\rm s}$ places an upper limit on the magnitude
of the vacuum energy density. Although the time required habitability
is not known, a minimum of $\sim1$ Gyr is often invoked
\cite{chyba,mckay,orgel} (see also the discussion in Section
\ref{sec:stars}). This choice leads to a bound on negative vacuum 
energy of the form  
\be
|\rhov| \simless 100\,\rho_{\rm crit} \approx 100\,
\rho_{\scriptscriptstyle{\Lambda}0} \approx (0.01\,\,{\rm eV})^4\,, 
\ee
where $\rho_{\rm crit}$ is the critical density in our universe today. 
Negative values of the cosmological constant are thus more constrained 
than positive values (compare with Figure \ref{fig:qvlambda}). 

\bigskip 
\section{Big Bang Nucleosynthesis} 
\label{sec:bbn} 

In order for a universe to become habitable, it must successfully
produce complex nuclei. In our universe, nucleosynthesis occurs in
multiple settings, including the early universe, stellar cores,
supernova explosions, and spallation in the interstellar medium. In
the first few minutes of its history, our universe processes about one
fourth of its baryonic material into helium-4 during an epoch known as
Big Bang Nucleosynthesis (BBN).  Although stars also make helium, this
early processing period is significant because BBN produces an order
of magnitude more helium than the stars --- and such a large helium-4
abundance is observed.  In addition, BBN produces trace amounts of
other light nuclei, especially deuterium, helium-3, and lithium-7.
These light elements are generally not synthesized in stars, but
rather are destroyed in stellar cores. By explaining the observed
abundances of these isotopes, BBN thus provides an important
confirmation of Big Bang theory \cite{kawanocode,wagoner,walker}
(see also \cite{coctwo} for a recent assessment). 

On the other hand, helium-4 is readily produced in stars, and none of
other nuclear species produced during BBN are known to be crucial for
the eventual development of life. As a result, BBN does not represent
a necessary constructive ingredient for a successful universe.
Instead, BBN provides a constraint: It is important that the early
universe does not emerge from its BBN epoch with an unacceptable
chemical composition. The universe could end up sterile if it
processes all of its protons and neutrons into heavier nuclei, thereby
leaving no hydrogen behind to make water. Unfortunately, the minimum
mass fraction of hydrogen required for habitability is not known.  
In this section, we review how different realizations of the basic
parameters lead to different nuclear compositions following the BBN
era.

The preceding discussion implicitly assumes that water is a necessary
ingredient for the development of life. For completeness, one should
keep in mind that that alternate solvents for biological operations
are possible \cite{lunine,scharf}, where ammonia (NH$_3$) is a leading
candidate \cite{haldane}. Ammonia is a polar molecule, is made of
common elements, supports numerous chemical reactions, and can
dissolve most organic molecules. Methane and other hydrocarbons are
also sometimes considered in this role \cite{mckaysmith}. However,
ammonia, methane, and other hydrocarbons all require hydrogen
(protons) as basic building blocks, so that the constraints on BBN
remain, namely, to leave behind an ample supply of free protons.

\subsection{BBN Parameters and Processes} 
\label{sec:bbnbasics} 

We begin with a brief assessment of the parameters that play a role in
BBN. The universe at this early epoch is radiation dominated, with the
expansion governed by equation (\ref{adotovera}) with $\rho=\rhor$ and
$k=0$. The gravitational constant $G$ (equivalently $\alpha_G$)
determines the expansion rate, where $H={\dot a}/a$ $\propto G^{1/2}$. 
Baryons are a trace constituent of the universe at this time and are
specified by the baryon to photon ratio $\eta\ll1$ (where
$\eta\approx6\times10^{-10}$ in our universe). The remaining
parameters determine the nuclear reaction rates. The weak interactions
play an important role by maintaining equilibrium early in the process
($T\simgreat1$ MeV, $t\simless1$ sec), and by allowing free neutrons
to decay ($t\sim1000$ sec). At the simplest level of description, the
weak rates are set by the Fermi constant $G_F$. A full specification
of BBN parameters requires the cross sections $\sigma_{jk}$ for all
of the relevant nuclear reactions and the binding energies $B_k$ for
all of the relevant nuclear species. The interaction rates ultimately
depend on the strengths of the strong, weak, and electromagnetic
forces.  Unfortunately, however, no simple transformation exists
between the fundamental parameters of the Standard Model (see
Section \ref{sec:particlephys}) and the quantities that appear in 
numerical treatments of BBN \cite{burst,kawanocode,wagoner,walker}. 

Although a detailed treatment of the BBN epoch requires the inclusion
of dozens of nuclear reactions, and hence a full numerical
treatment \cite{kawanocode,wagoner}, it is instructive to consider a
simplified description \cite{kolbturner}. At early times, the
background temperature of the universe is above that of nuclear
binding energies, and the weak interaction rate is fast enough to
maintain equilibrium. As result, the nuclear species are in Nuclear
Statistical Equilibrium (NSE), which determines their abundances as a
function of temperature. The ratio of neutron density to proton density
is of particular importance, and can be written in the form
\be
{n \over p} \equiv {n_n \over n_p} =  
\exp\left[-{\Delta{m}\over T} \right]\,,
\label{npnse} 
\ee
where the mass difference $\Delta{m}=m_{\rm n}-\mpro\approx1.29$ MeV
is a key parameter (see also Section \ref{sec:dneutpro}).  Note that
the chemical potentials of the electron and neutrino are assumed to be
small so that $(\mu_e-\mu_\nu)/T\ll1$.  

Once the cosmic age reaches $t\sim1$ sec and the temperature falls
below $T\sim1$ MeV (in our universe), the weak interaction rate
becomes slower than the expansion rate, and the nuclear species fall
out of equilibrium. In general this transition occurs when the weak
interaction rate $\Gamma_{\rm w}$ becomes smaller than the expansion
rate $H$. In the regime $T\gg\Delta{m}\sim\emass$, the rates become
equal \cite{kolbturner} when $\Gamma_{\rm w}=H$, i.e.,  
\be
{7\pi\over60}(1+3g_A^2)\,G_F^2T^5 = 1.66 g_\ast^{1/2}T^2/\mplanck 
\qquad \Rightarrow \qquad 
T_{\rm wf} \approx 1.3 \,G_F^{-2/3} \mplanck^{-1/3}\,. 
\label{weakfreeze} 
\ee
In our universe, the temperature scale $T_{\rm wf}\sim1$ MeV. 
After the temperature falls below this threshold, the neutron to
proton ratio stops tracking its NSE value. At this juncture, the
abundance ratio $n/p\sim1/6$, but it continues to decrease slowly due
to continued weak interactions and neutron decay.

With its large binding energy and low atomic number, helium-4 is the
most abundant nuclear species produced during BBN. As a result, the
most likely scenario where BBN leaves behind a sterile universe is for
essentially all of the baryons to be processed into helium-4.  Because
the neutrons are outnumbered by the protons (in our universe) and
interact readily due to their lack of charge, almost all of the free
neutrons become locked up in helium-4 during the BBN epoch.  A good
estimate for the corresponding helium mass fraction $Y_4$ is thus
given by 
\be
Y_4 \approx {4n_4 \over n_1} \approx 
{2 (n/p)_{\rm bbn} \over 1 + (n/p)_{\rm bbn}} \,\,\to\,\, 
{2 \over 1 + (n/p)_{\rm bbn}} \, {\rm min} \, 
\bigl\{ (n/p)_{\rm bbn}, 1 \bigr\} \,,  
\label{xhelium} 
\ee
where the subscripts refer to atomic mass numbers. In this expression,
the neutron to proton ratio is evaluated at the time/temperature when
most of the helium is produced. In our universe, for example,
$T\sim0.1$ MeV at this epoch, and the ratio $(n/p)_{\rm bbn}\sim1/7$ 
is slightly smaller than its earlier value $(n/p\sim1/6)$ when NSE was
compromised. The corresponding helium mass fraction $Y_4\approx0.25$. 
Note that the final version of this expression (\ref{xhelium})
generalizes the more familiar form \cite{kolbturner} to include
scenarios where the neutron abundance is greater than the proton
abundance \cite{grohsweakless}.

The simplest recipe for making a sterile universe thus requires
$Y_4\to1$, which in turn requires $(n/p)_{\rm bbn}\to1$. Since the
densities of neutrons and protons follow their NSE values until weak
interactions become slower than the cosmic expansion rate, and since
$n\approx p$ at high temperatures, a universe will be left with more
neutrons --- and greater helium abundance --- as the weak interaction
is made weaker \cite{weakless}. Notice also that smaller values of the
mass difference $\Delta{m}$ will also lead to more nearly equal
populations of protons and neutrons (equation [\ref{npnse}]). However,
the condition $(n/p)_{\rm bbn}\to1$ is necessary but not sufficient
for a universe to process all of its baryons into helium and heavier
elements. The interaction rates must be high enough for helium
production to proceed to completion before the universe becomes too
cool and diffuse. As a result, a viable regime of parameter space
always exists for sufficiently small values of the baryon to photon
ratio $\eta$ (see equation [\ref{etanuke}] below).

It is useful to identify the approximate conditions required for BBN
to operate in a regime similar to that of our universe, where some
light nuclei are produced, but not all of the protons are transformed
into helium and heavier elements \cite{carr}. Given the equilibrium
$n/p$ ratio from equation (\ref{npnse}) and the approximate helium
yield of equation (\ref{xhelium}), the interesting regime of BBN
requires that the weak interaction rate competes with the cosmic
expansion rate (see equation [\ref{weakfreeze}]) at a temperature
roughly comparable to the mass differences between the light
nuclei. The mass difference $\Delta{m}$ between the proton and
neutron, as well as the binding energies of deuterium and helium-3,
are of order 1 MeV, so that $T \sim 1$ MeV $\sim \emass$. The
requirement for interesting BBN thus becomes 
\be
G_F^2 \mplanck \emass^3 \sim 1 \,. 
\label{weakbalance} 
\ee
In our universe, the left hand side of the expression has a value
$\sim0.22$, relatively close to unity.  For stronger versions of the
weak interaction (larger $G_F$), decoupling of weak interactions
occurs later (at lower temperature).  The neutron to proton ratio
follows NSE for a longer time and the fraction of neutrons becomes
lower. With a diminished supply of neutrons, the universe supports
fewer nuclear reactions, and BBN yields decrease in general 
\cite{howeweakful}. In this limit, the universe could emerge from 
BBN with little nuclear processing, but this condition does not
preclude stellar nucleosynthesis at later epochs, and the universe 
can remain viable. 

In the opposite limit, where the weak force is less effective, the 
$n/p$ ratio is frozen out early and approaches unity. The neutron 
lifetime is long, so that neutron decay does not occur until well 
after the BBN epoch. In this regime, the helium fraction (from 
equation [\ref{xhelium}]) will approach unity if the nuclear reaction 
rate is fast enough. To prevent overproduction of helium, a universe
needs sufficiently low density given by the constraint 
\be
\eta n_\gamma \langle\sigma v\rangle \simless H \,,
\label{etanuke} 
\ee
where all quantities are evaluated at the temperature (time) when the
light elements are made. As a result, universes can avoid processing
all of their neutrons into helium if the baryon to photon ratio is
small enough. Even a universe with no weak interactions can remain 
viable \cite{grohsweakless,weakless,howeweakful}, with 10\% of the 
nucleons left over, provided that $\eta\simless10^{-10}$ (see Section
\ref{sec:bbnweakless}). In this weakless limit, one needs
$\eta\sim4\times10^{-12}$ to maintain the same hydrogen content as our
universe (where $\eta\approx6\times10^{-10}$).

For completeness, note that most treatments of BBN consider the
universe to be completely homogeneous during the relevant span of
time. This assumption is reasonable, given that our universe must
evolve into its present state at a much later epoch. In other
universes, however, inhomogeneities could be present during BBN
\cite{witten} and would change the abundance patterns
\cite{applegate,jedamzik}.  Although this present discussion focuses
on results for homogeneous BBN, the possibility of density variations
should be kept in mind.


\subsection{BBN Abundances with Parameter Variations} 
\label{sec:onevariable} 

This section considers the abundances of the light elements produced
during Big Bang Nucleosynthesis as the input parameters are varied.
This ensemble of results illustrates the difficulty in rendering the
universe sterile by using up all of the available protons (leaving
none behind for water production). Here we present results calculated
using the Kawano code \cite{kawanocode} as well as the more recent
{\sl BURST} code \cite{burst,burst2}, both of which are descendants of
the pioneering work of Wagoner \cite{wagoner}.

The numerical codes are set up so that variations can be made for the
baryon to photon ratio $\eta$, the gravitational constant $G$, and the
neutron lifetime $\tau_n$. One of the key results from BBN studies is
the determination of the baryon to photon ratio, which has a value of
$\eta\approx6\times10^{-10}$ for our universe. In addition, the
expansion rate can be varied and is equivalent to changes in the value
of the gravitational constant $G$ and/or the number of light neutrino
species. These codes also allow for variations in the neutron
lifetime $\tau_n$ because its value (in our universe) was not well
measured until recently. Finally, we also consider variations in the
fine structure constant $\alpha$. As outlined below, this latter
result is more approximate.

\subsubsection{Variations in the Baryon to Photon Ratio} 
\label{sec:bbneta} 

\begin{figure}
\centering 
\includegraphics[width=.95\textwidth,trim=0 150 0 150,clip]{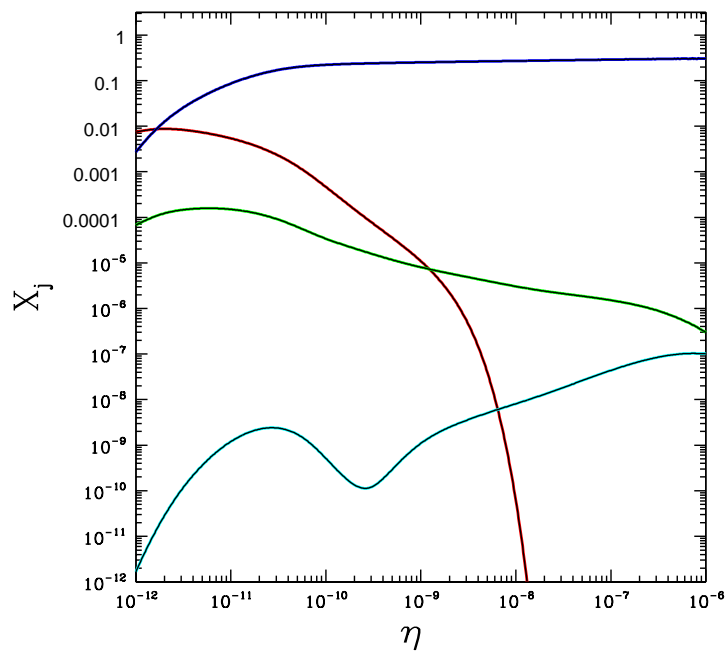}
\caption{BBN yields as a function of the baryon to photon ratio $\eta$.
The curves show the resulting mass fraction at the end of the BBN epoch 
for helium-4 (blue), as well as the corresponding abundances $X_j$ for 
deuterium (red), helium-3 (green), and lithium-7 (cyan). }
\label{fig:bbneta} 
\end{figure}

First we consider variations in the baryon-to-photon ratio $\eta$,
where the results are shown in Figure \ref{fig:bbneta}.
Historically, plots of this type have been made with the goal of
determining the value of $\eta$ that is consistent with the primordial
abundances of all of the light elements shown. The current estimated
value is $\eta\approx6\times10^{-10}$, although some uncertainty
arises due to the abundance of lithium.  In this context, 
we are interested in what input parameters could potentially disrupt 
the future operations of the universe. As illustrated in Figure 
\ref{fig:bbneta}, however, the baryon to photon ratio can vary over 
many orders of magnitude and still allow the universe to have an
acceptable chemical composition. 

Figure \ref{fig:bbneta} shows that for sufficiently small values of
$\eta\sim10^{-12}$, the mass fraction of helium-4 becomes smaller than
that of deuterium. As $\eta$ increases, the nuclear reaction rates
increase, and the mass fraction of helium-4 increases accordingly.
The abundances of deuterium and helium-3 decrease with increasing
$\eta$, as they are burned into helium-4.  Note that in the limit of
large $\eta$, the mass fraction of helium-4 reaches a limiting value
of $Y_4\sim0.3$. This limit corresponds to the regime where
essentially all of the available neutrons are burned into helium-4.
Notice also that the abundances of deuterium and helium-3 decrease
with increasing $\eta$. With most of the neutrons incorporated into
helium-4, few are left for the remaining light elements. Increasing
values of $\eta$ also facilitate the burning of the lighter elements
into heavier ones, which in turn leads to a slow increase in the
abundance of lithium-7. Even with a thousand-fold increase in $\eta$,
however, lithium remains a trace element after the BBN epoch. 

\subsubsection{Variations in the Gravitational Constant} 
\label{sec:bbngrav} 

\begin{figure}
\centering
\includegraphics[width=.95\textwidth,trim=0 150 0 150,clip]{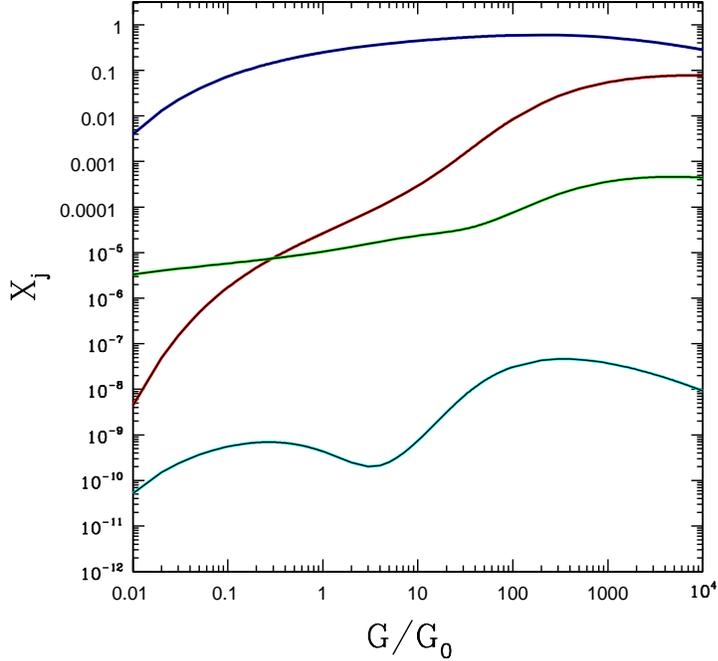}
\caption{BBN yields as a function of the gravitational constant,
$G/G_0$ (scaled to the value in our universe).  The curves show the  
resulting mass fraction at the end of the BBN epoch for helium-4
(blue), as well as the corresponding abundances $X_j$ for 
deuterium (red), helium-3 (green), and lithium-7 (cyan). }
\label{fig:bbngravity} 
\end{figure}

Variations in the gravitational constant result in corresponding
changes in the expansion rate of the universe, where $H \propto$
$G^{1/2}$. Figure \ref{fig:bbngravity} shows the results for varying the
gravitational constant over a range of six orders of magnitude, from
100 times smaller than in our universe to $10^4$ times larger. For
most of the range shown, the abundances of all of the light elements
increase with $G$.  For small values of $G$, with a slow expansion
rate, the freezing out of weak interactions occurs later in cosmic
history. As a result, protons and neutrons remain in NSE longer and
the $n/p$ ratio is smaller. As $G$ increases, the expansion rate
increases, freeze-out occurs earlier, and the $n/p$ ratio is larger.
Since most of the neutrons are processed into helium-4, its abundance
grows with increasing $G$. The abundances of deuterium and helium-3
also increase. For sufficiently large values of $G$, however, the
expansion rate is so fast that not all of the neutrons can be made 
into helium-4. As a result, the mass fraction of helium-4 has a 
maximum value of $Y_4\sim0.53$, which occurs at $G/G_0\sim100$. 

\subsubsection{Variations in the Neutron Lifetime} 
\label{sec:bbntau} 

Next we consider variations in the neutron lifetime $\tau_n$, 
which can be written in the form 
\be
\tau_n^{-1} = {G_F^2 \over 2\pi^3} \left( 1 + g_A^2 \right) 
m_e^5 \lambda_0\,,
\label{neutronlife} 
\ee 
where $G_F$ is the Fermi constant, $m_e$ is the electron mass, 
$g_A \approx 1.26$ is the axial-vector coupling for nucleons, 
and the dimensionless parameter $\lambda_0 \approx 1.636$ 
\cite{kolbturner}. Variations in the neutron lifetime are thus 
equivalent to variations in the Fermi constant, which has a 
value of $G_F \approx$ (293 GeV)$^{-2}$ in our universe. 

\begin{figure}
\centering
\includegraphics[width=.95\textwidth,trim=0 150 0 150,clip]{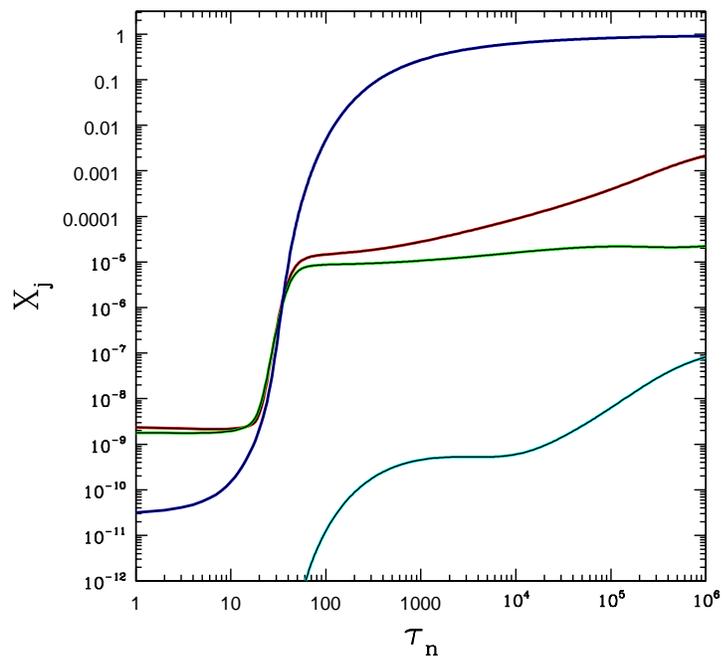}
\caption{BBN yields as a function of neutron lifetime $\tau_n$ (given 
by the horizontal axis in seconds).  The curves show the resulting
mass fraction at the end of the BBN epoch for helium-4 (blue),
as well as the corresponding abundances $X_j$ for deuterium (red), 
helium-3 (green), and lithium-7 (cyan). }
\label{fig:bbntau} 
\end{figure}

Figure \ref{fig:bbntau} shows the resulting abundances of the
light elements as a function of $\tau_n$, which varies from 1 to
$10^6$ sec (recall that $\tau_n \approx 885$ sec in our universe). As
a rule, the abundances of all of the relevant nuclear species increase
with increasing $\tau_n$, which corresponds to decreasing strength of
the weak force. In the limit of small $\tau_n$, which corresponds to
the weak force being stronger than in our universe, all of the
abundances are low. In this limit, the weak force competes with the
expansion of the universe down to low temperatures where the neutron
to proton ratio (at freeze-out) becomes extremely small. The universe
thus contains few neutrons due to both thermodynamic considerations
and due to their rapid decay. With low neutron abundance, the universe
can make few nuclei.

In the limit of long neutron lifetime $\tau_n$, the weak force is
unable to keep up with the expansion of the universe even at early
times and the freeze-out of weak interactions occurs at high
temperature $T\gg\Delta m\sim1$ MeV. In this limit, the freeze-out
value of the neutron to proton ratio approaches unity. Since the
neutron lifetime is long, nearly all of the neutrons can be
incorporated into helium-4 before they decay. Moreover, because
$p\approx{n}$ and $A=2Z$ for helium-4, nearly all of the protons are
used up as well. Keep in mind that the abundances shown in Figure 
\ref{fig:bbntau} are plotted on a logarithmic scale. Although 
$Y_4$ becomes large for the largest value $\tau=10^6$ sec considered
here, the mass fraction of helium-4 is only $Y_4 \approx 0.90$. Such a
universe still retains 10\% of its mass in protons, which are
available for making water and for undergoing nuclear fusion in stars
at later epochs (see also the discussion of Ref. \cite{hallpinner}).
This issue of overproduction helium-4 arises in the weakless
universe \cite{weakless}, where $G_F\to0$ and $\tau_n\to\infty$. As 
pointed out by the authors, universes with no weak interactions (or
extremely long $\tau_n$) can avoid fusing all of their nucleons into
helium-4 if they have a much smaller baryon to photon ratio $\eta$
(see Section \ref{sec:bbnweakless}). 

The abundance trends shown in Figure \ref{fig:bbntau} exhibit a
relatively sharp transition between universes with little nuclear
processing (left side of the figure) and those that emerge from the
BBN epoch with a primarily helium composition (right side). This
transition occurs near $\tau_n\sim100$ sec, which is close to the
observed value in our universe ($\tau_n=885$ sec). This proximity has
been found previously \cite{hallpinner}, which also notes that the
boundary can be expressed in terms of the value of $G_F$ or the mass
of the $W$ boson. Notice also that the value of the Fermi constant
$G_F=1/(\sqrt{2}{\cal V}^2)$, where ${\cal V}$ is the vacuum
expectation value of the Higgs. Changes in ${\cal V}$ can lead to
changes in the masses of the fundamental particles, which in turn
alter the structure of nuclei. The results shown in Figure
\ref{fig:bbntau} correspond to changes in the neutron lifetime 
only, but many other scenarios are possible (e.g., with different 
values of the Yukawa couplings that determine the masses of quarks 
and leptons, especially $\umass$, $\dmass$, and $\emass$). 

\subsubsection{Variations in the Fine Structure Constant} 
\label{sec:bbnalpha} 

This section considers the effects of varying $\alpha$.  Note that
variations in the fine structure constant have two effects on
nucleosynthesis. First, the value of $\alpha$ determines the size of
the coulomb repulsion factors that appear in the nuclear reaction
rates. The second effect is that, at the fundamental level, the value
of $\alpha$ will determine, in part, the binding energy of the proton
and neutron, and hence the mass difference $\Delta m$. This mass
difference, in turn, affects the proton to neutron ratio during the
BBN epoch.  Unfortunately, it remains difficult to calculate this mass
difference from first principles at the present time. The value of
$\alpha$ also affects the binding energies of the composite nuclei
being produced (such as helium-4). In our universe, the strong force
has the dominant contribution to the binding energies. If we use the
semi-empirical mass formula as a guide, then $\alpha$ must be
increased (relative to the strong force) by an order of magnitude to
compromise nuclear structure.

Unfortunately, no comprehensive studies of BBN currently include all
of the effects outlined above. In the absence of such results, here we
illustrate part of the issue by including variations in $\alpha$ into
the coulomb repulsion factors in the nuclear reaction rates. In this
treatment, the mass difference $\Delta m$, the nuclear binding
energies, and the coefficients in the reaction cross sections all
remain constant.  

\begin{figure}
\centering
\includegraphics[width=.95\textwidth,trim=0 150 0 150,clip]{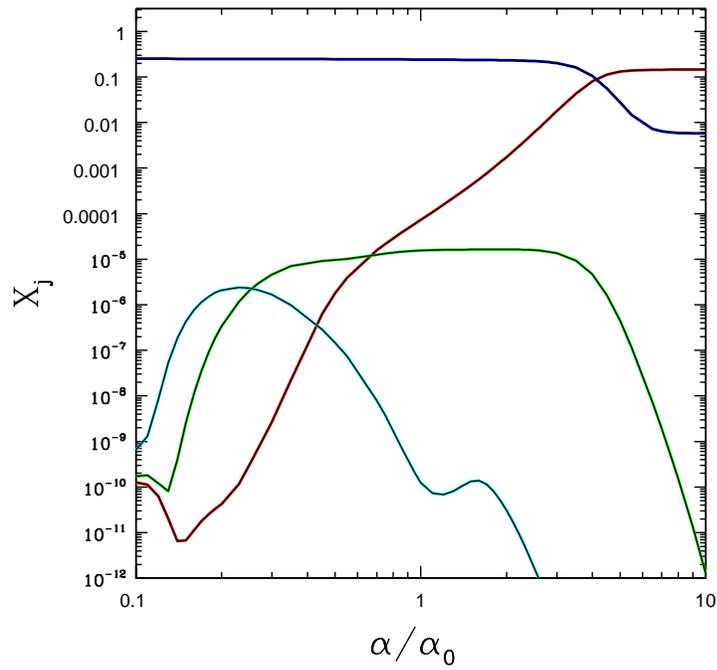}
\caption{BBN yields as a function of the fine structure constant
$\alpha/\alpha_0$ (scaled to the value in our universe). The curves 
show the resulting mass fraction at the end of the BBN epoch for
helium-4 (blue), as well as the corresponding abundances $X_j$ for 
deuterium (red), helium-3 (green), and lithium-7 (cyan). Note that 
the variations in $\alpha$ are included only for the coulomb 
repulsion factor, so that the binding energies and cross sections 
are held fixed (see text). }
\label{fig:bbnalpha} 
\end{figure}

The effects of variations in the fine structure constant are shown
in Figure \ref{fig:bbnalpha}, which plots the light element abundances
over a range in $\alpha$.  For large values of $\alpha$, the coulomb
repulsion effects are strong, so that the abundances of all of the
light elements except deuterium are suppressed. The deuterium
abundance grows with increasing $\alpha$ because its production
channel ($n+p\to d +\gamma$) is unchanged, whereas the reactions that
lead to its destruction are suppressed.  In fact, the deuterium
abundance becomes much larger than that of helium-4 for sufficiently
large values of the fine structure constant, namely
$\alpha/\alpha_0>7$.  As $\alpha$ is decreased, thereby removing the
coulomb barrier for nuclear reactions, the helium-4 abundance
increases. However, it reaches an asymptotic value $Y_4\sim1/3$, 
which corresponds to essentially all of the available neutrons being 
incorporated into helium-4. Significantly, over the range of values 
shown in Figure \ref{fig:bbnalpha}, all of the universes remain viable. 

\begin{figure}[tbp]
\centering 
\includegraphics[width=.99\textwidth,trim=0 150 0 150,clip]{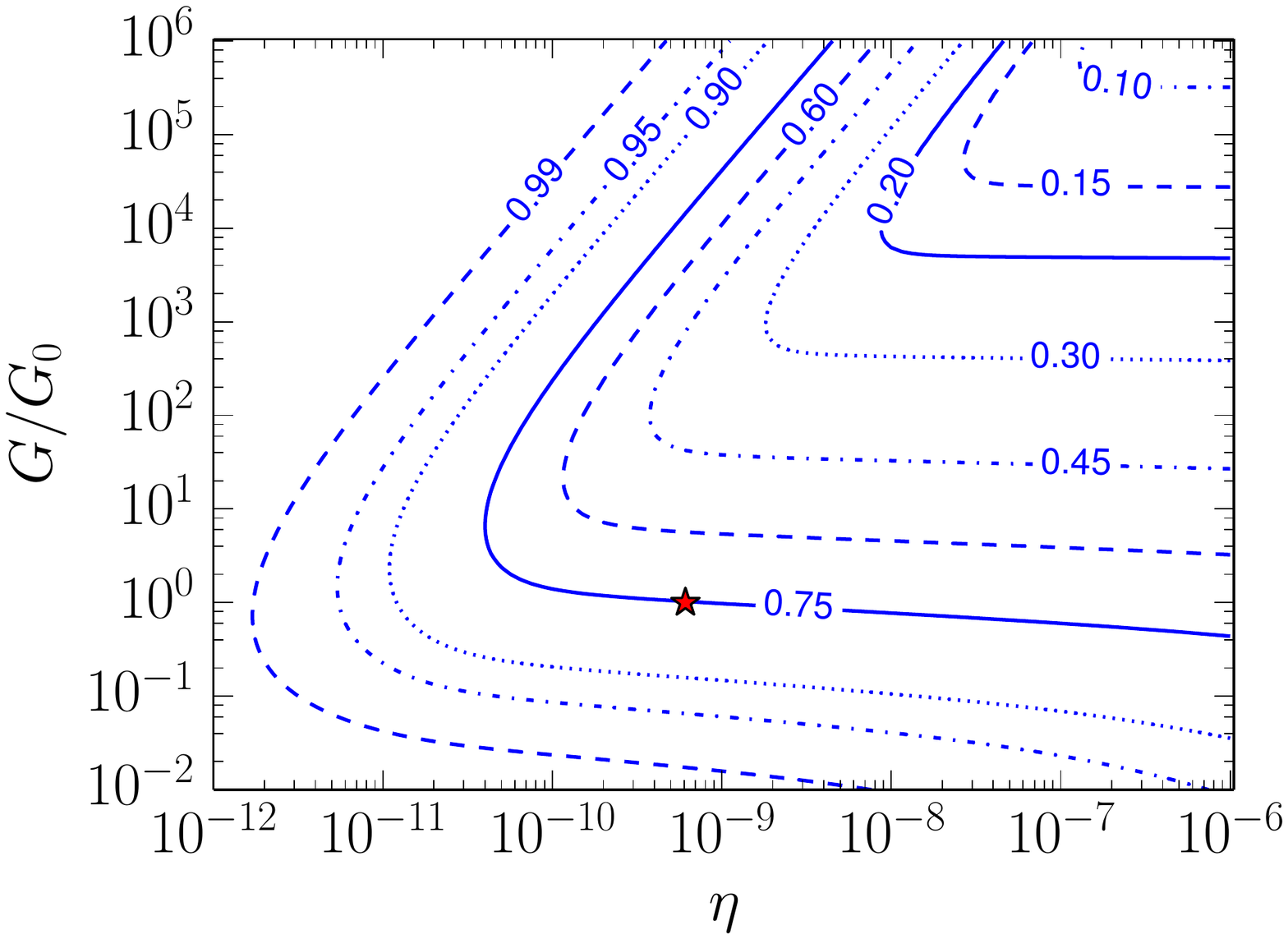}
\caption{Hydrogen mass fractions after the BBN epoch for varying values 
of the baryon to photon ratio $\eta$ and the gravitational constant $G$ 
(from \cite{adamsrhovac}).  Contours mark the abundance levels, where
the accounting includes all isotopes of hydrogen. The heavy contour
labeled with $X=0.75$ corresponds to hydrogen mass fractions comparable 
to that of our universe. The red star marks the location of our
universe in the diagram. The mass fraction of hydrogen falls below
10\% only in the upper right portion of the diagram, where
$G/G_0\simgreat10^6$ and $\eta\simgreat10^{-6}$. BBN thus leaves the 
universe with free protons over most of parameter space. }
\label{fig:bbnplane} 
\end{figure} 

\subsubsection{Variations in both $G$ and $\eta$} 
\label{sec:bbnetagrav} 

The discussion thus far has considered the variation of only one
parameter at a time. To illustrate the effects of multiple parameters,
consider the case where both the baryon to photon ratio $\eta$ and the
gravitational constant $G$ are allowed to vary. In order for the
universe to remain habitable, the abundance of protons cannot become
too small.  We thus consider contours of constant hydrogen (proton)
mass fractions in the $\eta$-$G$ plane. Figure \ref{fig:bbnplane}
shows the result (from \cite{adamsrhovac}) for an ensemble of BBN
simulations performed with the {\sl BURST} code \cite{burst,burst2}.  
The plane of parameters in the figure spans a factor of a million in
$\eta$ and a factor of 100 million in $G$. The proton mass fraction
remains above 10\% over almost the entire plane, with the exception of
the extreme upper right corner (where $G\simgreat3\times10^5\,G_0$ and 
$\eta\simgreat10^{-7}$). 
 
\subsection{BBN without the Weak Interaction} 
\label{sec:bbnweakless} 

As outlined above, universes can remain habitable in the face of
extreme variations in the baryon to photon ratio $\eta$
(Figure \ref{fig:bbneta}), the gravitational constant $G$
(Figure \ref{fig:bbngravity}), the neutron lifetime $\tau_n$
(Figure \ref{fig:bbntau}), and the fine structure constant $\alpha$
(Figure \ref{fig:bbnalpha}). The only parts of parameter space where
the proton abundance becomes dangerously low is for extremely large
values of $\eta$ and for long neutron lifetimes $\tau_n$. The latter
regime arises when the weak force becomes substantially weaker than
that of the Standard Model. Here we consider separately the case of
the ``weakless'' universe \cite{weakless}, where the weak interaction
is absent so that $\tau_n\to\infty$. By adjusting the parameters of
particle physics and cosmology, Ref. \cite{weakless} shows that one
can obtain universes with properties that are roughly similar to
our own (cf. \cite{clavelli}). 

\begin{figure}[tbp]
\centering 
\includegraphics[width=.95\textwidth,clip]{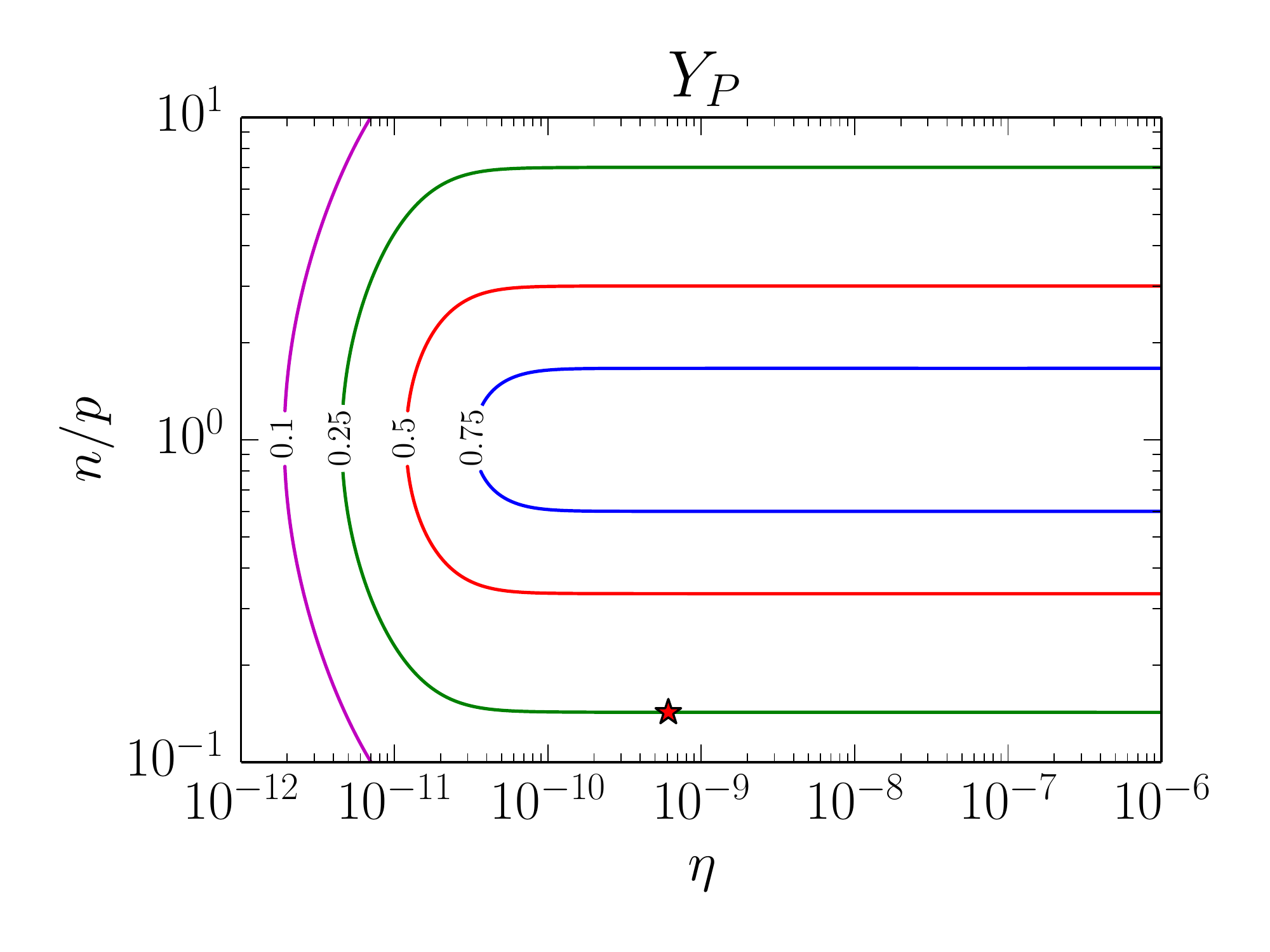}
\caption{Big Bang Nucleosynthesis in universes without the weak 
interaction (from \cite{grohsweakless}). The curves show the contours 
of constant helium mass fraction $Y_P$ in the $(\eta,n/p)$ plane, where
$\eta$ is the baryon to photon ratio and $n/p$ is the primordial 
neutron to proton ratio (in a weakless universe, this quantity is 
set well before the BBN epoch). The location of our universe is plotted 
as the red star symbol. For our universe, however, the ratio $n/p\sim1$ 
at early times, and approaches the value plotted at the start of the 
BBN epoch. The helium mass fraction $Y_P$ approaches unity only in 
the extreme center-right portion of the diagram, so that BBN leaves 
the universe with free protons over most of parameter space. }  
\label{fig:weaklessbbn} 
\end{figure} 

The BBN epoch plays out somewhat differently in a universe without
weak interactions. In standard cosmology, weak interactions occur
rapidly enough to maintain equilibrium at early times, and then later
become slow compared to the expansion rate. In a weakless universe, by
definition, weak interactions can never maintain equilibrium.  As a
result, the neutron to proton ratio is not determined by NSE (as in
equation [\ref{npnse}]) and is specified by other physics. If equal
numbers of up and down quarks are produced during the earlier epoch of
baryogenesis, then one expects $n/p\approx1$ (see \cite{weakless} for
further discussion). In this case, all of the nucleons can be readily
incorporated into helium-4, provided that the strong interactions are
sufficiently rapid. In order to avoid overproduction of helium-4, and
hence an accompanying hydrogen shortage, a smaller baryon abundance is
necessary.  For example, the value of $\eta\approx4\times10^{-12}$
produces a mass fraction of helium-4 that is comparable to that of our
universe, $Y_4\approx0.25$ \cite{grohsweakless,weakless}.

The possible ranges of the helium yields from Big Bang Nucleosynthesis
in the weakless universe are shown in Figure \ref{fig:weaklessbbn}
(from \cite{grohsweakless}) over a wide parameter space. The figure
shows contours of constant helium mass fraction $Y_P$ in the
$(\eta,n/p)$ plane, where $\eta$ is the baryon to photon ratio and
$n/p$ is the neutron to proton ratio.  Note that for the weakless
universe, the ratio $n/p$ is determined by processes in the early
universe, well before the start of the BBN epoch. One must thus
consider a range of values as shown. The helium abundance arising from
BBN has an acceptable level over most of the plane. For a universe
that is both weakless and symmetric ($n/p\sim1$), the preferred value
of the baryon to photon ratio $\eta$ is somewhat smaller than the
value in our universe. Although the baryon to photon ratio $\eta$ 
required to produce the helium mass fraction of our universe is
only $\eta\approx4\times10^{-12}$, any value of 
$\eta\simless10^{-10}$ will allow for $Y_P\simless0.90$ 
and hence a working universe. 

In the weakless universe, the mass fraction of helium-4 is an
increasing function of the baryon to photon ratio $\eta$.  More
specifically, the weakless universe emerges from its BBN epoch with
mass fraction $X_4\approx$ 0.45, 0.90, and 0.99 for $\eta$ =
$10^{-11}$, $10^{-10}$, and $10^{-9}$. The remaining baryonic mass is
(relatively) evenly distributed among free protons, free neutrons, and
deuterium \cite{grohsweakless}, with trace quantities of tritium and
helium-3. Weakless universes thus produce substantial amounts of
deuterium, with mass fractions of order $10^{-1}$ (compared to
$\sim10^{-5}$ for standard BBN). In addition, the free neutrons that
survive the BBN epoch can fuse with protons in dense regions of the
interstellar medium, thereby producing more deuterium.  At later
times, the resulting inventory of deuterium provides an important fuel
for stars: Without the weak force, free protons cannot fuse to make
deuterium (through $pp\to{d}e^+\nu_e$), so that the standard $p$-$p$
chain for stellar nucleosynthesis is unavailable. Instead, stars must
either burn free protons {\it and} any remaining free neutrons into
deuterium or fuse the deuterium into helium-4 (see 
\cite{grohsweakless} and Section \ref{sec:stars}).

\bigskip 
\section{Galaxy Formation and Large Scale Structure} 
\label{sec:galaxies} 

Galaxies are vital constituents of a habitable universe. They collect
and organize the interstellar gas that is required to form stars,
which provide the energy and heavy nuclei necessary for biological
development. Of equal importance, galaxies must be massive enough to
retain the heavy elements produced by stellar nucleosynthesis. In
order to fulfill these roles, galaxies must first be able to form. 

Galaxy formation involves a number of physical processes (see
\cite{benson2010} for a recent review and \cite{loebbook} for a 
general overview). In our universe, the mass budget is dominated by
weakly interacting dark matter. The first step is for the dark matter
to decouple from the expanding background universe and begin to
collapse into halos \cite{davis1985}. As outlined in Section 
\ref{sec:rhovac}, the collapse process must become sufficiently 
advanced before the epoch of vacuum domination, where this requirement
leads to an upper limit on $\rhov$ for a given matter inventory of the
universe (see equation [\ref{genweinbound}], Figure \ref{fig:qvlambda}, 
and Refs. \cite{adamsrhovac,efstathiou,martel,mersini,weinberg87,barneslss}).  
The next requirement is that the baryonic component of the universe
must be able to cool \cite{gunngott,reesost,whiterees}, so that the
gas condenses into structures analogous to galactic disks. These cool
and compact configurations are necessary for the subsequent formation
of stars.  Along with the initial collapse of dark matter into halos
and condensation of cooling gas, dynamical processes sculpt the
structure of the forming galaxies. Smaller halos merge together to
form larger ones, while individual galactic structures undergo
dynamical relaxation. The net result of these dynamical interactions
is to produce dark matter halos with a nearly universal form 
\cite{nfw}.  The baryonic component, consisting of both stars and
gas, forms spherical and disk-like structures \cite{binmer,bintrem}
that live at the halo centers.

In order for galaxy formation to be successful, the resulting
structures must meet a number of constraints. The density fluctuations
that produce the galaxies must become nonlinear, and detach from the
expanding background, before the universe becomes vacuum dominated.
The galaxies must be dense enough so that gas can cool, condense, and
eventually form stars, yet remain rarefied enough that solar systems
can survive. Finally, the galactic mass scales must be large enough to
form many stars and have deep enough gravitational potential wells to
retain the heavy elements produced by stellar nucleosynthesis. To 
assess the viability of galaxies in other universes, we thus need to
determine their masses, densities, and other properties as a function
of the cosmological parameters.

\subsection{Mass and Density Scales of Galaxy Formation} 
\label{sec:galaxymass} 

Density fluctuations are produced in the ultra-early universe, but 
do not begin to grow into galaxies until the energy density becomes
dominated by matter. At this epoch, when matter and radiation have 
equal energy densities, the cosmic background temperature is given by  
\be
\tempeq = \eta \mpro {\omegam \over \omegab} \,,
\ee
and the corresponding age of the universe is 
\be
\timeeq = {1\over8} \left({3\over\pi}\right)^{1/2}
{\mplanck \over a_R^{1/2} \tempeq^2} = 
{\mplanck \omegab^2 \over 8 a_R^{1/2} (\eta\mpro\omegam)^2}\,.
\ee
The total mass contained within the cosmological horizon at this 
epoch represents an important scale and can be written in the form 
\be
M_{\rm eq} \approx {\mplanck^3 \over 64 \tempeq^2} \approx
10^{72} \,{\rm GeV}\, \approx 10^{15} M_\odot\,,
\label{horizonmass} 
\ee
where we have used $\eta=6\times10^{-10}$ and $\omegam/\omegab=6$
to obtain the numerical values. 

All of the mass scales smaller than the horizon mass $M_{\rm eq}$ 
begin to grow at the epoch of equality and become nonlinear at 
a later epoch denoted here as $\timecol$. The age of the universe 
at this time is given approximately by the expression 
\be
\timecol \approx \timeeq f_{\rm vir} Q^{-3/2} \,, 
\ee
where $Q$ is the amplitude of the primordial density fluctuations
($Q\approx10^{-5}$ in our universe). The parameter $f_{\rm vir}<1$ is
a dimensionless factor of order unity \cite{tegmarkrees}, and varies
slowly with the halo mass (for $M\le M_{\rm eq}$).  If density
fluctuations on all scales were to grow exactly linearly with the
scale factor and start at the moment of matter domination, then we
would have $f_{\rm vir}=1$.  Notice also that larger mass scales
collapse at later times given by $t\approx\timecol(M/M_{\rm eq})$ 
\cite{tegmarkrees,tegmark}.

To leading order, the density of the galactic halos is determined by
the background density of the universe at the time $\timecol$ when 
the halo collapses. After the halo detaches itself from the cosmic
expansion taking place in the background, its subsequent collapse
proceeds in collisionless fashion, and results in a quasi-virialized
structure with characteristic density that is larger than its 
pre-collapse value by a factor of $f_{\rm col}\sim18\pi^2\sim200$ 
(e.g., see \cite{kolbturner,peacockbook,peebles67,schecter}). The 
characteristic density of the halo thus takes the form 
\be
\rho_{\rm c} = 18\pi^2 \rhoeq \left({\timeeq\over\timecol}\right)^2 
= 36\pi^2{a_R \over f_{\rm vir}^2} \tempeq^4 Q^3  
= {12 \pi^4 \over 5 f_{\rm vir}^2} \mpro^4 \eta^4 
\left({\omegam\over\omegab}\right)^4 Q^3 \,. 
\label{rhoscale} 
\ee
The characteristic density scales with the cube of the amplitude of
primordial density fluctuations. However, as emphasized by many
authors \cite{martel,tegmarkrees,tegmark}, the scaling depends on
additional parameters $\rho_{\rm c}\propto\tempeq^4 Q^3$, so that
$\rho_{\rm c}=\rho_{\rm c}(Q,\eta,\omegam,\omegab)$. Inclusion of 
the factor $f_{\rm vir}$ leads to the characteristic density
$\rho_{\rm c}$ increasing slowly with decreasing halo mass. On the
other hand, larger halos with $M>M_{\rm eq}$ collapse later and will
have lower characteristic densities, with the approximate scaling
$\rho_c\propto(M_{\rm eq}/M)^2$.

\subsection{Structure of Dark Matter Halos}
\label{sec:halostructure} 

Both numerical simulations and cosmological observations indicate that
--- to leading order --- galactic halos have a nearly universal form.
The pioneering theoretical study of this universality \cite{nfw} found
that the density profiles of these halo structures have the form
$\rho\propto r^{-3}$ at large radii. However, subsequent work indicates 
a somewhat steeper slope \cite{busha2005,busha2007,nagamine}, which 
arises from extending the analysis farther out in radius and evolving 
the halos into the future when they reach an asymptotic form (see also 
\cite{boily}). A good working model for galactic halos is provided by 
the Hernquist profile, originally put forth as a description for
galactic bulges, which are analogous collisionless systems. The
density distribution of this model has the simple form 
\be
\rho(r) = {\rho_0 \over \xi (1+\xi)^3} \qquad {\rm where} \qquad 
\xi = {r \over r_0} \,.
\label{hernquist} 
\ee
The profile is thus characterized by a density scale $\rho_0$ and a
length scale $r_0$. The density profile of equation (\ref{hernquist}) 
can be integrated to obtain a finite mass $M_0=2\pi\rho_0r_0^3$. 
All dark matter halos with masses $M<M_{\rm eq}$ will virialize at 
approximately the same epoch and will have characteristic densities 
given by equation (\ref{rhoscale}). As a result, we can make the 
identification 
\be
\rho_0 = \rho_{\rm c} \,.
\label{identify} 
\ee
For a given halo mass $M$, the corresponding length scale $r_0$ 
is then given by 
\be
r_0 = \left({M \over 2\pi\rho_{\rm c}}\right)^{1/3} = 
\left({M \over M_{\rm eq}}\right)^{1/3} 
{\mplanck \over 8\pi\tempeq^2}
\left({f_{\rm vir}^2 \over 9 a_R}\right)^{1/3} Q^{-1}\,.
\label{rscale} 
\ee

In addition to determining halo properties $(\rho_0,r_0)$ as functions
of the cosmological parameters (equations [\ref{rhoscale}],
[\ref{identify}] and [\ref{rscale}]), we also need to estimate the
structure of the baryonic component of galaxies. The density profile
of equation (\ref{hernquist}) provides a good working approximation to
the final collapsed state of a wide variety of collisionless systems
\cite{cannizzo}, including galactic bulges \cite{hernquist}, dark 
matter halos \cite{busha2005,nagamine,nfw}, and even young embedded
star clusters \cite{proszkow}. Moreover, elliptical galaxies have
similar morphologies to bulges \cite{binmer,bintrem}. As a first 
approximation, one can thus use equation (\ref{hernquist}) as a model
for the baryonic component of the galaxy, including stars, albeit with
different values of the density and radial scales \cite{coppess}. 
Since baryons dissipate energy, the baryonic component collapses to
higher densities, so one can write
\be
\rho_{\rm b0} = F \rho_0 \,,
\label{enhancement} 
\ee
where one expects $F={\cal O}(10)$. The baryonic component of the 
galactic mass budget is a fraction of the total and is given by 
$M_{\rm b} = (\omegab/\omegam)M$. With the density scale and mass 
specified, the radial scale for the baryonic component of the 
galaxy takes the form 
\be
r_{\rm b} = \left({M_{\rm b} \over 2\pi\rho_{\rm b}}\right)^{1/3}
= \left({\omegab\over\omegam F}\right)^{1/3} r_0 
\sim {r_0\over4}\,,
\ee
so that the density profile for the baryonic component of the galaxy 
can be written in the form 
\be
\rho_{\rm b} = {\rho_{\rm b0} \over \xi_{\rm b} (1+\xi_{\rm b})^3}
\qquad {\rm where} \qquad \xi_{\rm b} \equiv {r \over r_{\rm b}}\,.
\label{rhobaryon} 
\ee
As a reference point, for the value $\omegam/\omegab=6$ found in 
our universe, the baryonic density scale becomes 
\be
\rho_{\rm b0} \approx 10^9 M_\odot {\rm pc}^{-3} Q^3 
f_{\rm vir}^{-2} \left({\eta\over\eta_0}\right)^4\,,
\ee
where $\eta_0=6\times10^{-10}$. For $Q\simgreat10^{-2}$, the density
scale $\rho_{\rm b}\simgreat10^3 M_\odot$ pc$^{-3}$, comparable to the
mean density of a globular cluster \cite{bintrem}. The inner region
$\xi<\xi_{\rm b}$, one fourth of the galactic mass, will have higher 
density and will thus be susceptible to black hole formation. 

\subsection{Bounds on the Amplitude of Primordial Fluctuations 
from Planet Scattering} 
\label{sec:scatterbound} 

Galaxies that are too dense cannot support habitable solar systems
because passing stars can strip planets out of orbit. As outlined
above, large values of the amplitude $Q$ of the primordial density
fluctuations lead to higher galactic densities, which in turn leads to
more disruption. As a result, the requirement that planetary orbits
must survive places a constraint on the allowed values of $Q$ for
viable universes.

\begin{figure}[tbp]
\centering 
\includegraphics[width=.95\textwidth,trim=0 150 0 150,clip]{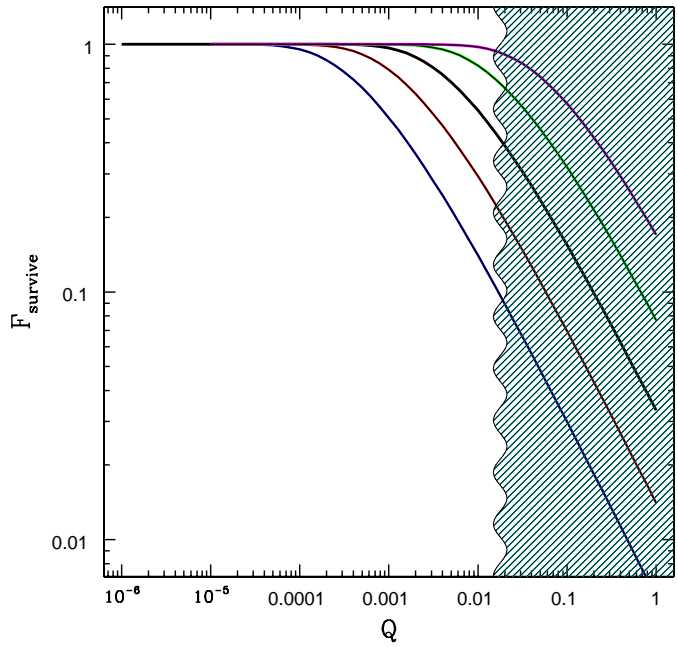}
\caption{Fraction of planets that survive disruption as a function 
of the amplitude $Q$ of the primordial density fluctuations (adapted 
from \cite{coppess}). Passing stars remove planets from their host
stars in regions where the galaxies are overly dense. The mean density
of galaxies and hence the rate of disruption is an increasing function
of $Q$. The survival fraction is a decreasing function of $Q$, as
shown here for galaxies with total masses $M/M_\odot$ = $10^{10}$
(blue), $10^{11}$ (red), $10^{12}$ (black), $10^{13}$ (green), and
$10^{14}$ (magenta). In the shaded region on the right side of
diagram, galaxies are likely to produce large black holes that disrupt
habitability \cite{bhrees}. This diagram was constructed using a 
baryon to photon ratio $\eta=10^{-9}$. The characteristic density 
scales as $\rho_{\rm c}\propto\eta^4$ (from equation [\ref{rhoscale}]), 
so that variations in $\eta$ can be included by taking the horizontal
axis variable to be ${\hat Q} = Q (\eta 10^9)^{4/3}$. }
\label{fig:galsurvive} 
\end{figure}  

The first assessment of this effect \cite{tegmarkrees} found that
solar system disruption starts to occur when the amplitude $Q$ is only
about an order of magnitude larger than that of our universe (see also
\cite{tegmark,liviorees}). These assessments make two assumptions: 
First, the cross section for disruption of habitable orbits is taken
to be the geometric cross section of the orbit, so that $\sigma_{\rm
p*}\sim\pi$(1 AU)$^2$. Second, the entire galaxy is assumed to have
the (single) characteristic density given by equations (\ref{rhoscale}) 
and (\ref{enhancement}). However, this calculation can be generalized
\cite{coppess} by allowing the galaxy to have a range of densities 
(e.g., with the profile of equation [\ref{hernquist}]) and by
calculating the cross sections for disruption of planetary orbits as a
function of encounter speed.

The cross section for disrupting Earth-like planets by stripping them
from orbit about their host stars can be calculated from a large
ensemble of numerical simulations \cite{laughlinadams,gdawg}. In this
type of calculation, numerical integrations follow either single stars
or binaries as they experience fly-by encounters with solar systems of
a given architecture. The passing stars sample the stellar initial
mass function as well as the observed distributions of binary orbital
parameters. The encounters sample a range of pre-encounter velocities,
impact parameters, and the angular variables necessary to specify the
geometry of the interaction.  All of the variables are then specified 
using a Monte-Carlo scheme, and numerical simulations are performed
for each realization of the variables. The results can then be
processed to find the the effective cross sections for any given
post-encounter result of interest. Here we need the cross section for
disrupting the orbit of a planet that is initially in a habitable
orbit, where disruption includes both ejection from the system and
increases in eccentricity beyond a specified threshold. The resulting
cross section for disruption can be written in the form \cite{coppess}
\be
\sigma_{\rm p*} = {\sigma_0 \over u (1+u)} \qquad 
{\rm where} \qquad u = {v \over 1000\,{\rm km}\,{\rm s}^{-1}} 
\qquad {\rm and} \qquad \sigma_0\approx 1\,{\rm AU}^2\,.
\label{csection} 
\ee  
Since dense galaxies have large velocity dispersion $\sim10^3$ km/s,
and the cross section decreases with encounter velocity, the value of 
$\sigma_{\rm p*}$ becomes significantly less than the geometric cross
section. As a result, more solar systems survive. The Monte Carlo 
scheme used to calculate the cross sections samples over a distribution 
of encounter speeds, centered on a characteristic value (denoted as $v_c$). 
The reported cross section is thus related to the velocity-averaged 
value such that $\sigma_{\rm p*} = \langle \sigma v\rangle / v_c$. 

With the interaction cross section specified, the requirement for 
solar system survival can be written in the form 
\be
n_* \, \langle\sigma{v}\rangle \, \tau < 1 \,, 
\label{tauscatter} 
\ee
where $n_*$ is the number density of stars, $v$ is the encounter
speed, and $\tau$ is the required lifetime of a habitable planet.
Although this latter quantity remains unknown, $\tau=1$ Gyr is often
taken as a benchmark value \cite{knoll,lunine,scharf}.  The number
density of stars can be determined from the density profile of
baryonic matter (e.g., equation [\ref{rhobaryon}]) by specifying the
star formation efficiency (denoted here as $\epsilon_{\rm sf}$). Note
that both the stellar density $n_*$ and the encounter velocity $v$
depend on the location within the galaxy.

Figure \ref{fig:galsurvive} shows the fraction of solar systems that
can survive --- not be disrupted by passing stars --- as a function of
the fluctuation amplitude $Q$. Survival curves are shown for galaxies
with a range of masses, from $M=10^{10}M_\odot$ (left side of the
diagram) to $M=10^{15}M_\odot$ (right side). Galaxies with masses
comparable to the Milky Way correspond to the central black curve.
The characteristic density of the galaxies increases with increasing
$Q$, and the survival fraction decreases accordingly. In addition, the
density decreases with increasing galactic mass. Although many solar
systems can be disrupted, the survival faction remains above 0.10
(10\%) over most of the range of parameter space. The most compromised
galaxies are those with the smallest masses, whereas the most
compromised universes are those with the largest amplitudes $Q$. The
region on the right side of the diagram is shaded to indicate that
universes with sufficiently large $Q$ tend to overproduce black holes,
which leads to inhospitable galaxies.  If the value of $Q$ approaches
unity, black hole formation can take place immediately following the
inflationary epoch \cite{greenliddle}; for somewhat smaller values, 
$Q=0.01-0.1$, a large fraction of the total mass can be incorporated 
into black holes during the process of galaxy formation \cite{bhrees}. 
Note that the survival fraction never reaches zero. Even in the
densest galaxies, the outer regions will be diffuse enough to allow
for some planets to survive in habitable orbits.

\subsection{Constraints from the Galactic Background Radiation} 
\label{sec:radbound} 

In addition to disruption by passing stars, habitable planets in dense
galaxies face another hazard. As a galaxy becomes denser, the
radiative flux provided by the background stars becomes more intense.
If this galactic background radiation flux is larger than that
received by a habitable planet from its host star, roughly comparable
to the solar insolation received by Earth, then planets will cease to
be habitable {\it in any orbit} \cite{coppess}. This type of radiative
disruption can occur in our universe in extreme environments,
including the centers of galaxies \cite{forbes} and sufficiently dense
star clusters \cite{thompson}.

Consider, for example, an elliptical galaxy where the baryonic
component can be modeled with a profile given by equation
(\ref{rhobaryon}). At the dimensionless radial location 
$a=\xi_{\rm b}$ within the galaxy, the background radiation 
flux $F_G(a)$ due to starlight has the form 
\be
F_G(a) = \rho_{\rm b0}\,\epsilon_{\rm sf}\,r_{\rm b}\,
{\langle L_\ast / m_\ast \rangle \over 2(a^2-1)} 
\left[ 1 - {2 \log a \over a^2 - 1} \right] \,,
\label{galaxyflux} 
\ee
where the angular brackets denote an average over the stellar 
population \cite{coppess}. In order for habitable planets to 
survive with temperate climates, the background radiation flux 
from the galaxy $F_G$ must be smaller than the flux received 
by a habitable planet from its host star, i.e., 
\be
F_G(a) \simless {L \over 4\pi\varpi^2} \sim S_\oplus \,, 
\label{radlimit} 
\ee
where $\varpi$ is the radius of the planetary orbit and $L$ is the
luminosity of the host star. In the second approximate equality,
$S_\oplus$ denotes the solar radiation flux received at Earth orbit.
Since the value of the flux directly determines the surface
temperature of the planet, the requirement of liquid water constrains
the flux from the host star ($L/4\pi\varpi^2$) to be roughly equal to
$S_\oplus$.  The galactic flux $F_G$ is a strictly decreasing function
of the radial variable $a$. The central regions of almost every galaxy
will support radiation fields that violate the constraint of equation
(\ref{radlimit}), leading to a situation analogous to that raised by
Olber's famous paradox \cite{harrison,wesson}. Conversely, the outer
regions of every galaxy will be cool and dark enough to allow for 
habitable planets. This compromise is similar to that found earlier
for disruption by scattering encounters.  The fraction of the solar
systems that survive in the face of strong radiation fields is
comparable to, but somewhat larger than, the fraction that survive
devastation by scattering events \cite{coppess}.

\begin{figure}[tbp]
\centering  
\includegraphics[width=1.0\textwidth,trim=0 150 0 150]{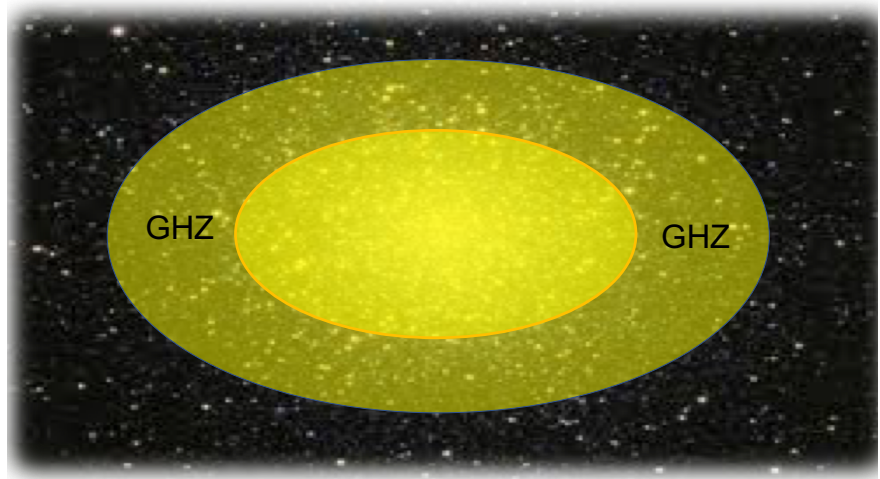}
\bigskip 
\caption{Schematic representation of galactic structure. In a universe 
with large amplitude $Q$ of the primordial density fluctuations,
galaxies are much denser than those in our universe. In the central
regions, both the intensity of the galactic background radiation and
the frequency of dynamical interactions are severe enough to
compromise habitability. The outer regions are diffuse, so that
habitable planets in suitable orbits survive as in our universe (note
that the galaxy extends far beyond the region depicted in the panel). 
In the intermediate regime, denoted here as the Galactic Habitable
Zone (GHZ), the galactic background radiation is as bright as the
daytime sky on Earth, so that planets are potentially habitable over 
a wide range of orbits (including unbound planets). }
\label{fig:ghz} 
\end{figure}

The presence of strong radiation fields in galaxies provides another
possible channel for planets to be habitable. Although radiation in
galactic centers will be too intense for planets to remain viable, and
the outer regions will be cold like in our universe, there exists an
intermediate zone where the background galactic flux is comparable to
the solar flux, i.e., $F_G \sim S_\oplus$. Any planets residing in
this intermediate regime can have appropriate surface temperatures for
nearly any orbit (see Figure \ref{fig:ghz} for a schematic depiction 
of galactic structure in this scenario).  The planets that reside too
close to their host stars will be too hot, but all other planets will
be warmed by the galactic background radiation and can have surface
temperatures that support liquid water environments. The size and
location of this Galactic Habitable Zone (GHZ) depend on the
dimensionless ratio 
\be
X \equiv {\rho_{\rm b0}\,\epsilon_{\rm sf}\,r_{\rm b}\,
\langle L_\ast / m_\ast \rangle \over 2 S_\oplus} \,. 
\label{xfactor} 
\ee
For values of the parameter in the range $X=1/2-100$, more than 10\%
of all solar systems reside in the GHZ. For the optimal case where
$X\approx3$, the GHZ contains approximately one fourth of the solar
systems in the galaxy.  If most stars have multiple planets, then
galaxies in such universes could support more habitable planets than
our own. As a result, our universe is not optimized for the support of
habitable planets.

The value of the parameter $X$ is primarily determined by the column
density of stars in the galaxy $N_\ast=n_\ast r_{\rm b}$, which must
depend on the amplitude $Q$ of the primordial density fluctuations.
The other quantities appearing in equation (\ref{xfactor}) are either
constant or slowly varying, and depend on stellar (rather than
galactic) properties. For galaxies like the Milky Way in our universe,
the column density of stars $N_\ast\sim10^2-10^3$. To allow the
parameter $X\sim3$, the column density must be much larger,
$N_\ast\sim10^9$.  Since galactic density scales with the fluctuation
amplitude $Q$ according to $n\sim\rho\sim Q^3$, and the galactic length
parameter scales as $r_{\rm b}\sim Q^{-1}$, the column density scales
as $Q^2$.  Universes that optimize their galactic habitable zones thus
need larger initial density fluctuations compared to our universe,
where the preferred value is $Q\approx10^{-2}$. As discussed above, 
the central regions of the resulting dense galaxies are susceptible 
to black hole formation. 

The fluctuation amplitude $Q$ that optimizes the GHZ corresponds to 
an enhancement by a factor of $\sim10^3$ if the other cosmological
parameters are fixed. The characteristic density for galaxies scales
as $\rho_{\rm c}\propto\eta^4Q^3$ (see equation [\ref{rhoscale}]). 
As a result, universes can achieve the same high galactic density 
if the baryon to photon ratio $\eta$ is larger than that of our 
universe by a factor of $\sim180$. 

Under the dense conditions of the GHZ, planets and their orbits are
subjected to a number of processes that are rare in our universe.
First, note that when the composite parameter $X\sim1$ (from equation
[\ref{xfactor}]), the optical depth for planetary disruption is also
of order unity (from equation [\ref{tauscatter}]). As a result, a
sizable fraction of the planets residing in the galactic habitable
zone will be freely floating, rather than in orbit about a particular
star. This condition does not necessarily preclude habitability, but
planets would not have the same astronomical cycles --- days and
seasons --- that characterize our calendar. Dynamical relaxation
provides another potential issue. With a dense field of potential
perturbers, planets (and stars) can random walk around the GHZ through
a large number of distant scattering encounters \cite{bintrem}. 
However, the dynamical relaxation time for stars/planets in the GHZ is
expected to be quite long, $t_{\rm relax}\sim10^5$ Gyr. Although this
time scale is much shorter than the relaxation time in our Milky Way
galaxy, it much longer than the canonical time scale $\sim1$ Gyr often
invoked for habitability. Most planets will thus stay within the GHZ
long enough for life to develop.

In addition to dynamical considerations, the dense environment of the
GHZ poses other hazards for habitable planets. The supernova rate is
expected to be comparable to that of galaxies in our universe.
However, habitable planets will be closer to the explosions due to the
enhanced stellar density. Since the galactic density scales with the
cube of amplitude of the initial cosmological fluctuations, 
$\rho\propto Q^3$, the distance $d$ to the nearest destructive
supernova will scale as $d\propto1/Q$.  Supernova can have a number of
detrimental effects on potentially habitable planets, including
stripping of the ozone layer from the atmosphere due to both cosmic
rays and gamma rays. Although the lethal distance for this process is
not well determined, estimates fall in the range $d$ = 8 -- 50 pc for
the atmosphere of Earth (see \cite{gehrels,melott,sloan} and references 
therein). Thicker atmospheres can tolerate greater exposure (see also 
\cite{cirkovicvukotic}). Gamma ray bursts present an analogous hazard 
\cite{thomas2005}. Other risks include large fluctuations in the 
galactic background radiation due to close stellar passages and
flaring activity on nearby stars \cite{lingamloeb}. These hazards 
arise due to the increased densities of galaxies. Note that galactic 
densities can also increase with the vacuum energy density $\rhov$, 
as only the denser perturbations can collapse \cite{totani}, which 
leads to analogous effects. 

For completeness, note that our universe had a brief epoch of
cosmic-wide habitability for redshifts $100\simless(1+z)\simless137$
when the temperature of the cosmic background radiation was in the
range $T$ = 273 -- 373 K. During this early epoch, suitable rocky
planets could have liquid water on their surfaces for virtually any
orbit \cite{loebcmb}, much like the case of the GHZ. In addition, the
concept of the Galactic Habitable Zone has been discussed previously
in the context of our Galaxy \cite{gonzalez,gzone,lineweaver}. This 
local version of the GHZ is based on the chemical makeup of the Galaxy
as a function of radial position, where the mean metallicity for
habitability is required to be within a factor of 2 of that of the
Sun. In practice, the metallicity $\metal$ decreases with
galactocentric radius, so that the outer regions of the galaxy have
low $\metal$ and are less suited for life. However, the metallicity
threshold for viable planets remains unknown. In addition, the degree
of disruption provided by the environment of our Galaxy is generally
much less severe than that considered above \cite{kaib}.

Supermassive black holes can provide another environment for a dense
concentration of habitable planets \cite{raymond}. In conventional
solar systems, relatively few planets can reside within the habitable
zone because orbits become dynamically unstable if they are too close
together. The minimum orbital spacing for long-term stability is
estimated to be $\sim10R_H$, where $R_H$ is the mutual Hill radius
given by
\be
R_H = \left({m_1+m_2\over3M_c}\right)^{1/3} {a_1+a_2\over2} \,.  
\label{hillrad} 
\ee
Here, $a_j$ are the semimajor axes and $m_j$ are the masses of the
orbiting bodies, and $M_c$ is the mass of the central object. For
bodies orbiting a supermassive black hole, the central mass $M_c$ can
be enormous (up to $\sim10^9M_\odot$) so that $R_H$ is small and the
orbits can be tightly spaced. Since black holes themselves have only
feeble radiative emission, another power source is required. One can
construct a hypothetical system in which many stars orbit the black
hole on interior orbits, with an extensive collection of planets
orbiting further out. The key feature of this scenario is that one can
place a large number of bodies, stars on the inside and planets on the
outside, in orbit about the same supermassive black hole.

\begin{figure}[tbp]
\centering 
\includegraphics[width=.95\textwidth,trim=0 150 0 150,clip]{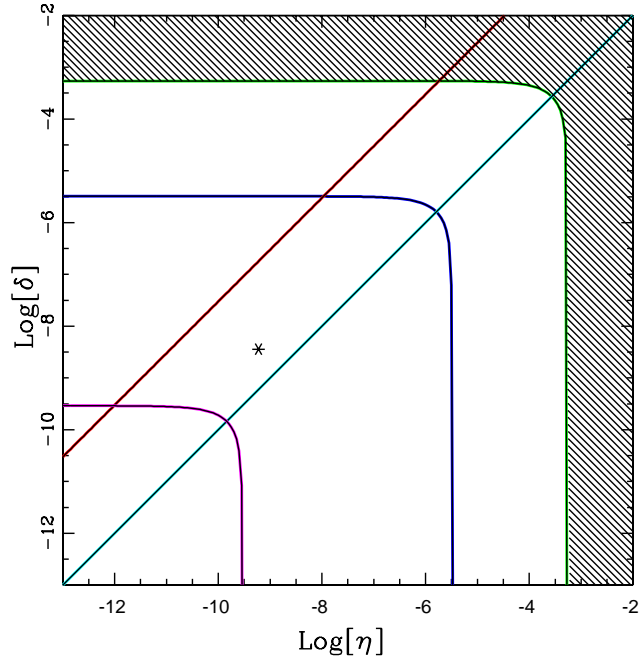}
\caption{Regions of parameter space for variations in the mass inventory 
of baryons and dark matter. The horizontal axis delineates the baryon
to photon ratio $\eta$, whereas the vertical axis delineates the
analogous parameter for dark matter $\delta=\eta\,\omegad/\omegab$. The
star symbol marks the location of our universe in the diagram. In the
region above the green curve, matter domination occurs before Big Bang
Nucleosynthesis; this region is disfavored and hence shaded. In the
region above the blue curve, some structures can become nonlinear
before decoupling. In the region below the magenta curve (lower left)
decoupling occurs before the epoch of matter domination. The red line
depicts the benchmark value $\delta=300\eta$; above this curve, the
larger abundance of dark matter could inhibit the fragmentation of
galactic disks and subsequent star formation. The cyan line depicts
the condition $\delta=\eta$; below this curve, baryonic matter is 
subject to enhanced Silk damping. } 
\label{fig:etadelta} 
\end{figure}  

\subsection{Variations in the Abundances of Dark Matter and Baryons}
\label{sec:massinventory} 

The relative amounts of dark matter and baryons in the universe can
affect the timing of galaxy formation and hence the resulting galactic
properties. Here we consider the matter content of the universe to
consist of two components, baryons and collisionless dark matter.  The
baryonic component is specified by the baryon to photon ratio
$\eta$. For the dark matter, we consider an analogous parameter
defined by $\delta=\eta\,\omegad/\omegab$. If the dark matter
particles had the same mass as baryons ($\mpro$), then $\delta$ would
be the dark matter to photon ratio (but the mass of the dark matter
particles is left arbitrary). In our universe, constraints from BBN
indicate that $\eta\approx6\times10^{-10}$ and a host of cosmological
measurements imply that $\delta\approx6\eta$ \cite{lahavliddle}. In
other universes, these quantities $(\eta,\delta)$ could have alternate
values, which would lead to different properties for galaxies, as
outlined below. (For completeness, however, we note that 
\cite{boussohall} provides an argument for comparable densities of
baryons and dark matter; see also \cite{wilczek}.)

This discussion roughly follows the arguments presented in \cite{tegmark} 
(see also references therein). The first consideration is that if the
total mass content of the universe is too large, then the epoch of
matter domination could occur before Big Bang Nucleosynthesis. As a
rough estimate, the energy scale of BBN can be taken to be comparable
to the electron mass $\emass$, so that these two epochs coincide when 
\be
(\eta + \delta)\,\mpro = \emass\,. 
\label{bbn_eq_matter} 
\ee
For earlier matter domination, both BBN and galaxy formation would
proceed in dramatically different fashion. Nonetheless, this regime of
parameter space is not necessarily ruled out. The previous section
shows that the abundances of light nuclei coming out of BBN are
generally not problematic, even for values of $\eta$ much larger than
that of our universe.

For somewhat smaller matter abundances, large scale structures (e.g.,
dark matter halos) can detach from the expanding background of the
universe and become nonlinear before the epoch of decoupling. The
first structures with roughly galactic mass scales become nonlinear
for temperatures $T\sim9Q(\eta+\delta)\mpro$ (see \cite{tegmark} for
further detail). To leading order, the temperature of recombination is
determined by solution to the Saha equation, which can be written in
the form
\be
{1-x\over x^2} = {2 \zeta(3) \over \sqrt{\pi}} \eta \alpha^3 
\left( {T\over B_H} \right)^{3/2} \exp\left[ B_H/T \right] \,,
\label{saha} 
\ee
where $x$ is the ionization fraction and $B_H=\emass\alpha^2/2$ is the
binding energy of Hydrogen. This expression makes a number of
simplifications \cite{kolbturner,peacockbook}, including the assumption 
of ionization equilibrium and the neglect of Helium. Because the baryonic
component is small $\eta\ll1$, the relevant solutions correspond to 
low temperature $T\ll B_H$, so that equation (\ref{saha}) is dominated
by its exponential term. As a result, the recombination temperature is
slowly varying and is given approximately by the expression 
\be
T_{\rm rec} \approx B_H/50 = {\alpha^2 \emass \over 100} \,. 
\label{rectemp} 
\ee
Using this result in conjunction with the requirement for structures
to become nonlinear, the boundary in parameter space where structure
grows before recombination takes the form
\be
\delta + \eta = {\alpha^2 \beta \over 900 Q}\,. 
\label{nonlin_eq_combine}
\ee

In our universe, the epochs of recombination and photon decoupling
occur after that of matter-radiation equality (the start of the matter
dominated era). If the order is reversed, then the initial growth of
structure will proceed differently. Instead of dark matter collapsing
before the baryons (as in our universe), both matter components would
collapse together starting from the epoch of equality. The boundary in
parameter space where this crossover occurs has the approximate form 
\be
(\eta+\delta)\,\mpro \approx {\alpha^2 \beta \mpro \over 100} \,. 
\label{rec_eq_matter} 
\ee 
Although the details of galactic collapse change with the ordering 
of the epochs of recombination and matter domination, collapse 
occurs either way. 

The above considerations constrain the total matter content of the
universe. The properties of galaxies can also be different if the
relative contributions of baryons and dark matter are different. If
the dark matter dominates the density of baryons by a large factor,
then galactic disks will be stabilized. This trend acts to suppress
the formation of molecular clouds and hence star formation.
Unfortunately, there is not a clean estimate for the maximum value of
the ratio $\delta/\eta$ that allows for star formation. However, the
working estimate of \cite{tegmark} suggests that galactic disks will
evolve differently when the dark matter parameter exceeds the limit 
\be
\delta \simgreat 300 \eta \,. 
\label{nodiskfrag} 
\ee
Similarly, if the dark matter content is much lower than in our 
universe, then baryons dominate the process of galactic formation. 
Although galaxies are still able to form, they will be subject to 
enhanced levels of Silk damping \cite{silkdamp}, and will wind up 
with different characteristics. The threshold for this type of 
behavior is approximately given by 
\be
\delta \simless \eta\,. 
\label{moresilkdamp} 
\ee

The constraints of equations (\ref{bbn_eq_matter}--\ref{moresilkdamp})
divide up the possible parameter space for the matter components into
distinct regions. The result is shown in Figure \ref{fig:etadelta}. In
the regime of parameter space above the green curve, matter domination
occurs before or during BBN. Although such universes may be able to
support life, they would be quite different from our own, and this
extreme region is shaded in the figure. The blue curves marks the
boundary where structures can become nonlinear before the epoch of
decoupling. Above this curve, forming galaxies remain coupled to the
background radiation field, at least for the first part of their
evolution. The magenta curve marks the boundary where decoupling
occurs before the epoch of matter domination. The two diagonal lines
in Figure \ref{fig:etadelta} correspond to constant ratios of baryonic
mass to dark matter $\eta/\delta$. Above the red curve, dark matter
dominates (by a factor of 300). In this regime, dark matter halos
provide stability to galactic disks (which are baryonic) and star
formation is suppressed. Below the cyan curve (where $\eta=\delta$),
baryonic matter dominates and Silk damping \cite{silkdamp} will be
enhanced. In this regime, the collapse of smaller scale structures
will be suppressed. Although galaxies and their dark matter halos will
have different properties over the parameter space delineated in 
Figure \ref{fig:etadelta}, the values of both $\eta$ and $\delta$ can
vary by many orders of magnitude and still allow for habitable
galaxies.

\subsection{Gravitational Potential of Galaxies} 
\label{sec:galpotential} 

One important role played by galaxies is that they retain and organize
the heavy elements produced by their stellar components.  If the heavy
nuclei synthesized through stars are lost to the intergalactic medium,
they cannot be incorporated into subsequent generations of stars,
planets, and possible life forms. In our universe, the smallest dwarf
galaxies lose interstellar gas through a combination of tidal and ram
pressure stripping, as well as internal feedback from stars. This mass
loss leads to inefficient star formation, and lower production rates
of heavy elements. Moreover, the efficacy of galactic mass loss
depends on the depth of the gravitational potential well.

For the halo density profiles of equation (\ref{hernquist}), 
the corresponding gravitational potential is given by 
\be
\Psi = {\Psi_0 \over 1 + \xi} \qquad {\rm where} \qquad 
\Psi_0 = 2\pi G \rho_0 r_0^2 \,,
\ee
where the scales $\rho_0$ and $r_0$ are defined in Section
\ref{sec:halostructure}. Using equations (\ref{rhoscale}), 
(\ref{identify}), and (\ref{rscale}) to specify the parameters 
$(\rho_0,r_0)$, the depth of the potential well for a halo becomes 
\be
\Psi_0 = {3\pi\over8} \left({\pi^2\over45}\right)^{1/3} 
\left({M \over f_{\rm vir} M_{\rm eq}}\right)^{2/3} Q 
\sim 0.7 Q \,.
\ee
The last approximate equality holds for the largest halos near 
the mass of the horizon at equality. Since the potential scale 
$\Psi_0$ defines the depth of the potential well and hence the 
escape speed from the galactic center, we have the result 
\be
v_{esc} \approx 0.8 Q^{1/2} c \,. 
\ee
In our universe, the escape speed from galaxies is typically
$v\approx300$ km/s or $v/c\approx10^{-3}$. The escape speeds from
stellar surfaces are also of this order. This latter quantity sets 
the scale for stellar winds, supernovae, and other stellar processes
involving the dispersal of heavy elements throughout the galaxy (and
beyond). As a result, successful retention of heavy elements requires 
a minimum escape speed and hence a minimum value of $Q$. As a starting 
point, we take the minimum escape speed to be $\sim100$ km/s, so that 
the minimum value of the fluctuation amplitude becomes 
\be
Q_{\rm min} \approx 10^{-7} \,. 
\label{qlimitgrav} 
\ee
This value is comparable to, but somewhat smaller than, the estimated
minimum value of $Q$ based on cooling considerations \cite{tegmarkrees}, 
as addressed in the following subsection. 

Notice that the bound of equation (\ref{qlimitgrav}) is not sharp:
Stellar processes will produce a distribution of ejection velocities,
and gas stripping will operate with a range of efficiencies, 
so that decreasing values of $Q$ result in a larger fraction (but not
all) of the heavy elements escaping. In addition, galaxies form with a
distribution of masses and escape speeds, so that the largest
structures could retain gases in universes with smaller amplitudes
$Q$. Yet another complication is that ejected material is not
necessarily gone forever: Exiled gas could cool and condense back into
the galaxy at some later epoch, or even be accreted by neighboring
galaxies.  Nonetheless, for sufficiently small fluctuation amplitude
$Q\ll Q_{\rm min}$, most heavy elements are expected to be lost,
leading to a commensurate loss of habitability.

\subsection{Cooling Considerations} 
\label{sec:galaxycooling} 

Another constraint on the properties of galaxies arises from the
requirement that galactic gas must cool promptly in order to make
stars. In addition to providing an estimate for the masses of viable
galaxies \cite{reesost}, this requirement of substantial cooling has
been used to place a lower bound on the amplitude $Q$ of primordial
fluctuations \cite{tegmarkrees,tegmark} with the result that
$Q\simgreat10^{-6}$. If one makes the further assumption that the gas
not only cools on the free-fall collapse time scale, but also turns into
stars on the same time scale, then an estimate for the star formation
rate can be found \cite{bousso2009,bousso2010}.

The cooling function for primordial (zero metallicity) gas involves a
number of physical processes \cite{abel,gallipalla}.  The dominant
mechanism varies with the temperature and density of the system.  At
high temperatures, bremsstrahlung is the most important process,
whereas a full accounting for lower temperatures includes line cooling
by neutral hydrogen, helium, and any heavier elements that are
present. Both bremsstrahlung and line cooling become inefficient below
a temperature $T\sim10^4$ K. Molecular cooling lowers this temperature
scale, but not substantially \cite{abel,gnedin,haiman,tegmarkbd}.  As
a result, gas generally cools down to this benchmark temperature
relatively quickly, but further cooling is much slower.

In order to illustrate the effects of cooling, and obtain analytic
expressions for the resulting constraint, we simplify the treatment
to include only bremsstrahlung. For this extreme limiting case, the
basic cooling rate has the form 
\be
{dE \over dt dV} = n_e n_P \Lambda = 
n_e n_P \langle \sigma v \rangle \epsilon\,,
\ee
where each scattering interaction cools the gas 
by the energy increment 
\be
\epsilon = {4 e^2 \over \lambda} = {2 \over \pi} \alpha \emass \,.
\ee
The cooling time can then be written in the form 
\be
t_{\rm cool} = {3 T \over 2 n \Lambda} = 
{3 T \over 2A n \sigma_T v_s \epsilon}\,,
\ee
where $v_s=(T/\emass)^{1/2}$ is the thermal speed of electrons,
$\sigma_T$ is the Thomson cross section, and $A$ is a dimensionless
constant of order unity (e.g., see \cite{carr}). The dynamics 
of the baryonic component of galaxies depends on the ratio of this 
cooling time scale to the gravitational collapse time 
\be
t_{\rm grav} = (G\rho)^{-1/2}\,.
\ee
The density in this expression is the characteristic density for
collapsed structures found previously. Equating the cooling rate 
with the collapse rate implies
\be
{2A\over3} {n \sigma_T v_s \epsilon \over T} = 
(G\rho)^{1/2} \,.
\ee
The number density can be expressed in the form 
$n=\eta\rho/(\eta+\delta)\mpro$, where the density $\rho$ is specified
through equation (\ref{rhoscale}).  After writing the cross section
$\sigma_T$, the cooling energy increment $\epsilon$, and the sound
speed $v_s$ in terms of fundamental constants, we find 
\be
Q = \left\{ {3f\over64A\pi^2} 
{\emass\over\mplanck} {1\over\eta(\eta+\xi)} \right\}^{2/3}
\left({15T\emass\over\mpro^2}\right)^{1/3}\,\alpha^{-2}\,\,. 
\ee
If we evaluate this expression by assuming that $f=0.01$
(corresponding to one of the denser galaxies) and that the 
temperature $T=10^4$ K (because cooling processes become much 
less effective for lower temperatures), the lower limit on 
the fluctuation amplitude becomes 
\be
Q > Q_{min} = 6 \times 10^{-7} \,. 
\label{qlimitcool} 
\ee
This limiting value of $Q$ is comparable to that determined using the
full cooling curves \cite{tegmarkrees,tegmark} and is slightly more
stringent that the limit obtained by requiring that galaxies retain
some fraction of their heavy elements (equation [\ref{qlimitgrav}]).
In approximate terms, we find that galaxies that can successfully
cool, condense, and potentially form stars will also have deep 
enough gravitational potentials to keep the metals produced by 
stellar nucleosynthesis.

Although successful cooling is necessary for star formation to take
place, we note that stars can be produced under a wide range of
initial conditions. Stars form in galaxies with markedly different
properties, with masses varying by factors of $\sim10^8$ and mean
densities varying over a corresponding range. In our universe, the
star formation rate became substantial at redshift $z\sim10$ and
peaked at $z\sim2$ when the universe was about 3.5 Gyr old 
\cite{madau}.  Although its current rate is smaller by an order of
magnitude, star formation takes place readily at the present
cosmological epoch \cite{mckeeost,sal1987}, and is expected to
continue (at a highly attenuated rate) for perhaps trillions of
years \cite{al1997}. As a result, star formation is not overly
sensitive to the conditions provided by the background universe.

\bigskip 
\section{Stars and Stellar Evolution} 
\label{sec:stars} 

Stars play two important roles regarding the habitability of our
universe, and presumably others.  First, they provide most of the
energy \cite{fukugita} that is available to support biospheres on any
conveniently situated planets. Second, they forge most of the heavy
nuclei necessary for the development of complex structures
\cite{timmes,trimble}, ranging from planets themselves all
the way down to biological entities.

\subsection{Analytic Model for Stellar Structure} 
\label{sec:starmodel} 

In order to estimate the range of parameter space that allows for the
existence of stars, it is useful to have a working semi-analytic model
for stellar structure. Toward this end, we utilize the model of
\cite{adams,adamsnew}, which solves the equations of stellar
structure \citep{chandra,phil,kippenhahn,hansen} subject to a number
of approximations. As one simplification, the physical structure of
the star is considered to be a polytrope with index $n$, which varies
from $n=3/2$ for fully convective stars to $n=3$ for high-mass
(radiative) stars. This approach uses a single reaction rate, which
allows for only one nuclear reaction chain at a time. The resulting
model \cite{adams,adamsnew} reproduces the properties of stars in our
universe to a reasonable degree of approximation (tens of percent),
but allows for variations in the fundamental constants over a wide
range of values (ten orders of magnitude).

Since a full description is given elsewhere \cite{adams,adamsnew},
this section presents only an outline of the model. The pressure 
is a function of the density so that 
\be
P = K \rho^\Gamma \qquad {\rm where} \qquad \Gamma = 1 + {1 \over n} \,.
\ee 
The density can be expressed in terms of a dimensionless function 
$f(\xi)$ such that 
\be
\rho = \rho_c f^n\,, \quad \xi \equiv {r\over R}\,, 
\quad {\rm and} \quad R^2 \equiv 
{K\Gamma \over 4\pi G(\Gamma-1)\rho_c^{2-\Gamma}} \,.
\ee
The function $f(\xi)$ is a solution to the Lane-Emden equation 
\cite{chandra,phil}. The mass of the star can then be written in 
terms of the dimensionless integral parameter $\mzero$, 
\be
M_\ast = 4\pi\rho_c R^3 \mzero \qquad {\rm where} \qquad 
\mzero \equiv \int_0^{\xi_\star} \xi^2 f^n d\xi\,, 
\ee
where the dimensionless radius of the star $\xi_\star$ is of order
unity. The temperature $T(r)$ within the star specified using the
change of variable 
\be
\Theta \equiv \left( {E_G \over 4 k T} \right)^{1/3} 
\qquad {\rm where} \qquad E_G = \pi^2 \alpha^2 Z_1^2 Z_2^2
{2 m_1 m_2 \over m_1 + m_2} c^2 \,, 
\ee
where $m_j$ and $Z_j$ are the masses and charges of the reactants.
Recall that the Gamow energy $E_G\approx493$ keV for hydrogen fusion
in our universe. 

For a single nuclear reaction chain, the cross section can be 
separated into factors through the ansatz 
\be
\sigma(E) = {S(E) \over E} \exp[-E_G/E]\,,
\ee
where $S(E)$ is a slowly varying function of energy. 
The model then defines a nuclear reaction parameter that 
encapsulates all of the nuclear physics in the star, 
\be
\conlum \equiv {8 \langle\Delta E\rangle S(E_0) \over 
\sqrt{3}\pi\alpha m_1 m_2 Z_1 Z_2 m_R c} \,,
\label{conlumdef} 
\ee
where $\langle\Delta E\rangle$ is the mean energy generated per
nuclear reaction and $E_0$ is the energy (temperature) where the
reaction rate is maximized for a given star \cite{kippenhahn,phil}.
In our universe, the parameter $\conlum\approx2\times10^4$
cm$^5$ s$^{-3}$ g$^{-1}$ for fusion via the proton-proton chain under
typical conditions in stellar interiors. With the nuclear parameters
specified, the stellar luminosity is given by 
\be
L_\ast = 4\pi R^3 \rho_c^2 \conlum I(\thetacen) 
\qquad {\rm where} \qquad I(\thetacen) = 
\int_0^{\xi_\star} f^{2n} \xi^2 \Theta^2 \exp[-3\Theta]d\xi\,,
\label{lumintegral} 
\ee
where $\thetacen=\Theta(\xi=0)$. In the integral, the function
$\Theta=\thetacen f^{-1/3}$, so that the expression is specified 
by the solution to the Lane-Emden equation up to the value of
$\thetacen$. 

Note that the definition (\ref{conlumdef}) of the nuclear parameter
$\conlum$ includes a factor of the fine structure constant. This
approach thus implicitly assumes that nuclear yield and reduced cross
section scale with $\alpha$. This choice of scaling is conservative,
in that it results in a smaller region of allowed parameter space than
the alternate scaling (without the factor of $\alpha$): For values of
$\alpha$ larger than in our universe, the total nuclear reaction rate
is exponentially suppressed by the coulomb factor, so this scaling
choice has little effect on the allowed parameter space. For smaller
values of $\alpha$, the value of $\conlum$ is smaller than it would be
otherwise, so that the allowed parameter is smaller. 

Using the energy generation equations, in conjunction with the
equations of energy transport, one can solve for the central
temperature of the star, or equivalently the central value $\thetacen$
of the temperature parameter. This quantity is determined by the
solution to the integral equation
\be 
I(\thetacen) \thetacen^{-8} = {2^{12} \pi^5 \over 45} 
{1 \over \betacon \kappa_0 \conlum E_G^3 \hbar^3 c^2} 
\left( {M_\ast \over \mzero} \right)^4 
\left( {G \mbar \over n+1 } \right)^7 \, . 
\label{tcsolution} 
\ee  
In this expression, $\mbar$ is the mean mass of the particles that
make up the star and $\kappa_0$ is the benchmark value of the stellar
opacity. In addition, $\betacon$ is a dimensionless parameter of order
unity and is determined by the luminosity integral over the structure
of the star, as characterized by the polytropic index $n$.  Note that
both sides of the equation are dimensionless. For the typical
parameter values in our universe, the right hand side of this equation
has a value of approximately $10^{-9}$.

Equation (\ref{tcsolution}) illustrates the parameters that determine
stellar structure solutions. In the crudest approximation, the opacity
scale $\kappa_0 \sim \sigma_T / \mpro$, where $\sigma_T$ is the
Thomson cross section, so that $\kappa_0 \sim \alpha^2$.  The
polytropic index $n$ varies over a small range $(3/2-3)$ as the
stellar mass varies by a factor of 1000 (in our universe).  Moreover,
the dimensionless integrals $\mzero$ and $\betacon$ are always of order
unity. If we fix the particle masses, then the remaining parameters
are the fine structure constant $\alpha$, the gravitational constant
$G$ (equivalently $\alpha_G$), and the composite nuclear parameter
$\conlum$. For a given stellar mass $M_\ast$, we thus have a three
dimensional parameter space $(\alpha,G,\conlum)$ that determines
stellar properties across the multiverse.

With the central temperature $\thetacen$ determined through equation
(\ref{tcsolution}), the equations of stellar structure specify the
remaining properties of the star. The stellar radius $R_\ast$ is
given by 
\be
R_\ast = {G M_\ast \mbar \over k \tcent} 
{\xi_\star \over (n+1) \mzero} \, . 
\label{rstar} 
\ee 
The stellar luminosity $L_\ast$ takes the form  
\be
L_\ast = {16 \pi^4 \over 15} {1 \over \hbar^3 c^2 \betacon \kappa_0 \thetacen} 
\left( {M_\ast \over \mzero} \right)^3 \left( 
{G \mbar \over n + 1 } \right)^4 \, . 
\label{lumstar} 
\ee 
The photospheric temperature $T_\ast$ of the star is then determined from 
the outer boundary condition so that 
\be
T_\ast = \left( {L_\ast \over 4 \pi R_\ast^2 \sigma_{\rm sb}} \right)^{1/4} \,,
\label{tphoto} 
\ee   
where $\sigma_{\rm sb}$ is the Stefan-Boltzmann constant. 

The above solutions specify stellar properties
$(T_c,R_\ast,L_\ast,T_\ast)$ for a given stellar mass $M_\ast$. For a
given set of fundamental parameters $(\alpha,G,\conlum)$, stars can
exist over a finite range of masses (see \ref{sec:massscales}). If a
star has too little mass, then its central temperature cannot become
hot enough to sustain nuclear fusion. If a star has too much mass,
then its equation of state is dominated by radiation pressure and the
structure becomes unstable. As a result, only part of the parameter
space allows for stable, long-lived nuclear burning stars. This
parameter space is determined by the values of the parameters
$(\alpha,G,\conlum)$ for which equations (\ref{tcsolution}
-- \ref{tphoto}) have solutions.

\begin{figure}[tbp]
\centering 
\includegraphics[width=.95\textwidth,trim=0 150 0 150,clip]{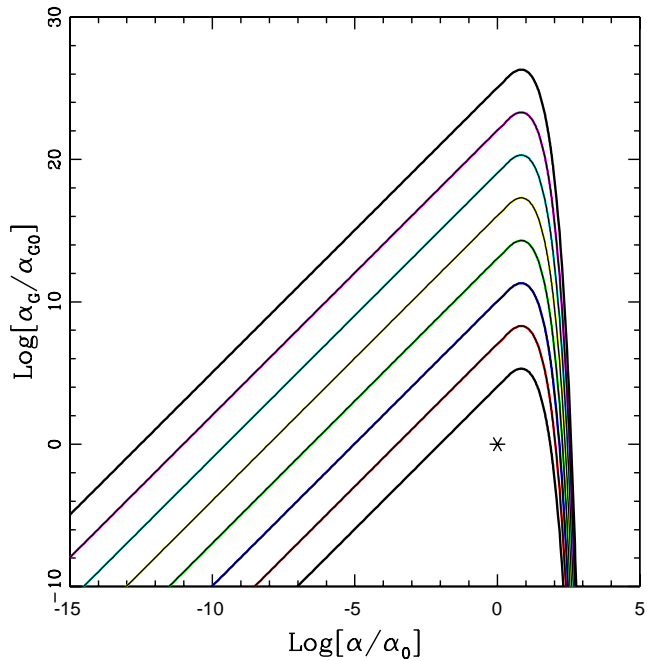}
\caption{Allowed plane of parameter space for the existence of stars. 
The area under each curve represents the allowed values of the
structure constants $\alpha$ and $\alpha_G$ for a given value of the
composite nuclear parameter $\conlum$.  The lower black curve 
corresponds to the value of $\conlum$ in our universe for the p-p 
chain of nuclear reactions. The other curves represent values of 
$\conlum$ that are larger by factors of $10^3$ (red), $10^6$ (blue),
$10^9$ (green), $10^{12}$ (yellow), $10^{15}$ (cyan), $10^{18}$
(magenta), and $10^{21}$ (upper black curve). The location of our 
universe is denoted by the star symbol. The magenta curve, for
$\conlum=10^{18}\conlum_0$, roughly corresponds to stellar nuclear
reactions taking place through the strong force (e.g., in universes
with stable diprotons or a large abundance of deuterium). }
\label{fig:staragc} 
\end{figure} 

The allowed parameter space for working stars is shown in 
Figure \ref{fig:staragc} (adapted from \cite{adams}; see also  
\cite{adamsnew,barnes2015}).  For a given value of the composite
nuclear parameter $\conlum$, the gravitational structure constant
$\alpha_G$ must lie below the curve in order for stable stars to
exist. Curves are shown in the figure for a wide range of values for
$\conlum$, from that in our universe (lower black curve) to a value
$10^{21}$ times larger (upper black curve). Note that the vertical
axis spans 40 orders of magnitude. Moreover, the largest value
$\conlum/\conlum_0=10^{21}$ is larger than that expected for a
universe in which diprotons are stable and nuclear burning can take
place through the strong interaction (see Section \ref{sec:twonucleon}). 
The intermediate curves correspond to values of $\conlum$ that
increase by factors of 1000, from bottom to top. The value of the
nuclear burning parameter $\conlum$ thus acts to change the scale in
the plot, but a large region of parameter space allows for working
stars.

\subsection{Minimum Stellar Temperatures}  
\label{sec:startemp} 

In addition to having sustained nuclear reactions, stars must also be
hot enough to drive chemical reactions and thereby host potentially
habitable planets \cite{kasting,kastingcatling}. This requirement
places constraints on the allowed range of parameters for viable
universes.

The surface temperature $T_{\rm P}$ of a planet is determined by
balancing the incoming radiation from the star and the exhaust heat
from the planet, 
\be
\sigma_{\rm sb} T_{\rm P}^4 = f_T {L_\ast \over 16 \pi d^2} \,,
\label{plantemp} 
\ee
where $d$ is the radius of the planetary orbit (taken to be circular).
The efficiency factor $f_T$ takes into account both the radiation
reflected away from the planet and the heat retained by the
atmosphere. 

The surface temperature $T_{\rm P}$ of the planet must be larger than
the temperature required to drive chemical reactions and hence
support biological operations. This required temperature (energy
scale) is a small fraction of the atomic energy scale and can be
written \cite{bartip} in the form  
\be
E_{\rm chem} = \epsilon_{\rm c} E_{\rm atom} =
\epsilon_{\rm c}\alpha^2 \emass c^2\,,
\label{biotemp} 
\ee
where $E_{\rm atom}=\alpha^2\emass c^2$ is the characteristic energy 
scale for atoms. In our universe, $E_{\rm atom}\approx27$ eV, whereas
chemical reactions take place at room temperature where $E_{\rm chem}$
= $kT \approx 0.026$ eV, so that the chemical conversion factor 
$\epsilon_{\rm c}\sim10^{-3}\ll1$.  Using the requirement that the
planet is warm enough, $kT_{\rm P} \ge E_{\rm chem}$, in conjunction
with the requirement that the orbit must lie outside the star, 
$d \ge R_\ast$, one can derive the constraint
\be
{L_\ast \over R_\ast^2} \simgreat {16 \pi \sigma_{\rm sb} \over f_T} 
\left( {\epsilon_{\rm c} \alpha^2 \emass c^2 \over k} \right)^4  
\propto \alpha^8 \,. 
\label{ratbound} 
\ee

Using the solutions for the stellar radius and stellar luminosity from
equations (\ref{rstar}) and (\ref{lumstar}), the ratio $L_\ast/R_\ast^2$ 
can be evaluated for a given stellar mass, 
\be
{L_\ast \over R_\ast^2} = 
{16 \pi^4 \over 15} {1 \over \hbar^3 c^2 \betacon \kappa_0 \thetacen} 
\left( {M_\ast \over \mzero} \right) 
\left( {G \mbar \over n + 1 } \right)^2  
\left( {k \tcent \over \xi_\star} \right)^2 \,.
\label{ratstar}
\ee
Because the right hand side of this expression increases with mass
$M_\ast$, we can derive a constraint by evaluating it using the
maximum stellar mass for a given universe.  

We thus need to determine the maximum stellar mass for a given set of
fundamental constants. As the mass of a star increases, the fraction
of its internal pressure that is provided by radiation pressure
(instead of gas pressure) increases. Let $\bcon$ denote the fraction
of the pressure provided by the ideal gas law, so that $(1-\bcon)$ is
the fraction provided by radiation. The star becomes unstable when the
radiation pressure dominates \citep{phil}. If we use the critical
value $\bcon\approx1/2$ to specify the pressure fraction for which the
stellar mass is maximum, we find 
\be
M_{\ast {\rm max}} = \left( {18 \sqrt{5} \over \pi^{3/2} } \right) 
\left( {1 - \bcon \over \bcon^4} \right)^{1/2}  
\left( {\mpro \over \mbar} \right)^2 \, \alpha_G^{-3/2} \mpro 
\approx 50 \alpha_G^{-3/2} \mpro \,.
\label{maxmass}
\ee 

The minimum temperature constraint from equation (\ref{ratbound}) can
now be evaluated using equation (\ref{ratstar}) to determine the ratio
$L_\ast/R_\ast^2$ and using equation (\ref{maxmass}) to specify the
maximum stellar mass. The resulting bound becomes 
\be
{\pi^3 \over 15} {1 \over \hbar \betacon \kappa_0 \thetacen^7} 
\left({G \over \hbar c} \right)^{1/2} 
\left( {50 \over \mzero} \right) 
\left( {\mbar \over \mpro (n + 1)} \right)^2
\left( {E_G \over 4 \xi_\star} \right)^2 > 
{\sigma_{\rm sb} \over f} 
\left( {\epsilon_{\rm c} \alpha^2 \emass c^2 \over k} \right)^4 \,.
\ee
This expression includes the temperature parameter $\thetacen$
evaluated at the stellar center, where this quantity is determined by
the stellar structure solution (\ref{tcsolution}) evaluated using the
maximum stellar mass (\ref{maxmass}), 
\be 
I(\thetacen) \thetacen^{-8} = {2^{12} \pi^5 \over 45} 
{\hbar^3 c^4 \over \betacon \kappa_0 \conlum E_G^3} 
{(50)^4 \over \mzero^4} 
\left( {G \mbar^7 \over (n + 1)^7 } \right) \mpro^{-8} \,.
\ee  

Now we can simplify the expressions further. Let $\mbar=\mpro$, 
$n=3/2$, and use the definition of $E_G$, so that the central 
temperature is given by 
\be 
I(\thetacen) \thetacen^{-8} = {2^{23} \pi^5 \over 9} 
{\hbar^3 c^4 \over \betacon \kappa_0 \conlum E_G^3} 
{G \over \mzero^4} \mpro^{-1} \, 
\ee 
and the constraint takes the form 
\be
{\pi^3 \over 30} 
{E_G^2 \over \hbar \kappa_0 \thetacen^7} 
\left( {G \over \hbar c } \right)^{1/2} 
\left( {1 \over \betacon \mzero \xi_\star^2} \right) > 
{\sigma_{\rm sb} \over f} 
\left( {\epsilon_{\rm c} \alpha^2 \emass c^2 \over k} \right)^4 \,.
\ee
This constraint on the fundamental constants is required for 
stars to have surface temperatures hot enough to support 
viable biospheres. 

\subsection{Stellar Lifetime Constraints} 
\label{sec:lifetime} 

For a universe to be habitable, at least some of its stars must live
long enough for biological evolution to take place. Stellar lifetime
increases as stellar mass decreases, so we can constrain the
fundamental parameters by considering stars with the lowest mass. The
minimum mass necessary for sustained nuclear fusion to take place has 
been derived previously \cite{adams,hansen,kippenhahn,phil} and takes 
the form 
\be
M_{\ast{\rm min}} = 6 (3\pi)^{1/2} \left({4\over5}\right)^{3/4} 
\left({kT_{\rm nuc} \over \emass c^2}\right)^{3/4} \alpha_G^{-3/2} \mpro\,.
\label{minmass} 
\ee 
If we invert equation (\ref{minmass}), it determines the maximum
temperature $T_{\rm nuc}$ that can be obtained with a star of a given
mass, where this temperature is an increasing function of stellar
mass. By using the minimum stellar mass from equation (\ref{minmass})
to specify the mass in equation (\ref{tcsolution}), we obtain the
minimum value of the stellar ignition temperature. This central
temperature, or equivalently the value of $\thetacen$, is determined
by solving the following equation
\be
\thetacen I(\thetacen) = \left( {2^{23} \pi^7 3^4 \over 5^{11} } \right) 
\left( {\hbar^3 \over c^2} \right) \left( {1 \over \betacon \mzero^4} \right) 
\left( {1 \over \mpro \emass^3} \right) \left( {G \over \kappa_0 \conlum} \right) \, . 
\label{iprofile}
\ee
The quantities on the right hand side of the equation have been
grouped to include pure numbers, constants that set units,
dimensionless quantities from the polytropic solution, particle
masses, and finally the stellar parameters that depend on the
fundamental constants. Note that this expression has been simplified
by setting $\mbar=\mion=\mpro$ and by using the polytropic index $n$ =
3/2. 

The stellar lifetime $t_\ast$ is determined by the available supply 
of nuclear fuel and can be written in the form  
\be
t_\ast = {f_{\rm c} \effish M_\ast c^2 \over L_\ast} = 
{9375 \over 256 \pi^4} f_{\rm c} \effish \hbar^3 c^4 \betacon \mzero^3 
\kappa_0 \thetacen M_\ast^{-2} 
\left( G \mbar \right)^{-4} \, , 
\label{startime} 
\ee
where $f_{\rm c}$ is the fraction of the stellar material that is
available for fusion and where $\effish$ is the nuclear conversion
efficiency (recall that $\effish \approx 0.007$ in our universe). 
Stars with masses comparable to the Sun have access to only a fraction
$f_{\rm c}\approx0.1$ of their nuclear fuel during the main sequence
phase, whereas smaller stars have larger $f_{\rm c}$
\citep{mdwarf,al1997}.

If the fine structure constant changes, then atomic structure and
atomic time scales are different. We thus want to measure stellar
lifetimes in units of the the time scale for atomic reactions, where
this latter quantity is given by 
\be
t_{\rm A} = {\hbar \over \alpha^2 \emass c^2} \,. 
\ee
This atomic time scale has the value $t_{\rm A}\sim2\times10^{-17}$
sec in our universe. On the other hand, the time required for
biological evolution to develop complex life forms (observers) on
Earth was $\sim1$ Gyr, which corresponds to $\sim10^{33}$ atomic time
units. Given our current sample size of one, the expected
characteristic time scale for biological evolution has enormous
uncertainty \cite{orgel,spiegel}, and many interpretations are
possible \cite{carter1983,carter2008}. For the sake of definiteness,
this treatment considers the terrestrial value of 1 Gyr as a fiducial
time scale, although it does not represent a definitive limit (see
also \cite{chyba,mckay}). In other words, we use the reference value
of $10^{33}$ atomic time units as a starting point.

The ratio of the stellar lifetime to the atomic time scale can be 
written in the form 
\be
{t_\ast \over t_{\rm A}} = 
{9375 \over 256 \pi^4} f_{\rm c} \effish \hbar^2 c^6 \betacon \mzero^3 
\kappa_0 \alpha^2 \emass \thetacen M_\ast^{-2} 
\left( G \mbar \right)^{-4} \, . 
\ee
In other universes, the largest possible value of this ratio, 
corresponding to the smallest, long-lived stars, is thus given by 
\be
\left( {t_\ast \over t_{\rm A}} \right)_{\rm max} = 
\left( {5^{13/2} \over 9 \pi^8 2^{10}} \right) 
\left( {c^3 \over \hbar} \right) 
\left( f_{\rm c} \effish \betacon \mzero^3 \right) 
\left( {\emass^{5/2} \mpro^{5/2} \over \mbar^4} \right)
\left( {\kappa_0 \over G \alpha} \right) 
\thetacen^{11/2} \,,
\label{timerat} 
\ee
where we have grouped the various factors as before.  Note that
equation (\ref{timerat}), as written, depends on the temperature
parameter $\thetacen$, which is specified via equation
(\ref{iprofile}). We can thus combine equations (\ref{iprofile})
and (\ref{timerat}) to solve for the ratio of time scales, and 
set it equal to the minimum required for life to develop (here 
we use $t_\ast/t_{\rm A} > 10^{33}$ as described above).   

\begin{figure}[tbp]
\centering 
\includegraphics[width=.95\textwidth,trim=0 150 0 150,clip]{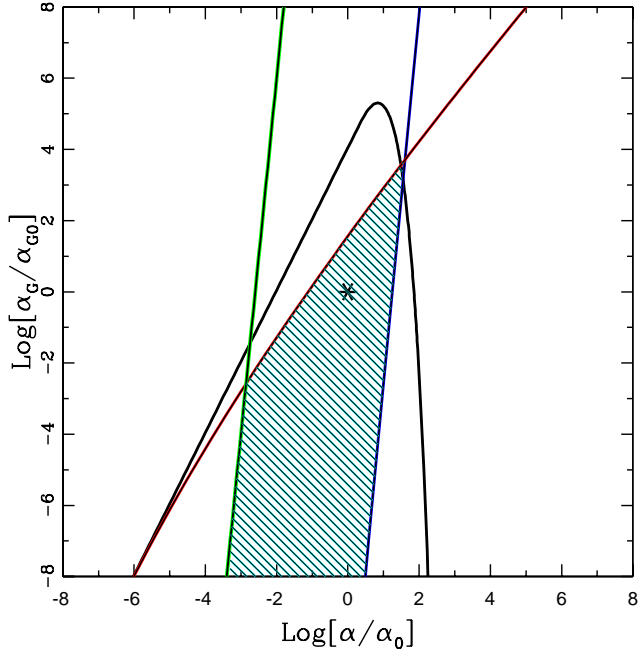}
\caption{Allowed plane of parameter space for the existence of stars 
with differing values of the structure constants $\alpha$ and
$\alpha_G$ \cite{adamsnew}. The shaded region delineates the portion
of the plane that remains after enforcing the following constraints:
The black curve shows the requirement that stable stellar
configurations exist. The blue curve shows the requirement that the
stellar temperature is high enough to allow habitable planets. The red
curve shows the constraint that stars live long enough for biological
evolution to occur ($10^{33}$ atomic time scales). Finally, for stars 
to have smaller masses than their host galaxies, $\alpha$ must fall to 
the right of the green curve. The location of our universe is depicted
by the star symbol in the center of the diagram. }
\label{fig:starplane} 
\end{figure} 

The requirements of a minimum stellar lifetime (this section, 
\ref{sec:lifetime}) and a minimum photospheric temperature 
(Section \ref{sec:startemp}) place further constraints on the allowed
parameter space for stars. This trend is illustrated in Figure 
\ref{fig:starplane}, which shows the viable regime of parameter
space for stars with the value of the nuclear burning parameter
$\conlum$ for hydrogen burning stars in our universe. The black curves
delineates the portion of the $\alpha$-$\alpha_G$ plane that allows
for working stars. In order for the stars to live long enough (using
the benchmark value of $10^{33}$ atomic time scales), the parameter
space is limited to lie below the red curve. In order for the stars to
have a sufficiently hot surface temperature, the parameter space must
fall to the left of the blue curve. Finally, the figure enforces an
additional requirement, that the stellar mass scale must be less than
the characteristic mass scale for a galaxy (see \ref{sec:massscales}).
This constraint limits the parameter space to the right of the green 
curve. These additional constraints thus remove about half of the
original parameter space shown in the diagram.

Figure \ref{fig:starplane} also provides constraints on the ratio of
the strengths of the gravitational and electromagnetic forces. In our
universe, this ratio is notoriously small, with
$\alpha_G/\alpha\sim10^{-36}$. In order for working stars to exist
(black curve), this ratio can be larger by a factor of $\sim10^4$. 
If we also require that the stars live for $10^{33}$ atomic time
scales (red curve), then the ratio can only be larger by a factor of
$\sim100$. Both of these factors are small compared to the observed
ratio.  As a result, viable universes are constrained to have an
extreme hierarchy of force strengths, provided that the nuclear
burning parameter $\conlum$ is comparable to that in our universe.
This degree of hierarchy between gravity and electromagnetism can be
made much less extreme for much larger values of $\conlum$, as shown
in Figure \ref{fig:staragc}. Specifically, if the parameter $\conlum$
is increased by a factor of $10^{21}$, which corresponds to nuclear
reactions occurring via the strong force instead of the weak force,
then the ratio $\alpha_G/\alpha$ can be larger by a factor of 
$\sim10^{24}$ (see also Section \ref{sec:twonucleon}).  

\subsection{The Triple-Alpha Reaction for Carbon Production} 
\label{sec:triplealpha} 

In our universe, the production of carbon takes place through a
somewhat convoluted process known as the triple alpha reaction
\cite{clayton,kippenhahn}. Because of the importance of carbon 
to life forms and the delicate nature of this requisite reaction,
carbon production is often used as an example of fine-tuning 
\cite{aguirre,barnes2012,bartip,carr,hogan}. Even for our own 
universe, the exact determination of the reaction rate and its 
dependence on fundamental parameters have been elusive (see 
\cite{deboer} for a recent review). 

The basic complication for carbon production, and hence the necessity
of the triple alpha reaction, arises because our universe has no
stable nucleus with atomic mass number $A=8$. After a star has
processed the hydrogen in its central core into helium, it adjusts its
internal structure by condensing so that the central temperature and
density increase. The seemingly obvious next step would be to fuse the
helium nuclei (alpha particles) into $^8$Be nuclei. Unfortunately,
this channel is unavailable because $^8$Be is unstable. In the absence
of this natural stepping stone, helium fusion must take place through
the triple alpha process \cite{clayton,kippenhahn},
\be
3\,\left(^4{\rm He}\right) \to \, {}^{12}\rm{C} + \gamma \,, 
\ee
where three helium nuclei combine to forge carbon. This reaction
relies on the formation of a transient population of $^{8}$Be
\cite{salpeter}. More specifically, helium nuclei continually combine 
to make $^8$Be, which decays back into its constituent particles with
a half-life of about $\tau\sim10^{-16}$ sec. Because they are mediated
by the strong force, these reactions occur rapidly enough to maintain
nuclear statistical equilibrium (NSE). This small and transient
population of $^8$Be is large enough to support the additional reaction 
\be
^4{\rm He} \, + {}^8{\rm Be} \to \, {}^{12}\rm{C} \,.
\label{hebe2c} 
\ee
In order for this latter reaction to operate fast enough, however, it
must take place in a resonant manner, which requires the $^{12}$C
nucleus to have a resonance at a particular energy. The existence and
energy level of this resonance was predicted by Hoyle \cite{hoyle}. 
Subsequent experiments measured the resonance in the laboratory 
\cite{dunbar} and provided a remarkable confirmation of this 
paradigm for carbon synthesis (see the review of \cite{fowler}).

The resonance in question is the $0^+$ excited state of the $^{12}$C
nucleus and has an energy of 7.6444 MeV, where this energy lies just
{\it above} that of a $^8$Be nucleus and an alpha particle considered
separately (given by the left side of equation [\ref{hebe2c}]). The key
question is thus how precisely specified this energy level must be in
order for stars to produce a significant amount of carbon. To address 
this issue, we start with some definitions: 

Although the reaction $^4$He + $^4$He $\to$ $^8$Be is not
energetically favored, unstable $^8$Be can be formed, with the
reaction rate (e.g., see \cite{epelbaum2013}) controlled by the energy
difference 
\be
(\Delta E)_b \equiv E_8 - 2 E_4 \,,
\ee
where $E_8$ and $E_4$ are the ground state energies of $^8$Be and $^4$He. 
The ground state of the carbon nucleus is denoted here as $E_{12}$ and
the excited state is $E_{12}^\star$. The energy difference between the
excited carbon nucleus and the reactants in equation (\ref{hebe2c}) 
is then given by
\be
(\Delta E)_h = E_{12}^\star - E_8 - E_4 \,. 
\ee
Finally, one can define the energy scale $E_R$ 
of the resonant reaction according to 
\be
E_R \equiv (\Delta E)_b + (\Delta E)_h = E_{12}^\star - 3E_4 \,. 
\ee
The currently measured value of this energy level is
$(E_R)_0\approx379.5$ keV.  With the above definitions, 
the resonant reaction rate $R_{3\alpha}$ is given by  
\be
R_{3\alpha} = 3^{3/2} n_\alpha^3 
\left( {2\pi\hbar^2 \over |E_4|kT} \right)^3 
{\Gamma_\gamma \over \hbar} 
\exp \left[ - {E_R\over kT} \right] \,,
\label{trialpharate} 
\ee
where $\Gamma_\gamma$ is the radiative width of the Hoyle state. Given
that the resonance energy $E_R$ appears in the exponential term, its
value determines the net reaction rate for the entire process. Here 
we let the resonance level vary up or down by an increment $\Delta E_R$ 
given by 
\be
\Delta E_R \equiv E_R - (E_R)_0 \,,
\label{defresdif} 
\ee
where the subscript denotes the value in our universe. 

If the energy level of the $^{12}$C resonance is higher, 
$\Delta E_R>0$, then carbon production is suppressed. At a given
temperature, a higher energy level for the resonance leads to a lower
reaction rate for helium burning (carbon production), so the stellar
core adjusts to a higher temperature in order for nuclear reactions to
support the star.  At the same time, however, this higher temperature
allows any extant carbon nuclei to fuse into oxygen through the
reaction 
\be
^4{\rm He} \, + {}^{12}\rm{C} \to \, {}^{16}\rm{O} \,.
\label{hec2o} 
\ee
The energy level of $^{16}$O lies at 7.1187 MeV, which is below the
combined energy of the reactants $^{12}$C and $^{4}$He (with energy
7.1616 MeV). The basic problem is that if the rate of burning carbon
into oxygen becomes large compared to the rate of making carbon, then
any carbon produced by stellar nucleosynthesis will immediately be
transformed into oxygen. No carbon would be left for making life forms
and other structures. 

Although the general trend of nuclear evolution in stars is to forge
ever larger nuclei (like burning carbon into oxygen), the case of the
triple alpha reaction is on a different footing. In general, the
production of larger nuclei involves larger coulomb barriers, which in
turn requires higher temperatures in the stellar core. Because carbon
production is suppressed by the requirement of passing through
unstable $^8$Be, the temperature required to make carbon is higher
than it would be otherwise and is comparable to the temperature
required to destroy it (e.g., through equation [\ref{hec2o}]).
This coincidence renders carbon production sensitive to the exact
energy levels of the $^{12}$C nucleus.

The sensitivity of the energy level of the carbon resonance has been
explored in a number of papers. One straightforward approach is to
vary the energy level of the $^{12}$C nucleus, but keep all other
parameters the same. At the fundamental level, changes in the excited
state of nuclei are determined by changes in the strengths of the
fundamental forces, especially the strong and electromagnetic
interactions \cite{epelbaum2011,epelbaum2012,epelbaum2013,epelbaumtwo,
lahde,meissner,meissner2}. Changes in these interaction strengths 
would affect all nuclear structures, not just the energy level of the
$^{12}$C resonance of interest here.  Nonetheless, variations in the
resonance energy provide a good starting point for understanding the
sensitive dependence of carbon production on its value.

It is important to note that carbon production also occurs through
a non-resonant reaction. The non-resonant contributions are larger 
than the resonant terms \cite{nomoto} for temperatures lower than 
$T\sim3\times10^7$ K (for $\alpha+\alpha\to$ $^{8}$Be) and
$T\sim7\times10^7$ K (for $\alpha$ + $^{8}$Be $\to$ $^{12}$C).  
As a result, if the resonance level is changed so that carbon is not
efficiently produced, the non-resonant reactions become important.

The first such treatment \cite{livio} considered the evolution of
stars with $M_\ast=20M_\odot$ as well as shell helium burning in
Asymptotic Giant Branch (AGB) stars. They found that increasing the
energy of the $0^+$ energy level of $^{12}$C by $\Delta{E}_R=+60$ keV
does not affect the amount of carbon produced in stellar interiors,
but a larger increase of $\Delta{E}_R=+277$ keV leads to significant
changes in nuclear burning patterns and relatively little carbon is
produced. In AGB stars, an increase in the resonance level also leads
to reduced carbon production, but the strength of the thermal pulse is
increased.  Stronger pulsations can increase the amount of (nuclear
processed) material that is transferred to the outer layers of the star
and ultimately distributed into the interstellar medium. Finally, if
the resonance level of the carbon nucleus is decreased, then the
amount of carbon produced can be increased significantly.

Subsequent studies \cite{csoto,oberhummer,schlattl} find similar
results to those outlined above, although stars with different masses
show somewhat different dependence on the value of $\Delta{E}_R$. For
example, stars with masses of $M_\ast$ = 15 and 25 $M_\odot$ were
considered in \cite{schlattl}. For an increase in the resonance level
of $\Delta{E}_R=+100$ keV, the carbon yield from 15 (25) $M_\odot$
stars decreased by a factor of 7 (17). On the other hand, for a
decrease in the energy of the resonance, $\Delta{E}_R=-100$ keV,
carbon production increased by a factor of $\sim3$ (7.5).

\begin{figure}[tbp]
\centering 
\includegraphics[width=.95\textwidth,trim=0 150 0 150,clip]{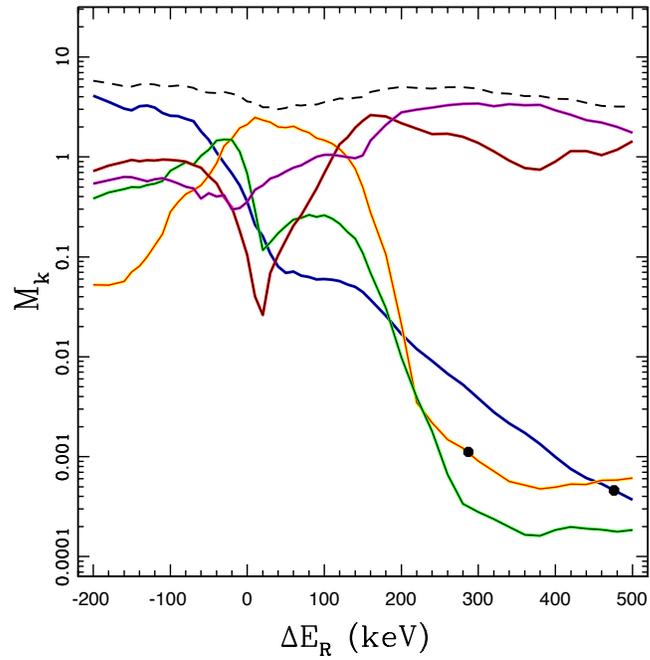}
\caption{Alpha element yields in massive stars as a function of the 
$0^+$ resonance energy of the carbon nucleus (adapted from \cite{huang}).
The resonance energy is specified on the horizontal axis by the
difference $\Delta{E}_R$ from the value in our universe (in keV).
Results are shown for calculations using the {\sl MESA} stellar
evolution package for stellar masses in the range $M_\ast$ = 
$15-40M_\odot$. The yields (in $M_\odot$) for each element are the
expectation values of the mass contained in that species produced 
by the stars (where the mean value is taken by integrating over the
stellar mass range weighted by the initial mass function). Yields are
shown for carbon (blue), oxygen (orange), neon (green), magnesium
(red), and silicon (magenta). The upper dashed black curve shows the
sum of the mass in these five elements. The black symbols on the
curves for carbon and oxygen show the initial values for the starting
metallicity $\metal=10^{-4}$. }
\label{fig:carbonyield} 
\end{figure}

Figure \ref{fig:carbonyield} provides an overview of results from more
recent stellar evolution calculations \cite{huang} that determine the
abundances of intermediate alpha elements, including carbon and
oxygen, as a function of the change $\Delta{E}_R$ in the resonance
energy. The figure shows results calculated using the {\sl MESA}
stellar evolution code \cite{mesaone,mesatwo} for stellar masses in
the range $M_\ast=15-40M_\odot$. The curves show the expectation
values for the mass in a given element produced per star during the
course of stellar evolution. These values are obtained by integrating
the yields over the stellar mass range, weighted by a stellar initial
mass function of the form $dN/dm\sim m^{-2.3}$ \cite{salpeterimf}.
Results are shown for yields of carbon (blue), oxygen (orange), neon
(green), magnesium (red), and silicon (magenta). The sum of the mass
in these elements is given by the black dashed curve. The stellar
models depicted in the figure began with metallicity $\metal=10^{-4}$
for computational convenience.  The black symbols on the curves for
carbon and oxygen show that expectation values of the mass in those
elements at the start of the simulations.

Several trends are evident from Figure \ref{fig:carbonyield}: The
carbon yields are a steeply decreasing function of the resonance
energy for values near that of our universe. Nonetheless, the carbon
yields are larger than the starting values (for $\metal=10^{-4}$) for
energy increments as large as $\Delta{E}_R\approx+480$ keV.  As a
result, somewhat larger changes in the energy level ($\Delta{E}_R$
$\simgreat+500$ keV) can seriously compromise carbon production. In
contrast, decreasing the energy level results in an {\it increase} in
mean carbon production, with $\Delta{E}_R=-200$ keV providing a factor
of $\sim10$ enhancement. Carbon production continues down to energy
increments $\Delta{E}_R\approx-300$ keV, where carbon production
ceases to be energetically favorable (in addition, the temperature
dependence of the triple alpha reaction rate ceases to allow for
stellar stability at similar values of $\Delta{E}_R$ \cite{huang}).
The oxygen yields are roughly anti-correlated with those of carbon.
Decreasing the resonance energy level $\Delta{E}_R$ results in less
oxygen production. Increasing the resonance level up to about
$\Delta{E}_R\sim+100$ keV leads to enhanced oxygen yields, but for
even larger increases in the resonance level the oxygen production
plateaus and then declines. The oxygen abundance falls below its
starting value in these simulations for $\Delta{E}_R\simgreat+280$
keV. Oxygen, which is necessary to make water, becomes scarce for
larger values.

For the five alpha elements shown in Figure \ref{fig:carbonyield},
each species has the maximum abundance for a range of $\Delta{E}_R$.
For increasing values of the resonance increment, the peak values
occur for carbon, neon, oxygen, magenta, and then silicon. With the
exception of neon, these peaks are found in order of increasing atomic
number. As the resonance energy (given by $\Delta{E}_R$) increases,
the operating temperature of the stellar core increases, and larger
elements can be synthesized. Notice also that the sum of the mass
(black dashed curve) contained in these five alpha elements is nearly
constant: Most of the mass in metals outside the iron core (which
becomes either the neutron star or black hole produced by the ensuing
supernova explosion) is contained in these five species, and their
mass contributions show nearly zero-sum behavior.

In approximate terms, these results can be summarized as follows:
Changing the carbon resonance energy by increments of $\sim300$ keV
changes the carbon and oxygen yields (in opposite directions) by
roughly an order of magnitude. The total range in the triple alpha
resonance energy for viable carbon production is given by $-300$ keV
$\simless\Delta{E}_R\simless+500$ keV, with the range for oxygen
somewhat smaller. Notice that the nuclear yields depend on stellar
mass (not shown -- see \cite{huang}).  Although the stellar mass
function is remarkably robust within our universe, it could vary in
other universes, and such variations would affect the masses in the
chemical species averaged over the stellar population.  Finally, note
that the mean yields of carbon and oxygen are equal at a resonance
level just below that of our universe (for $\Delta{E}_R\approx-35$
keV).

As illustrated in Figure \ref{fig:carbonyield}, stellar evolution
calculations indicate that the ratio of carbon yield to oxygen yield
is of order unity for massive stars in our universe. This finding is
consistent with the carbon-to-oxygen (C/O) ratio of [C/O]$\approx0.44$ found
in the Galaxy, the ratio [C/O]$\approx0.67$ observed for the Solar Wind
\cite{vonsteiger}, and the value [C/O]$\approx0.51$ measured for the Solar 
System as a whole \cite{cameron1973}.  Similarly, the C/O ratio has
been measured for nearby stellar populations in our universe,
including all stars in the Solar neighborhood \cite{brewer} and those
that are known to host extrasolar planets \cite{nissen}. The observed
distribution is relatively wide, with C/O ratios varying by at least a
factor of $\sim3$, and falling in the range [C/O]$\approx0.25-0.75$.
Although these values produce a consistent picture, one should keep in
mind that intermediate mass stars also provide a substantial
contribution to the carbon inventory of the universe.

In contrast to observed stellar abundances, the C/O ratio for Earth is
anomalously low, with estimates ranging from [C/O]$\approx0.01$ 
\cite{allegre} down to [C/O]$\approx0.002$ \cite{marty}. Earth is thus 
severely depleted in carbon relative to the Sun.  Although increasing
the carbon-12 resonance energy lowers both the carbon abundance and
the C/O ratio, the change must be larger than $\Delta{E}_R\approx+300$
keV in order to make these values smaller than the observed carbon
abundance and C/O ratio of Earth. As a result, the range of resonance
energy increment $-300$ keV $\simless\Delta{E}_R\simless+300$ keV
produces cosmic carbon abundances that are larger than those of Earth
--- which is the only place in the universe where life is known to
exist.  Unfortunately, the minimum carbon abundance required for
habitability remains unknown, but these considerations suggest that
the corresponding maximum value of the resonance energy must be larger
than $\Delta{E}_R\approx+300$ keV. A related unresolved issue is the
relationship between cosmic abundances, stellar abundances, and the
chemical compositions of planets \cite{bergin,meyer}, which form
within circumstellar disks associated with forming stars (see also
Section \ref{sec:planets}). For example, the chemical compositions of
planetessimals (building blocks of planets) are affected by both the
C/O ratio of the system and by their radial location within the
disk \cite{tvjohnson}. In our Solar System, the planet Venus has a
relatively low C/O ratio comparable to that of Earth \cite{venus},
whereas the majority of asteroids are carbonaceous (C-type), with a
carbon-rich composition close to solar \cite{asteroid} (after
accounting for the depletion of hydrogen, helium, and other
volatiles).

The above considerations indicate that the change $\Delta{E}_R$ in the
energy level of the carbon-12 resonance can vary over a total range of
$\sim800$ keV and still allow the universe to be viable. The next
question is whether this range is large or small.  The criterion for
an enhanced nuclear cross section is that the particle energies are
near a resonance, where the resonances are given by the energy levels
of the nucleus. In general, excited nuclear states are spaced at
intervals of order $\sim1$MeV \cite{caurier}. More specifically, the
carbon nucleus has excited states at $E$ = 4.44, 7.65, 9.64, 12.7, and
15.1 MeV, so that the energy intervals are about 3 MeV. Given the
allowed range of $\Delta E_R$, the chance of being sufficiently near
resonance is about 1 part in 4.

In addition to carbon and oxygen, the abundances of other elements
will be affected by changes to the carbon-12 resonance level. As one
example, the radionuclides $^{26}$Al and $^{60}$Fe, with half-lives of
order 1 Myr, provide an important energy source during the epoch of
planet formation. The abundances of these radioactive species are
particularly sensitive to variations in $\Delta{E}_R$ \cite{tur}. 

Another class of research considers the time variation of the 
constants of nature in our universe \cite{langacker,uzan,uzantwo},
including constraints from carbon production of Population III stars
\cite{eckstrom}. Possible variations in the constants of physics lead 
to corresponding variations in stellar properties \cite{vieira}.  In
this setting, however, the constants are constrained to have only
small variations. Over the age of the universe, the fine structure
constant has variations smaller than
$(\Delta\alpha)/\alpha\sim10^{-5}$. Such variations are thus too small
to affect carbon production at earlier epochs in our universe.

The above discussion indicates that changes in the carbon resonance
over a range of $\sim800$ keV allow for stars to produce carbon and
oxygen at acceptable levels, within an order of magnitude of values in
our universe, but larger changes are potentially problematic. As a
reference point: In order to move the resonance energy by an increment
of 100 keV, the corresponding change in the nuclear force strength
(e.g., \cite{schlattl}) is estimated to be $\sim0.5\%$ and/or the
change in the electromagnetic interaction is $\sim2-4\%$ (see also
\cite{coczero,cocone,eckstrom,epelbaum2013,epelbaumtwo}). On the other hand, 
the $^8$Be nucleus only fails to be bound by 92 keV. As a result, the
degree of change to nuclear physics required to compromise carbon
production is significantly larger than that needed to allow for
stable $^8$Be.  Detailed calculations using Lattice Chiral Effective
Field Theory confirm that bound states of $^8$Be require the strengths
of the nuclear and/or electromagnetic forces to vary by only about
$\sim1-2\%$ \cite{epelbaum2013}. Moreover, a stable isotope with mass
number $A=8$ removes the need for the triple alpha process altogether
\cite{agalpha,higa}. In universes with stable $^8$Be, stars can burn 
helium into beryllium, and later burn the beryllium into carbon. With
stable $^8$Be, the necessary carbon producing reactions can take place
at later evolutionary stages within the same star or in different
stars in later generations. Although stellar evolution is modified in
universes with stable $^8$Be, the epoch of Big Bang Nucleosynthesis
will continue to produce light nuclei with the usual abundances 
\cite{coc2007,coczero,scherrer2}. The conditions of low density, high
entropy, and short evolutionary time scale prevent BBN from producing
elements of larger atomic numbers.

If stars fail to make carbon through the triple alpha reaction, they
tend to produce other alpha elements with larger atomic numbers. In
this case, some carbon can still be synthesized through spallation
reactions with high energy cosmic rays \cite{spallreview}. 
Typical reactions include 
\be
^{16}{\rm O}+p\to ~^{12}{\rm C}+3p+2n
\qquad {\rm and} \qquad 
^{16}{\rm O}+p \to ~^{13}{\rm C}+3p+n\,,
\label{spallation}
\ee 
and many others. The cross sections for such reactions are of order
$\sigma\sim100$ mb \cite{spallupdate,spallcross}. For the cosmic ray 
flux of our Galaxy, the rate at which oxygen is converted into carbon
in the interstellar medium is only of order $3\times10^{-8}$ per
Hubble time per target nucleus. These rates can be enhanced in
supernova remnants \cite{spallsnr}, in planetary atmospheres 
\cite{spallatmos}, and by the inclusion of high energy photons 
\cite{spallradiation}. Since stars in other universes readily 
synthesize alpha elements, and supernovae are expected to accelerate
cosmic rays, spallation will always produce nonzero abundances of
carbon. Nonetheless, the expected mass fractions are much lower than
those observed in our universe (unless the cosmic ray flux is 
significantly enhanced). 

For completeness, we note that if the universe fails to make enough
carbon, another --- {\sl highly speculative} --- channel for life
could exist. The reason for low carbon abundance is not that it cannot
be synthesized, but rather that it is immediately processed into
heavier elements (first oxygen, then larger nuclei). The stars of
sufficiently high mass are thus likely to make appreciable amounts of
silicon (Figure \ref{fig:carbonyield}).  The idea of silicon based
life has long been the subject of science fiction, but scientists have
not ruled out the possibility.  The molecular diversity of life is
becoming understood \cite{pace1997} and life is known to thrive in
extreme environments \cite{grotzinger,rothschild}.  Recently, some
steps toward realizing silicon-based life have been carried out 
\cite{siliconlife}.  Both carbon and silicon can produce large
molecules that can (in principle) carry large amounts of biological
information \cite{pace2001}. On the other hand, carbon readily
produces chemical bonds with many other atoms, whereas silicon tends
to interact with many fewer species. As a result, carbon allows for
much greater chemical versatility, which in turn increases its
efficacy as a basis for biology. In addition, chemical reactions with
silicon are generally slower than those with carbon and the chemical
bonds (e.g., Si-Si) are weaker. Nonetheless, if life could exist with
a silicon architecture, then observers could still arise in
carbon-poor universes, and life would not depend on the triple alpha
process.

\subsection{Effects of Unstable Deuterium and Bound Diprotons on Stars} 
\label{sec:twonucleon} 

Another standard example of possible fine-tuning arises for nuclei
with atomic mass number $A=2$. In our universe, deuterium is stable,
whereas diprotons and dineutrons do not have stable bound states. A
common assertion in the existing literature of fine-tuning constraints
is that small changes in the strength of the strong nuclear force, in
either direction, would render the universe lifeless (for example, see
\cite{bartip,davies1972,dentfair,donoghuethree,dyson1971,hogan,pochet,
reessix,schellekens,tegmarktoe,tegmark} and references therein). The 
required change in the strong force is estimated to be $\sim15\%$. 

If the strong force were stronger, then diprotons and dineutrons could
be bound, and nuclear reactions in stars could take place through the
strong interaction, in contrast to the case of our universe where weak
interactions play an important role. Given the much larger reaction
rates, the concern is that stellar lifetimes could become too short.
On the other hand, if the strong force were somewhat weaker, then
deuterium would not be bound, and no stable $A=2$ nuclei would exist.
In our universe, both BBN and stellar nucleosynthesis rely on stable
deuterium as a stepping stone on the path to larger nuclei. Here the
concern is that no heavy nuclei can be made in the absence of stable
deuterium. In spite of the ubiquity of these arguments, recent
calculations of stellar structure and evolution show that stars can
serve as power sources for habitability in universes with stable 
diprotons \cite{barnes2015,bradford2009} (see also
\cite{adams,adamsnew}) and in universes with unstable deuterium 
\cite{agdeuterium,barnes2017}. These scenarios are discussed in 
the following subsections.

\subsubsection{Universes with Stable Diprotons} 
\label{sec:stabledip} 

In universes with stable diprotons or dineutrons, nuclear reactions
can be mediated by the strong interaction (without requiring the weak
interaction). This behavior stands in contrast to our universe, where
nuclear reactions in stars must convert four protons into a helium
nucleus.  The net reaction, proceeding through either the p-p chain or
the CNO cycle \cite{clayton,kippenhahn,phil}, necessarily involves the
transformation of two protons into two neutrons and hence requires the
weak force. Nuclear reactions involving only the strong force
generally have much larger reaction rates, and can potentially alter
the course of stellar evolution.

The magnitude of the change, from weak to strong nuclear reactions,
can be illustrated by the case of deuterium burning in stars in our
own universe. The nuclear reaction parameter $\conlum$ (see equation
[\ref{conlumdef}]) for deuterium burning can be written in the form
\be
\conlum \approx 2.1 \times 10^{17} \left({X_H \over 0.71}\right)^2 
\left({{\rm [D/H]}\over2\times10^{-5}}\right) 
\,{\rm cm}^5\,{\rm s}^{-3}\,{\rm g}^{-1} \,,
\label{conlumdeut} 
\ee
where the value is scaled to the deuterium abundance in our universe
\cite{barnes2015}. This value should be compared to that of ordinary 
hydrogen burning ($\conlum\approx2\times10^4$ in the same units), which
involves the weak interaction. Although nuclear physics in alternate
universes allows for a range of possible binding energies and reaction
cross sections for $A=2$ nuclei, a good starting assumption is that
nuclear burning rates in universes with stable diprotons are roughly
comparable to the case of deuterium. If the mass fraction of nuclear 
fuel is of order unity, comparable to hydrogen in our universe, then 
the nuclear reaction parameter for diprotons becomes $\conlum$ 
$\sim10^{22}$ cm$^5$ s$^{-3}$ g$^{-1}$, which is larger than the value
for the p-p reaction (in our universe) by a factor of $\sim10^{18}$.

Before analyzing the effect of changing the nuclear reaction rate by
such a large factor, we note that an analogous process takes place in
our universe. Star formation does not involve nuclear reactions, so
that stars are not born with the proper configurations required for
fusion to take place \cite{sal1987}. Instead, most stars (those with
masses $M_\ast\simless7M_\odot$) are born with large stellar radii
($\sim3-4$ times larger than their main sequence values) and central
temperatures that are too low to sustain hydrogen burning reactions 
\cite{stahlerbirth,stahlershutaam}. Young stars derive their energy 
from gravitational contraction and evolve over millions of years
without hydrogen fusion. As the stars condense, their central regions
eventually become hot enough ($T_{\rm c}\sim15\times10^6$ K) for
hydrogen burning to occur. However, well before the central
temperature reaches this benchmark value, deuterium fusion takes place
when the central temperature becomes $T_{\rm c}\sim10^6$ K (with
$\conlum$ given by equation [\ref{conlumdeut}]). Nothing disastrous
occurs when stars reach the configuration where deuterium burning
takes place. They briefly derive additional energy from this nuclear
process, which briefly delays their evolution. Because of the low
deuterium abundance, this phase is short compared to other stellar
timescales (of order 0.1 -- 1 Myr), and the stars subsequently
continue their contraction toward the main sequence.

In universes with stable diprotons, a number of different nuclear
reaction chains could be realized. The reaction networks will depend
on the values of the cross sections, the binding energies, and the
abundances of protons, diprotons, deuterium, and the helium isotopes
(in the stellar interior). As one example, consider the care where the
universe emerges from its BBN epoch with protons as the dominant
nuclear species. The first stage of nuclear burning in stars involves
the production of diprotons ($p+p\to$ $^2$He + $\gamma$), which takes
place through the strong interaction. The diprotons can then capture
free electrons to become deuterium, and subsequent reactions are the
same as those in the standard p-p chain \cite{clayton,hansen}. Note
that electron capture involves the weak force, but proceeds more
rapidly than the usual reaction (equation [\ref{ppchain1}]) due to the
absence of a coulomb barrier. Moreover, the starting reaction takes
place at a lower temperature than for stars in our universe (as
outlined above). As a result, the final step of the p-p chain ($^3$He
+ $^3$He $\to$ $^4$He $+2p$) does not occur promptly: The star first
builds up a abundance of $^3$He while the core operates at 
$T_{\rm c}\sim10^6$ K, and then produces $^4$He in a later stage of
nuclear burning. Other scenarios are possible, but in general the
nuclear burning parameter $\conlum$ is expected to be greatly
enhanced.

Stellar structure is relatively insensitive to the value of
$\conlum$. Figure \ref{fig:staragc} shows the allowed regions of the
$(\alpha,\alpha_G)$ parameter space for stars with a wide range of
nuclear burning parameter $\conlum$, extending up to values $10^{21}$
times larger than that of p-p burning in our universe. Stable,
long-lived stellar configurations are thus possible over an enormous
range of the parameter $\conlum$. In fact, as the value of $\conlum$
increases, the allowed region of the $(\alpha,\alpha_G)$ parameter
space becomes {\it larger}. Stellar structure solutions thus exist for
stars operating through the strong interaction.

The insensitivity of stars to the nuclear parameter $\conlum$ arises
for a number of (coupled) physical reasons: Stars are supported by
ordinary gas pressure, which balances gravity for any given stellar
mass, largely independent of the energy generation mechanism.  The
internal structure of the star adjusts itself so that the nuclear
reaction rate is whatever it needs to be to provide the required
pressure. If the nuclear parameter becomes larger (smaller), then the
star compensates by making its central temperature and density lower
(higher) to maintain the same pressure support. In addition, the
nuclear reactions take place via quantum mechanical tunneling. The
normal operating state of the star is such that the central
temperature is far lower than that required for protons to directly
overcome their mutual coulomb barrier (and thereby drive nuclear
reactions directly). If such conditions did not prevail, the star
would resemble a nuclear bomb rather than have a stable long-lived
structure. In any case, the nuclear reaction rate is exponentially
sensitive to the central temperature, so that small changes in 
$T_{\rm c}$ can compensate for large changes in $\conlum$.

This robust nature of the stellar structure solutions is exemplified
by equation (\ref{tcsolution}), which specifies the central
temperature required for sustained nuclear reactions. The
dimensionless integral $I(\thetacen)$ that determines the nuclear
burning temperature is a decaying exponential function of the
temperature parameter $\thetacen\propto{T_{\rm c}}^{-1/3}$, and can be
approximated with a fitting function of the form \cite{adamsnew} 
\be
I(\thetacen) \approx 0.83 \thetacen^{1.3} \exp[-3\thetacen] \,,
\label{ifit} 
\ee
where the numerical values correspond to stellar models with
polytropic index $n=3/2$ and radiative energy transport. The
combination of equations (\ref{tcsolution}) and (\ref{ifit}) shows
that the central temperature parameter $\thetacen$ depends only
logarithmically on the value of the nuclear burning parameter
$\conlum$. Increasing $\conlum$ by a factor of $10^{17}$ thus only
increases $\thetacen$ by a modest factor (an increment of
$\sim17\log(10)/3\approx13$), enough to lower the central temperature
of the star down to $T_{\rm c}\approx10^6$ K.

In addition to sustaining nuclear fusion reactions, stars in universes
with stable diprotons (large $\conlum$) must have sufficiently long
lifetimes in order to support habitable planets.  The relative
insensitivity of stellar structure to the value of $\conlum$ indicates
that stars in such universes can indeed be long-lived. As shown in
equation (\ref{startime}), the stellar lifetime $t_\ast$ depends only
linearly on the value of the central temperature parameter $\thetacen$
(which depends logarithmically on $\conlum$), but depends inversely on
the square of the stellar mass ($t_\ast\propto M_\ast^{-2}$).
Moreover, the range of stellar masses extends down to lower values as
the parameter $\conlum$ increases. As a result, the longest-lived
stars in a universe with bound diprotons have main sequence lifetimes
measured in trillions of years, comparable to the longest-lived stars
in our universe (this result follows from equations [\ref{iprofile}],
[\ref{startime}], and [\ref{ifit}]; see also Ref. \cite{barnes2015}).
For completeness, note that the smallest stars in our universe will
outlast the Sun by factors of $\sim10^3$ \cite{mdwarf}. Stars in
universes with stable diprotons thus live long enough to support 
habitability.

\begin{figure}[tbp]
\centering 
\includegraphics[width=.95\textwidth,trim=0 150 0 150,clip]{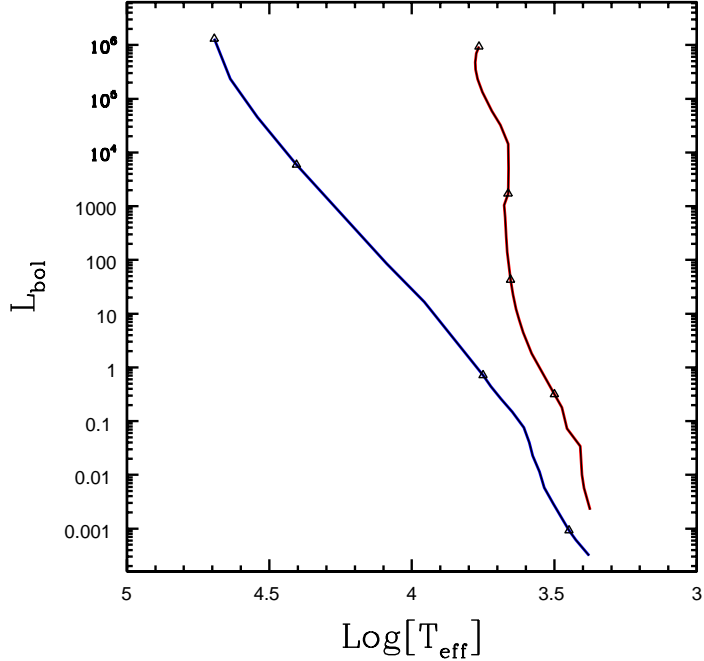}
\caption{Main sequence for the first stage of hydrogen burning in 
universes with stable diprotons. The red curve shows the main sequence
for the burning of protons into (stable) diprotons, which then capture
electrons to become deuterium, and interact with protons to make
$^3$He. For comparison, the blue curve shows the zero-age main
sequence for stars in our universe. The triangular points on both
curves mark benchmark locations for stellar masses $M_\ast$ = 100, 10,
1, and 0.1 $M_\odot$ (top to bottom). For stars with stable diprotons,
the surface temperatures are lower and nearly constant; the allowed
mass range is larger than in our universe, but the range in luminosity
is roughly comparable. }
\label{fig:msdiproton} 
\end{figure} 

To illustrate the effects of stable diprotons on stellar structure 
and evolution, we have used the {\sl MESA} stellar evolution code 
\cite{mesaone} to explore the case where the nuclear reaction
rates are enhanced by a factor of $10^{15}$. In this scenario, the
stars start with the same chemical composition as the Sun. The first
stage of nuclear processing converts free protons into diprotons,
which capture electrons to become deuterium, and then add protons to
make $^3$He. Because of the lower central temperatures ($\sim10^6$ K),
the resulting $^3$He does not burn promptly, but rather accumulates in
the stellar core. After the protons are exhausted in the central
regions, the stars adjust their structure and process the $^3$He into
$^4$He. This second stage of evolution completes the p-p chain of
nuclear reactions. For larger stars ($M_\ast\simgreat1M_\odot$), the
CNO cycle occurs alongside the aforementioned nuclear processes. For
sufficiently small stars ($M_\ast\simless0.03M_\odot$), the central
cores cannot become hot enough to support the second stage of nuclear
burning, and the stellar cores end with a $^3$He composition.  The
main sequence for these stars is shown as the red curve in Figure 
\ref{fig:msdiproton}. For comparison, the blue curve shows the
zero-age main sequence using the standard nuclear reaction rates.
Compared to hydrogen burning stars in our universe, these stars have a
larger range of masses, redder photospheres, and a comparable range of
luminosities.  The longest-lived stars have lifetimes of trillions of
years, comparable to those in our universe, and much longer than the
current cosmic age.

\subsubsection{Universes with Unstable Deuterium} 
\label{sec:unstabledeut} 

In universes without stable deuterium, the standard nuclear reaction 
chains in both BBN and stellar interiors must be altered. For example, 
in the p-p chain that powers most stars, the production of helium 
takes place by first combining two protons into a deuterium nucleus, 
\be
p + p \to d + e^{+} + \nu_e \,,
\label{ppchain1} 
\ee
which then reacts further to produce helium through the reactions
\be d + p \to {}^3{\rm He} 
\label{ppchain2} 
\ee
$$ {}^3{\rm He} + {}^3{\rm He} \to {}^4{\rm He} + p + p \,,$$
along with other branches \cite{clayton,hansen,kippenhahn,phil}.
Without the first step of deuterium production in equation
(\ref{ppchain1}), the subsequent reactions of the p-p chain cannot
take place. This difficulty has been noted by many authors
\cite{bartip,donoghuethree,hogan,pochet,reessix,schellekens,
tegmarktoe,tegmark}. Although the absence of stable deuterium 
compromises the p-p chain, stars have access to other sources 
of energy generation and other channels of nucleosynthesis. 

Even in the absence of any nuclear processing, stars can still
generate energy through gravitational contraction. As outlined above,
most stars are formed with somewhat extended configurations and with
central temperatures that are too cool to sustain nuclear reactions.
Young stars thus convert gravitational potential energy into
luminosity, grow smaller in radius, and their central cores become
hotter and denser. In our universe, this pre-main-sequence contraction
phase continues until the temperature at the center reaches the proper
nuclear ignition temperature for the given stellar mass (typically
about $T_{\rm c}\approx1-2\times10^7$ K). The onset of nuclear
reactions halts further contraction, and the star subsequently derives
its energy from nuclear power rather than from gravity. In the absence
of nuclear reactions, this contraction phase continues far longer. 
The longest-lived stars have a luminosity comparable to that of the 
Sun and can live for up to a billion years \cite{agdeuterium}, 
which could be long enough for biological evolution to take place 
on favorably situated planets. 

\begin{figure}[tbp]
\centering 
\includegraphics[width=.95\textwidth,trim=0 150 0 150]{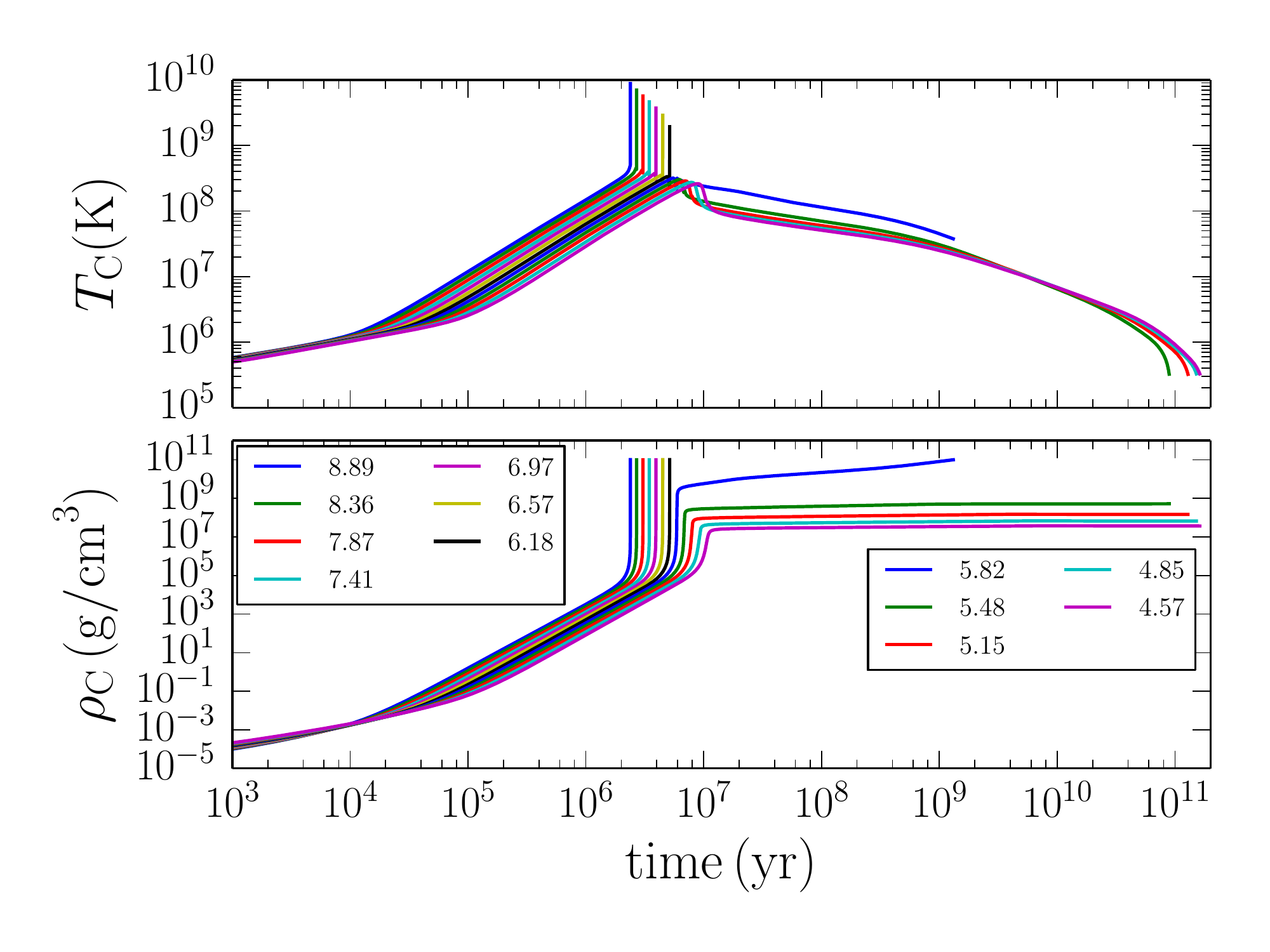} 
\vskip1.0truein
\caption{Central temperature and central density as a function of time 
for stars powered by gravitational contraction with no nuclear reactions
(from \cite{agdeuterium}).  The curves show the results for a range of
stellar masses, as labeled in solar units. The time evolution shows
different behavior above and below the Chandrasekhar mass, which
corresponds to $M_\ast=5.6M_\odot$ for stars with a pure hydrogen
composition. }
\label{fig:exnuke} 
\end{figure} 

The final fate of gravitationally contracting stars depends on the
stellar mass. Stars with sufficiently small masses can be supported by
the degeneracy pressure of non-relativistic electrons and end their
lives as white dwarfs. The maximum stellar mass that can become a
white dwarf given by the Chandrasekhar mass \cite{chandra}, where
$M_{\rm ch}$ $\approx5.6M_\odot$ for stars with a pure hydrogen
composition (as expected in the absence of nuclear reactions). Stars
with progenitor masses above the Chandrasekhar mass cannot be
supported by electron degeneracy pressure and continue to their
contraction until their cores reach enormously high temperatures 
and densities, as shown in Figure \ref{fig:exnuke} (from  
\cite{agdeuterium}).  The stellar models shown in the figure
(calculated using a modified version of the {\sl MESA} stellar
evolution package \cite{mesaone,mesatwo}) are evolved until the
central temperature reaches $T_{\rm c}\sim10^{10}$ K and the central
density reaches $\rho_{\rm c}\sim10^{11}$ g cm$^{-3}$. In actuality,
the stars condense further to even higher temperatures and densities.
Under such extreme conditions, nuclear reactions are no longer
suppressed: recall that $T\sim10^{10}$ K corresponds to $\sim1$ MeV,
the typical energy scale for nuclear reactions. At such temperatures,
protons have enough energy to overcome their coulomb barrier and can
undergo nuclear reactions without tunneling. As a result, many types
of nuclear reactions take place readily in the final death throes of
these stars. This process of explosive nucleosynthesis can supply the
universe with heavy nuclei, even in the absence of stable deuterium.

Universes without stable deuterium have additional channels for
stellar nucleosynthesis. If explosive nucleosynthesis --- or any other
process --- can produce a small amount of carbon, then stars can
operate through the CNO cycle. In this process, carbon acts as a 
catalyst to synthesize helium through a chain of reactions: 
$$
^{12}{\rm C}+p \to \, ^{13}{\rm N}+\gamma
\qquad \qquad 
^{13}{\rm N} \to \, ^{13}{\rm C} + e^+ + \nu_e
$$
\be
^{13}{\rm C}+p \to \, ^{14}{\rm N}+\gamma
\qquad \qquad 
^{14}{\rm N}+p \to \, ^{15}{\rm O}+\gamma
\label{cnocycle} 
\ee
$$
^{15}{\rm O} \to \, ^{15}{\rm N} + e^+ + \nu_e 
\qquad \qquad 
^{15}{\rm N}+p \to \, ^{12}{\rm C}+ \, ^4{\rm He} \,.
$$
Although the Sun generates only a few percent of its power through the
the CNO cycle, this nuclear process becomes dominant for somewhat
larger stars (with $M_\ast\simgreat1.5M_\odot$). Other branches of the
CNO cycle exist, but equation (\ref{cnocycle}) accounts for most of
the helium production. Notably, none of the branches of the CNO cycle 
involve deuterium, so that its instability is not an impediment to 
the production of heavier elements. 

\begin{figure}[tbp]
\centering 
\includegraphics[width=.95\textwidth,trim=0 150 0 150]{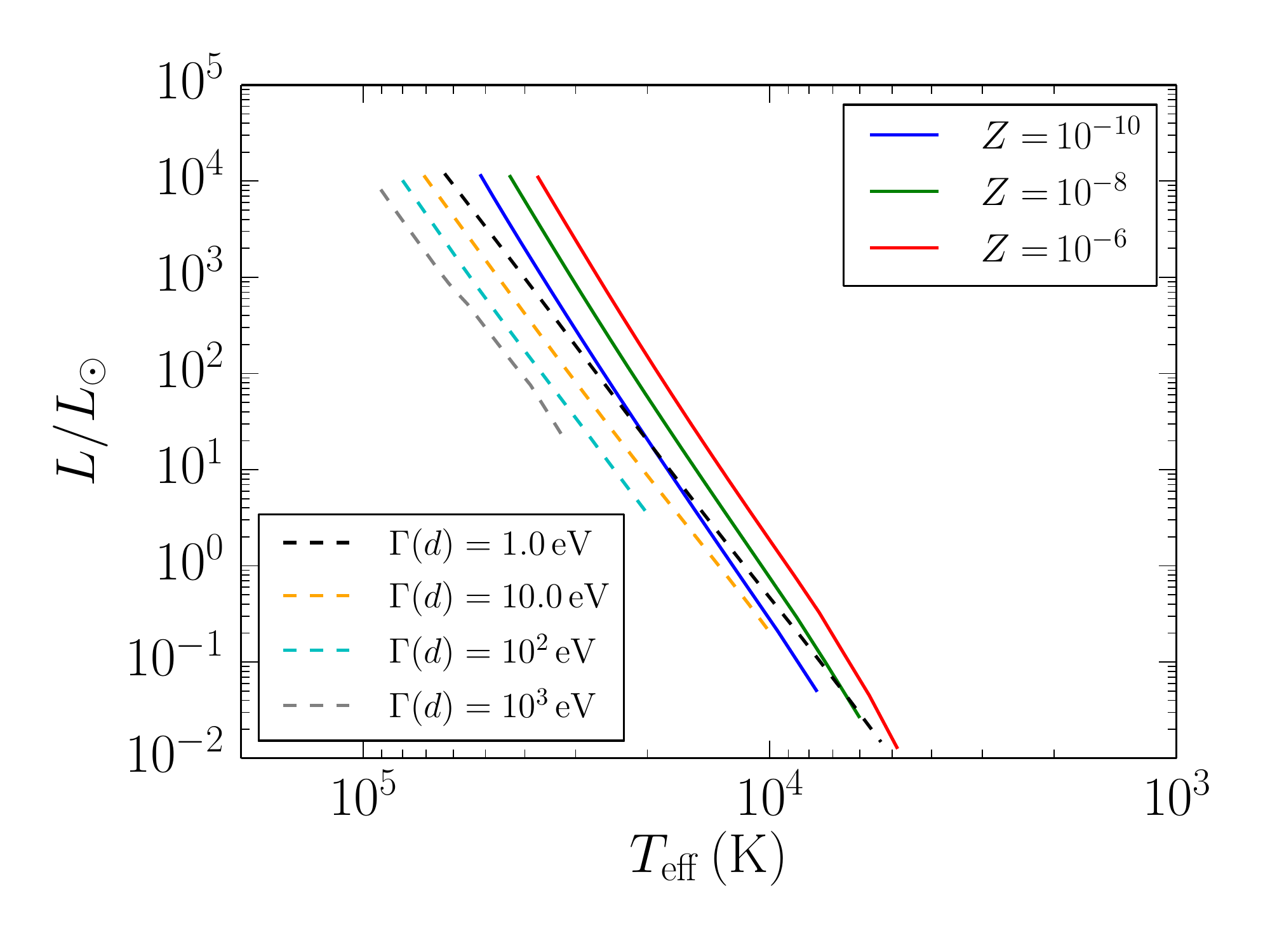} 
\vskip1.0truein
\caption{H-R diagrams for main sequence stars in universes without 
stable deuterium (from \cite{agdeuterium}). The solid curves show 
the hydrogen burning main sequence (luminosity as a function of
photospheric temperature) for stars generating power through the CNO
cycle. The curves correspond to metallicities in the range
$\metal=10^{-10}-10^{-6}$ (from left to right). The dashed curves show 
the main sequences for stars generating power through the triple
nucleon process, with varying values of the decay width for 
unstable deuterium: $\Gamma(d)$ = $1-10^3$ eV (from right to left). }
\label{fig:hrnod2} 
\end{figure}  

Given that stars can generate power through the CNO cycle, the key 
question is how much carbon is required to start the process. Figure 
\ref{fig:hrnod2} shows the main sequences for stars burning hydrogen 
via the CNO cycle for low metallicities in the range $\metal$ =
$10^{-10}-10^{-6}$ (shown as the solid curves in the diagram). These
main sequences were calculated \cite{agdeuterium} using the {\sl MESA}
stellar evolution package \cite{mesaone,mesatwo} with the p-p chain
removed from the nuclear reaction network. For the sake of
definiteness, the nuclear composition is taken to be the same as in
our universe, except for the overall normalization set by the
metallicity $\metal$.  Compared to the case of our universe, these
stars are somewhat hotter and brighter, and the allowed range of
stellar masses is truncated at the low-mass end.  These trends become
more pronounced as the metallicity becomes lower. Nonetheless, even
for $\metal=10^{-10}$, the main sequence appears relatively normal.
In addition, the stellar lifetimes for these stars (not shown; see 
\cite{agdeuterium}) are as long as $t_\ast\sim10^{11}$ yr or about
ten times the current age of our universe. As a result, stars with
such low metallicity can live long enough to support habitable
planets.  For even lower values of metallicity,
$\metal=10^{-12}-10^{-14}$, only high mass stars can sustain nuclear
burning and the main sequence becomes shorter. However, subsequent
stellar generations will attain higher metallicities, so that the
stellar population will behave similarly to that of our universe.

Stars without stable deuterium can generate energy and heavier nuclei
through yet another process involving three nucleons. This triple
nucleon process is roughly analogous to the triple alpha process that
drives carbon production in our universe, with some differences. Even
though the deuterium nucleus is unstable, the reaction $p+p\to d$ that
produces deuterium can still take place, even though the product will
fall apart a short time later. If the deuterium producing reactions
were fast enough, they could maintain nuclear statistical equilibrium,
and the star would have a standing population of transient deuterium
(analogous to the standing population of unstable $^8$Be that leads to
the triple alpha process). The reaction that produces deuterium
involves the weak force and does not operate quickly enough to
maintain equilibrium. When the deuterium nuclei are unstable, however,
they generally first decay into protons and neutrons, instead of the
original two protons. The neutron is also unstable, but its half-life
is of order ten minutes, which is enormous compared to the short time
scales of nuclear reactions in the stellar core. The star thus
maintains a standing population of neutrons, which can then interact
quickly enough to make heavier elements through the chain of reactions
\be
p + p \to d + e^+ \to p + n + e^+
\label{triplenuke} 
\ee
$$
n + p + p \to {}^3{\rm He} + \gamma \,. 
$$
Note that the final reaction takes place entirely through 
the strong force. The triple nucleon reaction thus leads 
to the net process 
\be
3p \to {}^3{\rm He} + e^+ + \gamma \,. 
\label{triplenet} 
\ee
The products interact further to make helium, 
\be
{}^3{\rm He} + {}^3{\rm He} \to {}^4{\rm He} + 2p \,,
\label{ontohelium} 
\ee
where this final reaction is the same as that utilized 
in the p-p chain. 

Figure \ref{fig:hrnod2} shows the main sequences for stars burning
hydrogen into helium through the triple nucleon process (shown as the
dashed curves). These calculations were carried out \cite{agdeuterium}
using the {\sl MESA} code \cite{mesaone,mesatwo}, where the CNO cycle
reactions have been removed and the p-p reactions are replaced by the
triple nucleon reactions described above. All of the reaction rates
and yields are the same as in our universe, except that deuterium is
unstable and decays on a short time scale given by the decay width
$\Gamma(d)$. Results are shown in Figure \ref{fig:hrnod2} for decay
widths in the range $\Gamma(d)=1-1000$ eV. The lower end of this range
corresponds to a half-life of about $10^{-16}$ sec, which is
comparable to that of unstable $^8$Be in our universe (in analogy to
the triple alpha process). A number of trends are evident from this
H-R diagram. As the decay width increases (so that the half-life of
deuterium decreases), the stars become increasingly hotter, and the
main sequence become shorter as the minimum stellar mass required for
nuclear ignitions increases. We note that working stars exist for even
larger decay widths $\Gamma(d)=10^4-10^5$ eV (not shown), although
only the most massive stars are operational. 

\subsection{Stellar Constraints on Nuclear Forces} 
\label{sec:weakstar} 

In our universe, the weak interaction plays an important role in
stars: Most of the stellar energy is generated by fusing together four
protons into a helium nucleus, which requires the conversion of two
protons into neutrons via the weak interaction. Nonetheless, stars can
still function in universes where the weak interaction is either
weaker or stronger than in the Standard Model \cite{howeweakful},
including scenarios where the weak interaction is absent altogether
\cite{weakless,grohsweakless} (see Section \ref{sec:starweakless}). 
On the other hand, the weak interaction cannot become too strong
without compromising stellar evolution; an upper bound is derived in
Section \ref{sec:starweaklimit}. The weak interaction is also
constrained \cite{carr} by the requirement that neutrinos must be 
optically thick in supernova explosions (Section \ref{sec:supernova}). 

\subsubsection{Stellar Evolution without the Weak Interaction} 
\label{sec:starweakless} 

Another scenario for stellar evolution arises in universes where the
weak interaction is absent \cite{grohsweakless,weakless}, or much
weaker than that of our universe \cite{howeweakful}. As outlined in
Section \ref{sec:bbn}, the weakless universe emerges from its BBN
epoch with a substantial mass fraction in helium-4, and with the
remainder of its baryons distributed among free protons, free
neutrons, and deuterium.  Given this chemical composition and the
absence of weak interactions, stellar evolution must rely on
unconventional nuclear reaction chains. Nonetheless, stars can still
function in such universes, and their properties are roughly similar
to those of ordinary stars.

In our universe, with its weak interaction, free neutrons decay to
protons, so that stars begin their evolution with a store of hydrogen
fuel (helium does not play a role in early evolution). The first step
of the reaction chain is to synthesize deuterium through the reaction
$p+p\to{d}+e^{+}+\nu_e$, which involves the weak force. This reaction
cannot take place in a weakless universe.  However, if free neutrons
remain, stars can make deuterium through the reaction $p+n\to{d}$,
which involves only the strong force. Some fraction of the primordial
free neutrons can be used up, through this same reaction, before the
neutrons are incorporated into stars. Whether the neutrons are fused
before star formation, or in stellar cores, the net result is a
substantial supply of deuterium, in addition to that produced during
BBN. This deuterium is available for the next steps of the reaction
chain, including $p+d\to$ $^3$He $+\gamma$, $d+d\to$ $^3$He $+n$, and
$d+d\to$ $^4$He $+\gamma$, and others. The first of these reactions is
the standard next step in the $p$-$p$ chain in ordinary stars
\cite{clayton,hansen,kippenhahn}. The final two reactions are highly
suppressed in our universe due to the small primordial abundance of
deuterium and its rapid reaction rate in stars. Given the enhanced
abundances of deuterium in the weakless universe, stars can generate
energy primarily through deuterium burning.

Numerical simulations have recently been carried out to illustrate 
how stellar evolution takes place in a weakless universe
\cite{grohsweakless}. These calculations use the {\sl MESA} stellar 
evolution package \cite{mesaone,mesatwo}, which has been modified to
incorporate the required nuclear reaction chains. For example,
interactions involving the weak force have been removed, whereas a
standing population of free neutrons and their corresponding reaction
pathways have been added.  Additional nuclear reactions that are rare
in our universe (given its chemical composition) are also included, 
primarily those involving deuterium (which is more plentiful in a 
weakless universe).

\begin{figure}[tbp]
\centering 
\includegraphics[width=.95\textwidth,clip]{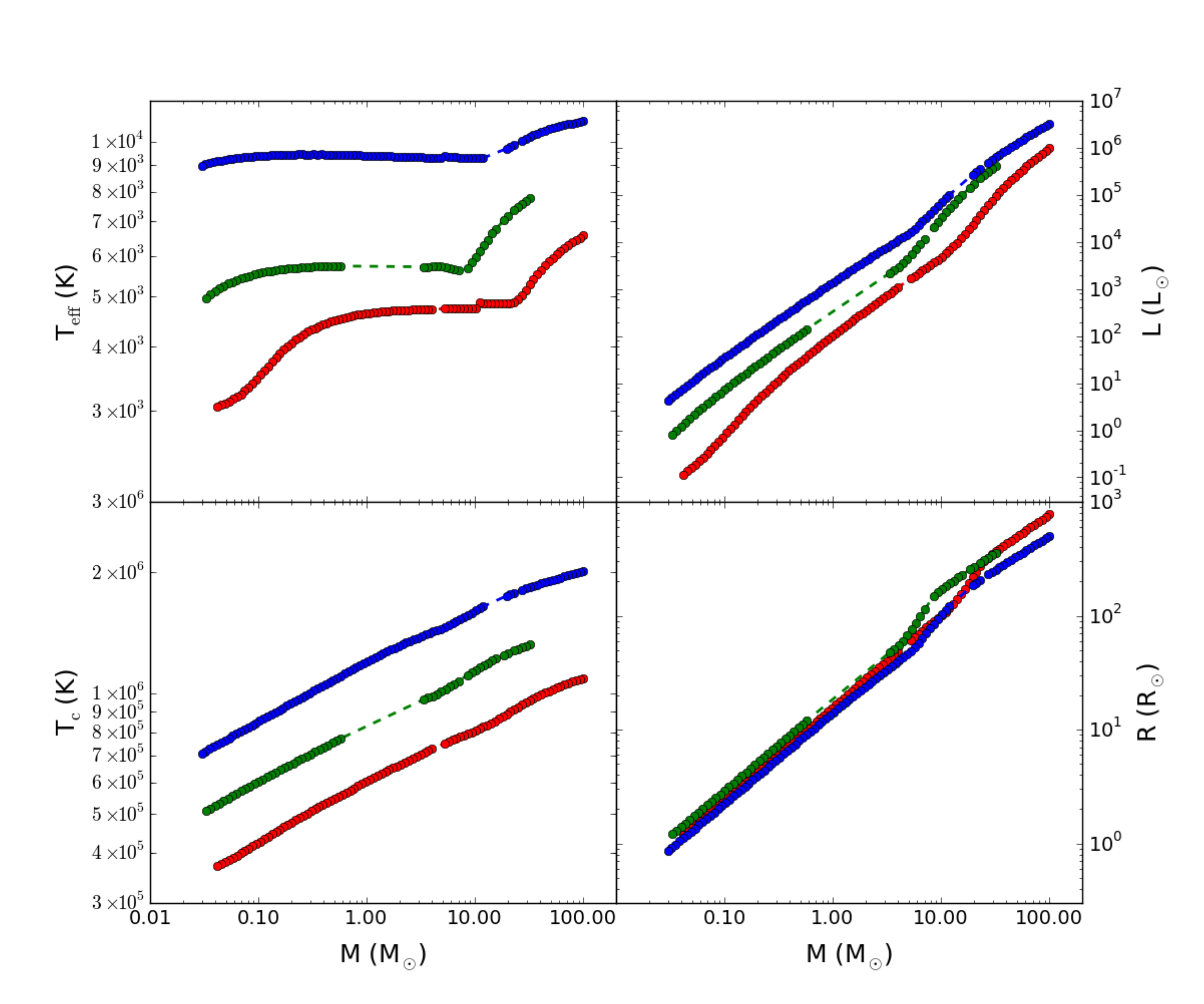}
\caption{Stellar evolution in universes without the weak interaction
(from \cite{grohsweakless}). Each panel shows properties of weakless
stars on the zero-age main sequence as a function of stellar mass,
where the range of stellar masses is the same as in our universe.  
The three curves correspond to different chemical compositions, i.e.,
those produced by BBN for $\eta$ = $10^{-11}$ (red), $10^{-10}$
(green), and $10^{-9}$ (blue).  The stellar properties include the
effective photospheric temperature (top-left panel), the central core
temperature (bottom-left panel), the luminosity (top-right panel), and
the stellar radius (bottom-right panel). Mass, luminosity, and radius 
are given in Solar units; temperature is given in Kelvin. The dashed 
portions of the curves correspond to stellar models that did not fully 
converge, so that these results are more approximate than the rest 
of the curves. } 
\label{fig:weakstar} 
\end{figure} 

The results of stellar evolution calculations for the weakless
universe are shown in Figure \ref{fig:weakstar}. Stellar properties
are plotted for three different compositions emerging from BBN (see
Section \ref{sec:bbnweakless}), corresponding to baryon to photon
ratios $\eta$ = $10^{-11}$ (red curves), $10^{-10}$ (green), and
$10^{-9}$ (blue). The initial neutron to proton ratio is taken to be
unity, $n/p=1$, for all cases considered here.  As $\eta$ increases,
the helium abundance increases, whereas the mass fractions of
deuterium and other nuclear species decrease.

The upper left panel shows the effective surface temperature of the
stars as a function of stellar mass. Over the range of compositions
shown, and the entire range of stellar masses, these temperatures fall
in the interval 3000 K $\simless T_{\rm eff}\simless$ 10,000 K (recall
that the Sun has $T_{\rm eff}\approx5800$ K). Compared to stars in
our universe, these objects have comparable surface temperatures, but
span a somewhat narrower range.  The lower left panel in the Figure
shows the central temperature of the stars versus mass. As expected,
the central temperature is a slowly increasing function of stellar
mass, and also increases with decreasing deuterium abundance. The most
significant feature is that the central temperatures are of order
$10^6$ K, characteristic of deuterium burning. For comparison, the
central temperature of the Sun is $\sim1.5\times10^7$ K, the value
characteristic of hydrogen burning.

The upper right panel of Figure \ref{fig:weakstar} shows the stellar
luminosity as a function of mass. The luminosity displays the
approximate power-law form $L_\ast\propto M_\ast^2$, which is less
steep than the dependence in our universe for hydrogen burning stars.
The range of luminosities is somewhat smaller than in our universe: 
The luminosity for massive stars is comparable, but that for low-mass
stars is larger by factors of 10 -- 100.  The lower right panel shows
the stellar radius as a function of mass, which displays a nearly linear
dependence $R_\ast\propto M_\ast$. This finding is consistent with the
outer boundary condition for stars, $L_\ast$ = $4\pi R_\ast^2$ 
$\sigma_{\rm sb} T_{\rm eff}^4$, given the relatively constant 
surface temperatures and the mass-luminosity relation. For massive
stars, the stellar radii are somewhat larger than those of our
universe, consistent with their cooler photospheric temperatures. 
As a result, massive stars in the weakless universe look much like  
the red giants of our universe. 
 
In spite of their non-standard nuclear reactions, stars in a weakless
universe have roughly standard properties, including luminosities and
lifetimes. One can understand this similarity as follows: The net
result of nuclear burning --- both with and without the weak
interaction --- is the production of helium-4, which has a binding
energy of about 28 MeV. Instead of starting the process with protons,
stars in the weakless universe begin with deuterium, which has a
binding energy of 2.2 MeV. The energy supply in weakless stars is 
smaller than that of ordinary stars, but only by $\sim10\%$. 

\subsubsection{Stellar Constraint on the Weak Interaction} 
\label{sec:starweaklimit} 

The previous section showed that stars can function and universes can
remain viable in the absence of the weak interaction. The requirement
of working stars can also be used to place an upper bound on the
strength of the weak interaction. Nuclear burning in stellar interiors
produces a large flux of neutrinos, which freely stream out of the
stars. If the weak interaction cross section were much larger, then
the neutrinos could be optically thick, and stellar structure could be
altered accordingly \cite{howeweakful}. It remains possible for stars
to function with optically thick neutrinos, but their evolution would 
change significantly. 

In this setting, the optical depth for neutrinos can be
written in the form
\be
\tau_\nu = \int_0^{R_\ast} n \sigma_\nu dr \approx 
\sigma_\nu \langle n \rangle R_\ast \,, 
\ee
where the integral is taken over the extent of the star and 
$\langle n \rangle$ is defined via the mean value theorem.  
The cross section for neutrino interactions is of order
$\sigma_\nu\sim G_F^2 E_\nu^2$, where $E_\nu$ is the energy of 
the stellar neutrinos (typically $E_\nu\sim1$ MeV for stars in 
our universe \cite{bahcall}).  The mean number density of 
particles in the star can be written $\langle n \rangle \approx$ 
$3M_\ast/(4\pi\mpro R_\ast^3)$. The stellar radius has the form 
given by equation (\ref{rstar}), so that $R_\ast\approx$
$GM_\ast\mpro/\tcent$, where $\tcent$ is the central nuclear burning
temperature (in natural units). Finally, the stellar mass can be
expressed in terms of the fundamental mass scale for stars
(see \ref{sec:massscales}) so that $M_\ast=X\alpha_G^{-3/2}\mpro$, 
where $X$ is a dimensionless parameter of order unity.  The
combination of these factors allows the optical depth to be 
written in the form 
\be
\tau_\nu = {3 \over 4\pi X}  G_F^2 E_\nu^2 \tcent^2 
\alpha_G^{-1/2} \,.
\ee
The requirement that neutrinos can freely stream out of 
stellar interiors thus implies a constraint on the weak 
interaction of the form 
\be
G_F \simless \alpha_G^{1/4} E_\nu^{-1} \tcent^{-1} \approx 
0.28 {\rm GeV}^{-2} \approx 2.4\times10^4 (G_F)_0\,,
\ee
where we have used $E_\nu$ = 1 MeV and $\tcent$ = 1 keV 
to evaluate the bound and where $(G_F)_0$ = (292 GeV)$^{-2}$ 
is the value in our universe. 

\subsubsection{Supernova Constraint on the Weak Interaction} 
\label{sec:supernova} 

In order for neutrinos to play a role in supernova explosions, 
they must have a sufficiently high interaction rate. In order of 
magnitude, the neutrino interaction rate must be comparable to the 
free-fall collapse time for the inner region of the star \cite{carr}. 
This condition can be written in the form 
\be
n \sigma_\nu v = c_1 (G \rho)^{1/2} \,,
\ee
where $n$ is the number density of particles that the neutrinos
interact with, $\rho$ is the total mass density of collapsing stellar
material, and $c_1$ is a dimensionless constant of order unity. The 
parameter $c_1$ determines how closely the time scales for neutrino 
interactions and stellar collapse must match (and $c_1$ could have 
a range of values). The speed $v=c=1$ and the cross section can be 
written in the form 
\be
\sigma_\nu = c_2 G_F^2 E_\nu^2 \,, 
\ee
where $E_\nu$ is the neutrino energy and the dimensionless constant
$c_2$ depends on the types of particles in the stellar material. The
density is close to nuclear densities, so we can express it in the
form $\rho=c_3\mpro\mpion^3$, where $c_3$ is another dimensionless
constant, and $\mpion$ is the pion mass. The neutrino energies are 
of order 1 MeV, so we can write $E_\nu=c_4\emass$. Combining these
expressions thus yields the requirement that must be met in order for
neutrinos to affect supernovae,  
\be
G_F^4 = {c_1^2 \over c_3 c_2^2 c_4^4} 
{\mpro \over \mplanck^2 \emass^4 \mpion^3} \,. 
\ee
If we define $\alpha_{\rm w}=G_F\mpro^2$ and use the values of the 
constants $c_j$ corresponding to equality in our universe, then we 
find the approximate constraint 
\be
\alpha_{\rm w} \sim 20 \alpha_G^{1/4} \beta^{-1} \,. 
\label{snconstraint} 
\ee

In order for neutrinos to help enforce supernova explosions, the
approximate equality of equation (\ref{snconstraint}) must be met. If
the weak interactions are ineffective, then all of the neutrinos can
leave the system without interacting and the explosion stalls. On the
other hand, successful detonation seems to require that the neutrinos
escape the stellar core and reach the outer layers that are driven off
by the explosion. For neutrino interactions that are too strong, the
neutrinos scatter many times before leaving the core and distribute
their momenta among many nuclei.  If this momentum deposition occurs
deep in the gravitational potential, then Type II supernovae could
still fail. Moreover, two-dimensional and three-dimensional effects,
along with stellar mass and composition, also influence the explosion.
Unfortunately, our current understanding core-collapse supernovae
remains incomplete (see \cite{janka,jankarev} for recent reviews), so
that we do not know how closely the two sides of equation
(\ref{snconstraint}) must approach equality.

For completeness, we note that supernovae from massive stars are not
the only sources of elements heavier than iron. In our universe,
neutron star mergers provide the primary environments for the rapid
neutron-capture process required to synthesize such large nuclei 
\cite{frebelbeers,freiburghaus}, and the gravitational radiation 
from such events can now be observed \cite{ligonsns}.  Explosive
nucleosynthesis can also take place in white dwarfs, which result from
the deaths of lower mass stars. Explosion of these objects takes
when they accrete enough mass to exceed the Chandraskehar limit 
\cite{iwamoto,nomoto1997}, and more rarely via collisions. In any 
case, core-collapse supernova explosions are not strictly necessary
for heavy element production, although they do provide the requisite
neutron stars for mergers in our universe.  On the other hand, neutron
stars can also be produced by mergers of white dwarfs \cite{wdmerger},
so that many evolutionary channels are possible in principle.

\subsubsection{Supernova Constraints on the Nucleon Potential} 
\label{sec:supercore} 

The successful launch of supernova explosions depends on a number of
factors. As discussed above, one requirement is that neutrinos must be
sufficiently optically thick to help drive out the explosion. Another
possible constraint arises from the form of the nuclear potential. In
addition to the general Yukawa form described by equation
(\ref{yukawa}), the nucleon-nucleon potential has a repulsive core
that acts on short distance scales because the wavefunctions of the
nucleons do not readily overlap \cite{epelbaum2009}. If the nuclear
potential had a somewhat different form, then supernovae could be
compromised.

After a massive star produces a degenerate iron core, its central
regions rapidly collapse and reach densities comparable to that of
nuclear material.  In successful explosions, the repulsive core of the
nucleon-nucleon potential prevents the central region of the star from
becoming so dense that the material falls within its event horizon and
becomes a black hole. Although some supernovae produce black holes in
our universe, the majority result in explosions that distribute heavy
elements into the galaxy and leave behind neutron stars. In other
universes, if the nuclear potential has different properties, more (or
perhaps all) core-collapse supernovae could result in failed
explosions and black hole formation. Such universes would still
produce heavy elements via AGB stars, Type-Ia supernovae, and
collisions within the diminished population of neutron stars, but 
they could have much lower metallicity than ours.

The short-range repulsive force can be modeled by the exchange of
vector mesons. The relevant particles are the $\omega$ and $\rho$
mesons, both with masses $m\sim770-780$ MeV, corresponding to a
Compton wavelength of $\sim0.25$ fm. This scale is comparable to 
the repulsive core of the nucleon potential, which has an estimated 
size $\sim0.4-0.5$ fm \cite{epelbaum2009}. The vector mesons can be
described in terms of their constituent quarks, e.g., the $\omega$
particle can be written in the form  
\be
\omega \sim {1\over\sqrt{2}} \left( u{\bar u} + d{\bar d} \right) \,.
\label{vectormeson} 
\ee
Larger quark masses would thus imply heavier vector mesons and a
smaller spatial extent of the repulsive core of the nuclear potential. 
In a universe with such properties, when massive stars collapse at the
end of their lives, they would reach higher central densities and
would be susceptible to black hole formation.

A detailed assessment of the properties of the nuclear potential
necessary for supernova explosions has not yet been carried out. In
approximate terms, we know that supernovae generally result in nuclear
densities and the production of neutron stars. Since neutron stars
have masses $M\sim1M_\odot$ and radii $R\sim10$ km, whereas black
holes of the same mass have radii $R_{\rm bh}=2GM/c^2\sim3$ km, a
density enhancement of $\sim20-30$ during stellar collapse should be
sufficient to favor black hole production.  This enhancement would
result from a factor $\sim3$ decrease in the spatial extent of the
repulsive core in the nuclear potential, and hence a factor of $\sim3$
increase in the masses of both the vector mesons and their constituent
quarks. This constraint on the light quark masses is roughly
comparable to those found in Section \ref{sec:quarklimit}. 

The description given here in terms of meson exchange is highly
approximate, and a full treatment using effective field theory 
\cite{epelbaum2009,meissner2} should be carried out. The change in the 
length scale of the repulsive core can also be thought as a change in
the equation of state of dense nuclear matter. If the equation of
state becomes sufficiently soft \cite{shapteuk}, then black hole
formation is enabled. Moreover, near the densities where the repulsive
core becomes important, nucleons begin to break up into their
constituent quarks. The quarks are not truly free until they reach
much larger densities, so a proper treatment must include the physics
of the transition and its complications. For reference, note that
recent lattice QCD calculations find that the quark/hadron phase
transition occurs at energies $T_c\approx173$ MeV \cite{qhtrans}.

\bigskip 
\section{Planets} 
\label{sec:planets} 

Planets represent the smallest astrophysical objects that are
necessary for the development of life (as we know it). In order for a
given universe to become habitable, the laws of physics must allow for
the production of planets with a number of basic properties, as
outlined in this section. These bodies must be small enough in mass so
that they are not degenerate.  This requirement is also necessary (but
not sufficient) for the planet to have a solid surface \cite{ikoma}.
On the other hand, the planets must have enough mass to retain a
gaseous atmosphere and must have enough particles to support a
biosphere with sufficient complexity. One also expects planets to be
smaller in mass than their host stars and galaxies.  These
considerations place constraints on the allowed range of parameter
space $(\alpha,\alpha_G)$.

In addition to constraints on the properties of planets themselves,
the existence of habitable planets also involves a number of
environmental constraints. Potentially habitable planets in our
universe are subject to well-known requirements: Planets must be the
right distance from their host stars to allow for liquid water, the
planet mass must be roughly comparable to Earth, and the stars must 
have sufficiently long lifetimes  
\cite{kastingcatling,lammer,lunine99,lunine,scharf}.
Additional requirements must be enforced in other universes: As
discussed above, planets must reside in galaxies that are dense enough
to cool and make stars and planets, yet remain diffuse enough to allow
for the survival of habitable orbits (Section \ref{sec:galaxies}).
Planets must be made of heavy elements, which require successful
stellar nucleosynthesis (Section \ref{sec:stars}), along with mild
constraints on Big Bang Nucleosynthesis (Section \ref{sec:bbn}). At
the more fundamental level, nuclei themselves must be stable (Section
\ref{sec:particlephys}). In addition, if habitable planets require 
particular nuclear structures, such as iron being the most stable
nucleus and/or particular types of radioactivity, then even tighter
constraints on the fundamental parameters (e.g, $\alpha$) can be
derived \cite{sandora2016,sandora2017}.

\subsection{Mass Scale for Non-Degenerate Planets} 
\label{sec:planetnondegen} 

In order for an astronomical body to function as a planet, it must be
small enough in mass to remain non-degenerate. This requirement allows
the planet to have a solid surface --- or liquid in the case of
so-called water worlds.

The standard way to formulate this constraint on planetary mass is to
require electromagnetic forces to dominate gravitational forces on the
scale of the planet \cite{bartip,presslight,weisskopf}. Equivalently, 
the electromagnetic energy $E_{\rm em}$ must be larger than that of
self-gravity $E_{\rm g}$.  Consider a planet with mass $M_P$ and
radius $R_P$.  This constraint can be written in the form 
\be
E_{\rm em} = N {e^2 \over \ell} > f_p {GM_P^2\over R_P} = 
E_{\rm g} \,,
\ee
where $N$ is the number of atoms in the planet, $\ell$ is the
effective distance between charges, and $f_p$ is a dimensionless
constant of order unity that depends on the internal density
distribution of the planet \cite{chandra}, where typical values fall
in the range $f_p=3/5-3/4$. The mass of the planet and the number of
atoms are related so that 
\be
M_P = A N \mpro\,,
\ee
where $A$ is the mean atomic weight for the constituent atoms. 
With these specifications, the mass of the planet must obey the 
upper limit 
\be
M_P < A^{-3} \left({3\over4\pi f_n^3}\right)^{1/2} 
\left( {\alpha\over\alpha_G}\right)^{3/2} \mpro\,. 
\label{planmassdegen} 
\ee
This scale is comparable to the mass of Jupiter, where
$M_J\sim300M_\oplus$. In our universe, planets with such large masses
are observed to have low density and are inferred to be primarily made
of gaseous material. In order for a planet to be rocky, and have a
solid surface, the mass must be significantly smaller, with current
observations indicating $M\sim10M_\oplus$ \cite{wolfgang}. As a
result, the limit of equation (\ref{planmassdegen}) is necessary, but
not sufficient. In addition to hard surfaces, habitable planets must
also have surface gravities weak enough to allow for the survival of
flora and fauna \cite{huygens}.  Both of these considerations suggest
that the expected masses for habitable planets are closer to the lower
bound considered in the following section.

\subsection{Mass Scale for Atmospheric Retention} 
\label{sec:planetatm} 

The surface gravity of a planet must be sufficiently strong in order
for it to retain its atmosphere \cite{bartip,press1980,presslight}.
At the same time, the surface of the planet, and hence the atmosphere, 
must be hot enough to support the chemical reactions necessary for life. 
This latter requirement can be written in the form 
\be
kT > E_{\rm chem} = \epsilon_{\rm c} \alpha^2 \emass c^2 \,,
\label{minchem} 
\ee
where $T$ is the surface temperature and $E_{\rm chem}$ is the 
energy required for chemical reactions to occur (from equation
[\ref{biotemp}]). Recall that the chemical conversion factor 
$\epsilon_{\rm c}\sim10^{-3}$. 

In order for a planetary atmosphere to remain intact, its 
constituent molecules cannot evaporate on short timescales. 
As a result, the surface temperature must correspond to an 
energy scale less than the gravitational binding energy of 
an atmospheric molecule, i.e., 
\be
kT < {GM_P (A_{\rm atm}\mpro) \over R_P} \,,
\label{noevap} 
\ee
where $A_{\rm atm}$ is the mean atomic weight of the molecules (recall
that $A_{\rm atm}\approx29.6$ for terrestrial air).  The constraint of
equation (\ref{noevap}) is approximate: The evaporation of planetary
atmospheres depends on a number of additional factors, including the
planetary structure (which determines the surface gravity), the
planetary magnetic field, the location within its solar system, and
spectral properties of the host star \cite{owenwu}. 

The surface temperature of the planet must be hot enough to support
chemistry (equations [\ref{minchem}, \ref{biotemp}]) and cold enough 
to suppress evaporation (equation [\ref{noevap}]). In order to evaluate 
the planetary radius, one can assume that the atoms in the (solid) 
planet subtend a volume $\sim a_0^3$, where $a_0=(\alpha\emass)^{-1}$
is the Bohr radius. These combined constraints then result in a 
lower bound for the planetary mass 
\be
M_P > \left({\epsilon_{\rm c}\over A_{\rm atm} A}\right)^{3/2} 
\left({\alpha\over\alpha_G}\right)^{3/2}\,\mpro\,. 
\label{planmassatm} 
\ee
Keep in mind that the atomic weight of the atmosphere $A_{\rm atm}$
can be different from that of the bulk of the planet $A$. 

Note that the maximum mass for non-degenerate planets from equation
(\ref{planmassdegen}) is significantly larger than the minimum mass
from equation (\ref{planmassatm}). If we let $A_{\rm atm}$ = $A$, then
the two mass scales are related such that $M_{P{\rm min}}$ $\approx$
$\sqrt{3}\epsilon_{\rm c}^{3/2}M_{P{\rm max}}$. Since the chemical
conversion factor $\epsilon_{\rm c}\sim10^{-3}\ll1$, one finds that
$M_{P{\rm min}} \ll M_{P{\rm max}}$, as required. 

For completeness, notice also that planetary bodies with too little
mass will not have enough self-gravity to become quasi-spherical. In
sufficiently small bodies, like most of the asteroids and Kuiper Belt
objects in our Solar System, imperfections in the planetary surface
(effectively mountains) can be as large as the planet itself. In
general, the requirement of retaining an atmosphere is stronger than
that of maintaining a spherical shape \cite{bartip,weisskopf}.

\begin{figure}[tbp]
\centering 
\includegraphics[width=.95\textwidth,trim=0 150 0 150,clip]{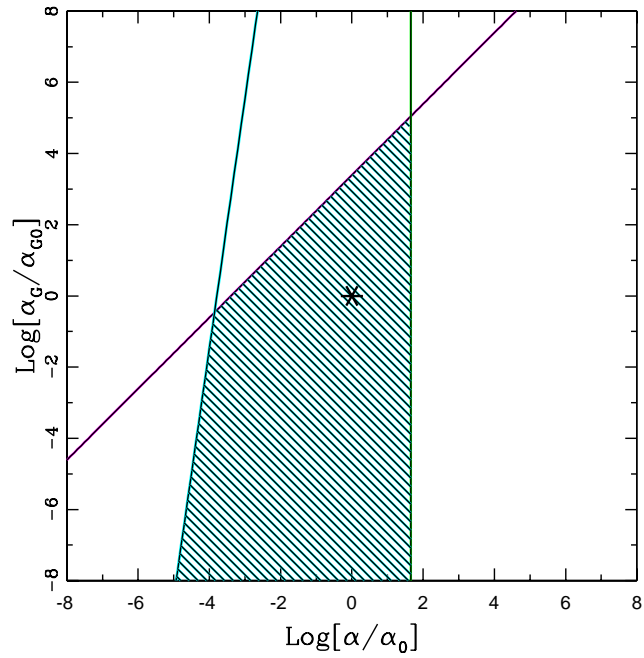}
\caption{Allowed parameter space for the existence of planets with 
differing values of the structure constants $\alpha$ and $\alpha_G$. 
The shaded region delineates the portion of the plane that remains
after enforcing the following constraints: For planets to be smaller
than stars, the fine structure constant $\alpha$ must lie to the left
of the vertical green line. For planets to be smaller in mass than
their host galaxies, $\alpha$ must fall to the right of the cyan
curve. For planets to carry enough information content to support a
biosphere, and remain non-degenerate, the parameters must fall below
the diagonal magenta line. The star symbol in the center of the 
diagram marks the location of our universe. } 
\label{fig:planplane} 
\end{figure} 

\subsection{Allowed Range of Parameter Space for Planets}
\label{sec:planetrange} 

With the above constraints specified, this section estimates the
allowed range of parameter space that supports potentially habitable 
planets.  We first note that planets must be larger than the mass
scale of equation (\ref{planmassatm}) to retain their atmospheres and
smaller than the mass scale of equation (\ref{planmassdegen}) to
remain non-degenerate.  Both of these mass limits can be written as
dimensionless parameters times the mass scale
\be
M_P = A^{-3} \left({\alpha\over\alpha_G}\right)^{3/2} \mpro\,, 
\label{planmassone} 
\ee
where $A$ is the atomic weight of the planetary material. (Note
that we can define a single characteristic planetary mass scale 
by setting $A=1$; see equation [\ref{massplanet}]). 
The lower limit (\ref{planmassatm}) is smaller than the upper limit
(\ref{planmassdegen}) by a factor of 
$\epsilon_{\rm c}^{3/2}\sim10^{-9/2}\ll1$, so that both limits can be
simultaneously satisfied. 

Notice also that the scale of equation (\ref{planmassone}) that
characterizes planetary masses differs from the corresponding mass
scale for stars (see \ref{sec:massscales} and equation 
[\ref{starmassform}]) by a factor of $\alpha^{3/2}$. If one requires
that planets are less massive than their host stars, then the fine
structure constant obeys the constraint $\alpha\simless1$.

The planetary mass scale from equation (\ref{planmassone}) must be
larger than the minimum mass required for a planet to become
habitable. One specific constraint is that planets must be large
enough to support a working biosphere \cite{coppess}. Since we do not
know the minimum information content of a biosphere, this limit is
necessarily speculative. As a starting point, we can use the values in
our universe to obtain a working estimate: The biosphere of Earth has
500 to 800 billion tons of carbon, which implies a possible
information content of $Q_B\approx4\times10^{40}$ bits. The
corresponding mass of our biosphere is
$M_B=12Q_B\mpro\approx10^{-10}M_\oplus$. In general, the host planet
must be larger in mass than its biosphere by a large but unknown
factor $f_{\rm bio}$. If we require the planetary mass scale to be
larger than that necessary to support a biosphere, the following 
constraint must be met 
\be
\left({\alpha\over\alpha_G}\right) > 
(f_{\rm bio} 12 Q_B)^{2/3} A^2 \approx 10^{30} f_{\rm bio}^{2/3}\,.
\label{biolimit} 
\ee
Although the limiting value of $f_{\rm bio}$ is not known, only the
outer layers of a planet are hospitable for biology.  On our planet,
the biosphere has a thickness of order 10 km, whereas the planet has
radius $R_\oplus\approx6400$ km. As a result, we expect 
$f_{\rm bio}\gg1000$.

One can derive another (relatively weak) constraint on the structure 
constants by requiring that planetary masses are smaller than the masses 
of their host galaxies. This constraint is necessary, but not sufficient, 
for planet formation to take place. Using the planetary mass scale 
(\ref{massplanet}) and galactic mass scale (\ref{massgalcool}) from 
\ref{sec:massscales}, this constraint takes the form
\be
\alpha_G < \alpha^7 \beta^{-1}\,,
\label{gplimit} 
\ee
where $\beta$ is the electron-to-proton mass ratio. 
Since $\alpha_G\ll\alpha$, this constraint only becomes 
significant for extremely small values of the fine
structure constant $\alpha$. 

The constraints described above limit the allowed parameter space for
universes with differing values of the structure constants $\alpha$
and $\alpha_G$, as shown in Figure \ref{fig:planplane}. In order for
planets to be smaller in mass than their host stars, the fine
structure constant $\alpha$ must be small, as delimited by the
vertical green line. In order for planets to be smaller in mass than
their host galaxies, $\alpha$ must be larger than the limit given by
the cyan line (from equation [\ref{gplimit}]). Finally, the magenta
line shows the limit required for planets to simultaneously be small
enough to remain non-degenerate and contain enough mass to support a
biosphere of sufficient complexity (from equation [\ref{biolimit}],
where the planet is assumed to be $f_{\rm bio}=10^4$ times more
massive than its biosphere).  The star symbol shows the location of
our universe in the diagram.  The allowed range in $\alpha$ spans six
orders of magnitude, whereas the range in $\alpha_G$ spans more than
eleven orders of magnitude (although one should keep in mind that the
upper limit on $\alpha_G$ depends on biosphere requirements and is
thus uncertain).  In any case, the existence of planets does not
require highly specialized values of $(\alpha,\alpha_G)$. Notice also
that the allowed range of parameter space for planets (from Figure 
\ref{fig:planplane}) is larger than the corresponding range for
the existence of working stars (from Figure \ref{fig:starplane}).

\subsection{Planet Formation} 
\label{sec:planetform} 

The discussion thus far has considered the structure of planets -- and
the their prospects for survival -- but not the processes that form
these bodies. We have thus implicitly assumed that if planets can
exist, they can somehow be made. Although a comprehensive theory of
planet formation remains under construction, current observations of
extra-solar planets suggest that planet formation takes place readily.
Here we briefly review the current state of the field \cite{armitage}
and identify the basic ingredients necessary for planet formation to
occur.

The field of exoplanets has developed rapidly over the past two
decades. The first confirmed planet in orbit about another star was
discovered in 1992 associated with a pulsar \cite{alex}. The detection
of the first planet orbiting a main sequence star, 51 Pegasi, was
reported in 1995 \cite{mayorqueloz}, and planetary discoveries have
steadily piled up. Exoplanets have now been discovered by radial velocity
methods, transits of their host stars, microlensing, and by direct
imaging. The first planet with a mass comparable to Earth and an orbit
consistent with liquid water was reported in 2014 \cite{quintana}, and
about twenty so-called habitable planets are now known \cite{kanehabit}.
The inventory of exoplanets is thus enormous, with thousands of
detections already reported \cite{kepler}. Moreover, current
projections indicate that the Galaxy contains more planets than stars
and that the fraction of Sun-like stars with Earth-like planets is
sizable, namely $\eta_\oplus\approx0.1$ \cite{etaearth}.

Planets form in the nebular disks that are found ubiquitously around
young stars. This nebular hypothesis for the origin of planets 
\cite{kant,laplace,wright} predates by centuries the actual observations
of circumstellar disks. These structures were unambiguously discovered
in the 1980s through their infrared radiation signatures and other
indirect evidence. They can now be imaged directly using submillimeter
interferometry and their properties can be measured with exquisite
precision (see \cite{hartmann} for a recent review). These disks have
roughly solar system size scales, with typical outer radii in the
range 10 -- 100 AU. The total mass in the disk represents the material
available for planet formation and varies with time. Current
observations indicate that disk mass can be as large as 10 percent of
the stellar mass at the end of the star formation process and drops 
rapidly over the next 3 -- 10 Myr \cite{jesus}. 

Within these disk environments, planets can form through two
conceptually different channels. First, planet formation can take
place from the top down, through gravitational instabilities in
massive circumstellar disks \cite{boss,cameron,rafikov2005}.  Second,
planet formation can take place from the bottom up, through the
accumulation and agglomeration of rocky building blocks into ever
larger bodies \cite{pollack1996}.  If this process occurs fast enough,
these growing rocky cores can accrete large amounts of gas from the
nebula and become giant planets (like Jupiter). In other cases, the
gas in the disk can be removed before runaway accretion takes place,
and the planet is left with a rocky core, along with ices and other
volatiles. This latter scenario produces planets like Uranus and
Neptune, and apparently occurs $\sim$ten times more often than the
successful production of Jovian planets.

Although both of the aforementioned mechanisms are likely to take
place, the core accretion (bottom up) scenario is thought to dominate.
This paradigm accounts for the observed mass distribution of planets,
which is weighted toward bodies of lower mass. Moreover, many of the
observed Jovian planets are close to their stars, where they can be
observed using both radial velocity and transit techniques. As a
result, their mass and radius can be measured, and hence their
densities, and their internal structure can be inferred. This program
indicates that many Jovian planets have high metallicity, consistent
with the presence of large rocky cores. These considerations suggest
that sufficiently high metal content is a necessary ingredient for
planet formation. In fact, in the current theoretical calculations of
giant planet formation, the surface density of solid material is the
most important variable in the problem \cite{ikoma2000}. 

In considering the potential viability of a universe, it is important
to keep in mind that the chemical composition of planets is not
necessarily the same as that of the host star --- or the universe as a
whole. Moreover, all of the planets in a given system will generally
not have the same composition \cite{bergin,tvjohnson,meyer}. In our
Solar System, for example, the giant planets are enriched in carbon
relative to the Sun, whereas Earth is significantly depleted 
\cite{allegre,marty}.  A related issue is that the composition of 
the planetary surface layers, which are most relevant for biospheres,
is generally not the same as the bulk composition. Planetary surfaces
are often sculpted during their late formation stages, leading to both
chemical enrichment and depletion. Relevant processes include
continuing impacts of minor bodies, which provide one channel for
water delivery to otherwise dry planets, and cataclysmic collisions,
thought to be the mechanism that formed our moon \cite{agnor,benz}. 

Our universe produces planetary bodies with a wide range of sizes by
utilizing a variety of processes (as outlined above). Even our own
Solar System contains many objects that are too large to have rocky
surfaces (the four giant planets) and a multitude of objects that are
too small to retain atmospheres (including $\sim10$ dwarf planets and
thousands of smaller bodies). Since planets of all sizes are readily
produced, it is likely for some to fall within the mass range required
for habitability. In other words, universes that support planetary 
structures (see Figure \ref{fig:planplane}) are likely to produce 
some bodies with favorable properties. 

\subsection{Planets and Stellar Convection} 
\label{sec:convection} 

Another potential fine-tuning constraint arises from the requirement
that both convective and radiative stars are necessary for a
successful universe. If all stars were convective, then massive stars
would mix all of their nuclear burning layers as they evolve, and
thereby erase the classic ``onion skin'' structure found during the
advanced stages of nuclear burning. Such mixing would cycle all of the
elements with intermediate mass numbers (such as carbon and oxygen)
into the stellar core, where they would be processed into larger
nuclei. This scenario could potentially leave the universe with little
carbon and oxygen. On the other hand, if all stars were radiative,
then some authors have claimed (starting with \cite{carter1974}) that 
planet formation would be compromised. This section re-examines the
argument that  both convective and radiative stars are necessary 
for a universe to be viable. 

As a general rule, low mass stars remain convective over much of 
their lifetimes, whereas larger stars develop a radiative structure 
\cite{clayton,hansen,kippenhahn,phil}. The boundary in stellar mass 
between these two regimes can be expressed in terms of fundamental 
constants \cite{bartip,carr} and has the form 
\be
M_{\rm r/c} \approx \alpha_G^{-2} \alpha^{10} \mpro = 
\alpha_G^{-1/2} \alpha^{10} \starmass \,, 
\label{radiconv} 
\ee 
where $\starmass$ is the mass scale of a typical star. If one requires 
that the mass threshold $M_{\rm r/c}$ is comparable to the typical 
stellar mass $\starmass$, then the following approximate equality 
must hold: 
\be
\alpha_G \sim \alpha^{20} \,. 
\label{radconlimit} 
\ee
Because of the large exponents on $\alpha$ in equations
(\ref{radiconv}) and (\ref{radconlimit}), modest changes in the fine
structure constant could raise or lower the mass threshold and lead to
a universe where all stars are either convective or radiative.

Although suggestive, the mass threshold in equation (\ref{radiconv})
is not sharp. Most stars --- including the Sun --- are convective in
their early phases of evolution and develop radiative cores later on.
As a result, the question of whether stars have convective versus
radiative structure depends on time.  In addition, when massive stars
enter into their advanced stages of nuclear burning, various layers of
the star become convective and radiative as the stars evolve (this
complicated behavior is readily seen in stellar evolution calculations
of intermediate and high mass stars using modern computational methods 
\cite{mesaone,mesatwo}). In any case, it is overly simplistic to use 
a back of the envelope estimate such as equation (\ref{radconlimit}) 
to delineate the boundary between convective and radiative energy 
transport in stellar interiors. 

Putting aside the above complications, suppose that the mass threshold
$M_{\rm r/c}$ is raised, so that all stars become convective. In the
worst case scenario, all of the massive stars would remain completely
convective over their entire lifetimes.  After being produced during
intermediate evolutionary stages, carbon and oxygen would be mixed
deep into the stellar interior and processed into even larger nuclei.
By the time the iron core is fully developed and the star explodes as
a supernova, most if not all of the carbon and oxygen could be
depleted.

Although this scenario would lead to a universe different from our
own, with different cosmic abundances of the elements, habitability is
not necessarily compromised. Stars with intermediate masses, roughly
comparable to the Sun and somewhat larger, burn their helium into
carbon and oxygen, but do not produce heavier elements. In universes
with convective massive stars, these intermediate mass stars could
provide enough carbon and oxygen to support habitability. In fact, 
in our universe, a substantial fraction of the carbon inventory is
produced during the Asymptotic Giant Branch (AGB) phases of
intermediate mass stars \cite{agbstars,gustafsson,henry}. AGB stars in
other universes could thus provide the necessary carbon, even if more
massive stars maintain convective cells and burn through their supply.

If the mass threshold of equation (\ref{radiconv}) is lower, then all
stars would be radiative. In spite of the original claim 
\cite{carter1974} (see also \cite{bartip,carr,tegmarktoe,tegmark} and 
others), planet formation does not rely on stars having a convective
internal structure. As outlined above, planets form in circumstellar
disks, which are found nearly ubiquitously around young stars, and
which are decoupled from the internal structure of the star. Both
mechanisms of planet formation, core accretion and gravitational
instability, rely on physical processes that take place within the
circumstellar disk, and do not depend on the stellar interior. Taking
an observational perspective, we have now detected planets around
stars without convective zones, including higher mass stars (see 
\cite{planbigstar} and many others), degenerate white dwarfs 
\cite{veras,zuckerman}, and even pulsars \cite{alex}. As a result, 
both observations and theoretical considerations definitively show
that convective stars are not necessary for planet production.

\section{Exotic Astrophysical Scenarios} 
\label{sec:exotica} 

Our local bubble of parameter space is suitable for habitability, and
the discussion thus far has focused on delineating the boundaries of
this region.  Given the wide range of particle physics and
cosmological parameters that could be realized across the multiverse,
it becomes possible for other universes to utilize unconventional
pathways and power sources. As a result, additional bubbles of
habitability could exist with cosmic properties markedly different
from our own. This section explores some of these possibilities:

Dark matter annihilation provides a promising channel of detecting
this elusive material \cite{cirelli}, but the expected contribution to
the galactic energy budget is minimal.  In other universes, however,
with denser galaxies and/or different dark matter properties, this
source of energy can be substantial (Section \ref{sec:astrohalos}).
Similarly, dark matter is expected to collect and annihilate within
stellar bodies in our universe, especially degenerate stellar remnants
such as white dwarfs and neutron stars. This channel of energy
production could also be enhanced in universes with denser galaxies
and more interactive dark matter (Section \ref{sec:wdcapture}). Note
that the dark matter abundance is determined in the early universe at
age $t\sim1$ sec and depends on the cross section for
self-interactions \cite{kolbturner}.  In order for other universes to
have comparable dark matter abundances but different cross sections,
other cosmological parameters must vary, or the dark matter abundance
must be determined by non-thermal (out of equilibrium) processes
\cite{baer,jungman,kaneluwatson}. Next we note that black holes can
provide a source of power through Hawking evaporation. Although this
process is completely negligible in our universe, it could become
important in the regime where gravity is much stronger relative to the
electromagnetic force (Section \ref{sec:bholestars}). If the weak
force is less effective, then compact objects composed of degenerate
dark matter can play the role of stars (Section \ref{sec:dmstar}).
Finally, even in the absence of nuclear reactions, universes can still
produce dark matter halos, galaxies, and even stars that shine via
gravitational contraction (Section \ref{sec:nukefree}). These latter
universes would not be habitable, but would nonetheless bear an eerie
resemblance to our own.

\subsection{Dark Matter Halos as Astrophysical Objects} 
\label{sec:astrohalos} 

Dark matter halos are essentially inert in our present-day universe,
as their evolutionary time scales are much longer than the current
cosmic age \cite{al1997}. Here we consider the case where dark matter
can be more interactive than in our universe, either through enhanced
densities or larger cross sections. For the sake of definiteness, we
consider the dark matter halos to have the form of a Hernquist profile
(see equation [\ref{hernquist}], Section \ref{sec:halostructure}, and
Refs. \cite{hernquist,nfw}).

In our universe, primordial density fluctuations, inferred from
observed inhomogeneities in the Cosmic Background Radiation, have
amplitude $Q \approx 10^{-5}$. These fluctuations could be larger in
other universes, with the consequence that galaxies can form earlier
and become denser (see Section \ref{sec:galaxies} and references
therein).  Here we define the relative amplitude
\be
q \equiv {Q \over Q_0} \,,
\ee
where $Q_0\approx10^{-5}$ is the value in our universe. The parameters
$\rho_0$ and $r_0$ that specify the properties of dark matter halos
vary with the fluctuation amplitude that specifies the initial
conditions.  For dark matter halos with density profiles given by
equation (\ref{hernquist}), the dependence of $\rho_0$ and $r_0$ on
the amplitude $Q$ has been derived previously (see equations
[\ref{rhoscale}] and [\ref{rscale}]), where these results are based on
the standard paradigm for galaxy formation \cite{whiterees}. The
resulting scaling laws take the form 
\be 
\rho_0 \propto q^3 
\qquad {\rm and} \qquad 
r_0 \propto q^{-1} \,. 
\label{qscale} 
\ee 
The fluctuation amplitude can be larger by more than a factor of
$\sim1000$ (Section \ref{sec:scatterbound}), so that the halo densities 
can be enhanced by many orders of magnitude. 

\subsubsection{Power from Dark Matter Annihilation} 
\label{sec:power} 

The annihilation rate per particle $\Gamma_1$ at a 
given radial location within the halo has the form 
\be
\Gamma_1 = n \sigdark = 
{\rho_0\over \xi (1+\xi)^3} {\sigdark \over \mdm} \,,
\ee
where $\sigdark$ is the cross section for dark matter annihilation and
$\mdm$ is the mass of the particle.  The total annihilation rate and 
hence the luminosity are determined by integrating over the entire 
halo. The luminosity due to dark matter annihilation is given by 
\be
L_{\rm dm} = {4\pi\over5} \rho_0^2 c^2 r_0^3 {\sigdark \over \mdm} \,. 
\label{dmluminosity} 
\ee
Typical values in our universe for Milky-Way-like galaxies are
$\rho_0\sim10^{-25}$ g cm$^{-3}$ and $r_0\sim65$ kpc
\cite{binmer,bintrem}, whereas typical dark matter properties are
$\sigdark\sim10^{-27}$ cm$^3$ s$^{-1}$ and $\mdm\sim100\mpro$
\cite{feng,jungman}. For these parameters, the luminosity
(\ref{dmluminosity}) evaluates to $L_{\rm dm}\approx500L_\odot$.
Since the luminosity of the halo scales as $q^3$, other universes can
have an enhancement of order $\sim10^9$ (using $q$ = 1000).  The dark
matter luminosity thus becomes $L_{\rm dm}\sim5\times10^{11}$
$L_\odot$, comparable to the stellar luminosity of a moderate-sized
galaxy in our universe.  The luminosity $L_{\rm dm}$ results from
annihilation of dark matter and corresponds to a mass loss rate given
by 
\be
{dM \over dt} = {d \over dt} \left( 2\pi \rho_0 r_0^3  \right) 
= - {4\pi\over5} {\rho_0^2 \sigdark \over \mdm} r_0^3 \,.  
\label{mdot} 
\ee

The radiation flux at a location $\xi=a$ within the galaxy 
can be written in the form 
\be
F_G(a) = 2\pi {\rho_0^2 c^2 \sigdark \over \mdm} r_0
\int_0^\infty {d\xi \over (1+\xi)^6} 
\int_{-1}^1 {d\mu \over \xi^2 + a^2 - 2a\xi\mu} \,.
\label{dmfluxone} 
\ee
The angular integral can be evaluated to obtain 
\be
F_G(a) = \rho_0^2 c^2 r_0 {\sigdark \over \mdm} 
{\pi \over a} I(a) = {5 L_{\rm dm} \over 4\pi r_0^2} 
{\pi \over a} I(a) \,, 
\label{dmfluxtwo} 
\ee
where we have defined a dimensionless integral function 
of the position $a$, i.e., 
\be
I(a) \equiv \int_0^\infty 
{\left\{ \log\left[\xi^2 + a^2 + 2a\xi\right] - 
\log\left[\xi^2 + a^2 - 2a\xi\right] \right\}
\over \xi (1+\xi)^6} d\xi \,.
\label{aintegral} 
\ee
Near the galactic center, where $a\ll1$, the integral 
approaches the form $I\to-\log(a)$. In the outer parts 
of the halo where $a\gg1$, the integral $I\sim a^{-1}$, 
so that the flux $F_G\sim a^{-2}$. 

The typical value of 
$F_G \sim L_{\rm dm}/(4\pi r_0^2)\sim4\times10^{-12}$ 
erg s$^{-1}$ cm$^{-2}$ $\sim 3 \times 10^{-18} S_\oplus$ 
(where $S_\oplus$ is the radiation flux received by Earth 
from the Sun).  The radiation flux from dark matter annihilation
scales as $F_G \sim q^5$, so that dense galaxies in other universes
can be enhanced by a factor of $10^{15}$ or more. Inner parts of 
such galaxies, or somewhat denser galaxies resulting from 
$q\approx3200$, can thus have background radiation fields equal 
to the value $S_\oplus$ due to Solar irradiance on Earth.  

\subsubsection{Time Evolution of Dark Matter Halos} 
\label{sec:evolution} 

For a given mass, the halo properties are determined by specification
of the parameters $\rho_0$ and $r_0$.  In order to provide a
quantitative description of the time evolution of the dark matter
halo, we need another constraint on the halo structure in addition to
the equation (\ref{mdot}) that determines the mass loss rate. Here we
make the approximation of adiabatic compression \cite{sellwoodmag},
which assumes that the specific angular momentum of orbits (here, for
dark matter particles) remains constant as the mass changes. In this
context, the approximation is equivalent to assuming that that
composite parameter 
\be
\rho_0 r_0^4 = constant\,.
\label{adcomp} 
\ee
This constraint, in conjunction with the specification of mass 
loss through equation (\ref{mdot}), determines the time evolution 
of the halo properties. We first define a dimensionless time 
variable 
\be
\tau = {t \over t_{\rm dm}} \qquad {\rm where} \qquad 
t_{\rm dm} \equiv {M_0 c^2 \over L_0} \,,
\label{taudef} 
\ee
where $M_0$ and $L_0$ are the initial mass and luminosity of the halo
respectively. The second equality defines the characteristic time
scale for time evolution of dark matter halos. For typical parameters
$\sigdark \approx 10^{-27}$ cm$^3$ s$^{-1}$ and $\mdm \approx 100 m_P$, 
this time scale become $t_{\rm dm}\approx3\times10^{21}$ yr for a dark 
matter halo with initial properties similar to those of the Milky Way.

However, the characteristic timescale for the evolution of dark matter
halos scales as $t_{\rm dm} \sim q^{-3}$.  Using the larger value
$q=1000$, the time scale can be shorter by a factor of $\sim10^{9}$.
Given the fiducial time scale for halo evolution of $\sim10^{22}$ yr,
denser halos in other universes could evolve on time scales of
$\sim10^{13}$ yr without changing the properties of the dark matter.

The time-dependent density, potential, and radial scale are given by  
\be
\rho = {\rho_{0(t=0)} \over (1 + 4\tau) \xi (1+\xi)^3} \,,
\qquad \Psi = {\Psi_{0(t=0)} \over (1+4\tau)^{1/2} (1+\xi)} \,, 
\qquad r_0 = r_{0(t=0)} (1 + 4\tau)^{1/4}\,, 
\label{halosolution} 
\ee
and the corresponding solutions for the time evolution of 
the halo mass and luminosity have the forms 
\be
M_{\rm dm} (\tau) = {M_0 \over (1+4\tau)^{1/4}}
\qquad {\rm and} \qquad 
L_{\rm dm} (\tau) = {L_0 \over (1+4\tau)^{5/4}}\,.
\label{othersolution} 
\ee



The column density of the dark matter halo, integrated from spatial 
infinity to a radial location $\xi$, is given by the expression 
\be
N(\xi) = \rho_0 r_0 \left\{ 
\log\left[ {1 + \xi \over \xi} \right] - 
{2\xi + 3 \over 2(\xi+1)^2} \right\} \, . 
\ee
The optical depth of the halo to its radiation field (that 
generated by the dark matter annihilation) is thus of order 
\be
\tau = \rho_0 r_0 {\sigma_{\rm rad} \over \mdm} \,,
\ee
where $\sigma_{\rm rad}$ is the cross section for interactions 
between the annihilation photons and the remaining dark 
matter particles. For values in our universe, the optical 
depth $\tau \sim 10^{-16}$. As a result, the halo does not 
have a photosphere -- it is optically thin to the radiation 
it generates, so that photons freely stream outwards. 

On the other hand, the halo could contain a gaseous component. Here we
assume that the mass in baryons is a fraction $f$ of the dark matter
density and that the gas has the same form for its density
distribution (see equation [\ref{hernquist}]). The optical depth 
of the baryons to the radiation produced via dark matter 
annihilation is given approximately by 
\be
\tau_{\rm gas} = f \rho_0 r_0 {\sigma_T \over \mpro} \approx 
0.008 f \,, 
\ee
where the numerical estimate assumes properties comparable to those of
our Galaxy.

The optical depth scales as $q^2$, and hence can be larger by a factor
of $\sim10^6$. Even with this level of enhancement, the annihilation
products (gamma rays) would have an optical depth that is much less
than unity for interactions with dark matter. However, the halo would
become optically thick due to its baryon content. In this case, the
annihilation products interact with hydrogen gas and the radiation
field would be processed to longer wavelength (lower energy), which is
more compatible for supporting habitable planets.

\begin{figure}[tbp]
\centering  
\includegraphics[width=.90\textwidth,trim=0 150 0 150]{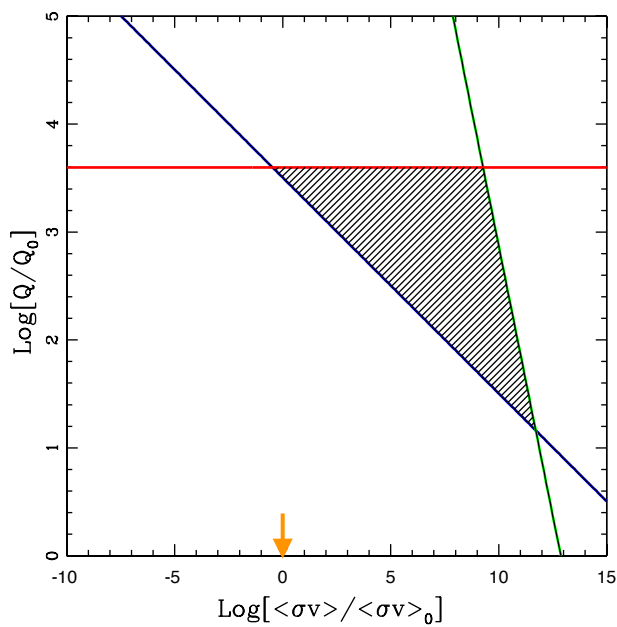}
\caption{Parameter space for which dark matter halos provide enough 
radiation for habitability. The dark matter annihilation cross section
is given on the horizontal axis and the amplitude of the primordial
density fluctuations is given on the vertical axis. These quantities
are scaled relative to benchmark values $\sigdark$ = $10^{-27}$ cm$^3$
s$^{-1}$ and $Q_0=10^{-5}$.  In order for the annihilation flux to be
larger than the Solar irradiance, the value of $Q$ must lie above the
blue curve. In order for the lifetime of the halo to be long enough to
support habitability, $Q$ must fall below the green curve. The 
horizontal red curve marks the maximum value of $Q$ beyond which black
hole formation becomes a serious issue. The orange arrow at the origin
marks the location of our universe, which is relatively far from the
delineated region of parameter space. Note that the horizontal axis
(cross section $\sigdark$) spans many more orders of magnitude than
the vertical axis (fluctuation amplitude $Q$). }
\label{fig:exhalo} 
\end{figure} 

Figure \ref{fig:exhalo} shows the allowed parameter space where dark
matter halos produce enough radiation to support habitable planets,
where this energy is produced via dark matter annihilation. The plane
of parameter space includes the amplitude $Q$ of the primordial
density fluctuations and the annihilation cross section for dark
matter particles.  For larger $Q$, galaxies form earlier and become
more compact, thereby leading to enhanced radiation flux due to
annihilation (see equation [\ref{dmfluxtwo}]). Similarly, larger cross
sections lead to enhanced radiation fields. The flux due to dark
matter annihilation is larger than the radiation received by Earth
from the Sun for the parameter space above the blue curve.  If dark
matter annihilation is too efficient, however, dark matter halos will
evolve too quickly. The green curve shows the boundary where the
potential well of the halo becomes too shallow ($v<100$ km/s) over a
time scale of 10 Gyr, which is comparable to the current age of our
universe. If the amplitude $Q$ is too large, above the red curve in
the figure, then black hole formation can proceed catastrophically.
The figure shows that a wide range of parameters allows for dark
matter to provide the energy necessary for habitability. On the other
hand, this regime is far removed from the parameters of our universe,
which is marked by the orange arrow at the origin.

Notice that Figure \ref{fig:exhalo} does not include an upper limit on
the annihilation cross section for dark matter, although $\sigdark$
cannot be made arbitrarily large.  This cross section depends on the
strength of the weak force. In our universe the strong force dominates
on the scale of atomic nuclei, so that the weak force provides only a
perturbative effect on nuclear structure. Increases in the cross
section $\sigdark$ correspond to increases in the weak coupling
constant, which cannot become too large without compromising nuclear
reactions in stars and even the existence of bound states. In the low
energy limit, the weak coupling constant $\alpha_{\rm w}$ = 
$G_F \mpro^2 \sim10^{-5}$, which is much smaller than the strong
coupling constant $\alpha_{\rm s}\sim10$ (see Section
\ref{sec:standardmodel}). As a result, the Fermi constant $G_F$ cannot
increase by more than a factor of $\sim10^6$, and the interaction
cross section $\sigdark\propto G_F^2$ cannot increase by more than a
factor of $\sim10^{12}$. More stringent bounds are likely, but require
an in-depth analysis of nuclear structure. These considerations thus
limit the parameter space shown in Figure \ref{fig:exhalo}. This
effect is roughly comparable to the requirement that the halos live
long enough, as depicted by the green curve.

\subsection{Dark Matter Capture and Annihilation in White Dwarfs} 
\label{sec:wdcapture} 

White dwarfs provide another channel through which dark matter can 
be processed in a galactic halo \cite{al1997}. These dense stellar
remnants can accrete dark matter particles, which accumulate in the
stellar core. Once the population is sufficiently large, the system
reaches a steady state where the rate of annihilation in the stellar
core is balanced by the rate of particle accretion from the halo. 
Although this process is expected to produce (at most) modest
luminosities in our universe, this channel of energy generation can 
be significant in alternate universes with larger $Q$ and denser 
galaxies.  Notice also that current experiments for the direct
detection of dark matter \cite{akerib,aprile} put tight constraints 
on interactions between dark matter and baryons in our universe, but
the relevant cross sections could be different in other universes.

White dwarfs are expected to be optically thick to dark matter
particles. Here we assume that cross section of interaction between
dark matter particles and the baryonic stellar material is
$\sigone\sim10^{-38}$ cm$^2$. The density of a white dwarf is of order
$\rho_{\rm wd} \sim 10^6$ g cm$^{-3}$, which corresponds to a number
density $n_{\rm wd}\sim10^{30}$ cm$^{-3}$. With the radius of a white
dwarf, $R_{\rm wd}\sim10^8$ cm, we find the optical depth to be 
\be
\tau_{\rm wd} \sim n_{\rm wd} \sigone R_{\rm wd} \sim 1 \,. 
\ee
The rate of capture of dark matter particles by the star 
is thus given by 
\be
\Gamma_{\rm cap} = n_{\rm dm} \sigma_{\rm wd} v_h \,,
\ee
where $\sigma_{\rm wd}$ is the cross section of the star for the 
capture of dark matter. The cross section is enhanced over 
the geometric cross section of the star through gravitational
focusing so that 
\be
\sigma_{\rm wd} = \pi R_{\rm wd}^2 
\left(1 + {GM_{\rm wd}\over R_{\rm wd} v_h^2} \right) 
\approx 10^{18}\,{\rm cm}^2\, 
\left(1 + \left[{3000\,{\rm km/s}\over v_h}\right]^2\right)\,.
\ee
The dark matter particles will collect inside the star until 
the annihilation rate and the capture rate become equal, so 
that a steady state is reached. The resulting luminosity of 
the star, produced by the capture and subsequent annihilation 
of dark matter, is thus given by 
\be
L_{\rm \ast dm} = n_{\rm dm} \mdm c^2 \sigma_{\rm wd} v_h =
\rho_{\rm dm} c^2 \sigma_{\rm wd} v_h \,.
\ee
This luminosity is about $10^{17}$ Watt for white dwarfs in 
our galaxy. Including the billions of white dwarfs in the galaxy, 
the total luminosity produced through this channel is of order 
1 $L_\odot$. 

The power generated by dark matter capture and annihilation in 
white dwarfs is comparable to that generated via direct (particle 
on particle) annihilation. This approximate equality arises due 
to the similarity in opacities (cross section per unit mass) of 
the two processes, i.e., 
\be
{\sigone \over \mdm} \sim {\sigma_{\rm wd} \over M_{\rm wd}} = 
{\pi R_{\rm wd}^2 \over M_{\rm wd}} 
\left(1 + {GM_{\rm wd}\over R_{\rm wd} v_h^2} \right) \approx 
{\pi G R_{\rm wd} \over v_h^2} \,. 
\ee
With typical values for the parameters, both opacities are of 
order $10^{-14}-10^{-13}$ cm$^2$ g$^{-1}$. 

In other universes, with larger values of $Q$, the galactic halos will
be denser and the luminosity of individual white dwarfs generating
energy via this mechanism will be larger by a factor of $q^3$. As
discussed above, the initial fluctuations could be enhanced by a
factor as large as $q=1000$, thereby increasing the luminosity to
$L_{\rm \ast dm}\sim10^{-2}L_\odot$. The total radiation flux from the
galactic background will thus be given by equation (\ref{galaxyflux}),
where the luminosity per star is taken to be $L_{\rm \ast dm}$. Since
this luminosity scale is smaller (on average) than that of
main sequence stars, the radiation fields from ordinary stars would
dominate as long as they are actively burning nuclear fuel. In the 
absence of nuclear power, however, the luminosity of the galaxy is 
only smaller by a factor of $\sim10^2-10^3$, so that the prospects 
for habitability are not completely diminished. 

For completeness, we note that the dense conditions found in white
dwarf interiors are conducive to pycnonuclear reactions, where nuclear
fusion takes place at low temperature via quantum tunneling
\cite{cameronpycno,schrammpycno,salpeterpycno}. In this process, which
is too slow to be important in our universe \cite{shapteuk}, the
quantum mechanical zero point energies of the nuclei allow them to
overcome their mutual Coulomb repulsion and interact. Although
hydrogen can fuse in white dwarfs and thereby produce some additional
helium, larger nuclei are not synthesized, even over time scales that
vastly exceed the current age of the universe \cite{al1997}. In other
universes with different fundamental parameters, however, such
pycnonuclear reactions could compete with dark matter annihilation as
an additional energy source for white dwarfs. This issue should be
considered further.

\subsection{Black Holes as Stellar Power Sources} 
\label{sec:bholestars} 

Black holes are expected to exist in any universe. Given that our
universe forms black holes on both stellar and galactic scales, one
also expects that the formation of black holes will not be completely
suppressed. As a result, black holes will provide a power source
through Hawking evaporation \cite{hawking}. Such radiation is
completely negligible in our universe at the present epoch, but could
play a role in universes with stronger gravity and other favorable
parameters. Black hole properties depend primarily on the
gravitational constant $G$ (equivalently $\alpha_G$), whereas
habitability involves atomic energies and time scales, and thus
depends on the fine structure constant $\alpha$.

Black holes can exist over a wide range of masses, but they require a
production mechanism. For this illustrative treatment, we focus on the
case of stellar black holes, where the mass scale is approximately
given by $M_{\rm bh}\approx\starmass=\alpha_G^{-3/2}\mpro$, comparable 
to the Chandrasekhar mass (see equation [\ref{starmassform}]). Black 
holes have effective temperatures given by 
\be
T_{\rm bh} = {1 \over 8 \pi GM_{\rm bh}} = 
{\alpha_G^{3/2} \over 8 \pi G \mpro} \,, 
\label{bholetemp} 
\ee
where $M_{\rm bh}$ is the mass of black hole, which is taken to 
be $\starmass$ in the second equality. This temperature 
must be larger than that required to drive chemical reactions 
(from equation [\ref{biotemp}]). Black holes must satisfy the 
constraint 
\be
T_{\rm bh} > E_{\rm chem} \qquad \Rightarrow \qquad 
\alpha_G > \left( 8 \pi \epsilon_{\rm chem} \beta \alpha^2 \right)^2\,. 
\label{bhtempconst} 
\ee

The lifetime of a black hole with initial mass $M_{\rm bh}$ 
takes the form 
\be
\tau_{\rm bh} = {2650\pi \over g_\ast} G^2 M^3_{\rm bh} = 
{2650\pi \over g_\ast} \alpha_G^{-5/2} \mpro^{-1} \,,
\label{bholelife} 
\ee
where $g_\ast$ is the number of effective degrees of freedom in the
radiation field produced by Hawking evaporation and where the second 
equality assumes $M_{\rm bh}=\starmass$. In order for black 
holes to serve as engines of habitability, their lifetime must be 
sufficiently long. If we measure time in terms of atomic time 
scales, and require the benchmark number of such time units 
$N_{\rm bio}=10^{33}$, we find the following constraint 
\be
\alpha_G < \alpha^{4/5} \beta^{2/5} \left[ 
{2650\pi \over g_\ast N_{\rm bio}} \right]^{2/5}\,. 
\label{bhtimeconst} 
\ee

\begin{figure}[tbp]
\centering 
\includegraphics[width=.95\textwidth,trim=0 150 0 150,clip]{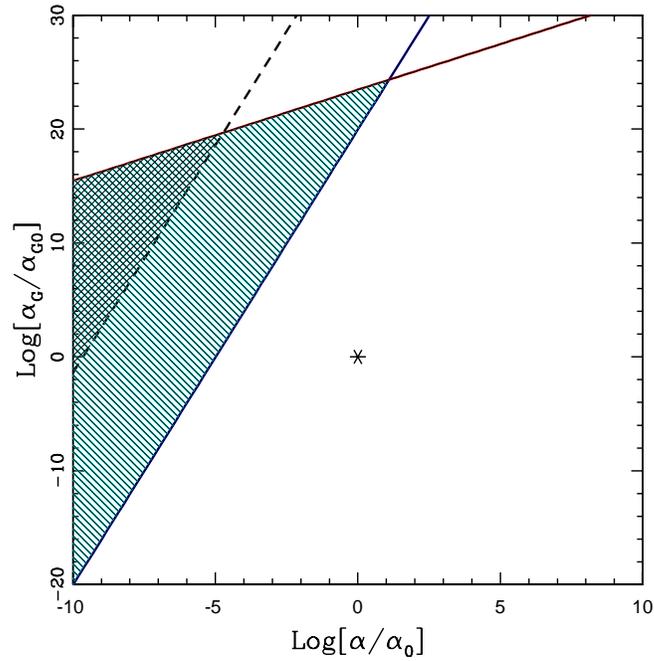}
\caption{Allowed parameter space for black holes to play the role of 
stars in other universes with different values of the structure
constants $\alpha$ and $\alpha_G$. For illustrative purposes, the 
black hole mass is taken to be the stellar mass scale 
$M_{\rm bh}=\starmass=\alpha_G^{3/2}\mpro$.  In order for the black
hole radiation to drive chemical reactions, $\alpha_G$ must lie above
the blue curve. In order for the black hole lifetime to be long enough
to support biological evolution, $\alpha_G$ must fall below the red
curve. The shaded region delineates the parameters that satisfy both
constraints. The dashed black curve shows the minimum value of
$\alpha_G$ for which the total power output is larger than a minimum
benchmark value $P\sim10^{17}$ erg/sec. Our universe is located at the
origin in the figure, as marked by the star symbol. }
\label{fig:bholeplane} 
\end{figure} 

In general, these stellar black holes have a relatively small
luminosity, given by 
\be
L_{\rm bh} = {1 \over 15360\pi G^2 M_{\rm bh}^2} = 
{\alpha_G \mpro^2 \over 15360\pi} \,.
\label{bholepower} 
\ee
If we require that the total power output is larger then a fiducial
minimum value $L_{\rm min}$, then we obtain an additional constraint.
Unfortunately, the value of $L_{\rm min}$ remains unknown. For the
sake of definiteness, we take the (arbitrary) value $L_{\rm min}$ =
$10^{17}$ erg/sec and scale the result for other universes. This
benchmark value corresponds to 10 gigawatts, enough power to run a
relatively large city. For universes with varying $\alpha$, we scale
the minimum power $L_{\rm min}$ as follows: The energy levels of atoms
vary $\sim\alpha^2$ and the atomic time scales vary according to
$t_A\sim\alpha^{-2}$.  In order for the black hole luminosity to
provide the same number of atomic reactions over the lifetime of the
system, the scaling law becomes $L_{\rm min}$ = 
$L_{{\rm min}0}(\alpha/\alpha_0)^4$, where the subscript denotes
values in our universe. 

The allowed region of parameter space that satisfies both the
temperature constraint (\ref{bholetemp}) and the lifetime constraint
(\ref{bholelife}) is shown in Figure \ref{fig:bholeplane}. The red
line provides the upper limit on $\alpha_G$ by requiring that the
black holes live for $N_{\rm bio}=10^{33}$ atomic time scales, taken
here to be the time required for biological evolution. The blue curve 
provides the lower limit on $\alpha_G$ by requiring that the black
hole temperature is high enough to support chemical reactions and
hence life. The shaded portion of the diagram depicts the parameter
space for which black holes are both hot enough and sufficiently
long-lived to play the role of stars. The dashed line represents 
the additional requirement that the power output of the black holes 
must be larger than the fiducial minimum value $L_{\rm min}$. This 
constraint requires $\alpha_G$ to lie above the dashed curve. Our 
universe, marked by the star symbol, falls well outside the region 
where black holes are an important power source. 

As illustrated in Figure \ref{fig:bholeplane}, the allowed range of
parameters for which black holes can play the role of stars spans many
orders of magnitude and thus appears reasonably large. However, this
diagram assumes particular values for the time scales necessary for
biological evolution (red curve) and the minimum luminosity for a
biosphere (dashed curve). One should keep in mind that these choices
are uncertain, and could be either much larger or smaller than the
fiducial values used here. Moreover, the viable region for black holes
shown in Figure \ref{fig:bholeplane} does not overlap with the region
of the plane corresponding to working stars from Figure
\ref{fig:starplane}. This mismatch indicates that universes are
unlikely to have both nuclear burning stars and stellar black holes 
providing power sources for biology.

The discussion thus far has focused on black holes with masses
comparable to stars $M_{\rm bh}\sim\starmass$. Another possibility 
is for universes to produce primordial black holes with masses much
smaller than stars \cite{carrhawk}. To illustrate how such objects
could play the role of stars, we define a fiducial mass scale
\be
m_{15} \equiv 10^{15} {\rm g} = 5\times10^{-19} M_\odot \,,
\ee
which represents the minimum mass for a black hole to live for the
current age of the universe (comparable to the time required to
support habitability). The black hole temperature (\ref{bholetemp}),  
lifetime (\ref{bholelife}), and luminosity (\ref{bholepower}) can 
be written in the forms 
\be
T_{\rm bh} = {1.2 \times 10^{11} {\rm K} \over g\,m_{15}} \,, 
\quad \tau_{\rm bh} = (10^{10} {\rm yr})\, g^2 m_{15}^3\,, 
\quad {\rm and} \quad 
L_{\rm bh} = {3 \times 10^8 \, {\rm W} \over g^2 m_{15}^2} \,, 
\ee
where $g=G/G_0$ is the gravitational constant scaled to the 
value in our universe. Next we impose the constraints that the 
surface temperature must be sufficiently hot, $T_{\rm bh}>300$ K, 
and that the objects must live longer than $10^{10}$ yr. The 
mass and scaled gravitational constant then obey the constraints 
\be
gm_{15} < 3 \times 10^8 \qquad {\rm and} \qquad g^2 m_{15}^3 > 1 \,. 
\ee
The smallest masses leads to the largest luminosities.  For the
gravitational constant in our universe ($g=1$), the maximum black hole
luminosity compatible with these constraints is $L_{\rm bh}\approx360$
MW, enough power to run a small city. Even larger luminosities are 
possible for weaker gravity (smaller $g$). Primordial black holes 
thus provide another channel for alternate universes to generate power.

\subsection{Degenerate Dark Matter Stars} 
\label{sec:dmstar} 

This section considers the possibility that an alternate universe can
produce stellar objects composed entirely of dark matter and supported
by the degeneracy pressure of the constituent particles.  Such dark
matter stars can generate energy through the process of dark matter
self-annihilation. Unlike the case of dark matter halos (Section
\ref{sec:astrohalos}) where the interaction rate is too slow in our
universe to play a significant role, degenerate dark matter stars are
so dense that their annihilation rates are too rapid. Although the
formation of these types of stars is problematic, this section
outlines the parameter space necessary for degenerate dark matter
stars to play the role of hydrogen burning stars in our universe
(see also \cite{adams}). 

The properties of these stellar bodies can be determined using
arguments analogous to those used for white dwarfs. The equation of
state for a degenerate star is that of an $n=3/2$ polytrope, where 
the leading coefficient $K$ in this setting is given by
\be
K = (3 \pi^2)^{2/3} {\hbar^2 \over 5 \mdm^{8/3} } \, , 
\ee 
where $\mdm$ is the mass of the dark matter particle. 
The mass-radius relation has the form 
\be
M_\ast R_\ast^3 = \xi_\star^3 \mzero {9 \pi^2 \over 128} 
\hbar^6 \mdm^{-8} G^{-3}  \,, 
\ee 
and the central density is given by 
\be
\rho_{\rm c} = {32 \over 9 \pi^2 \mzero^2} 
{G^3 \mdm^8 M_\ast^2 \over \hbar^6} \,. 
\ee
These stars will have a maximum mass that can be supported by the
degeneracy pressure of the dark matter particles. This mass scale is
the analog of the Chandrasekhar mass and can be written in the form 
\be
M_{\rm ch} = \mzero { (3 \pi)^{1/2} \over 2} 
\left( {\hbar c \over G \mdm^2} \right)^{3/2} \mdm \,, 
\label{chandradm} 
\ee 
where the parameter $\mzero\approx2.714$ for an $n=3/2$ polytrope
(Section \ref{sec:stars}). Note that this expression does not include
general relativistic corrections \cite{shapteuk}.  For dark matter
particles with mass $\mdm=100\mpro$, the Chandrasekhar mass scale
$M_{\rm ch}\approx0.0007M_\odot$. Notice also that if we consider
stars that are a fraction of the Chandrasekhar mass, so that
$M_\ast=XM_{\rm ch}$, then the central density is given by $\rho_{\rm
  c} \sim X^2 \mdm^4$ (in natural units).

The luminosity of the stars is determined by the annihilation rate 
of the constituent dark matter particles. The annihilation rate per
particle $\Gamma_1$ is given by 
\be
\Gamma_1 = n \sigdark \,, 
\ee
where $n$ is the number density of particles in the star. 
The corresponding total annihilation rate $\Gamma_T$ integrated
over the volume of the star is given by 
\be
\Gamma_T = {\gamma_\star \over \mzero} 
\left({M_\ast \over \mdm}\right) 
{\rho_{\rm c} \over \mdm} \sigdark 
\qquad {\rm where} \qquad \gamma_\star \equiv 
\int_0^{\xi_\star} \xi^2 f^{3} d\xi \,, 
\ee
where $\gamma_\star\approx1.128$. As a result, the total
annihilation rate is given approximately by $\Gamma_T\sim0.4$
$N_T\Gamma_1$, where $N_T$ is the number of particles in the star 
and $\Gamma_1$ is evaluated at the stellar center.
The corresponding stellar luminosity is then given by
\be 
L_\ast = {\gamma_\star \over \mzero} 
\left(M_\ast c^2\right) 
{\rho_{\rm c} \over \mdm} \sigdark = 
{32 \over 9 \pi^2} {\gamma_\star \over \mzero^3} 
G^3 M_\ast^3 {\mdm^7 c^2 \over \hbar^6} \sigdark \, . 
\ee
For the parameters of our universe, this luminosity is enormous, 
and these degenerate stars will be short-lived.  In order for 
this scenario to produce stellar objects that are useful for 
habitability, the annihilation cross section must be smaller, 
as discussed below. 

We want to consider stars, with masses below the Chandrasekhar mass 
$M_{\rm ch}$, that satisfy two constraints. The first requirement 
is that the star is has a high enough surface temperature to drive 
chemical reactions on suitably situated planets, which implies that 
\be 
{L_\ast \over R_\ast^2} > 16\pi\sigmasb 
\left( { \epsilon_{\rm c} \alpha^2 \emass c^2 \over k} \right)^4 \,,
\ee
where the efficiency $\epsilon_{\rm c}\sim10^{-3}$ (see Section 
\ref{sec:stars}). After some simplification, this constraint can 
be written in the form 
\be
G^5 M_\ast^{11/3} \mdm^{37/3} \sigdark 
> B_1 \epsilon_{\rm c}^4 \alpha^8 \emass^4 c^4 \hbar^7 \,,
\ee
where the numerical constants have been combined into a single
constant $B_1\approx52,000$. We can write the mass of the star 
in terms of the fundamental mass scale for degenerate stars with
particle mass $\mdm$, i.e., 
\be
M_\ast = X \left({\hbar c \over G \mdm^2}\right)^{3/2} \mdm \,. 
\ee
With this ansatz, the constraint for surface temperature becomes 
\be
X^{11/3} \left({\hbar c \over G \mdm^2}\right)^{3/2} G \mdm^8 
\sigdark > B_1 \epsilon_{\rm c}^4 \alpha^8 \emass^4 \hbar^3 \,.
\label{darktemp}
\ee

Next we require that the stellar lifetime is sufficiently long. 
If the degenerate star starts its evolution with initial mass $M_0$
and later has a mass $M_\ast(t)\ll M_0$, then its age $t_\ast(M_\ast)$ 
is related to its current mass $M_\ast$ through the expression
\be
t_\ast (M_\ast) = {M_\ast c^2 \over 2 L_{\ast} } \, , 
\label{dmtime} 
\ee
where $L_{\ast}$ is the luminosity of the star when it has mass
$M_\ast$. We want the stellar age $t_\ast$ to exceed the time required
for biological evolution. This constraint can be written in the form 
$t_\ast>N_{\rm bio}t_A$, where $t_A$ is the time scale for atomic
processes and where we expect $N_{\rm bio}\sim10^{33}$. We thus 
obtain the requirement  
\be
{9\pi^2 \over 64} {\mzero^3 \over \gamma_\star} 
\hbar^5 \emass c^2 \alpha^2 > N_{\rm bio} 
\left[G^3 M_\ast^2 \mdm^7 \sigdark \right]\, .
\ee
As before, we write the stellar mass in terms of the 
fundamental mass scale and we combine the dimensionless 
constants into a single composite $B_2$, so that we obtain 
\be
\hbar^2 \emass \alpha^2 > B_2 N_{\rm bio} X^2 (c) \mdm^3 \sigdark \,.
\label{darktime}
\ee

Both constraints of minimum temperature (equation [\ref{darktemp}])
and long stellar lifetime (equation [\ref{darktime}]) must be met in a
viable universe. More specifically, these expressions define the
parameters necessary for dark matter stars to play the role of
ordinary stars. To explore this parameter space, we set $X=0.1$,
corresponding to stars that are comparable to, but smaller than, the
maximum mass limit. The cross section for dark matter annihilation 
can be written in the approximate form 
\be
\sigdark \sim G_F^2 \mdm^2\,,
\ee
where the Fermi constant $G_F=1/(\sqrt{2}{\cal V}^2)$ (and where 
$G_F\approx$ (293 GeV)$^{-2}$ in our universe). For a given 
stellar mass, the parameter space is specified by the fine structure 
constant $\alpha$, the strength of the weak force specified by 
${\cal V}$, and the mass $\mdm$ of the dark matter particles. 

\begin{figure}[tbp]
\centering 
\includegraphics[width=1.0\textwidth,trim=0 150 0 150]{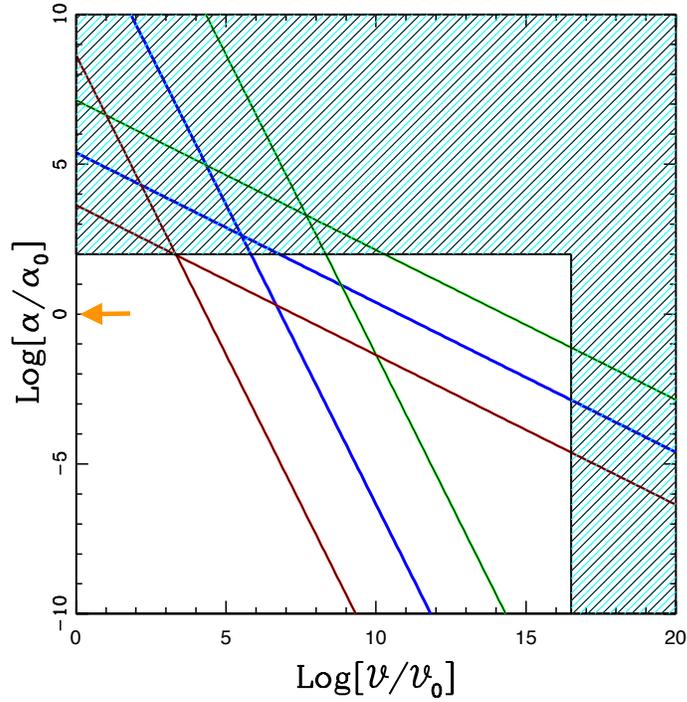}
\caption{Parameter space for degenerate dark matter stars to play the 
role of hydrogen burning stars in our universe. The horizontal axis 
shows the weak energy scale ${\cal V}$ relative to the value in our 
universe; the vertical axis shows the scaled fine structure constant 
$\alpha$. Two curves are shown for three masses of the dark matter: 
$\mdm$ = 0.01 $\mpro$ (red), 1 $\mpro$ (blue), and 100 $\mpro$ (green).
In order for the surface temperature to be high enough, the parameters 
must fall below the curves of shallow slope; in order for the stellar 
lifetimes to be long enough, the parameters must fall to the right 
(above) the curves of steeper slope. The weak scale becomes larger 
than the Planck scale on the right side of the diagram, whereas the 
fine structure constant becomes larger than unity in the top part 
of the diagram; these regions are disallowed. The location of our 
universe is marked by the orange arrow. }  
\label{fig:dmstar} 
\end{figure}  

Figure \ref{fig:dmstar} shows the constraints derived above in the
plane of parameter space $({\cal V},\alpha)$ for different masses of
the dark matter: $\mdm$ = 0.01 $\mpro$ (red curves), 1 $\mpro$ (blue
curves), and 100 $\mpro$ (green curves). The curves with shallow slope
depict the constraint that the stellar surface temperature is high
enough to allow suitably situated planets to have temperatures that
support biological (chemical) processes. Viable universe must fall
below these curves (for a given mass $\mdm$). The curves with steeper
slope depict the constraint that the stellar lifetime is long enough
to allow for biological evolution, where we use the benchmark value
$N_{\rm bio}=10^{33}$. All of the curves shown use stellar masses with
$X=0.1$, roughly comparable to, but smaller than the Chandrasekhar
mass. The shaded region in the diagram is not allowed for viable 
universes. It is bounded on the right by the constraint that the 
weak scale parameter must be smaller than the Planck scale 
${\cal V}<\mplanck$. The region is bounded from above by requiring 
that the fine structure constant $\alpha\ll1$. 

This analysis shows that a sizable region of parameter space allows
for degenerate dark matter stars to live long enough and have
sufficiently high luminosities to serve as engines of biological
evolution. However, the formation mechanism(s) for these bodies
remains unknown. In principle, dark matter can have self-interactions,
which allow for energy dissipation and the formation of bound
structures \cite{dietl,narain}. On the other hand, the annihilation
cross sections can be large enough to prevent the stars from achieving
hydrostatic equilibrium \cite{stojkovic}.  Another possibility is for
bound structures to be produced through phase transitions in the early
universe \cite{witten}. Nonetheless, the formation of dark matter
stars represents a formidable obstacle.

\subsection{Nuclear-Free Universe} 
\label{sec:nukefree} 

As another non-conventional scenario, it is interesting to consider
universes in which no nuclear reactions can take place
\cite{liviorees2018,rees2016}. In this case, the universe is assumed
to include the strong force so that it supports protons as bound
states of quarks.  Unlike our universe, however, no further nuclear
processing can take place, so the universe retains a pure hydrogen
composition. Such a universe would be far too simple to support life,
or anything resembling biochemistry. On the other hand, when viewed
from an astronomical perspective, this specter universe would look
superficially much like our own:

On the largest scales, galaxy formation takes place through the
collapse of dark matter to form halos \cite{reesost,whiterees}. The
interplay between cosmic expansion and gravitational collapse is
largely independent of nuclear considerations, so that large scale
structure would be essentially unchanged. After the dark matter halos
form, baryonic gas must cool and condense, and cooling processes
depend on chemical composition. In our universe, however, most of
galaxy formation takes places with primordial abundances of the
elements. As a result, the main difference between the nuclear-free
universe and our own is the presence of primordial helium. However,
the cooling from higher temperatures is dominated by bremsstrahlung
processes \cite{abel,gallipalla}, so that heavy nuclei (and molecules)
provide only higher order effects.

After galaxies are in place, stars are produced on smaller scales.
The star formation process itself does not depend on nuclear
considerations \cite{mckeeost,sal1987}. On the other hand, the initial
conditions for star formation depend on cooling processes in the dense
interstellar medium, where molecules and dust play an important role,
and these starting conditions affect the distribution of stellar
masses \cite{bastian}.  The cooling processes of the nuclear-free
universe will be much like those of our universe during the epoch when
the first generation of stars was forming \cite{abel,tegmarkbd}. We
thus expect the distribution of stellar masses to be similar to those
of Population III stars. Although the initial mass function for these
early stars is not known, it is thought to skew towards stars of
higher mass compared to the stellar mass distribution of the present
day \cite{bromm,schneider,susa}.

The vast majority of stars are born with large radii and relatively
cool central temperatures. Nuclear burning does not take place until
after an extended phase of pre-main-sequence contraction, which
depends on stellar mass and typically takes millions of years
\cite{clayton,hansen,phil}. In the absence of nuclear reactions, the
gravitational contraction phase continues much longer --- until the
stars reach an end state analogous to either white dwarfs or black
holes \cite{agdeuterium}, depending on the initial mass of the object.
Note that the cooling time for white dwarfs is longer than the current
age of our universe \cite{wood}, so that these stellar remnants can
continue to provide power over times scales $t\simgreat10$ Gyr after
the end of the contraction phase.

Just as in our universe, the starting state of the star formation
process is expected to have substantial amounts of angular momentum.
As collapse proceeds, most of the infalling material gets channeled
onto a circumstellar disk, which subsequently transfers material onto
the star. The formation of this disk is significant, as it allows for
planets to form. As outlined in Section \ref{sec:planetform}, planet
formation can take place via two channels. The core accretion paradigm
requires heavy elements to make large rocky cores \cite{pollack1996}
and would not operate in a nuclear-free universe. However, planets
could still form in principle through gravitational instabilities in
the circumstellar disks \cite{boss,cameron,rafikov2005}. The natural
mass scale of secondary objects formed in this manner is roughly
$M_S\sim10M_{\rm Jup}$, significantly larger than most planets in our
universe. By default, these planets would be composed of hydrogen, 
which would cool and take a solid form. 

For stars with masses less than the Chandrasekhar limit, stellar
evolution ends with all of the energy leaking out of the star, which
eventually reaches the radius of a zero-temperature white dwarf. The
total energy radiated over the course of the stellar lifetime is thus
given by 
\be
E_T = f_{\rm wd} {GM_\ast^2 \over R_{\rm wd} } - 
f_1 {GM_\ast^2 \over R_1} \approx 
f_{\rm wd} {GM_\ast^2 \over R_{\rm wd} } \,,
\ee
where we expect the dimensionless constant $f_{\rm wd}\approx3/7$ in
the long-time limit where the star becomes an $n=3/2$ polytrope.  The
final equality holds because the final radius $R_{\rm wd}$ (comparable
to the radius of Earth) is much smaller than the starting radius $R_1$
(several times the radius of the Sun). The total energy $E_T$ produced 
is thus the binding energy of a white dwarf, which can be written in 
the form 
\be
E_T \approx E_{\rm wd} = 4f_{\rm wd} 
\left({2 \over 9\pi^2 \mzero}\right)^{1/3} 
{M_\ast^{7/3} \emass \mpro^{5/3} G^2 \over \hbar^2} \,. 
\ee
If we write the stellar mass in the form $M_\ast = X \starmass$ 
(see equations [\ref{starmassform}] and [\ref{xstarmass}]), 
the above expression can be written 
\be
E_T \approx 0.35 (M_\ast c^2) X^{4/3} \beta \,,
\label{estarlow} 
\ee
where $\beta=\emass/\mpro$. In our universe, the energy generated by a
star via hydrogen burning on the main sequence is $E_H \sim f_{core}
M_\ast \effish$, where $f_{core}\approx0.1$ is the fraction of the
mass in the stellar core, and where $\effish=0.007$ is the nuclear
efficiency for hydrogen fusion. Since $\beta=1/1836$, low mass stars
in the nuclear-free universe will produce about 10 times less energy
than those in our universe.

For non-rotating stars with masses above the Chandrasekhar limit,
gravitational contraction continues until the star is smaller than its
event horizon, and the object becomes a black hole. Over the course of
its evolution, a high mass star radiates an energy comparable to its
total mass energy so that
\be
E_T \sim M_\ast c^2 \,. 
\label{estarhigh} 
\ee
If the star has appreciable angular momentum, then not all of the mass
would immediately fall within the event horizon. Instead, the material
with the highest specific angular momentum would form an accretion
disk surrounding the collapsed central object. Note that the size of
this disk structure (many Schwarzschild radii, perhaps 10 -- 100 km) 
is much smaller than the sizes of circumstellar disks that form
planets (10 -- 100 AU).  This disk could then dissipate energy,
transfer angular momentum, and eventually channel more of the mass
into the black hole, with the total radiated energy approaching that
of equation (\ref{estarhigh}).

Compared to our universe, the nuclear-free universe thus has galaxies
and other large scale structures that are essentially the same, stars
that generate less total energy by factors of $\sim10$, and many fewer
planets that have larger masses. On the other hand, such a universe
would have no rocky planets, no nuclei heavier than protons, and
nothing approaching biological complexity. This specter universe would
thus be quite similar to ours on cosmological scales and completely
different on terrestrial scales.

\bigskip 
\section{Conclusion} 
\label{sec:conclude} 

An intricate network of constraints must be satisfied in order for a
given universe to be viable. This section provides a guide through 
the labyrinth by organizing the results of this review in several 
different ways: Section \ref{sec:resultsummary} presents a
straightforward summary of the most important results. Section
\ref{sec:trends} identifies general trends emerging from this 
collection of constraints. The relation to anthropic arguments is
addressed in Section \ref{sec:anthropic}, and parameter variations
that potentially allow universes to be more habitable are outlined in
Section \ref{sec:bestworld}. This overview concludes with a brief
discussion of open issues (Section \ref{sec:futureissues}) and the
insights gained from this endeavor (Section \ref{sec:insight}).

\subsection{Summary of Fine-Tuning Constraints} 
\label{sec:resultsummary} 

The first step in determining the degree of fine-tuning of our
universe --- and others --- is to delineate the range of parameter
space for which observers can arise. This issue is made difficult
because we have no definitive determination of what parameters are
allowed to vary and what requirements must be enforced to ensure
habitability. For this latter issue, this treatment considers a
universe to be viable if it can successfully produce complex
structures, including composite nuclei, planets, stars, and galaxies.
The genesis of these entities puts additional constraints on the
universe itself. Although a definitive assessment remains elusive,
this review suggests the following constraints:

\medskip 

The masses of light quarks are among the most constrained parameters
of the Standard Model of Particle Physics. The mass of the down quark
can only vary by a factor of $\sim7$, while the mass of the up quark
can vary by several orders of magnitude (see Figure \ref{fig:updown}).
The ranges are asymmetric, however, favoring lighter quarks, rather
than quarks with larger masses. Note that previous treatments 
\cite{agrawalprl,agrawal,aguirretegmark,damour,donoghuetwo,
donoghuethree,hogan} 
generally obtain even tighter constraints on the quark masses by
invoking the (unnecessary) constraints that deuterium must be bound
and diprotons cannot be bound. 

The mass difference between the proton and neutron $\Delta{m}$ =
$m_{\rm n}-\mpro$ depends on both the mass difference between the
light quarks $\delta{m}=\dmass-\umass$ and the fine structure constant
$\alpha$. Although $\delta{m}$ cannot be made too small, the parameter
space allows $\delta{m}$ to vary by a factor of $\sim10$ for the fine
structure constant in the range $0\le\alpha/\alpha_0\le2$ and 
$\Delta{m}=1-4$ MeV (see Figure \ref{fig:massalpha}). 

The fine structure constant $\alpha$ and the ratio $\beta$ of the
electron mass to the proton mass can vary by several orders of
magnitude while allowing for stable atoms and working stars
(Figure \ref{fig:abplane}). The corresponding allowed region for
$\alpha$ and its strong force counterpart $\alpha_{\rm s}$ is somewhat
smaller (Figure \ref{fig:asplane}), but still spans many orders of
magnitude.

In order for the universe to evolve to its present state, the initial
value of the total energy density $\Omega$ must be extremely close to
unity, so that the space-time of the early universe must be spatially
flat to one part in $\sim10^{60}$ (e.g., Figure \ref{fig:omegaevolve}). 
This type of tuning can be explained if the early universe experiences
an inflationary epoch \cite{guth2000} or its equivalent. Successful
inflation is far from guaranteed, however, and the initial conditions
required to achieve such a solution to the flatness problem can
introduce additional fine-tuning issues
\cite{bardeen,kaloper,tegmarkinflate,vilenkin95,vilenkin98}. 

The energy density $\rhov$ of the vacuum (equivalently, the
cosmological constant) can vary by many orders of magnitude (Figure
\ref{fig:qvlambda}) and still allow galaxies and clusters to form, 
contrary to many previous claims.  If the amplitude $Q$ of the
primordial density fluctuations varies over its allowed range, 
$\rhov$ can be larger than its observed value by a factor of 
$\sim10^{10}$. Universes with even larger values of $\rhov$ can produce
structure if the baryon to photon ratio $\eta$ increases. The bound is
proportional to $\eta^4$ (see equation [\ref{genweinbound}]) so that
the upper limit increases by an additional factor of $\sim10^{12}$. 
For universes with large $(\eta,Q)$, the resulting galaxies would 
be much denser than those in our universe, so only a fraction of the 
solar systems (residing in the outer galaxy) would remain viable. 

Big Bang Nucleosynthesis generally does not cause a universe to lose
its potential habitability. In order for BBN to render a universe
lifeless, the early universe must process almost all of its nucleons
into helium and heavier elements --- leaving no hydrogen behind to
make water.  However, the relevant cosmological parameters can vary
over many orders of magnitude without over-producing helium, including
the baryon to photon ratio (Figure \ref{fig:bbneta}), the
gravitational constant (Figure \ref{fig:bbngravity}), the neutron
lifetime (Figure \ref{fig:bbntau}), and the fine structure constant
(Figure \ref{fig:bbnalpha}). Even without the weak force, universes 
can emerge from BBN with viable compositions if the baryon to photon 
ratio is smaller than that of our universe \cite{grohsweakless,weakless}. 

The range of allowed amplitudes $Q$ for the primordial density
fluctuations is approximately given by $10^{-6}\simless Q \simless
10^{-2}$ (see Figure \ref{fig:galsurvive}). The density of forming
galaxies scale with the value of $Q$ such that $\rho_{\rm c}\sim Q^3$. 
For smaller values of $Q$, galactic gas has difficulty cooling, so
that star formation is suppressed (or at least delayed). For larger
values of $Q$, galaxies become so dense that planets can be stripped
out of their orbits by passing stars, background radiation fields
become more intense than the solar flux received by Earth, and
galactic black hole formation becomes problematic. Nonetheless, 
the allowed range of $Q$ spans about four orders of magnitude. 

The baryonic and dark matter inventories of the universe can also vary
by many orders of magnitude without disrupting habitability (Figure 
\ref{fig:etadelta}). For large values of $\delta/\eta$, which sets 
the mass ratio of dark matter to baryons, galactic disks become
stabilized, which slows down star formation; for smaller values of the
ratio $\delta/\eta$, Silk damping becomes important and suppresses
structure formation on small scales.  Although galaxies and their
universes can have different properties in the separate regimes
delineated within Figure \ref{fig:etadelta}, most of the parameter
space remains habitable.

Stable hydrogen burning stars can exist over a wide range of parameter
space, specified here by the fine structure constant $\alpha$, the
gravitational structure constant $\alpha_G$, and a composite parameter
$\conlum$ that specifies the nuclear reaction rate (Figure 
\ref{fig:staragc}). The allowed parameter space is reduced by
requiring stars to have sufficiently high surface temperatures and
long lifetimes, but still spans many orders of magnitude (Figure 
\ref{fig:starplane}). The ratio $\alpha_G/\alpha$ is notoriously 
small ($\sim10^{-36}$) in our universe. Within the range of the
$(\alpha,\alpha_G)$ plane allowed by working stars, this ratio can be
larger by a factor of $\sim10^4$, but still remains small compared to
unity ($\sim10^{-32}$). Planet properties do not depend on the nuclear
reaction parameter $\conlum$, and the allowed range of the
$(\alpha,\alpha_G)$ parameter space for viable planets is even larger
than that for working stars (Figure \ref{fig:planplane}).

In addition to operating over a wide range of parameters, stars and
stellar structure are less sensitive to nuclear considerations than
suggested by previous claims. Stars can continue to make substantial
amounts of carbon as long as the energy of the triple-alpha resonance
is not raised by more than $\sim500$ keV. If the energy of the
resonance is lowered, then stars actually make {\it more} carbon (see
Figure \ref{fig:carbonyield}). Moreover, the $^8$Be nucleus fails to
be bound by only 92 keV, so that changes to its binding energy of this
magnitude can lead to stable beryllium and remove the need for the
triple-alpha reaction altogether. Finally, the carbon found in the 
Earth is depleted relative to Solar and cosmic abundances by a factor
of $\sim100$ \cite{allegre,marty}, so that habitability does not 
necessarily require a full carbon inventory.  

Stars can also operate with both stable diprotons and unstable
deuterium, again contrary to many previous claims. If diprotons are
stable and nuclear reaction rates are enormously larger (by factors of
$\sim10^{16}$), then the central temperatures in stars decrease from
$\tcent\approx15\times10^6$ K down to about $\tcent\approx10^6$ K, but
stars otherwise function normally. As the nuclear reaction rates
increase, the allowed parameter space for the structure constants
$(\alpha,\alpha_G)$ increases substantially (Figure \ref{fig:staragc}). 
In the opposite case, where deuterium is unbound and cannot provide
the usual stepping stone toward larger nuclei, stars continue to
operate through a variety of processes.  Gravitational contraction is
sufficient as an energy source, and explosive nucleosynthesis can
produce heavy nuclei at the end of stellar lifetimes. A triple nucleon
process allows stars to burn hydrogen into helium in the absence of
stable deuterium.  Finally, the CNO cycle continues to operate, and
allows for working stars, even with metallicity as small as
$\metal\sim10^{-14}$. Through these four stellar processes, stars can 
provide both energy and nucleosynthesis in universes without stable 
deuterium. 

In universes with parameters that are significantly different from
those of our universe, new types of astrophysical processes can
contribute to the generation of energy (Section \ref{sec:exotica}).
For the case of dense galactic halos and larger cross sections for
weak interactions, the energy generated by dark matter annihilation
can compete with stellar radiation as a power source for habitable
planets (Figure \ref{fig:exhalo}). Dark matter can also collect inside
stellar remnants (such as white dwarfs) and subsequently annihilate.
This channel of power generation is also enhanced with denser halos
and larger weak interaction cross sections. In the regime of stronger
gravity and weaker electromagnetism (smaller $\alpha$), black holes can
be bright enough and sufficient long-lived to serve as hosts for
habitable planets (Figure \ref{fig:bholeplane}).

In order to delineate the allowed ranges for the fundamental and
cosmological parameters, we needed to express the mass scales of
astrophysical objects in terms of the fundamental constants. These
mass scales are collected in \ref{sec:massscales}. The number of
dimensions of space represents another physical property that 
could in principle vary from universe to universe  
\cite{barrow1983,ehrenfest1,ehrenfest2,reessix,tegmarkdimension,whitrow}.  
Although this quantity rests on a somewhat different footing from the
other parameters reviewed in this paper, a number of arguments suggest
that three spatial dimensions ($\dimnum=3$) are strongly preferred
(\ref{sec:dimensions}). Constraints on the fundamental parameters due
to requirements imposed by chemistry and working biomolecules are not
as developed as those arising from physics. Work to date suggests that
chemical constraints on the fine structure constant $\alpha$ and the
mass ratio $\beta$ are not as stringent as those arising from physics
(\ref{sec:chemistry}). Considerations of stellar structure can be used
to place global upper bounds on the gravitational constant and on the
ratio $\alpha_G/\alpha$. Although the hierarchy between gravity and
electromagnetism can be made smaller by several orders of magnitude,
it remains large (\ref{sec:globalbounds}). A complete description of
fine-tuning requires knowledge of the probability distributions from
which the parameters are sampled; this part of the problem is much
less developed than the constraints outlined above
(\ref{sec:probability}). Finally, variations in the fundamental
constants can lead to unstable nuclei and a smaller periodic table.
These constraints are outlined in \ref{sec:semfappend} using the
Semi-Empirical Mass Formula for nuclear structure.

The constraints reviewed in this paper are summarized in Table
\ref{table:summary}. For each quantity of interest, the table lists
the range of that quantity that allows for a viable universe, where 
the result is expressed in decades. If $x$ is a parameter, with 
minimum and maximum allowed values $x_{\rm min}$ and $x_{\rm max}$, 
then the range (in decades) is defined by 
\be
D(x)\equiv\log_{10}\left[{x_{\rm max}\over x_{\rm min}}\right]\,. 
\label{decadedef} 
\ee
Note that, in general, the ranges of allowed parameters are 
asymmetric with respect to the values in our universe. 

The table is organized into three sections. The top portion includes
fundamental parameters associated with particle masses, as specified
by the Standard Model of Particle Physics. The central portion
considers the dimensionless strengths of the four fundamental forces,
where the values in our universe are evaluated in the low energy
limit. The bottom portion includes the cosmological parameters.  Note
that the ranges of the up quark mass, the gravitational constant, the
strength of the weak force, and the energy scale of the vacuum energy
do not have well-defined lower limits.  The up quark mass and the
strength of gravity cannot vanish, however, so the results given in
Table \ref{table:summary} correspond to the parameter space presented
in the Figures, but the ranges could be even larger. On the other
hand, universes could be viable without the weak force and without any
vacuum energy; for these parameters, Table \ref{table:summary} lists
the number of decades for which the parameters could be larger. 

Notice also that the allowed range for the strong coupling constant
$\alpha_{\rm s}$ would be much smaller if one imposes the constraints
that diprotons remain unbound and/or deuterium remains bound. Although
these requirements are not necessary for functioning stars (see
Section \ref{sec:twonucleon}), they would limit variations in
$\alpha_{\rm s}$ to $\pm15\%$, corresponding to a range of only
$D(\alpha_{\rm s})\sim0.06$. The allowed range of the strong coupling
constant (and the fine structure constant) would also be much smaller
if one requires the triple alpha reaction (Section 
\ref{sec:triplealpha}).  The triple alpha constraint limits
variations in $\alpha_{\rm s}$ to $\sim2\%$, corresponding to a
range of only $D(\alpha_{\rm s})\sim0.01$. However, smaller variations
in the fundamental constants allow for stable $^8$Be nuclei, which
provide an alternate channel for carbon production and obviate the 
triple alpha constraint. 

With the allowed ranges of the fundamental and cosmological parameters
specified in Table \ref{table:summary}, we are left with the question
of whether these ranges of possible variations are large or small.
Given that the concept of fine-tuning evokes the image of tuning the
dials of a radio, as one point of comparison we consider the tuning
required to capture a radio station. The AM carrier frequencies lie in
the range $f_{\scriptscriptstyle AM}$ = 535 to 1605 kHz, where the
spacing between stations is 10 kHz. To find a particular radio station,
the electronics must be accurate to about 1 part in 100.  Similarly,
the FM radio band extends from $f_{\scriptscriptstyle FM}$ = 88 to 108
MHz, where the central carrier frequencies for individual stations are
assigned at 200 kHz intervals.  Finding an FM station thus requires
tuning the frequency to 1 part in 500.  The final two lines of Table
\ref{table:summary} give the ranges (in decades) required for
successful AM and FM radio reception, $D(f)$ = 0.0043 and 0.00087, 
respectively.  These ranges are much smaller than those required of
the fundamental parameters. In fact, the allowed ranges for all of the
parameters listed in the table are enormously larger than the entire
range of radio frequencies. Tuning a radio thus requires far more 
precision than tuning a universe. 

\bigskip 
\begin{table} 
\centering 
{\bf Ranges of Parameter Values for Viable Universes} 
\medskip
\def\arraystretch{1.15} 
\begin{tabular}{lccc}
\\
\hline 
\hline 
quantity & symbol & observed value & range (decades) \\
\hline 
\hline 
Up quark mass & $\umass$ & 2.3 MeV & $>3$ \\ 
Down quark mass & $\dmass$ & 4.8 MeV & 0.85 \\ 
Electron-proton mass ratio & $\beta$ & 1/1836 & 5\\ 
Up-down quark mass difference & $\delta{m}$ & 2.5 MeV & 1 \\ 
\hline 
Gravitational constant & $\alpha_G$ & $6\times10^{-39}$ & $>10$\\ 
Weak coupling constant & $\alpha_{\rm w}$ & $10^{-5}$ & $6+$\\ 
Fine structure constant & $\alpha$ & 1/137 & 4\\
Strong coupling constant & $\alpha_{\rm s}$ & 15 & 3\\ 
\hline 
Fluctuation amplitude & $Q$ & $10^{-5}$ & 4 \\ 
Baryon to photon ratio &$\eta$&$6\times10^{-10}$& 6\\
Dark matter abundance & $\delta$ & $3\times10^{-9}$ & 6\\
Vacuum energy scale & $\lambda$ & 0.003 eV & $10+$\\ 
\hline 
\hline 
AM Radio & $f_{\scriptscriptstyle AM}$ & 535 -- 1605 kHz & 0.0043 \\
FM Radio & $f_{\scriptscriptstyle FM}$ & 88 -- 108 MHz & 0.00087 \\ 
\hline 
\hline 
\end{tabular}
\caption{Table of the fundamental and cosmological parameters and  
their allowed ranges. The ranges are expressed in terms of decades 
of allowed variation, as defined by equation (\ref{decadedef}). 
The ranges expressed for the mass of the up quark and the gravitational 
constant are lower limits on the full range. The ranges listed for the 
weak force and the energy scale of the vacuum correspond only to 
variations to greater values (any lower values are allowed). For 
comparison, the final two lines specify the amount of tuning 
required for radio reception (see text). } 
\label{table:summary} 
\end{table}  

Limits on the cosmological density parameter $\Omega$ are notably
absent from Table \ref{table:summary}. As discussed in Section 
\ref{sec:cosmology}, the value of $\Omega$ must be either (a) tuned 
exquisitely to one part in $10^{60}$ at the Planck epoch, or (b)
driven to its observed value $\Omega\rightsquigarrow1$ by some
mechanism (e.g., an inflationary epoch in the early universe or its
analog). Although the universe must indeed be spatially flat to high
precision, almost all universes could have this property according
some measures on the space of initial conditions
\cite{carrollbook,carroll2010,carroll2014}. The allowed range of 
$\Omega$ in the early universe is thus extremely small, perhaps
$D(\Omega)\approx-60$, but its implications for fine-tuning depend on
the (as yet unknown) likelihood of the universe achieving successful
inflation or the equivalent, and/or on the resolution of the
cosmological measure problem \cite{corichi,gibbons,schiffrin}. 

Finally, note that solutions to specific fine tuning problems in
stellar astrophysics require additional physical processes and are
summarized in Table \ref{table:startable}.  The top part of the table
addresses the triple alpha problem for carbon production, which occurs
through a resonant reaction in our universe. Possible solutions
include varying the resonance level $\Delta{E}_R$ and/or increasing
the binding energy of beryllium-8. Spallation reactions can produce
some carbon by breaking down larger alpha elements, although the
abundances are small.  The table then lists processes relevant for
universes with unstable deuterium. Energy can be generated by
gravitational contraction via small stars, and heavy elements can be
produced through explosive nucleosynthesis in larger stars. In
addition, stars can burn hydrogen through triple nucleon reactions (if
the deuterium half-life $\tau_{1/2}(d)$ is long enough) and through
the CNO cycle (if the metallicity $\metal$ is high enough).  Stars can
function with the nuclear burning parameter $\conlum$ (see equation
[\ref{conlumdef}]) varying over more than 21 orders of magnitude,
including values appropriate for universes with stable
diprotons. Finally, in universes without the weak interaction, stars
can burn deuterium over $\sim$Gyr lifetimes (if the baryon to photon
ratio $\eta$ is small enough).

\bigskip 
\begin{table} 
\centering 
{\bf Alternate Processes in Stellar Astrophysics} 
\medskip
\def\arraystretch{1.2}  
\begin{tabular}{llll}
\\
\hline 
\hline 
Problem & Solution & Parameter Constraint \\ 
\hline 
\hline 
Triple alpha reaction & allowed resonance levels & 
--300 keV $\simless\Delta{E}_R\simless$ 500 keV \\ 
$\qquad\qquad\vdots$ & stable beryllium-8 & $B_8>0$ ~(92 keV change) \\ 
$\qquad\qquad\vdots$ & spallation & (large cosmic ray flux) \\ 
\hline 
Unstable deuterium & gravitational power & $M_\ast\simless0.8M_\odot$\\ 
$\qquad\qquad\vdots$ & explosive nucleosynthesis 
& $M_\ast\simgreat5.6M_\odot$\\ 
$\qquad\qquad\vdots$ & triple nucleon reactions 
& $\tau_{1/2}(d)\simgreat10^{-21}$ sec \\ 
$\qquad\qquad\vdots$ & CNO cycle & $\metal\simgreat10^{-14}$ \\ 
\hline 
Stable diprotons & (none required) 
& $1\le\conlum/\conlum_0\simless10^{21}$ \\
\hline 
No weak force & deuterium burning & $\eta\simless10^{-10}$ \\ 
\hline 
\hline 
\end{tabular}
\caption{Physical processes that can operate in alternate universes 
and alleviate stellar fine tuning issues. For each issue (left column)
and process (middle column), the right column lists the range of the
relevant parameter that allows for viable universes. These parameters
depend on the process under consideration and include the change in
the energy level of the carbon-12 resonance $\Delta{E}_R$, the
binding energy $B_8$ of the beryllium-8 nucleus, stellar mass
$M_\ast$, deuterium half-life $\tau_{1/2}(d)$, metallicity $\metal$, 
nuclear burning parameter $\conlum$, and primordial baryon to 
photon ratio $\eta$. }
\label{table:startable} 
\end{table}  

\subsection{General Trends}
\label{sec:trends} 

\noindent
Another way to summarize the degree of fine-tuning of the universe is
to organize the results described above into the following general
trends:

\medskip \noindent $\bullet$ 
{\sl The allowed parameter space is large.}  Most of the relevant
parameters can vary by several orders of magnitude and still allow for
the development of complex structures, from atoms to stars to
galaxies: The parameters that are allowed to vary include the 
masses of the light quarks and leptons $(\umass,\dmass,\emass)$, the 
structure constants $(\alpha,\alpha_{\rm s},\alpha_{\rm w},\alpha_G)$, 
and cosmological parameters $(\eta,\delta,\rhov,Q)$. 

\medskip \noindent $\bullet$ 
{\sl Particle physics is more sensitive than astrophysics.}  
Particle physics considerations are more constraining than
astrophysical considerations: The allowed mass range for the light
quarks (shown in Figure \ref{fig:updown}) is smaller than the
corresponding ranges for the astrophysical parameters, including the
fluctuation amplitude $Q$ (Figure \ref{fig:galsurvive}), the baryon to
photon ratio $\eta$ and its dark matter counterpart $\delta$
(Figures \ref{fig:bbneta} and \ref{fig:etadelta}), and the energy
density of the vacuum $\rhov$ (Figure \ref{fig:qvlambda}). Stars can
operate over a wide range of parameter space and can produce heavy
nuclei through many channels (Section \ref{sec:stars}). The key issue
is that stable nuclei (especially carbon) must exist (see 
\ref{sec:semfappend}). Finally, although only preliminary work has
been carried out, chemistry seems to be less confining than physics
(see \cite{kingchemistry} and \ref{sec:chemistry}).

\medskip \noindent $\bullet$ 
{\sl Large hierarchies remain $\rightsquigarrow$ gravity must be weak.}  
Even if the parameters of physics and cosmology can deviate from their
values in our universe by orders of magnitude, `unnaturally small'
ratios are still required: For example, the cosmological constant can
vary over a wide range, but must be small compared to the Planck scale
(Section \ref{sec:qvbounds}).  Similarly, the ratio $\alpha_G/\alpha$
of the gravitational structure constant to the fine structure constant
can vary by several orders of magnitude, but must remain small
compared to unity (\ref{sec:globalbounds}). Both of these ratios are
extremely small due to the required weakness of gravity, equivalently,
the large value of the Planck mass. These hierarchies for the physical
parameters ultimately lead to the enormous ranges of mass and size
scales observed in the universe (Figure \ref{fig:astroscales}). In
general, the universe exhibits more Hierarchical Fine-Tuning than it
does Sensitive Fine-Tuning. A summary of remaining hierarchies in the
universe is provided by Table \ref{table:hierarchy}.

\medskip \noindent $\bullet$ {\sl Multiple variations are important.} 
More possibilities for working universes arise if more than one
parameter is allowed to vary: For example, if all other parameters are
fixed, then the vacuum energy density $\rhov$ can only be larger than
its observed value by a modest factor. If the amplitude $Q$ of the
density fluctuations is larger, the allowed range in $\rhov$ increases
by a a factor of $\sim10^9$. If the baryon to photon ratio is larger,
the allowed range increases by a factor of $\sim10^{12}$. 

\medskip \noindent $\bullet$
{\sl Our universe does not lie at the center of parameter space.}  
The ranges for viable parameters are often asymmetric and are
sometimes constrained in only one direction. If the triple alpha
resonance level is raised, then carbon production decreases, but stars
produce {\it more} carbon if the resonance level is lower.  The fine
structure constant $\alpha$ cannot be too large without compromising
nuclear structure, but larger ratios of the strong-to-electromagnetic
force lead to a wider variety of stable nuclei. Stars no longer
function if the gravitational constant is increased by more than a
factor of $\sim10^6$, but $G$ can become arbitrarily small and working
stellar solutions still exist.  Similarly, large values of the
cosmological constant $\Lambda$ compromise structure formation, but
essentially all smaller values are allowed. Finally, limits on the
light quark masses $(\umass,\dmass)$ are asymmetric: Larger masses
lead to a shorter range for the strong force, and result in tight
constraints from nuclear structure. Smaller quark masses lead to a
longer range for the strong force, and the limits are not nearly as
stringent.

\medskip \noindent $\bullet$ {\sl Universes have multiple pathways.} 
Viable universes are not required to be exactly like our own --- they
can in principle achieve habitability through alternate routes: 
Although our universe has $\beta\ll1$, so that the electron mass is
much smaller than the proton mass, the opposite ratio $\beta\gg1$
could also allow for working atoms. If the nuclear parameters are
different so that carbon production via the triple alpha reaction
becomes compromised, then some universes can support stable
beryllium-8 nuclei and would not need the triple alpha process
\cite{agalpha}. Another possible scenario is that of a Cold Big 
Bang \cite{aguirre}, where the cosmological parameters
$(\eta,\delta,\Lambda,Q)$ differ from those in our universe by orders
of magnitude. Yet another alternate universe could have no weak
interactions \cite{weakless,gedalia}; the weakless universe requires
smaller values of $\eta$ to avoid overproducing helium during BBN, but
can remain habitable. With sufficiently large departures of dark
matter properties and gravity, other universes can generate enough
energy to sustain habitable planets through dark matter annihilation
or even black hole radiation (Section \ref{sec:exotica}). These 
alternate pathways expand the range of parameter space for viable 
universes. 

\medskip \noindent $\bullet$ 
{\sl The fine structure constant must be small.} Although the value of
$\alpha$ can vary over a wide range, a number of independent lines of
argument indicate that the fine structure constant must remain much
less than unity: Stable long-lived stars burn their nuclear fuel
through a quantum mechanical tunneling process, which would shut down
if $\alpha$ is too large (Figure \ref{fig:staragc}). An even stronger
upper limit on $\alpha$ arises from the requirement that the
photospheric temperatures of stars are hot enough to support chemical
reactions (Figure \ref{fig:starplane}). If the value of $\alpha$ 
becomes of order unity, then large atomic nuclei would not exist
\cite{barrowetal2002,davies1972}, and the periodic table would be 
much smaller (see also \ref{sec:semfappend}). Even if the nuclei
remain bound, the electrons in atoms would become relativistic for
larger values of $\alpha$. Although such atoms can remain in
existence, contrary to many previous claims, the atomic energy levels
and hence the chemical properties would be markedly different
(see \cite{greiner,greinerschramm,liebyaualt,reinhardt} and references
therein).  The value of $\alpha$ must remain less than unity in order
for planets to have smaller masses than their host stars (equation
[\ref{planmassone}]). Finally, smaller values of $\alpha$ result in
relatively small changes to biomolecules, but larger values of
$\alpha$ are disallowed (\ref{sec:chemistry}
and \cite{kingchemistry}). All of these considerations constrain the
fine structure constant to be small, such that $\alpha\ll1$. As a
result, physicists in any viable universe should be able to understand
quantum electrodynamics, as QED will always lie in the perturbative
regime.

\medskip \noindent $\bullet$
{\sl Myths and unnecessary constraints:} A number of constraints that
are often enforced on viable universes are not as severe as some
previous work suggests: In universes with stable diprotons, where
nuclear reactions proceed through the strong force only, stars operate
with somewhat lower central temperatures, but do not burn through
their nuclear fuel in a catastrophic manner (Figure \ref{fig:staragc}).  
Similarly, in universes with unstable deuterium, stars can generate
energy and synthesize heavy elements through alternate channels,
including gravitational contraction, explosive nucleosynthesis, the
CNO cycle, and the triple-nucleon process (Figure \ref{fig:hrnod2}
and \cite{agdeuterium}). A possible constraint on the weak coupling
constant arises from the requirement that core collapse supernovae
must have optically thick neutrinos (so that $\alpha_{\rm w}^4$
$\sim\alpha_G$; see Section \ref{sec:supernova}), but Type Ia
supernovae can provide heavy elements \cite{weakless} even if all
massive stars collapse to form black holes. Another commonly invoked
constraint is the requirement that some stars must be convective in
order to produce planets \cite{carter1974}; we now know that planet
formation is independent of stellar convection (Section
\ref{sec:convection}).  The relaxation of these unnecessary
constraints results in a significantly larger parameter space for
working universes.

\bigskip 
\begin{table} 
\centering 
{\bf Hierarchies in the Universe} 
\medskip 
\def\arraystretch{1.1} 
\begin{tabular}{lccc}
\\
\hline 
\hline 
quantity & symbol & observed ${\cal R}$ & minimum ${\cal R}$ \\ 
$\,$ & $\,$ & (decades) & (decades) \\ 
\hline 
\hline 
Horizon size vs proton & $r_H/r_{\rm p}$ & 41 & 37 \\
Horizon mass vs proton & $M_H/\mpro$ & 80 & 76 \\ 
Electromagnetic force vs gravity & $\alpha/\alpha_G$ & 36 & 32 \\ 
Planck mass vs vacuum energy & $\mplanck/\lambda$ & 30.6 & 20 \\ 
Planck mass vs quark masses & $\mplanck/(\umass+\dmass)$ & 21.2 & 16.6 \\ 
\hline 
\hline 
\end{tabular}
\caption{Hierarchies of scales in cosmology and particle physics. 
For each pair of scales, the larger quantity is taken to be in the
numerator, so that the ratios ${\cal R}$ are large numbers. The values
are then expressed in decades $\log_{10}({\cal R})$. All of the 
hierarchies shown in the table result from the required weakness of 
gravity (large value of the Planck mass). }
\label{table:hierarchy} 
\end{table}   

\subsection{Anthropic Arguments} 
\label{sec:anthropic} 

Anthropic arguments are related to --- but not equivalent to --- the
fine-tuning discussion considered in this paper. Although anthropic
arguments have been put forth with a range of definitions (classic
references include \cite{bartip,carr,carter1974,carter1983}), the
basic consideration is that the universe must have the proper version
of physical law in order to develop observers. As discussed throughout
this review, the fundamental constants and cosmological parameters
must lie within specified ranges in order for the universe to develop
interesting structures such as galaxies, stars, and planets. Since the
universe must produce these astronomical entities in order to support
observers, at least those of familiar form, the finding that the
fundamental constants lie within the aforementioned specified ranges
is not by itself surprising. If the ranges of allowed parameters were
sufficiently small, then the finding that our universe has the proper
parameters becomes interesting. In the opposite limit, if the universe
could develop structure while its parameters vary over enormous
ranges, then the observed properties of the universe do not provide a
strong constraint. As a result, the range of parameters for viable
universes must be small in order for anthropic arguments to carry
weight. In other words, some fine-tuning of the universe is necessary
--- but not sufficient --- for the efficacy of anthropic arguments.

One point of contention is the degree to which anthropic arguments 
can predict the values of the fundamental parameters, while another 
difficulty is the extent to which such arguments can be falsified 
\cite{ellissilk}. Discussion of these issues is complicated by the 
vast literature on the subject, where some reviews describe at least
30 different definitions of the anthropic cosmological principle
\cite{bostrom,stenger}. A full review of this subject is beyond 
the scope of this contribution (for example, see \cite{balashov,
bartip,bostrom,carr,davies1983,liviorees,meissner,weinstein} and
references therein). Here we briefly outline the nature of anthropic
arguments and elucidate their relation to the degree of fine-tuning
(see also \cite{barrow1981} for a history of the subject). 

To illustrate the nature of anthropic arguments, let's consider a
generic example. Suppose for, example, that $X$ is a fundamental
constant, and the existence of stars requires that the value of $X$
lies in the range $X=X_0\pm\delta{X}$, where $X_0$ is the observed
value of the parameter in our universe. If the value of $\delta{X}$
is small, then we can say that the existence of stars provides an
anthropic argument for the value of $X$. If the value of $X_0$ is not
yet measured, then this argument provides an anthropic prediction for
the value of $X$. Note that in order for such an argument to carry
much weight, the allowed range of the parameter (and hence $\delta{X}$) 
must be small in some sense. In other words, the parameter $X$ must be
fine-tuned at some level. As a result, the fine-tuning of fundamental
parameters is a prerequisite for anthropic arguments. 

In spite of its seeming simplicity, the above class of arguments has 
a number of complications: First, we have no consensus on how small
$\delta{X}$ must be in order for an anthropic argument to be
meaningful (or for the parameter $X$ to be fine-tuned). The range can
be measured in absolute terms, as a relative change $\delta{X}/X_0$,
as a factor $f=(X_0+\delta{X})/X_0$ by which the variable can change,
and so on. In addition, more than one parameter could vary from
universe to universe. In some cases, for example, the range
$\delta{X}$ must be small if all of the other fundamental parameters
are held constant, but can be much larger for other (allowed) values
of those parameters.

Even if the anthropic argument is successful, so that the argument
confines the parameter $X$ to a small range, we are left with the
question of how much the argument actually explains: If, for example,
$X$ must be very close to its observed value in order for stars to
work, do we now know {\it why} $X$ has such a value? A more
fundamental argument that specifies the value of the parameter $X$
would clearly be preferable. The anthropic argument thus provides only
a partial explanation for the value of $X$. 

In spite of the complications outlined above, and many others,
anthropic arguments have been used to explain the observed values 
of both fundamental constants and cosmological parameters in our
universe.  These studies include constraints on the cosmological
constant, starting with the original work of Weinberg
\cite{weinberg87}, which has subsequently been generalized to 
include variations of additional parameters \cite{adamsrhovac,garriga2006,
kallosh,liviorees,mersini,hartle,peacock,vilenkin2004}. 
Additional studies have presented anthropic arguments for the energy
scale of electroweak symmetry breaking \cite{agrawalprl,barrkhan,jeltema}, 
big bang nucleosynthesis \cite{macdonaldmullan}, stellar nucleosynthesis
including the triple-alpha resonance \cite{jeltema,livio,meissner}, the 
parameters of aluminum-26 decay \cite{sandora2017}, the mass of the proton 
\cite{page}, the masses of neutrinos \cite{tegmarkneutrino}, the finite 
age of the universe \cite{cirkovic2000}, and for the existence of three 
generations in the Standard Model \cite{gould,ibe,schellekens}. 

To illustrate the difficulties faced by anthropic arguments, it is
useful to consider a concrete example. Perhaps the most widely
discussed anthropic argument is that invoked for the cosmological
constant. As outlined in Section \ref{sec:rhovac}, in order for
structure formation to successfully take place, the energy density of
vacuum must obey the bound of equation (\ref{genweinbound}). For
fixed values of the parameters $(Q,\eta,\omegam,\omegab)$, the
cosmological constant cannot be much larger than its observed value.
This argument was put forth \cite{weinberg87} a decade before
astronomical observations provided unambiguous evidence for a nonzero
value of $\rhov$ \cite{riess,riess2}. This example is often considered
not only as an explanation for the observed value of $\rhov$, but also
as a successful anthropic prediction. On the other hand, the upper
bound on $\rhov$ is proportional to the product $\eta^4Q^3$, and both
of these parameters can be larger than the values realized in our
universe by several orders of magnitude (Figure \ref{fig:qvlambda}).
To summarize the situation: An anthropic bound on $\rhov$, derived
before the observations, provides an interesting upper limit if all 
of the other parameters are held fixed. For alternate values of 
$(\eta,Q)$ that allow for viable universes, the bound is weaker by 
a factor of $\sim10^{21}$. For values of $\rhov$ smaller than that 
observed, universes are not only viable, but perhaps even ``better''
than our own (see the following section). Finally, constraints on the
allowed range of $\rhov$ are not the same as a fundamental explanation
for its value.

The Anthropic Cosmological Principle, which requires our universe to
contain observers, could be considered at odds with the Copernican
Principle, which holds that no physical locations are privileged.  The
Copernican Revolution is generally considered an important milestone
in the history of science \cite{kuhncopernicus,kuhnparadigm}.  
Considerations of the multiverse suggest that our universe could be
one out of many, so that our place in the cosmic order must be
reassessed. Nearly five centuries ago, Copernicus argued that Earth
does not occupy a privileged location within the Solar
System \cite{copernicus}.  Subsequent astronomical discoveries have
continually degraded our status. The Solar System does not lie at the
center of the Galaxy.  Given that the observable universe is
homogeneous and isotropic, the Milky Way does not occupy a special
location. Continuing this trend, the idea of a Copernican Time
Principle \cite{al1997} suggests that the current cosmological epoch
does not have special significance.  With the emergence of the
multiverse paradigm, with the possibility that far-away disconnected
regions of space-time exist, our universe no longer occupies a
privileged location within the cosmic archipelago. The multiverse thus
extends the Copernican Principle.  At the same time, the existence of
multiple universes allows for Anthropic Selection, which provides a
mechanism for our universe to develop observers and hence be
privileged. However, whether or not our universe is special depends on
how the assessment is made: While Earth does not lie at the center of
the Solar System, it {\sl does} reside the proper distance from the
Sun to allow for liquid oceans on its surface. In a similar vein, our
universe is not special in an {\it a priori} sense, but it {\sl does}
exhibit the proper version of laws of physics to support the existence
of observers. It remains to be seen how this change of cosmic status,
that our universe could be but one of many, will affect the status of
the Copernican Principle.

\subsection{Is our Universe Maximally Habitable?} 
\label{sec:bestworld} 

Discussions of fine-tuning often implicitly assume that our universe
is optimized for the development of observers \cite{leibniz}.
However, it is interesting to revisit this panglossian assumption 
\cite{voltaire} and ask if different choices for the fundamental 
constants or the cosmological parameters could lead to universes that
are even more favorable to the development of life, or at least the
production of complex cosmic structure (see also \cite{naumann}).  
In the realm of extra-solar planetary systems, researchers are now  
considering whether or not Earth is the best prototype for a habitable
planet \cite{heller,pierrehumbert}.  Asking the same question on a
cosmic scale, we find that several parameter choices could lead to
possible improvements of the universe:

\medskip \noindent $\bullet$ {\sl Smaller vacuum energy:} 
The cosmological constant could vanish (equivalently, $\rhov=0$).  
In this case, the universe would never enter into a late-time phase of
accelerated expansion. This scenario would favor structure formation,
as the matter dominated era would never end. In addition, ever larger
cosmological structures would continue to enter the horizon, so that
large scale structures of ever larger masses could be constructed
\cite{al1997,cirkovic,dyson1979,krauss2000,nagamine}. Even if the 
energy density of the vacuum is nonzero, it could be much smaller than
the value realized in our universe, so that structure formation could
continue over a longer span of time.

\medskip \noindent $\bullet$ {\sl Larger primordial fluctuations:} 
The amplitude $Q$ of the primordial density fluctuations could be
larger. If some type of inflation occurs in the ultra-early universe,
then the probability of producing density fluctuations with a given
amplitude could be an increasing function of $Q$, so that larger
values of $Q$ are more natural \cite{garriga2006,susskind2005}.  
In addition, universes with larger values of $Q$ produce denser
galaxies \cite{tegmarkrees,tegmark}, and allow structure formation 
to occur with larger values of the cosmological constant 
\cite{liviorees,mersini,adamsrhovac}.  With the right choice of the 
amplitude, roughly $Q\sim10^{-2}$, galaxies can be dense enough that
starlight from the background galaxies allows for planets to have
habitable temperatures from almost any orbit \cite{coppess}, thereby
producing a Galactic Habitable Zone (see Figure \ref{fig:ghz}). This
scenario allows the universes in question to support even more
potentially habitable planets than our own. For fluctuation amplitudes
as large as $Q=10^{-2}$, galaxies are susceptible to the
overproduction of supermassive black holes \cite{bhrees}, so that the
optimum value of $Q$ is somewhat lower.

\medskip \noindent $\bullet$ {\sl More baryons:} The ratio $\eta$ 
of baryons to photons could be larger. Such enhancements lead to
earlier matter domination, which allows for large scale structures to
grow more easily. Such universes could tolerate a greater range of
fluctuation amplitude $Q$ or vacuum energy density $\rhov$; for some
parameter choices, they could produce denser galaxies and support more
habitable planets (analogous to increases in $Q$). 

\medskip \noindent $\bullet$ {\sl More stars:} The range of stellar 
masses in our universe spans a factor of $\sim1000$, but could be much
larger in other universes with other versions of the fundamental
constants. For example, if the fine structure constant is smaller, the
range of stellar masses is larger, with a maximum stellar mass range
of $\sim18,000$ for $\alpha\sim1/6300$ (see \ref{sec:massscales}).
Stars can also have longer nuclear burning lifetimes with other
choices of parameters (Section \ref{sec:stabledip}).

\medskip \noindent $\bullet$ {\sl Stable beryllium-8:} 
If the strong force were slightly stronger than in our universe, then
$^8$Be could be a stable isotope \cite{agalpha,epelbaum2013,hafstad}.  
In this type of universe, carbon can be produced through the reaction of
equation (\ref{hebe2c}) without the need for the triple alpha reaction
(which is more sensitive to stellar conditions 
\cite{clayton,kippenhahn} and the underlying parameters 
\cite{eckstrom,epelbaum2011,epelbaum2013,oberhummer,schlattl}).  
Carbon production could thus take place within the same star that
produced the $^8$Be, or much later in a different star \cite{agalpha}.
In addition to allowing for more branches of the nuclear reaction
network to produce carbon, this type of universe would have a more
orderly nuclear inventory. In our universe, the most abundant isotopes
beyond hydrogen are $^4$He, $^{16}$O, $^{12}$C, and $^{20}$Ne (in that
order). All of these nuclei are made up of alpha particles, with
numbers $N_\alpha$ = 1, 4, 3, and 5, respectively. The $N_\alpha=2$
nucleus is thus conspicuously absent, but a `more logical' universe
could make all of its most common isotopes with integer numbers of
alpha particles.

\medskip \noindent $\bullet$ {\sl Weaker gravity:} If the strength of
gravity is weaker than in our universe, then the cosmos would expand
more slowly, so that life would have more time to emerge and evolve
\cite{liviorees2018}. In addition, stars and planets would be larger
in size, allowing for larger animals (\cite{press1980}, equation
[\ref{masslife}]) and more interesting topography \cite{bartip}.
Perhaps more importantly, for weaker gravity, the range of values for
the fine-structure constant that allows for working stars would be
wider (see Figures \ref{fig:staragc} and \ref{fig:starplane}).

\subsection{Open Issues} 
\label{sec:futureissues} 

This review has discussed the possible parameters from particle
physics and cosmology that can vary across the multiverse and can
potentially affect the habitable properties of the constituent
universes. The main focus has been to delineate the allowed ranges for
these parameters that allow a universe to develop complex structures,
including nuclei, planets, stars, and galaxies. Although this
enterprise has made steady progress, and now has a vast literature,
many open issues remain: 

Although this review has discussed the parameters that are allowed to
vary, this determination is not definitive. At the present time, all
of the individual parameters that appear in the Standard Model of
Particle Physics are considered as independent. Many workers hope that
a more fundamental theory, such as string theory or its descendants,
would contain fewer fundamental parameters, so that the currently
considered quantities (masses, mixing angles, and coupling constants)
would be derived from a smaller set
\cite{donoghuethree,hogan,kaneperry,schellekens2008,schellekens,susskind}. 
Until a more fundamental theory is in place, however, we are left with
the rather large number of parameters outlined in
Section \ref{sec:particlephys}.

Similarly, the cosmological parameters have been allowed to vary
separately, as outlined in Section \ref{sec:cosmology}, but these
values could in principle be determined by physics beyond the Standard
Model. For example, the amplitude $Q$ of the primordial density
fluctuations could be determined by the properties on an inflaton
field, a high energy (perhaps near the GUT scale) scalar field that is
not yet discovered. The baryon to photon ratio $\eta$ is considered
here as a freely varying parameter, but its value must ultimately be
set by the physics of baryogenesis (which is driven by out of
equilibrium, CP and baryon number violating processes at high
energy). Similarly, the value of the vacuum energy density $\rhov$
could be specified by additional physics in the gravitational sector,
but remains undetermined in the current state of physics.

On a related note, the parameters considered here are allowed to vary
independently of each other. It remains possible --- perhaps even
likely --- that a more fundamental understanding of physics would
require parameter values to vary together \cite{kaneperry,schellekens}.
As one example, the strengths of the coupling constants might have
fixed ratios in a fundamental theory, but the overall strength could
scale up and down (since the coupling constants are also energy
dependent, this scaling would have to apply at a specific energy, or
in the limit zero temperature). Grand Unified Theories generally have
this property (starting with \cite{georgiguts}).  As another example 
\cite{bjorken}, the parameters $X_{sm}$ of the Standard Model could 
vary with the scale $\lambda$ of the vacuum energy according to a 
renormalization group equation, 
\be
-{1\over2} \lambda {\partial X_{sm} \over \partial \lambda} 
= p_{\scriptstyle X} \, X_{sm} \,, 
\label{renormalize} 
\ee
where the exponents $p_{\scriptstyle X}$ are chosen so that the flow
reaches a fixed point at the defining energy scale (e.g., the 
Planck scale). 

Many of the results presented herein are represented as allowed ranges
of parameters in a plane of two variables (see Figures \ref{fig:updown} 
-- \ref{fig:bholeplane}).  If the possible variations are coupled, as 
outlined above, universes would sample only a subset of the plane. The
viable range of parameter space would then correspond to a curve
passing through the planes presented here. On the other hand, most of
the current exploration of parameter space starts with the values
realized in our universe and changes the parameter values until some
failure point is reached. In principle, there could exist distant
islands of parameter space, far removed from values in our universe,
that allow for viable universes (e.g., Figure \ref{fig:abplane}). 

As discussed in Section \ref{sec:intro}, a full assessment of
fine-tuning requires not only a specification of the allowed parameter
space, but also the underlying probability distribution for universes
to realize a given set of parameters (\ref{sec:probability}). As
outlined in this review, the field has made substantial progress in
delineating the allowed ranges of parameters, in the context of
particle physics, cosmology, and stellar astrophysics. In contrast,
our understanding of the underlying probability distributions remains
in its infancy. An important challenge for the future is to develop
{\it a priori} determinations of the underlying distributions from
which the fundamental parameters are sampled.

Another remaining challenge is to develop a rigorous theory for the
creation of individual universes and the mechanism through which they
select their vacuum states, which then determine the laws of physics
for that region. The launch of a universe onto an expanding trajectory
--- one that separates itself from the background space-time of the
rest of the multiverse --- involves a full theory of quantum gravity,
which remains elusive. In addition, we need a rigorous assessment of
the possible vacuum states of the universe (compare 
\cite{banks2012,banksdinegorb,bena} with 
\cite{boussopolchinski,hogan2006,kachru,schellekens,susskind}),
and the manner in which an individual universe settles into such a
state. Eternal inflation provides one specific mechanism that allows
for the fundamental constants and cosmological parameters to vary from
universe to universe (Section \ref{sec:eternal}). Bouncing cosmologies
provide another example. In this latter scenario, the universe expands
and re-collapses, and then begins a new phase of expansion 
\cite{battefeld,lehners}. This process continues in cyclic fashion, 
where each reincarnation of the universe can result in a new
realization of the fundamental/cosmological parameters
\cite{alexanderbounce,barrow2004}. The idea of cosmological natural 
selection \cite{smolin} provides yet another mechanism for producing
multiple universes. In this hypothetical scenario, singularities at
black hole centers act as sources of new universes, which are hidden
from the parental universe by the event horizon and can have different
realizations of the fundamental parameters.

Another complication that arises in assessments of fine-tuning is the
sheer number of possibilities for different types of universes with
different types of physics. In addition to the variations discussed
above, many generalized versions of the laws of physics (and alternate
cosmological models) have been put forth for applications in our
universe. Although most of these scenarios are highly constrained by
experimental data, these generalizations could be realized in other
regions of space-time. For example, the constants of nature could vary
with time \cite{barrow2005,bergstrom,chiba,teller,uzan}, including
variations in the effective value of the gravitational constant 
\cite{bransdicke}. More extreme departures from the standard theory 
of general relativity are also possible, as described by a wide range
of modified gravity theories 
\cite{aguirregrav,capozziello,clifton,joyce,nojiri},
including the particular paradigm of Modified Newtonian Dynamics
\cite{bekenstein,mcgaugh,milgrom}.  The equivalence principle, which 
is well established experimentally, could be violated in other
universes \cite{damour2009}. In addition to taking on different
values, the cosmological `constant' could evolve over cosmological
history \cite{quintessence}, including being an increasing function of
time \cite{caldwellbigrip}. The vacuum energy contribution could also
conspire to produce a steady-state universe \cite{bondigold}.
Although the flux of magnetic monopoles in our universe is highly
constrained \cite{adtarle,parkerbound,turnerparker}, they arise 
naturally in unified gauge theories \cite{dirac1931,thooft1974}, are
readily produced cosmologically \cite{preskill}, and could be abundant
elsewhere. Other cosmological defects are highly constrained in our
universe \cite{planckstring}, but domain walls, cosmic strings, and
global texture could instigate structure formation in other universes
\cite{kibble,turok1989,vilenkin85}. The global geometry of space-time 
could also vary, including having a compact topology
\cite{barlev,cornish,lachieze,levin}. Neutrinos in our universe have 
relatively small masses and do not contribute significantly to the
current cosmic density, but larger neutrino masses and other
variations in the dark matter inventory are possible
\cite{pogosian,tegmarkneutrino}. One can also envision universes 
with different numbers of light quarks \cite{jaffe}, photons with 
nonzero mass \cite{goldhaber}, and/or different numbers of particle 
generations \cite{gould,ibe,schellekens}. Other generalizations 
could lead to additional (fifth) forces \cite{fischbach,niebauer},
which could have astrophysical consequences \cite{lizhao}. All of
these possible variations in the laws of physics --- and many others
--- are likely to affect the potential habitability of other universes
and should be addressed in the future.

\subsection{Insights and Perspective} 
\label{sec:insight} 

The consideration of counterfactual universes necessarily lies near
the boundary of science. Some authors have considered this enterprise
to lie safely within the scientific realm, while others have adamantly
taken the opposite point of view (e.g., see \cite{bostrom,carrellis,
carroll2018,davies2004,ellissilk,freivogel,friederich} for further 
discussion).  Instead of continuing the debate as to whether or not
fine-tuning arguments and alternate universes should be officially
considered as part of science, perhaps a better question is whether or
not the results are useful. Taking the affirmative position, we wrap
up this review with a brief summary of what assessments of fine-tuning
reveal about physics and astrophysics in our universe. By determining
the parameter variations necessary to render the cosmos devoid of
life, we gain a greater understanding of how the universe operates.

Stars are more robust than most people realize: The range of
fundamental constants that allow for working stars is enormous
(Section \ref{sec:stars}). Stars exist while the fine structure and
gravitational constants vary by many orders of magnitude. The nuclear
reaction rate can change by factors of $10^{21}$ and stars will still
function. One can eliminate the weak interaction altogether (Figure 
\ref{fig:weakstar}), make deuterium unstable (Figure \ref{fig:hrnod2}), 
or make diprotons stable (Figure \ref{fig:staragc}), and the stars 
will still shine. And not only will stars continue to operate as
stable nuclear-burning entities, their power output, surface
temperatures, and lifetimes are commensurate with values considered
appropriate for habitability (Figure \ref{fig:starplane}).  The
limiting feature for a viable universe --- including ours --- is not
the astrophysical engines that synthesize heavy elements and generate
energy, but rather the parameters of particle physics that allow for
the existence of stable nuclei (see Figures \ref{fig:updown} and 
\ref{fig:nukebind}).

Although they represent a significant problem for particle physics
theories, the observed hierarchies of the fundamental parameters are 
a distinguishing feature of the cosmos \cite{carr}, and extreme
hierarchies are {\it required} for any universe to be viable: The
strength of gravity can vary over several orders of magnitude, but it
must remain weak compared to other forces so that the universe can
evolve and produce structure, and stars can function. This required
weakness of gravity leads to the hierarchies of scale that we observe
in the universe (Figure \ref{fig:astroscales}). In addition, any
working universe requires a clean separation of the energies
corresponding to the vacuum, atoms, nuclei, electroweak symmetry
breaking, and the Planck scale (Figure \ref{fig:energyhist}).  
In other words, the hierarchies of physics (see also Figure  
\ref{fig:quarkhist}) lead to the observed hierarchies of
astrophysics, and this ordering is necessary for a habitable
universe. Moreover, these required hierarchies can occur with
reasonably high probability if the underlying distribution of energy
scales has a log-random (scale-free) form (\ref{sec:probability}).

Finally, in spite of its biophilic properties, our universe is not
fully optimized for the emergence of life. One can readily envision
more favorable universes (Section \ref{sec:bestworld}). Possible
improved cosmic properties include a larger primordial fluctuation
amplitude $Q$, larger baryon to photon ratio $\eta$, smaller fine
structure constant $\alpha$, stable beryllium-8, and less dark energy
(smaller $\rhov$). Such variations could lead to more galaxies, stars,
and potentially habitable planets, which provide more opportunities
for biological development. 

The universe is surprisingly resilient to changes in its fundamental
and cosmological parameters, whether such variations are realized in
other regions of space-time or are merely gedanken in nature.
Considerations of these possible variations thus improve our
understanding and alter our interpretation of observed aspects of
physics and astrophysics -- in our universe and others.

\bigskip
\bigskip
\noindent 
{\bf Acknowledgments:} This review benefited from discussions and
input from many colleagues. I would especially like to thank
Konstantin Batygin, Juliette Becker, Tony Bloch, Sean Carroll, Gus
Evrard, George Fuller, David Garfinkle, Evan Grohs, Alex Howe, Lillian
Huang, Dragan Huterer, Gordy Kane, Jake Ketchum, Martin Rees, Frank
Timmes, and James Wells. I also thank the anonymous referee for many 
useful suggestions. This work was supported by the University of
Michigan and in part by the John Templeton Foundation through grant
ID55112 {\it Astrophysical Structures in Other Universes}.

\bigskip
\bigskip 
\appendix 

\bigskip 
\appendix 
\section{Mass Scales in terms of Fundamental Constants} 
\label{sec:massscales}

This Appendix provides a summary of the astrophysical mass scales
discussed in the main text and introduces some additional masses (see
also the previous treatments from \cite{bartip,burost,carter1974,page,
pagenew,presslight,rees1972,weisskopf}). Each quantity is expressed 
in terms of fundamental constants.

We start the discussion with stars. A characteristic stellar mass
scale $\starmass$ \cite{phil} can be written in the form 
\be 
\starmass \equiv \alpha_G^{-3/2} \mpro = \left( {\hbar c \over G} 
\right)^{3/2} \mpro^{-2} \,,
\label{starmassform}  
\ee 
where the value of this quantity in our universe is $\starmass\approx$
$3.7\times10^{33}$ g $\approx1.85M_\odot$.  Although this scale mass
is relatively close to the mass of the Sun, it is important to keep in
mind that $\starmass$ is an order of magnitude larger than the typical
mass of stars in our universe. The stellar initial mass function has a
nearly log-normal form with a characteristic mass of about
$M_C\approx0.2M_\odot\sim\starmass/10$ \cite{chabrier}. It is also
important to keep in mind that the range of stellar masses spans a
factor of $\sim1000$ in our universe \cite{clayton,kippenhahn,phil}, 
and such a wide distribution of masses cannot be fully characterized 
by a single value. 

The minimum mass necessary to sustain nuclear fusion is determined by
the requirement that stars can achieve a sufficiently high central
temperature in the face of degeneracy pressure \cite{clayton,hansen,phil}. 
This minimum mass can be written in the form 
\be
M_{\ast{\rm min}} = 6 (3\pi)^{1/2} \left({4\over5}\right)^{3/4} 
\left({kT_{\rm nuc} \over \emass c^2}\right)^{3/4} 
\alpha_G^{-3/2} \mpro \,, 
\label{massminone} 
\ee
where $T_{\rm nuc}$ is the temperature required for sustained nuclear
fusion. For hydrogen burning reactions in our universe, the required
temperature $T_{\rm nuc}\approx10^7$ K, although a more general
estimate can be derived.  Nuclear reaction rates in stellar cores
depend on quantum mechanical tunneling of the particles through the
Coulomb barrier. With this requirement, in conjunction with the
Boltzmann distribution of thermal speeds, the reaction rate is
proportional to an exponential factor of the form $\exp[-3\Theta]$,
where $\Theta=(E_G/4kT)^{1/3}$ and $E_G$ is the Gamow energy
($E_G\approx\pi^2\alpha^2\mpro{c^2}$ for proton reactions).
Semi-analytic stellar models \cite{adams,adamsnew} show that the
parameter $\Theta$ must be of order unity, with a typical value 
in the range $4-7$.  Putting together these considerations, and 
setting $\Theta=\Theta_X$, the nuclear burning temperature can be 
written in the form 
\be
k T_{\rm nuc} = {\pi^2 \over 4 \Theta_X^3} \alpha^2 \mpro c^2 \,, 
\label{nuketemp} 
\ee
so that the minimum stellar mass scale becomes 
\be
M_{\ast{\rm min}} = 6 (3\pi)^{1/2} 
\left({\pi^2 \over 5 \Theta_X^3}\right)^{3/4}  
\left({\mpro \over \emass}\right)^{3/4} \alpha^{3/2} 
\alpha_G^{-3/2} \mpro \approx {1\over2} 
\left({\mpro \over \emass}\right)^{3/4} \alpha^{3/2} \starmass\,. 
\label{massminstar} 
\ee
The final approximate equality assumes $\Theta_X\approx6$. 

Stars become unstable if the pressure contribution from radiation
exceeds the gas pressure by a sufficiently large margin 
\cite{clayton,phil}. The maximum stellar mass is given by the
expression 
\be
M_{\ast {\rm max}} =  
\left( {18 \sqrt{5} \over \pi^{3/2} } \right) 
\left( {1 - \bcon \over \bcon^4} \right)^{1/2}  
\left( {\mpro \over \mbar} \right)^2 \, \alpha_G^{-3/2} \mpro 
\approx 50 \starmass \,, 
\label{massmaxstar}
\ee
where $\bcon\approx1/2$ denotes the fraction of the pressure 
provided by the ideal gas law. 

Next we consider the Chandrasekhar mass, which represents the maximum
stellar mass that can be supported by the degeneracy pressure of
non-relativistic electrons \cite{chandra}. This mass scale is the
upper limit for white dwarfs and can be written in the from 
\be
M_{\rm ch} \approx {1 \over 5} (2\pi)^{3/2} 
\left({Z \over A}\right)^2 \alpha_G^{-3/2} \mpro \approx 
{\,\,(2\pi^3)^{1/2} \over 10\,\,} \starmass \approx 1.4 M_\odot\,, 
\label{masschandra} 
\ee
where $Z$ and $A$ are the mean atomic number and atomic weights 
of the stellar material. The mass scale thus depends on chemical 
composition, where $Z\approx{A}/2$ for the carbon and oxygen 
composition expected for white dwarfs in our present-day universe. 

High mass stars ($M_\ast\simgreat8M_\odot$ in our universe) leave
behind neutron stars as remnants. To leading order, the expected mass
scale for neutron stars is comparable to the Chandrasekhar mass from
equation (\ref{masschandra}). The full story is more complicated,
however, and involves the subtleties of nuclear physics. In
approximate terms, the maximum mass for a neutron star is given by the
requirement that the object cannot exceed nuclear densities, where the
inter-particle spacing is determined by the Compton wavelength of the
pion, which sets the range of the strong force. These
considerations \cite{burost} result in the mass scale 
\be
M_{\rm NSmax} = \alpha_G^{-3/2} \mpro 
\left({\mpro \over 2f m_\pi}\right)^{3/2} = 
\left({\mpro \over 2f m_\pi}\right)^{3/2} \starmass \,, 
\label{nstarmax} 
\ee
where $m_\pi$ is the pion mass (about 140 MeV) and the dimensionless
parameter $f$ is the factor by which the star is larger than the
Schwarzschild radius for the same mass. 

The minimum mass accessible to a neutron star is given by the
requirement that the object has a lower energy state than the white
dwarf configuration of the same mass \cite{shapteuk}.  The neutron
star is smaller in radius than the equivalent white dwarf, so that its
gravitational binding energy is greater; in contrast, the neutrons in
the neutron star are free, rather than bound into nuclei, and this
freedom costs energy. Given that degenerate objects have larger radii
for lower masses, a crossover point occurs when the gravitational
binding energy per nucleon is equal to the nuclear binding energy per
nucleon. The resulting mass scale is given by 
\be
M_{\rm NSmin} = \alpha_G^{-3/2} \mpro \,\, \alpha_{\rm s}^{3/2} = 
\alpha_{\rm s}^{3/2} \starmass \,,  
\label{nstarmin} 
\ee
where $\alpha_{\rm s}\approx0.2$ is the analog of the fine-structure
constant for the strong force (so that the binding energy for a
nucleon is given $E_B\approx\alpha_{\rm s}^2\mpro$).

In order to make stars, interstellar gas must condense and fragment.
A number of studies have considered the minimum mass subject to
opacity limited fragmentation \cite{rees1976,silk1977}.  In this
scenario, parcels of gas continue to condense until they become
sufficiently optically thick that they can no longer radiate away the
energy released during contraction. This type of analysis results in a
variety of different expressions for the minimum fragmentation mass,
depending on the assumptions. If we find the point where a gas parcel
contains one Jeans mass, has optical depth unity, and can radiate away
is energy of self-gravity on a free-fall time scale, the resulting
fragmentation mass takes the form
\be
M_{\rm frag} = \alpha_G^{-10/7} 
\left({\emass\over\mpro}\right)^{2/7} \alpha^{-2/7} \mpro = 
\alpha_G^{1/14} \left({\emass\over\mpro}\right)^{2/7} 
\alpha^{-2/7} \starmass \,.  
\label{fragmass} 
\ee

The mass scale for rocky planets is determined, in part, by the
requirement that the body is supported by electromagnetic forces
rather than degeneracy pressure (see Section \ref{sec:planets}).  
This constraint leads to the characteristic mass scale for planets
\cite{adamsnew,bartip,burost}, which takes the form 
\be
M_{\rm P} = \left( {\alpha \over \alpha_G} \right)^{3/2} \mpro 
= \alpha^{3/2} \starmass \,. 
\label{massplanet} 
\ee
In our universe, this mass scale is roughly comparable to that of
Jupiter. Significantly, since $\alpha\ll1$ in our universe, and we
expect $\alpha\simless1$ more generally, planets supported by
electromagnetic forces have smaller masses than their host stars.

The mass scale of the cosmological horizon at the epoch of equality
between the matter and radiation components \cite{coppess,tegmarkrees} 
is given approximately by the expression 
\be
M_{\rm eq} = \left( {5 \over \pi} \right)^{1/2} {3 \over 64\pi} 
\alpha_G^{-3/2} \mpro \left( {\mpro c^2 \over kT_{\rm eq}} \right)^2 
\approx {1 \over 64 \eta^2} 
\left( {\omegab \over \omegam} \right)^2 \, \starmass \,.
\label{massequal} 
\ee
In this expression, $\eta$ is the baryon to photon ratio, whereas
$\omegab$ and $\omegam$ are the energy densities of baryons and
matter, respectively.

One characteristic mass scale for galaxies is given by the requirement
that the gas cooling time is comparable to the free-fall collapse time
for cosmological structures. In terms of the structure parameters, the
resulting mass scale \citep{reesost,tegmarkrees} can be written in the form 
\be
M_{\rm gal} = \alpha_G^{-2} \alpha^5 
\left( {\mpro \over \emass} \right)^{1/2} \mpro = 
\alpha_G^{-1/2} \alpha^5 \left( {\mpro \over \emass} \right)^{1/2} 
\, \starmass \,. 
\label{massgalcool} 
\ee 

For completeness, we note that mass scales for sub-galactic systems
can also be defined.  Here the result depends on the nature (and
abundance) of dark matter, and whether the primordial fluctuations are
adiabatic or isothermal \cite{carrrees1984}. In spite of these
complications, for our universe one generally finds 
$M_{\rm sub}\sim10^5-10^6M_\odot$, close to the Jeans mass of the
baryons at the epoch of recombination. Similarly, at the earlier 
epoch of mass-radiation equality, we can write the Jeans mass in the 
form 
\be
M_{J({\rm eq})} = {\pi\over6} \left({\pi kT \over \mpro G}\right)^{3/2} 
\rho^{-1/2} = {\pi\over6} 
\left[ 15\pi\omegab \over \eta\omegam \right]^{1/2} \starmass 
\sim \starmass \eta^{-1/2} \,, 
\label{mjeanseq} 
\ee
where the final expression ignores all of the dimensionless 
constants or order unity. 

One can also define a mass scale for entire universe, provided that it
contains a non-zero contribution of vacuum energy $\rhov=\lambda^4\ne0$.
At late times, the universe develops a horizon with size given by the 
inverse Hubble parameter, which approaches a constant (e.g., 
\cite{bjorken}). The mass scale of the universe thus becomes  
\be
M_{\rm univ} = {56 \over \pi^3} \sqrt{{5\over2\pi}} 
{\mplanck^3 \over \lambda^2} \approx 1.6 
\left({\mpro\over\lambda}\right)^2\,\starmass\,.
\label{massuniverse} 
\ee
Ignoring the dimensionless coefficient of order unity, the number
$N_\star$ of `stellar mass units' in the universe is given by 
$(\mpro/\lambda)^2$. In our universe, $\lambda\approx0.0023$ eV, 
so that $N_\star\approx10^{23}$. 

We can also derive a mass scale for a matter dominated universe, with
no dark energy, although the result is a function of time.  From the
Friedmann equation (\ref{adotovera}), the density of the universe is
given by $\rho=3H^2/(8\pi G)$, where $H$ is the Hubble parameter. The
inverse $H^{-1}$ is essentially the horizon distance, so that the
corresponding mass scale at time $t$ becomes 
\be
M_{\rm univ} \approx {3 \over 8\pi GH} \to 
{3 c^3 t \over 8\pi G} \,.  
\ee
A viable universe must live for a sufficiently long time, which places
a lower limit of the value of $t$. Here we require the universe to
live for as long as the minimum stellar lifetime imposed in Section
\ref{sec:stars}, so that $t>10^{33}t_{\rm A}$, where $t_{\rm A}$ = 
$\hbar/(\alpha^2\emass{c^2})$ is the atomic time scale and where 
the numerical factor corresponds to the equivalent of 1 Gyr. 
The mass scale of a viable matter dominated universe is thus 
bounded from below such that 
\be
M_{\rm univ} > {(3\times10^{33})\hbar c \over 8\pi G\alpha^2\emass} 
\approx 10^{32} {\mpro \over \beta \alpha_G \alpha^2} = 10^{32} 
{\alpha_G^{1/2} \over \beta \alpha^2}\,\starmass\,.  
\label{massunivbound} 
\ee 

A crude estimate for the mass scale of life forms can be found by
invoking the following requirements \cite{press1980,presslight}: 
The life forms are assumed to (a) be composed of molecules, 
(b) reside on planets with the masses and radii discussed in 
Section \ref{sec:planets}, and (c) become as large as possible 
without breaking apart if they fall down on the planetary
surface. This set of constraints leads to a mass estimate 
of the form 
\be
M_{\rm zoo} \sim \epsilon_{\rm c}^{3/4} 
\left({\alpha\over\alpha_G}\right)^{3/4} \mpro = 
\left(\epsilon_{\rm c}\,\alpha\,\alpha_G\right)^{3/4} \starmass\,,
\label{masslife} 
\ee
where $\epsilon_{\rm c}$ is the chemical conversion factor for life
forms introduced in Section \ref{sec:stars}.  In order to describe
observed energy scales of chemical reactions, the conversion factor
$\epsilon_{\rm c}\sim10^{-3}$. In this context, however, equation
(\ref{masslife}) provides a better description for animal life on 
our planet if the efficiency $\epsilon_{\rm c}$ is of order unity 
(see also \cite{giraffe}). Analogous considerations for plant life 
limit the height of trees to $\sim100$ meters on Earth \cite{treeheight},
about 4 times larger than a blue whale. 

\begin{figure}[tbp]
\centering 
\includegraphics[width=.95\textwidth,trim=0 150 0 150,clip]{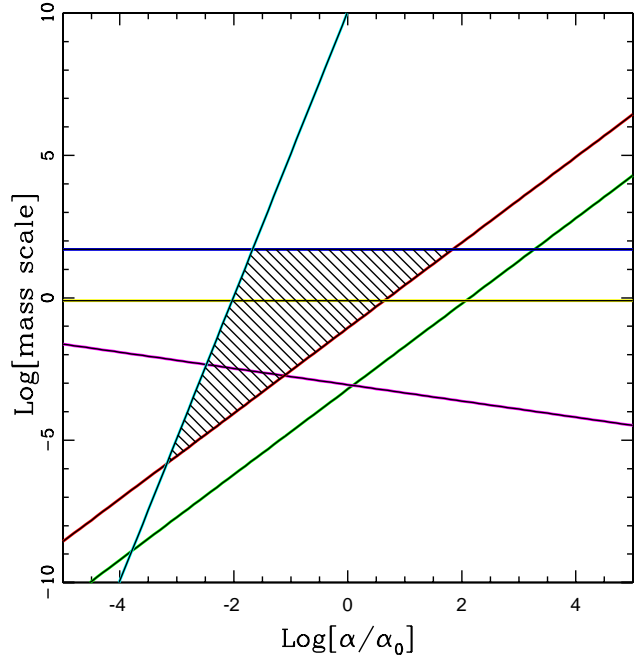}
\caption{Mass scales in the universe plotted versus the fine structure 
constant $\alpha$ (with all other parameters fixed to their values in
our universe). Masses are given in units of the fundamental stellar
mass scale $\starmass$ = $\alpha_G^{-3/2}\mpro$ from equation
(\ref{starmassform}). The curves show the mass scales for galaxies
(cyan), maximum stellar mass (blue), Chandrasekhar mass (yellow),
minimum stellar mass (red), minimum mass for opacity limited
fragmentation (magenta), and the planetary mass scale (green). 
Shaded region shows the allowed range of stellar masses, determined 
by the constraints that the minimum stellar mass is smaller than the
maximum stellar mass, and that stars are smaller in mass than their
host galaxies. }
\label{fig:mscales} 
\end{figure} 

The mass scales presented in this Appendix are plotted in Figure
\ref{fig:mscales} as a function of the fine structure constant
$\alpha$. The maximum stellar mass (blue curve) and the Chandrasekhar
mass (yellow curve) are independent of $\alpha$ and follow horizontal
lines in the figure. The minimum stellar mass (red curve) increases
with $\alpha$ and eventually becomes larger than the maximum stellar
mass for $\alpha/\alpha_0\approx68$ or equivalently
$\alpha\approx1/2$. The maximum planetary mass scale (green curve)
increases with $\alpha$ in parallel to the minimum stellar mass, but
the planetary mass is always much smaller. The galaxy mass scale is a
steeply increasing function of $\alpha$ (cyan curve). For sufficiently
small values of the fine structure constant,
$\alpha/\alpha_0<6\times10^{-4}$ or $\alpha<4.6\times10^{-6}$, the
galactic mass scale is smaller than the smallest star, so that such
universes are not viable. For completeness, the minimum fragment mass
is shown as the magenta curve, which decreases slowly with
$\alpha$. For $\alpha<1.6\times10^{-4}$, the smallest fragment mass 
is larger than the smallest star, but is unlikely to inhibit star 
formation (given that stars form in cloud cores that are much larger
in mass \cite{sal1987}). Note the the mass scale (\ref{massuniverse})
for the universe is much larger than the mass range depicted in Figure
\ref{fig:mscales}, whereas the mass scale for life forms (\ref{masslife})
is much smaller. 

The shaded region in Figure \ref{fig:mscales} depicts the allowed
range of stellar masses as a function of the fine structure constant
(see also Section \ref{sec:stars}). For large values of $\alpha$, the
range of stellar masses decreases and then vanishes altogether (due to
the lack of viable stellar structure solutions). The range of stellar
masses also narrows for sufficiently small values of the fine
structure constant. As $\alpha$ decreases, the mass scale for galaxies
decreases rapidly, and galaxies become smaller than stars, first only
the high mass stars, but eventually stars of all possible
masses. These trends underscore two important issues: First, the range
of $\alpha$ that allows for working stars is relatively wide, spanning
about five orders of magnitude.  Second, the value of $\alpha$ that
allows for the largest range of stellar masses is smaller than the
value in our universe by a factor of $\sim46$, i.e., for
$\alpha\sim1/6300$. For this value, stars span a range of masses 
corresponding to a factor of $\sim18,000$ (compared to only $\sim1000$
in our universe). As a result, our universe is {\it not} optimized to 
support the widest possible range of stellar masses. On the other hand, 
for universes with a wider range of stellar masses, the mass hierarchy 
between galaxies and stars is smaller. 

\bigskip 
\begin{table} 
\centering 
{\bf Summary of Mass Scales in the Universe} 
\medskip 
\def\arraystretch{1.25}
\begin{tabular}{lllc}
\\
\hline 
\hline 
quantity&symbol& $\,\,\,\,\mathfrak{M}\equiv M/\starmass$ &observed value\\
\hline 
\hline 
Vacuum dominated universe & $M_{{\rm univ}(\Lambda)}$ & 
$(56/\pi^3) \sqrt{5/2\pi} (\mpro/\lambda)^2$ & $3\times10^{23}$ \\
Matter dominated universe & $M_{{\rm univ}(m)}$ & 
$10^{32} \alpha_G^{1/2} \,\beta^{-1} \alpha^{-2}$ & $3\times10^{22}$ \\
Horizon mass at equality & $M_{\rm eq}$ & 
$(8\eta\omegam/\omegab)^{-2}$ & $10^{15}$ \\
Jeans mass at equality & $M_{J({\rm eq})}$ & 
$(\pi/6) (15\pi\omegab/\eta\omegam)^{1/2}$ & $4\times10^4$ \\
Galactic mass scale & $M_{\rm gal}$ & $\alpha_G^{-1/2}\alpha^5\,\beta^{-1/2}$ 
& $2\times10^{10}$ \\[3pt]
\hline 
\hline 
Maximum stellar mass & $M_{\ast {\rm max}}$ & 50 & 50 \\
Chandrasekhar mass & $M_{\rm ch}$ & $(1/5) (2\pi)^{3/2} (Z/A)^2$ & 0.8 \\
Minimum stellar mass & $M_{\ast{\rm min}}$ & 
$(1/2)\,\beta^{-3/4}\alpha^{3/2}$ & 0.09 \\ 
Maximum Neutron star & $M_{\rm NSmax}$ & $(\mpro / 2f m_\pi)^{3/2}$ & 1.2 \\
Minimum Neutron star & $M_{\rm NSmin}$ & $\alpha_{\rm s}^{3/2}$ & 0.09 \\
\hline 
\hline  
Opacity limited fragment & $M_{\rm frag}$ & 
$\alpha_G^{1/14} \,\beta^{2/7} \alpha^{-2/7}$ & $10^{-3}$ \\[3pt]
Planet mass & $M_{\rm P}$ & $\alpha^{3/2}$ & $6\times10^{-4}$ \\
Animal mass & $M_{\rm zoo}$ & $(\epsilon_{\rm c}\,\alpha\,\alpha_G)^{3/4}$ 
& $10^{-31}$ \\
\hline 
\hline 
\end{tabular}
\caption{Mass scales in the universe. For each astrophysical quantity 
discussed in the text, the table lists the value of its mass in terms
of the fundamental constants. The masses are scaled relative to the
stellar mass scale $\starmass=\alpha_G^{-3/2}\mpro$, which has a value
of $\starmass\approx1.85M_\odot$ in our universe (as a result, the 
masses are not given in solar masses). The fourth column lists the
values realized in our universe. The upper portion of the table
provides the mass scales of galaxies and cosmology, the middle portion
lists stellar mass scales, and the bottom portion includes smaller
entities. }
\label{table:masses} 
\end{table}  

Another way to summarize the mass scales of the universe is to write
all of the quantities in terms of the fundamental constants and in
units of the stellar mass scale $\starmass$. The resulting set of mass
scales is presented in Table \ref{table:masses}. For each quantity of
interest, the table lists the expression for the mass scale in terms
of fundamental and/or cosmological parameters, where the result is
scaled by $\starmass=\alpha_G^{-3/2}\mpro$. The top portion of the
table includes the masses of the universe and galaxies, which have
masses much larger than the stellar scale $\starmass$ for our
universe. Note that the weakness of gravity (small $\alpha_G$), and
to a lesser extent the small value of $\beta$, allows galaxies to
contain billions of stellar mass scales. Similarly, the small energy
scale of the vacuum $\lambda$ (compared to the proton mass $\mpro$)
allows the universe to contain a large number of stellar masses
(trillions of billions).  The central portion of the table includes
mass scales resulting from stellar evolution, which are comparable to
$\starmass$. Finally, the bottom portion of the table includes planets
and life forms, which have masses much less than $\starmass$. On these
scales, small values of the structure constants $(\alpha,\alpha_G)$
allow stars to host many planetary masses and for planets to host an
enormously large number of lifeforms.

\bigskip
\section{Number of Space-Time Dimensions} 
\label{sec:dimensions} 

The number of space-time dimensions represents another property of the
universe that could vary from region to region within the multiverse.
Developments in string theory and M-theory indicate that the
fundamental manifold of space-time must have at least 10 dimensions.
In most versions of the theory, these extra dimensions only manifest
themselves at sufficiently high energies and small size scales that
are currently outside the reach of experimental probes. In our universe,
only three of the spatial dimensions have become large, whereas the
remaining dimensions are compactified.  Given the possible existence
of alternate universes with higher numbers of dimensions, we are left
with the question of why our universe has only three large spatial
dimensions (and one temporal dimension). This Appendix briefly reviews
some of the arguments for the standard $3+1$ structure of space-time.

We first note that most authors agree that habitable universes 
should have only one time dimension 
\cite{barrow1983,borstnik,tegmarkdimension,tegmarktoe}. If space-time 
had more than one temporal dimension, then closed time-like loops
could be constructed. Such loops, in turn, allow for observers to
revisit the ``past'' and thereby affect causality. In addition to
violations of causality, multiple time dimensions can lead to
violations of unitarity, tachyons, and ghosts \cite{fostermuller}.
Although space-times with multiple time variables can be constructed,
the corresponding universes would have properties vastly different
from the types of potentially habitable universes discussed herein. 
As a result, they will not be considered further. 

Notice also that the question of the number of space-time dimensions
rests on a different footing than the other issues considered in this
review.  If one considers variations in the strength of the
gravitational constant or the mass of the electron, for example, the
basic form of the laws of physics would remain the same. Moreover,
many of the physical structures relevant for habitability --- such as
atoms and stars --- could retain their general form, albeit with
different specific properties.  If the number of dimensions varies,
however, more drastic changes to physics arise. The first obvious
change is to the number of phase space dimensions, which could be much
larger or much smaller than in our universe. As a result, the
structure of orbits depends sensitively on the number of dimensions
--- compare orbits in 1, 2, and 3 spatial dimensions in our universe.
These differences arise not only from the additional degrees of
freedom available in spaces of higher dimension, but also due to
changes in the functional form of force laws. As long as the forces,
such as gravity and electromagnetism, are described by a version of
the Poisson equation, or its generalizations, the force laws will
change with the number of dimensions.

One basic requirement for a viable universe is that the force laws
must allow for stable orbits. The orbits in question include both the
motion of planets around their host stars and the motion of electrons
around atomic nuclei. Although the latter orbits are more complicated
due to quantum mechanics, their stability leads to the same constraint
on the number of spatial dimensions. Ehrenfest is often credited with
starting this line of inquiry \cite{ehrenfest1,ehrenfest2} by
presenting constraints from the stability of classical orbits and wave
propagation. A host of additional considerations have since been invoked,
including the necessity of even numbers of dimensions based on
topological considerations \cite{whitrow}, the existence of quantum
mechanical solutions for atomic structure \cite{gurevich}, orbits in
Schwarzschild space \cite{tangherlini}, and the structure of white
dwarf stars \cite{chavanis}. Here we outline the basic arguments from
the stability of orbits (\ref{sec:orbits}) and the stability of atoms
(\ref{sec:atoms}). A number of previous reviews provide additional
detail (e.g., see \cite{barrow1983,bartip,tegmarkdimension,tegmarktoe}).

As outlined below, the stability of orbits and atoms places an upper
bound on the number of spatial dimensions, namely $\dimnum\le3$.
Nonetheless, these stability constraints would allow universes to have
lower numbers of dimensions $\dimnum$ = 1 or 2. The standard argument
against universes of lower dimension is that they are too simple to
allow for the complexity necessary for a universe to become habitable.
This argument has been given many times \cite{bartip,reessix,whitrow},
but is briefly summarized here: A universe with $\dimnum=2$ would be
confined to a plane, much like the classic satirical novella 
{\sl Flatland} \cite{abbott}.  Orbits of both planets and electrons
would also be confined to planes. The resulting atomic structures
(electron orbitals) would necessarily be much simpler than those in
our universe, so that chemistry would be compromised.  Another issue
arises with the construction of networks, such as neural pathways or
electrical circuits: A sufficiently complicated network, confined to a
two-dimensional surface, would result in wires crossing, whereas
higher dimensional spaces allow more possibilities. Universes with only
a single spatial dimension $\dimnum=1$ would be even simpler and hence
not viable.

Universes with $\dimnum=1,2$ are also disfavored by considerations 
of signal propagation. The properties of wave equations depend
sensitively on the number $\dimnum$ of spatial dimensions 
\cite{bartip,ehrenfest1,ehrenfest2}. For example, consider
a simple wave equation of the form 
\be 
{1 \over c^2} {\partial^2 S \over \partial t^2} = 
\nabla_\dimnum^2 \,S \qquad {\rm where} \qquad 
S = S(t, {\bf r}_\dimnum) \,, 
\label{dimwave} 
\ee 
and where $\nabla_\dimnum$ is the Laplacian operator in $\dimnum$
dimensions. For the case of one- and two-dimensional wave equations,
the properties of the wave solutions depend on the entire domain over
which the wave propagates. This property implies that disturbances can
propagate at any speed $v\le{c}$. Slower waves, emitted earlier, can
thus be overtaken by faster waves produced later, which acts to 
impede the transmission of sharply defined wave signals \cite{hadamard}.  
On the other hand, for $\dimnum=3$, only the boundary of the domain
(here the spherical surface at $r=ct$) determines the wave solution. 
As a result, waves in three spatial dimensions must propagate at the
speed $c$ appearing in the wave equation (\ref{dimwave}). Universes
with $\dimnum=3$ are thus distinguished from those of lower dimension
in terms of their capability for transmission of wave signals and
information processing.

\subsection{Stability of Classical Orbits} 
\label{sec:orbits} 

The stability of classical orbits has been studied in many contexts,
and places a constraint on the number of spatial dimensions $\dimnum$
such that 
\be
\dimnum \le 3 \,. 
\label{maxdim} 
\ee
One can illustrate this constraint as follows. Suppose that one
considers the force due to a point mass (or a point charge for
classical orbits mediated by the electric force).  For the case where
the force carrier is massless, the force law is given by a version of 
the Poisson equation and takes the familiar power-law form  
\be
F = {B \over r^q} \,.
\label{forcelaw}
\ee
In equation (\ref{forcelaw}), $F$ is the magnitude of the force
(assumed to be attractive), $B$ is a constant determined by the 
strength of the coupling constant, and the power-law index $q$ is 
related to the number of spatial dimensions $\dimnum$ according to 
$\dimnum=q+1$. For a classical orbit with a force law of the form 
(\ref{forcelaw}), the orbital frequency is given by 
\be
\Omega^2 = { B/m \over r^\dimnum} \,,
\ee
where $m$ is the mass of the test particle in question. One 
way to determine the stability of the orbit is to consider the 
epicyclic frequency $\kappa$, i.e., the frequency of oscillation 
of the particle's motion about the guiding center of the orbit 
\cite{bintrem,md1999}. This quantity is given by  
\be
\kappa^2 = {1 \over r^3} {d \over dr} \left( r^4 \Omega^2 \right) 
= (4-\dimnum) \Omega^2\,. 
\ee
This result shows that for $\dimnum>4$, the epicyclic frequency is
imaginary and orbits are unstable. For $\dimnum=4$, the epicyclic
frequency vanishes, but further analysis shows that the orbit is also
unstable. Stable orbits thus require $\dimnum<4$, consistent with
equation (\ref{maxdim}), which implies $\dimnum=$ 1, 2, or 3 (assuming
the number of dimensions is discrete).

The above result can be generalized to include massive force 
carriers and other departures from a power-law force law. We 
can write the force law in the generalized form  
\be
F = B {g(r) \over r^q} \,, 
\label{forcegeneral} 
\ee
where $g(r)$ is a dimensionless function of radius. For example, 
the function $g(r)\sim\exp[-\mu r]$ for the Yukawa potential. 
With this generalized form (\ref{forcegeneral}), the 
epicyclic frequency becomes 
\be
\kappa^2 = \Omega^2 \left[ (4-\dimnum) + {r\over g} {dg\over dr} \right]\,.
\ee
If the function $g(r)$ is a decreasing function of radius, then 
the constraint on the number of dimensions becomes more restrictive.

\subsection{Stability of Atoms: Bound Quantum States} 
\label{sec:atoms} 

Next we consider the stability of atoms. In the semi-classical limit,
the stability problem for electronic orbits reduces to an analog of
that considered in the previous section for classical orbits
\cite{ehrenfest1,ehrenfest2}, and one obtains the same constraint
$\dimnum<3$ (from equation [\ref{maxdim}]). Since atoms are quantum 
mechanical, semi-classical arguments do not provide an adequate proof
of stability or instability (see the discussion of \cite{barrow1983}), 
and one must start from basic quantum mechanical principles. 

The stability of a hydrogenic atom represents the simplest case, which
we outline here, both for three dimensional space and higher values of
$\dimnum$. The problem can be formally stated as follows:
The ground state energy of a quantum mechanical system is given by the
expectation value 
\be
E_0 = {\rm min}_{\textstyle \psi} \,
\bigl\{\,\,\langle\psi|H|\psi\rangle\,\,\bigr\}\,,
\label{energymin} 
\ee
where $H$ is the Hamiltonian operator and the minimum is taken over
the space of all properly normalized trial wave functions $\psi$ (so
that $\langle|\psi|^2\rangle=1$). The system is stable if the ground
state energy, as defined here, is finite, i.e., is bounded away from
negative infinity \cite{lieb1990}.

Historically, two different (but related) approaches to this problem
have been taken. The direct way to proceed is to start from the
Schr{\"o}dinger wave equation (SWE) for the problem and directly
construct the ground state wave function. Of course, the SWE results
from applying the calculus of variations to the problem posed in
equation (\ref{energymin}) subject to the constraint
$\langle|\psi|^2\rangle=1$.  Another option is to proceed indirectly
by using a series of inequalities to show that the ground state energy
$E_0$ is bounded. For the hydrogen atom in three dimensional space,
the SWE can be readily solved, and the direct method provides a simple
way to prove stability. For more complicated systems, including the
classic problem of the stability of bulk matter, solutions to the SWE
are generally not available, but the indirect method provides a way
forward \cite{dyson1967,lieb1975}. 

To illustrate the problem, consider a hydrogenic atom in
$\dimnum$-dimensional space. After suitable definitions, the SWE can
be written in dimensionless form, and the angular part of the wave
function can be separated. The remaining radial part of the wave
function $R(\xi)$ is then governed by the equation
\be
{d^2 R \over d\xi^2} + {\dimnum-1\over\xi} {dR\over d\xi} 
+ \left[ {Z \over \xi^{\dimnum-2}} - {\lambda \over \xi^2} 
+ \epsilon \right] R = 0 \,.
\label{genradial} 
\ee
If we take the dimensionless radius to be $\xi=r/r_0$, where $r_0$ is
a constant scale, the remaining dimensionless parameters include the
energy, 
\be
\epsilon = {2 m E r_0^2 \over \hbar^2} \,, 
\ee
the eigenvalue of the angular part of the Laplacian
operator in $\dimnum$ spatial dimensions, 
\be
\lambda = \ell (\dimnum - 2 + \ell) \,,
\ee
and the depth of potential 
\be
Z = {2m e^2 r_0^{4-\dimnum} \over (\dimnum - 2) \hbar^2}\,.
\ee
For $\dimnum\ne4$, one can define the radial scale $r_0$ so that
$Z=1$. For our universe with $\dimnum=3$ spatial dimensions,
$r_0=a_0/2$, where $a_0=\hbar^2/(me^2)$ is the Bohr radius.  For the
ground state with $\ell=\lambda=0$, equation (\ref{genradial}) admits
the well-known solution $\psi=\exp[-Z\xi/2]$; the corresponding ground
state energy eigenvalue is finite and has the value $\epsilon=-Z^2/4$.
For higher dimensions with $\dimnum>3$, however, the energy spectrum
extends to $\epsilon\to-\infty$, so that the energy has no lower
bound, and the hydrogen atom is unstable \cite{gurevich}. This result
indicates that universes with more than three (large) dimensions are
not habitable.

The result described above rests on the assumption that the electric
force in higher dimensions obeys Gauss's law, so that the potential
$V(\xi)\propto\xi^{2-\dimnum}$. For completeness, note that one could
assume that Maxwell's equations are no longer valid in higher
dimensions, and that the electric potential retains its form
$V\propto1/\xi$ in spaces with arbitrary dimension (here $\xi$ is the
radial coordinate in $\dimnum$ dimensional space). In that case, the
hydrogen atom could be stable in higher dimensions (see \cite{caruso} 
and references therein for further discussion).  Another option is to 
consider non-Euclidean spaces. For example, the higher dimensional
space $\mathbb{R}^3\times{S}^1$ allows for stable hydrogenic
atoms \cite{bures}.

Another way to consider stability is to write the energy eigenvalue 
of the ground state in the form 
\be
\epsilon = \int dV |\nabla \psi|^2 - 
Z \int dV \xi^{-(\dimnum-2)} |\psi|^2 \,, 
\ee
where $\dimnum>2$. To show that atoms are stable, one can show that
the energy $\epsilon$ is bounded from below. If the above expression
is finite for all trial wavefunctions $\psi$, then the ground state
energy must be finite (as codified in equation [\ref{energymin}]). 

For three dimensional space, one can use the Sobolev inequality
\cite{lieb1990} to write the constraint in the form 
\be
\epsilon > {\rm min}_{\textstyle \psi}\,\left\{ 
S \left[ \int dV\,|\psi|^6 \right]^{1/3} - 
Z \int dV \,{1\over\xi}\, |\psi|^2 \, \right\} \,,
\ee
where $S$ is a dimensionless constant. By defining the probability
density $\prob\equiv|\psi|^2$, and using H{\"o}lder's inequality to
rewrite the first term \cite{lieb1990}, one obtains the bound
\be
\epsilon > {\rm min}_{\textstyle \prob} \,\left\{ \, \int dV 
\left[ K\,\prob^{5/3} - {Z \over \xi} \prob \right] \,\,
\right\} \,,
\ee
where $K$ is another dimensionless constant.  Finally, applying the
calculus of variations to the integral, subject to the constraint
$\int\prob\,dV=1$, one finds a lower bound on the energy of the form
\be
\epsilon > - C Z^2 / K \,,
\label{lowerbound} 
\ee
where $C>0$ is a dimensionless constant. This argument shows that the
ground state energy is larger than some value (namely that given by
equation [\ref{lowerbound}]) by showing that the true expression for
the energy is larger than a series of energy functionals evaluated 
for {\it any} trial wavefunction (so the energy must be larger than 
the energy functional evaluated with the ``correct'' wavefunction). 

We can now show how this argument fails for higher dimensions: 
The energy eigenvalue in $\dimnum$-dimensional space can be written
in the form 
\be
\epsilon_{\dimnum} = \int dV_{\dimnum}\,|\nabla \psi|^2 - 
Z \int dV_{\dimnum}\,\xi^{-(\dimnum-2)} |\psi|^2 \,. 
\label{energeneral} 
\ee
Assume here that the ground state has zero angular momentum, 
so that the wavefunction has no angular dependence. The volume 
element can be written 
\be
dV_{\dimnum} = \Omega\,\xi^{\dimnum-1}d\xi \,. 
\ee
Now we consider a particular trial wavefunction of the form 
\be
\psi_\ast = A \exp[ - \mu \xi] \,,
\ee
where $A$ is the normalization constant. This form is motivated 
by the ground state of the hydrogen atom in three dimensional 
space. The parameter $\mu$ defines the effective width of the 
wavefunction and is left arbitrary for now. After finding the 
normalization constant and evaluating the integrals from 
equation (\ref{energeneral}), we find the following expression 
for the energy 
\be
\epsilon (\psi_\ast) = \mu^2 - 
Z {(2\mu)^{\dimnum-2} \over \dimnum!} = 
{1 \over L^2} - {Z 2^{\dimnum-2} \over \dimnum!} 
{1 \over L^{\dimnum-2}} \,,
\label{toymodel} 
\ee
where we have defined $L\equiv1/\mu$.  Equation (\ref{toymodel})
illustrates the basic issue of stability in higher dimensions. If
$\dimnum>4$, then the negative second term has a higher power of the
length scale $L$ in the denominator. As a result, the energy of the
trial wavefunction can become arbitrarily large and negative as $L$ is
made increasingly smaller. In contrast, for $\dimnum=3$, we have
$\epsilon \sim 1/L^2 - Z/6L$, so that the first positive term
dominates for small $L$. We can summarize this result with the 
following heuristic argument \cite{lieb1990}: The first term
represents the squeezing of the electron into a smaller volume, but
the wavefunction develops an effective ``pressure'' contribution 
($\propto1/L^2$) that is strong enough to overcome the tendency for
energy gain due to the electron going deeper into the potential. For
higher dimensions, however, the second term dominates, indicating that
this ``pressure'' is not sufficient to overcome the change in energy.
As a result, it becomes energetically favorable for electrons to fall
ever deeper toward/into the nucleus, the energy has no lower bound,
and atoms are unstable.

Note that the mathematical arguments presented in this Appendix are
abridged. One should refer to the primary references for a more
rigorous treatment. Finally, given the preference for three spatial
dimensions, the next step is to determine the manner in which an
initially higher dimensional universe relaxes into a configuration
with (only) three large dimensions and to assess its probability 
(e.g., \cite{brandvafa,karchrandall}). 
\bigskip 

\section{Chemistry and Biological Molecules} 
\label{sec:chemistry} 

The discussion thus far has considered fine-tuning issues in physics
by requiring that the fundamental constants and cosmological
parameters support the existence of galaxies, working stars, stellar
nucleosynthesis, habitable planets, complex nuclei, and stable atoms.
These constraints are necessary but not sufficient \cite{carrreestwo}.
Additional requirements are needed in order for a universe to actually
develop observers. Specifically, a viable universe must be able to
construct the essential molecules required for life \cite{elliskopel}.
Although the exact requirements are not known \cite{binderellis},
these biomolecules are likely to include water with suitable
properties, proteins, DNA, and RNA (or equivalent types of complex
molecules). We are thus excluding discussion of more exotic life forms
such as Hoyle's Black Cloud \cite{blackcloud}.  Life in its familiar
form also requires fatty acids, sugar, starch, and cellulose.  Unlike
physical considerations reviewed in the main text, the possible
fine-tuning in biologically relevant molecules is relatively
unexplored. This Appendix provides a brief overview of current work.

In order for life to exist, biological molecules must function
properly in the context of their host organisms. One recent summary 
of the requirements suggests that life requires proteins, metabolic 
networks, gene regulatory networks, and signal transmission networks
\cite{wagner}. Biology thus requires that particular molecules 
(e.g., proteins) function properly, which requires the right types of
chemistry. These chemical requirements, in turn, place constraints on
the laws of physics. At the fundamental level, the link between
chemistry and physics is provided by the Schr{\"o}dinger equation,
which can be written in the dimensionless form 
\be 
- \left[ {\emass \over 2 m} \nabla^2 + {\alpha \over r}\right] 
\Psi(r,t) = i {\partial \Psi \over \partial t} \,,
\label{swe} 
\ee
where $(r,t)$ are dimensionless variables defined by $r\to r/\ell$ 
and $t\to t/\tau$, where the reference values are defined via 
$\ell=\hbar/\emass c$ and $\tau=\hbar/\emass c^2$. In equation 
(\ref{swe}), the mass $m$ is that of the particle described by 
the wave equation, whereas the electron mass $\emass$ appears 
because it is used to define the dimensionless units. 

One important issue is the manner in which the sizes and other
properties of atoms change with the values of the fundamental
constants. Suppose we let $\alpha \to S\alpha$, where $S$ is a
dimensionless scaling factor. If the reference length is rescaled
according to $\ell\to \ell/S$, then both the potential and the kinetic
term in equation (\ref{swe}) are scaled by a factor of $S^2$, so that
the energy changes by this same factor. Different values of $\alpha$
thus result in changes to the atomic energy levels. If one includes
the hyperfine structure (which depends on the proton mass and hence
$\beta$), then the atomic structure changes as well. As a result, one
does not expect chemistry to operate in the same manner with different
values of $\alpha$ and $\beta$ (see \cite{atkins} for further
discussion).  Moreover, although it is called the ``fine structure
constant'', $\alpha$ is one of the most important parameters for the
specification of atomic structure: For quantum systems where 
$m=\emass$ and hyperfine corrections are ignored, the constant
$\alpha$ is the {\it only} parameter appearing in the dimensionless
Schr{\"o}dinger wave equation (\ref{swe}).

\begin{figure}[tbp]
\centering 
\includegraphics[width=.95\textwidth,trim=0 150 0 150,clip]{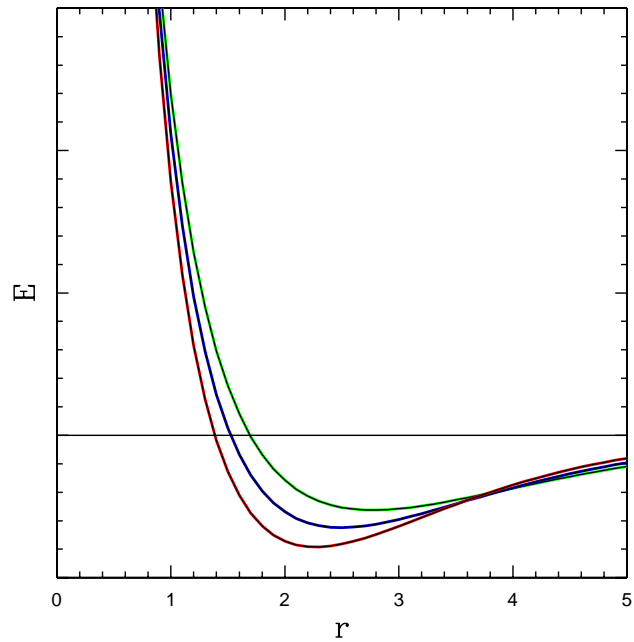}
\caption{Electronic energy for the Hydrogen molecule as a function of 
internuclear separation for different values of the fine structure
constant $\alpha$. The energy $E$ and separation $r$ are given in
dimensionless units (see text). The minimum of the energy curve
defines the size of the molecule. These energies are calculated using
the approximation of Linear Combination of Atomic Orbits \cite{atkins}. 
The three curves correspond to different values of the fine structure
constant: $\alpha$ = $0.9\alpha_0$ (red), $\alpha_0$ (blue), and
$1.1\alpha_0$ (green), where $\alpha_0$ is the value measured in our
universe. These 10\% variations in the fine structure constant result
in $\sim10\%$ variations in the effective size of the Hydrogen
molecule. }
\label{fig:molecule} 
\end{figure} 

To illustrate the difficulty of assessing fine-tuning in molecules,
one can consider the ionized Hydrogen molecule H$_2^+$, one of the
simplest cases with a single electron in orbit about two protons. Even
for this molecule, an analytic solution for the energy levels is not
available. One can proceed using the method of Linear Combination of
Atomic Orbitals, where the electron wave functions for the two protons
are found separately, and then combined, taking into account the
overlap integrals and the proper normalization (details are given in
most advanced physical chemistry texts -- see \cite{atkins}). Even in
this highly simplified case, the solutions for the energy levels and
the internuclear potential are rather cumbersome to write down. The
resulting energy is shown as a function of internuclear separation in
Figure \ref{fig:molecule} for three different choices of the fine
structure constant $\alpha$. The minimum energy state corresponds to
the equilibrium separation of the two protons and thus defines the
size of the molecule.  The blue curve shows the result for the value
of $\alpha$ in our universe. For comparison, the figure shows the
corresponding energies for $\alpha$ values that are 10\% larger (red
curve) and 10\% smaller (green curve). Figure \ref{fig:molecule} shows
that as the fine structure constant decreases, the molecular size
increases. More specifically, 10\% changes in the fine structure
constant $\alpha$ lead to corresponding $\sim10$\% changes in the
nuclear separation, as determined by the minimum of the energy curve.

Relatively little work has been carried out concerning fine-tuning 
of molecular structures larger than H$_2^+$.  One pioneering study  
\cite{kingchemistry} considers how variations in the fine structure 
constant $\alpha$ and the electron to proton mass ratio $\beta$ can
affect the structure of biologically important molecules (see also 
\cite{rthompson1975,rthompson2017}). Both of these dimensionless 
constants are small compared to unity. This study shows that the
quantum chemistry \cite{atkins} of hydrogen, carbon, nitrogen, and
oxygen correspond to the asymptotic limit where $\alpha,\beta\to0$. 
As a result, even smaller values of these two parameters leave
biologically important molecules, such as water, with properties
similar to those in our universe. For sufficiently larger values of
these parameters, however, the necessary biomolecules could cease to
function properly.

The water molecule, one of the essential ingredients for life as we
know it, provides a working example for how molecular structure
changes with the values of $\alpha$ and $\beta$. The equilibrium bond
angle in the water molecule decreases with increasing values of the
fine structure constant $\alpha$. If this angle becomes too small,
then water will no longer have its tetrahedral bonding capabilities
between hydrogen atoms. Since this property of water is essential 
for water to expand upon freezing, the loss of this property alters 
the characteristics of `liquid water environments' and hence changes 
the prospects for habitability. Increases in $\alpha$ also reduce the
polar nature of the water molecule and reduce the strengths of its
bonds.  If the mass ratio $\beta$ is increased, the energy for the
water producing reaction 
\be
{\rm O}_2 + 2{\rm H}_2 \to 2{\rm H}_2{\rm O}\,, 
\label{burnwater} 
\ee
increases accordingly. This increase in energy, in turn, increases the
thermodynamic stability of water, which partially offsets the negative
effects of increasing $\alpha$. 

Current work does not provide definitive bounds on $(\alpha,\beta)$
due to the constraint of working water molecules (and other molecules
such as H$_2$, O$_2$, and CO$_2$). However, as a benchmark
\cite{kingchemistry}, an increase in $\alpha$ by a factor of 7
decreases the strength of O-H bond in water by an increment of 7 kcal
mol$^{-1}$. An increase in $\beta$ by a factor of 100 increases the
strength of that same bond by 11 kcal mol$^{-1}$. For comparison,
burning one mole of hydrogen into water releases about 60 kcal. As a
rough summary of these findings: An increase in $\alpha$ by an order
of magnitude and/or an increase in $\beta$ by two orders of magnitude
change the energetics of the reaction (\ref{burnwater}) by $\sim10$
percent. Keep in mind that decreases in the parameters
($\alpha,\beta$) have much smaller effects. 

Relatively little work has been carried out to study the effects of
changes in the fundamental constants on larger molecules. Ultimately,
we would like to know how variations in the fundamental parameters
(e.g., $\alpha$ and $\beta$) alter the nature of biologically
important structures such as DNA \cite{elliskopel}.  These
biomolecules are complex and involve extremely long chains of base
pairs. Their successful operation requires precisely controlled
interactions that involve molecular recognition at the relevant
binding sites. The characteristic distance between adjacent sugars
and/or phosphates in DNA molecules is about $\ell\approx6$ \AA,
whereas working DNA chains require the distance $\ell$ to fall between
about 5.5 and 6.5 \AA $\,$ \cite{calladine}. This variation represents
a tolerance of about $\sim15\%$. However, if biological evolution were
to take place under different conditions where $\ell$ is significantly
larger or smaller, it is not known if DNA molecules could operate with
new spacings, or how far the distance $\ell$ could vary. Moreover, the
variations in the fundamental constants required to produce such
changes has not been calculated (although the required variations are
of order 10\% for the Hydrogen molecule, as shown in Figure
\ref{fig:molecule}).

In any case, sufficiently severe changes to the fundamental parameters
$(\alpha,\beta)$ would lead to significant changes to chemical
properties and reactivity, and hence to biology. However, as the
authors \cite{kingchemistry} state ``$\dots$ the broad
$(\alpha,\beta)$ sensitivity determined here for the chemistry of
life-supporting molecules is not as spectacular as the narrow
constraints on these fundamental constants established previously in
physics and cosmology''.

\bigskip 
\section{Global Bounds on the Structure Constants} 
\label{sec:globalbounds} 

This Appendix uses the equations of stellar structure to derive global
bounds on the fundamental parameters that determine the properties of
stars (see \cite{adamsnew,barnes2015,alexander}).  The relevant
variables for this problem include the fine structure constant
$\alpha$, the electron to proton mass ratio $\beta$, the nuclear
burning parameter $\conlum$, and the gravitational constant $G$
(equivalently $\alpha_G$). We can also find global bounds on the ratio
$\alpha_G/\alpha$.


These results follow from the analytic model of stellar structure
developed previously (see Section \ref{sec:stars} and
\cite{adams,adamsnew}). We start by writing all of the constraints in
dimensionless form.  The stellar mass can be written in terms of the
fundamental stellar mass scale $\starmass$ (from equation 
[\ref{starmassform}]) so that
\be
M_\ast = X \starmass = X \alpha_G^{-3/2} \mpro \,. 
\label{xstarmass} 
\ee 
The fine structure constant, electron to proton mass ratio, 
nuclear parameter, and gravitational constant can be written 
in terms of their values in our universe, 
\be
\alpha \equiv a \, \alpha_0\,, \qquad 
\beta \equiv b \,\beta_0 \,, \qquad 
\conlum \equiv \ck \,\conlum_0 \,,
\qquad {\rm and} \qquad 
\qquad G \equiv g \, G_0 \,,
\ee
where the subscripted quantities are the values in our universe. 

The equation that sets the central temperature of the star necessary
for a long-lived stable configuration can then be written in the form
\be
I(\thetacen) \thetacen^{-8} = A X^4 a^{-8} \ck^{-1} g \,,
\label{starstruck} 
\ee
where the dimensionless constant $A$ is given by 
\be
A = \left( {2^{19} \pi^5 \over 9 \cdot 5^8} \right) 
\left( {1 \over \betacon \mzero^4} \right) 
\left( {\hbar^3 \ck^4 \over E_G^3 \mpro} \right) 
\left( {G \over \kappa_0 \conlum} \right) 
\approx 5.23 \times 10^{-9} \,. 
\label{aconstant} 
\ee
All of the quantities in the above expression correspond to the 
values in our universe.

The condition that stars have a minimum temperature can be written 
\be
B X g^{1/2} > a^6 b^4 \thetacen^7 \,,
\label{temperature} 
\ee 
where the dimensionless constant $B$ is given by 
\be
B = \left( {\pi \over 25} \right) 
\left( {1 \over \betacon \mzero \xi_\star^2} \right) 
\left( {E_G^2 \over \kappa_0} \right) 
\left( {G \over \hbar c } \right)^{1/2} 
{(\hbar c)^2 \over \left( \epsilon \alpha^2 \emass \ck^2 \right)^4} 
\approx 5.70 \times 10^{10} \,.
\label{bconstant} 
\ee
The condition that stars have a sufficiently long lifetime 
takes the form 
\be
C a^4 b \, \thetacen > X^2 g \,,
\label{lifetime} 
\ee
where the constant $C$ is given by 
\be
C = \left( {9375 \over 256 \pi^4} \right) 
\left( f_{\rm c} \effish \betacon \mzero^3 \right) 
\left( {\emass \ck^3 \over \hbar} \right) 
\left( {\kappa_0 \alpha^2 \over G} \right) 
{1 \over N_{\rm life}} \approx 0.586 \, ,  
\label{cconstant} 
\ee
where $N_{\rm life}$ is the number of atomic time scales required
for a functioning biosphere (and where we use $N_{\rm life}$ = 
$10^{33}$ to obtain the numerical value). 

The maximum allowed value of the stellar mass defines a maximum value
of the parameter $X$, i.e.,
\be
X \le X_{\rm max} \approx 50\,. 
\label{xmax} 
\ee
Finally, the minimum stellar mass can be written in terms of 
the minimum value of $X$ such that 
\be
X \ge X_{\rm min} = D a^{3/2} b^{-3/4} \thetacen^{-9/4} \,,
\label{xmin} 
\ee
where the constant $D$ is defined by 
\be
D = 6 \left( 3\pi \right)^{1/2} 
\left( {\pi^2 \mpro \over 5 \emass} \right)^{3/4} 
\alpha^{3/2} \approx 5.36\,. 
\label{dconstant} 
\ee


The bounds for a minimum stellar temperature (equation
[\ref{temperature}]) and a minimum stellar lifetime (equation
[\ref{lifetime}]) can be combined and written in the form
\be
C a^4 b \thetacen > g X^2 > a^{12} b^8 \thetacen^{14} B^{-2} \,.
\label{alphabound} 
\ee
The outer parts of the composite inequality (\ref{alphabound})
lead to the constraint 
\be
C B^2 > a^8 b^7 \thetacen^{13} \,. 
\ee
In order for the stellar structure equation (\ref{starstruck}) for 
the central temperature to have a valid solution, the parameter 
$\thetacen$ must be bounded from below so that 
\be
\thetacen > (\thetacen)_{\rm min} \approx 0.869 \,. 
\ee
The previous two equations thus imply the bound 
\be
a b^{7/8} < 
\left( C B^2 \right)^{1/8} (\thetacen)_{\rm min}^{-13/8} 
\approx 574 \, . 
\label{abound} 
\ee


Next we derive an upper limit on the gravitational structure constant
$\alpha_G$. Using the same procedure, we also find a corresponding 
upper limit on the ratio $\alpha_G/\alpha$ of the structure constants. 
If we combine the stellar temperature equation (\ref{starstruck}) with
the minimum value of the stellar mass parameter $X$ from equation
(\ref{xmin}), we obtain the inequality 
\be
I(\thetacen) \thetacen^{-8} \ge A X_{\rm min}^4 a^{-8} \ck^{-1} g =
A D^4 a^6 b^{-3} \thetacen^{-9} a^{-8} \ck^{-1} g \,. 
\ee
This result can be simplified to obtain the form  
\be
\thetacen I(\thetacen) \ge A D^4 a^{-2} b^{-3} \ck^{-1} g \,. 
\ee
We also require that the minimum stellar mass is less than the 
maximum stellar mass, $X_{\rm min} \le X_{\rm max}$. This 
condition can be used to obtain a bound on the scaled fine 
structure constant $a$, 
\be
a^2 \le 50^{4/3} D^{-4/3} \thetacen^{3} b \,. 
\ee
Combining the previous two equations then yields the 
inequality 
\be
\thetacen I(\thetacen) \ge A D^4 a^{-2} b^{-3} \ck^{-1} g 
\ge A D^4 b^{-3} \ck^{-1} g 50^{-4/3} D^{4/3} \thetacen^{-3} b^{-1} \,,
\ee
which can be rewritten in the form 
\be
g b^{-4} \ck^{-1} \le A^{-1} D^{-16/3} 50^{4/3} 
\left[ \thetacen^4 I(\thetacen) \right]_{\rm max} 
\approx 4.5 \times 10^6 
\left[ \thetacen^4 I(\thetacen) \right]_{\rm max} \,. 
\label{gbound} 
\ee
Note that we have replaced the value of the function 
$\thetacen^4 I(\thetacen)$ with is maximum value. 

Similarly, we can make an analogous argument to find a limit 
on the ratio $g/a$, which results in the upper bound
\be
\left({g \over a}\right) b^{-7/2} \ck^{-1} \le 
A^{-1} D^{-14/3} 50^{2/3} 
\left[ \thetacen^{5/2} I(\thetacen) \right]_{\rm max} 
\approx 10^6 
\left[ \thetacen^{5/2} I(\thetacen) \right]_{\rm max} \,. 
\label{govabound} 
\ee

Using the definition of the integral function $I(\thetacen)$
\cite{adams}, we can find a bound on the function of the form 
\be
I(\thetacen) < J_0 \thetacen^2 \exp[-3\thetacen] \,,
\ee
where $J_0$ is given by the integral 
\be
J_0 = \int_0^{\xi_\star} \xi^2 d\xi f^{2n-2/3} \,,
\ee
where $f(\xi)$ is the solution to the Lane-Emden equation 
for polytropic index $n$. Note that we can also write the 
expression for $J_0$ in the form 
\be
J_0 = \int_0^{\xi_\star} \xi^2 d\xi f^n \left[ f^{n-2/3} \right] \,.
\ee
As long as the polytropic index $n > 2/3$, the factor in square 
brackets is less than unity, whereas the remaining part of the 
expression is just $\mzero$, so that we obtain the bound 
\be
J_0 < \mzero\,.
\ee
Given the upper limit on $I(\thetacen)$, we can find an upper 
limit on functions of the form 
\be
F (\thetacen) = \thetacen^k I(\thetacen) \,,
\ee
which is bounded by 
\be
F \le F_{\rm max} < \mzero \left({k+2 \over 3}\right)^{k+2} 
\exp[-(k+2)] \,. 
\ee
Using this result to evaluate the bounds of equations 
(\ref{gbound}) and (\ref{govabound}), we find the limits 
\be
g \simless 2 \times 10^6 \left(b^4 \ck\right) 
\qquad {\rm and} \qquad 
{g \over a} \simless 2 \times 10^5 \left(b^{7/2} \ck\right) \,. 
\ee
The bounds of $\alpha$ and $\alpha_G$ thus scale linearly with the
value of the nuclear burning parameter $\conlum$, and also depend on
the mass ratio $\beta$. As discussed in Section \ref{sec:alphabeta},
the electron to proton mass ratio is constrained to be much less than
unity, with an approximate bound $b\simless23$ or equivalently 
$\beta\simless1/3$ (see also Section \ref{sec:trends}). With this 
additional information, the global bounds can be written in the form 
\be
{g\over \ck} \simless 5.6 \times 10^{11} 
\qquad {\rm and} \qquad 
{g \over \ck a} \simless 1.2 \times 10^{10} \,. 
\ee
One should keep in mind that these are upper bounds on the structure
constants. Tighter limits can in principle be found.  Nonetheless,
although the ratio $g/a$ can be made larger by many orders of
magnitude, it always falls well short of the hierarchy of 36 orders 
of magnitude found in our universe. 

\bigskip 
\section{Probability Considerations} 
\label{sec:probability} 

The main focus of this review has been to delineate the ranges over
which parameters can vary and still allow a universe to be potentially
habitable. In order to make a complete assessment of the degree of
fine-tuning, however, the probability for universes to obtain
particular parameters must be considered. Unfortunately, the relevant
probability distributions are neither specified by theory nor measured
by experiments at the present time. As a result, discussion of this
topic, which is subject to much uncertainty, is relegated to this
Appendix.

Note that (at least) two additional probabilistic complications must
be taken into account \cite{barnesbayes,hartlesrednicki,tegmark}:
First, the allowed ranges for working parameters generally do not have
sharp boundaries.  For example, one constraint on the fine structure
constant is that stars must live `long enough'.  Although this
constraint is undoubtedly necessary, it is not known how long it takes
for life to evolve.  For the sake of definiteness, we have taken the
required time to be $10^{33}$ atomic oscillations, equivalent to 1 Gyr
in our universe, but other reasonable choices could be made. Moreover,
the time required for biological evolution probably does not have a
single minimum value. Instead, life is unlikely to evolve on short
time scales, and more likely to evolve if given more time.  In this
context, we would like to know the probability $P(t)$ that life can
arise on a suitable planet within a given time $t$. Unfortunately,
again, we do not know the probability distribution for this
eventuality. Of course, this probability distribution is likely to
depend on the properties of both the planet and its environment. 
Because of these complications, the allowed regions of the parameter
spaces (shown in the figures in this review) must be subject to some
(as yet unknown) quality weighting. One should also keep in mind that
the most favorable part of the parameter space (that with the highest
weighting) does not necessarily correspond to the parameter values in
our universe.

Second, in addition to the range of allowed values, which must be
weighted as outlined above, we need to know the {\it a priori}
probabilities for universes to obtain given values of the parameters.
At the present time, we do not even know (with certainty) what
parameters are allowed to vary in a fundamental theory, much less the
probability for a universe to realize them. As a result, the
probability distributions for a universe to attain given values of the
parameters considered herein currently remain highly uncertain. As one
example \cite{tegmark}, the {\it a priori} probability distribution
for the vacuum energy density $\rhov$ was assumed to be flat 
(independent of $\rhov$) because only a small portion of the possible
range can produce structure. Other choices are possible (e.g., see 
\cite{martel,sandora2019} and many others). 

Even though we do not know the probability distributions sampled by the
relevant parameters, we can consider some general trends. As noted in
Section \ref{sec:resultsummary}, for most of the parameters
traditionally considered in fine-tuning discussions, the allowed
ranges span a couple to several orders of magnitude (Figures 
\ref{fig:updown}, \ref{fig:abplane}, \ref{fig:asplane}, 
\ref{fig:qvlambda}, \ref{fig:bbnplane}, \ref{fig:galsurvive}, 
\ref{fig:etadelta}, \ref{fig:staragc}, \ref{fig:starplane}, and 
\ref{fig:planplane}). These ranges are summarized in Table 
\ref{table:summary}. For our universe at the present epoch, the 
full range of allowed energy/time/mass scales spans a factor of
$\sim10^{61}$, or 61 decades (see Section \ref{sec:cosconprob} and
equations [\ref{tplanck}] and [\ref{lplanck}]). Similarly, the range
of mass and size scales allowed by physical considerations spans about
$80\times15$ decades (Figure \ref{fig:astroscales}).  If the parameter
probabilities were distributed evenly in logarithmic space across a
comparable range (of order 100 decades --- see also the discussion
below), then the allowed ranges of parameter space correspond to a few
percent (maybe $\sim1-10\%$) of the total.  In contrast, if the
parameters were distributed uniformly over the full range --- and the
viable parameters do not lie at the upper end of that range --- then
the allowed range would represent an incredibly small fraction of the
total.

The cosmological constant provides the canonical example of this
probability issue. The observed value of energy density of the vacuum
$\rhov=\lambda^4\sim10^{-10}$ eV$^4$ is smaller than the maximum (and
perhaps expected) scale $\mplanck^4$ by $\sim120$ orders of magnitude.
The energy density $\rhov$ can be much larger than the observed value
(by a factor of $\sim10^{10}$ or more; see Section \ref{sec:rhovac})
and still allow for a working universe, but even the maximum allowed
energy density is $\sim100$ orders of magnitude smaller than the
benchmark. If the possible values of $\rhov$ are distributed
uniformly, then the chances of realizing such a small value are only 
$\sim10^{-100}$ and hence highly improbable. On the other hand, if the
possible values of $\rhov$ are distributed in a log-uniform manner,
the probability of obtaining a workable value of $\rhov$ is given by
\be
P_\Lambda \approx {10 - \log_{10}(\rho_{min}/\rho_{obs}) \over 
120 - \log_{10}(\rho_{min}/\rho_{obs})} > {1\over12}\,, 
\ee
where $\rho_{min}$ is the minimum allowed value of the vacuum energy
density. The numerical result on the right assumes that the minimum
value is equal to the observed value $\rho_{obs}$.  Any smaller choice
leads to a larger probability. For example, if the energy density
could be 120 orders of magnitude smaller than that observed (that
energy scale corresponds to a wavelength comparable to the
cosmological horizon size), then $P_\Lambda\sim13/24\sim1/2$. Even in
the worst case scenario, approximately 1 out of 12 universes would
inherit a value of the cosmological constant that is compatible with
habitability.

This same state of affairs applies to other fundamental constants and
cosmological parameters. The strengths of the four forces span a range
of about 40 orders of magnitude (note that because the coupling
constants run with energy, this value depends on the energy scale of
interest, and this estimate applies in the low energy limit). The
range of masses measured for the quarks spans only about 5 or 6 
orders of magnitude. However, if one considers masses for the full
collection of existing particles, from the Higgs mass at $125$ GeV
down to estimated neutrino masses of order $\sim10^{-3}$ eV, the range
becomes about 14 orders of magnitude. Cosmological parameters, such as
the baryon to photon ratio $\eta\sim10^{-10}$ and the fluctuation
amplitude $Q\sim10^{-5}$, could also vary over $\sim10$ orders of
magnitude. Given these ranges for the tunable parameters (tens of
orders of magnitude), in conjunction with the ranges of values that
allow for habitable universes (typically an order of magnitude or a
few -- see Table \ref{table:summary}), log-uniform distributions imply
that successful realizations would be chosen with probabilities of 
order $1-10\%$. Taken together, these results suggest that:
{\sl If the fundamental parameters are sampled from log-uniform} 
{\sl distributions, then hierarchical fine-tuning issues are largely}
{\sl alleviated}. On the other hand, if the parameters are sampled 
from a random distribution with a large high-end cutoff, then 
hierarchical fine-tuning problems remain troublesome, even 
pernicious. 

As is well known in probability theory \cite{sivia}, in the absence of
any additional defining information, the principle of maximum entropy
\cite{jaynes,shannon} implies that the underlying probability distribution 
should be flat or uniform: $P(x)\approx$ {\sl constant}. On the other
hand, we get a different result for problems where the relative change
in parameters is important, rather than their absolute values. In this
case, the preferred probability distribution has the form 
\be
P(x | I) \propto {1 \over x} \, , 
\ee
which corresponds to the probability being distributed in a
log-uniform manner.  This probability density function, sometimes
called the {\it Jeffreys prior} \cite{jeffreys}, arises in problems
that do not have a single well-defined fundamental scale (except for
possible cutoffs). More generally, power-law distributions arise in
problems that are scale-free or self-similar \cite{barenblatt}, and
are found in a wide variety of observed phenomena, including
earthquake magnitudes, crater sizes, intensity of solar flares, and
populations of cities \cite{newman}.

The key issue is whether the tunable parameters of physics and
cosmology are distributed in a uniform or a log-uniform manner. The
parameters could sample a more complicated distribution, of course,
but it is useful to consider this simple dichotomy as a starting
point. This choice is equivalent to asking whether the parameters are
governed by a single well-defined scale, or if they are essentially
scale-free.  Significantly, this question has no definitive answer at
the present time.

\begin{figure}[tbp]
\centering 
\includegraphics[width=1.0\textwidth,trim=0 150 0 150]{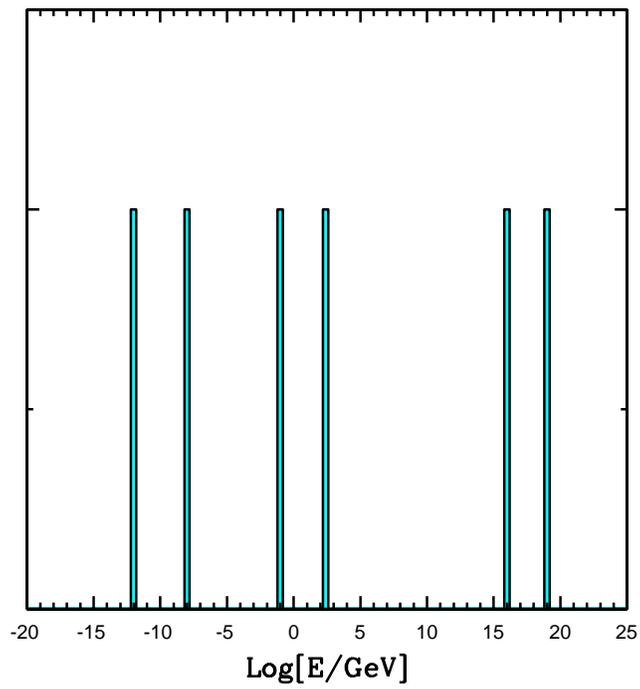}
\caption{Energy scales in the universe. The six spikes correspond to 
the energy scales of the dark energy (0.001 eV), atomic energy levels
(10 eV), the QCD scale (100 MeV), the electroweak scale (250 GeV), the
Grand Unified Scale ($10^{16}$ GeV), and finally the Planck scale at 
$10^{19}$ GeV (from left to right). }
\label{fig:energyhist} 
\end{figure}  

Nonetheless, one piece of evidence is provided by Figure 
\ref{fig:quarkhist}, which shows that the quark masses are in fact 
distributed in a (nearly) log-uniform manner (for further discussion,
see \cite{donoghuemass,jaffe}).  These data are consistent with a
log-uniform sampling of the particle masses. This finding, in turn,
argues for the lack of a well-defined mass scale and hence some
alleviation of hierarchical fine-tuning issues.

Another example of a possible log-uniform distribution is provided 
by the energy scales of the universe, as shown in Figure 
\ref{fig:energyhist}. This plot shows the spacing of six fundamental 
energy scales (values are taken from Table 1 of \cite{carroll2006} and
from \cite{particlegroup}; five of the scales are also discussed in 
\cite{bjorken}): The largest value is the Planck scale, where gravity 
exhibits quantum behavior ($E_{\rm pl}=\mplanck=10^{19}$ GeV). The
next step on the hierarchy is the Grand Unified Scale (taken here to
be $E_{\rm gut}=10^{16}$ GeV) where the other three forces become
unified. The Fermi scale for electroweak interactions is next 
($E_{\rm w}=G_F^{-1/2}\sim250$ GeV), followed by the QCD scale
($E_{\rm qcd}=\Lambda_{\rm qcd}\sim100$ MeV). Atomic energy scales are
set by the fine structure constant and are given approximately by
$E_{\rm atom}\sim\alpha^2\emass\sim10$ eV. Finally, the energy scale
of empty space, given by the dark energy or cosmological constant, is
the lowest scale on the diagram ($\rhov^{1/4}=\lambda\sim10^{-3}$ eV).
As shown in Figure \ref{fig:energyhist}, these scales span a wide
range of energies extending over more than 30 decades. Although the
spacing is not perfectly even, this collection of energy scales is
roughly consistent with a log-uniform distribution.

The biggest apparent difference between a log-uniform distribution and
that of the energy scales shown in Figure \ref{fig:energyhist} is the
large gap between the electroweak scale and the GUT scale. This
well-known feature is sometimes denoted as {\sl The Particle Desert}.  
But such a gap is not unlikely: If the six energy scales are sampled
from a log-uniform distribution over the range shown, an interval as
large as the observed desert will arise about 54 percent of the time
(note that this percentage will vary with the chosen range of possible
energy scales).  Nonetheless, the presence of an additional energy
scale within the gap would go a long way toward making the
distribution even closer to log-random. One candidate for such an
additional scale arises from the paradigm of supersymmetry, which
posits a basic relationship between fermions and bosons, where each
known particle has a superpartner of the opposite type 
\cite{haberkane,nilles}. With the current deficit of experimental
data, the scale of supersymmetry could lie anywhere within the energy
interval corresponding to the particle desert.  One collection of
arguments suggests an energy scale of $\sim1$ PeV \cite{wellspev}, 
while other models put the energy scale closer to that of
unification \cite{arkani}.

Finally, note that one can turn the argument around: If the tunable
parameters of physics are sampled from a uniform distribution, then
the probability of attaining certain small values required for a
successful universe (like the observed energy density $\rhov$ of the
vacuum) would become uncomfortably small. On the other hand, if the
underlying distributions are log-uniform, the probabilities for
realizing parameters consistent with a habitable universe are no
longer problematic. This result could thus be considered as evidence
in favor of scale-free and hence log-uniform distributions. Such
distributions are also suggested by renormalization group treatments
(see equation [\ref{renormalize}]). Nonetheless, the construction of
credible probability distributions for particles masses, energy
scales, and other fundamental parameters represents a formidable
challenge for the future. 

For completeness, note that the dark energy density $\rhov$ could be
negative, whereas this discussion and Figure \ref{fig:energyhist}
consider only the scale $\lambda=|\,\rhov|^{1/4}$, which takes on only
positive values. One must thus consider a more complicated treatment
(instead of using only a log-random distribution) to take into account
negative $\rhov$. In addition, the parameter value $\lambda=0$ is
allowed, but $\lambda=0$ lies an `infinite distance' away from the
observed $\lambda\ne0$ value according to the log-random distribution.

\bigskip 
\section{Nuclei and the Semi-Empirical Mass Formula}
\label{sec:semfappend} 

As outlined in the main text, the stability of nuclei is an important
consideration for the potential habitability of other universes. One
general trend emerging from the entire collection of constraints is
that the requirement that bound nuclear states exist can be more
restrictive than the requirement that those nuclei can be synthesized
in stars or during Big Bang Nucleosynthesis.  This Appendix explores
the possible ranges of allowed nuclear structures in of other
universes through the use of a standard version of the Semi-Empirical
Mass Formula or SEMF \cite{semf}.  This formalism has already been
invoked in Section \ref{sec:strongcon} to place limits on the strong
coupling constant (see equation [\ref{nofission}] and Figure
\ref{fig:asplane}). Here we consider the range of allowed nuclei.

Within the context of this model, the nuclear binding energy 
$E_{\rm b}$ can be written in the form 
\be
E_{\rm b} (A,Z) = V A - S A^{2/3} - C {Z(Z-1)\over A^{1/3}} - 
B {(A-2Z)^2 \over A} + \delta_{\rm qp}(A) \,. 
\label{semf} 
\ee
The corresponding mass $M(A,Z)$ of the nucleus is then given by 
\be
M(A,Z) = Z \mpro + (A-Z) m_{\rm n} - E_{\rm b} \,.
\label{nukemass} 
\ee
The coefficients $(V,A,C,B)$ and the function $\delta_{\rm qp}(A)$ are
chosen to provide a good fit to the observed nuclear binding energies
in our universe, but are based on physical considerations as outlined
below.  The contributions in equation (\ref{semf}) include the volume
term due to nearest neighbor interactions, the surface term which
corrects for the decrease in such interactions for nucleons near the
surface, the Coulomb term due to electromagnetic repulsion, the
asymmetry term which favors equal numbers of protons and neutrons, and
finally the quantum pairing term $\delta_{\rm qp}$ which favors even
numbers of nucleons. Different authors provide different fits for the
binding energy, and hence the coefficients, but the values do not vary
by wide margins. For the sake of definiteness, we use the following
choices \cite{basdevant}: 
\be
V = 15.75\,{\rm MeV}\,, \quad S = 17.80\,{\rm MeV}\,, \quad  
C = 0.7103\,{\rm MeV}\,, \quad B = 23.69\,{\rm MeV} \,. 
\label{semfcon} 
\ee

In the present context, we apply the SEMF across a range of universes
with different values of the fundamental parameters. To provide a
simplified treatment, we ignore the pairing term, which favors nuclei
with even numbers of neutrons and protons. This term models variations
in the binding energy curve as the values of $(Z,A)$ vary from odd to
even integers. Here we are interested in the gross properties of the
nuclear binding energy curves and thus ignore these ``smaller scale''
variations. 

Although the constants from equation (\ref{semfcon}) are fit to
experimental data, they have a well motivated theoretical basis.  
In particular, the Fermi gas model allows for a semi-quantitative
assessment of the terms appearing in equation (\ref{semf}).  The
coefficients of the volume term $V$, the surface term $S$, and the
asymmetry term $B$ are all proportional to the Fermi Energy 
$\varepsilon_F$, with coefficients of order unity \cite{basdevant}. 
The Fermi energy is given by  
\be
\varepsilon_F = {\hbar^2 \over 2m} (3\pi^2 n)^{2/3}\,,
\label{efermi} 
\ee
where $n$ is the number density of nucleons in the nucleus and $m$ 
is the nucleon mass.  In our universe, the benchmark nuclear density 
$n_0\sim0.15$ fm$^{-3}$ is set by the strength and range of the 
strong force. Here we can parameterize $n_0$ through the ansatz
\be
n_0 = \mu^3 \,, 
\ee
where the mass scale $\mu\approx106$ MeV will reproduce the observed
nuclear density $n_0$ in our universe. This scale is close to the pion
mass (as expected). The number density $n$ appearing in the definition
of the Fermi energy (\ref{efermi}) is the density of the protons, or
neutrons, considered as separate components of the Fermi gas. Under
the approximation that $Z\approx N\approx A/2$, the density 
$n\approx n_0/2$ and the Fermi energy can be written in the form
\be
\varepsilon_F = 2^{-5/3} (3\pi^2)^{2/3} 
{\mu^2\over{m}} \sim 3 {\mu^2\over{m}} 
\approx 36 \, {\rm MeV} \,.
\label{efermitwo} 
\ee
We can thus define scaled versions of the coefficients such that 
\be
V = a_V \varepsilon_F \approx 0.44 \varepsilon_F \,, \qquad 
S = a_S \varepsilon_F \approx 0.49 \varepsilon_F \,, \qquad 
B = a_B \varepsilon_F \approx 0.66 \varepsilon_F \,. 
\ee

The Coulomb term arises from the electromagnetic repulsion of the
constituent protons. This term is determined by the potential energy
of the charge distribution. For a uniform charge density, this
potential energy takes the form 
\be
U = {3\over5} {e^2 Z^2 \over R} \,, 
\ee
where $R$ is the size of the nucleus, which is generally written 
as $R = A^{1/3} r_0$, where $r_0 \sim \mu^{-1}$. The Coulomb 
term thus takes the form 
\be
C {Z(Z-1)\over A^{1/3}} \approx U \approx {3 \over 5} \, 
(r_0 \mu)^{-1} \, \alpha \mu {Z^2 \over A^{1/3}} \,. 
\ee
As a result, we can define a scaled version of the coefficient 
\be
C = a_C \alpha \mu \approx 0.92 \, \alpha \, \mu \,. 
\ee
The scaled version of the SEMF becomes 
\be
E_{\rm b} = \varepsilon_F \left[ a_V A - a_S A^{2/3} 
- a_C \left( {\alpha \mu \over \varepsilon_F} \right) 
{Z(Z-1)\over A^{1/3}} - a_B {(A-2Z)^2 \over A} \right] \,. 
\label{semftwo} 
\ee
When written in this form, the Fermi energy $\varepsilon_F$ defines
the overall energy scale for nuclear binding energies.  The scaled
coefficients $(a_V,a_S,a_C,a_B)$ are dimensionless quantities of order
unity. The only remaining parameter is the ratio ${\cal R}$ of the
Coulomb coefficient to the Fermi energy, 
\be
{\cal R} \equiv \left( {\alpha \mu \over \varepsilon_F} \right) 
\sim {\alpha m \over \mu} \,, 
\ee
which determines the shape of the nuclear binding energy curve. 

\begin{figure}[tbp]
\centering 
\includegraphics[width=1.0\textwidth,trim=0 150 0 150]{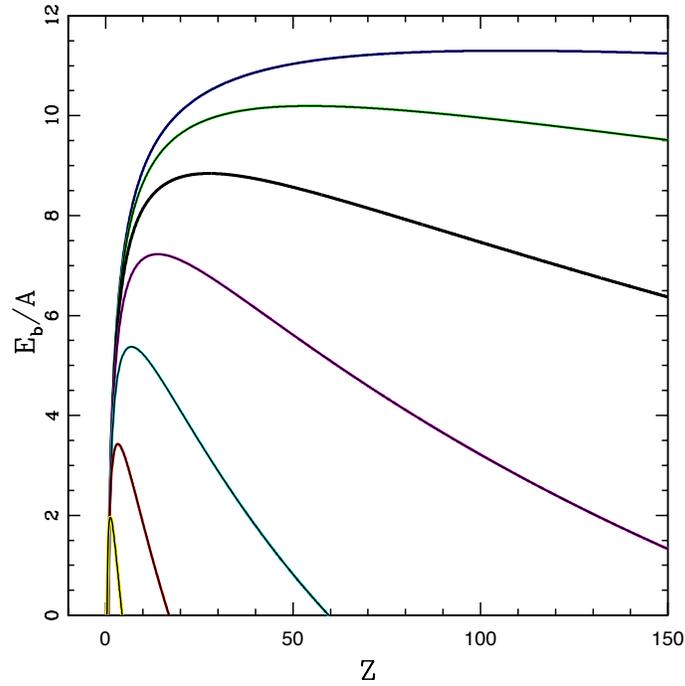}
\caption{Nuclear binding energy curves for universes with different 
values of the fundamental constants. The nuclear binding energy per
nucleon $E_{\rm b}/A$ (in MeV per nucleon) is plotted as a function 
of atomic number $Z$ for the most bound nucleus according to the
Semi-Empirical Mass Formula. In this simplified treatment, the quantum
pairing term is neglected, so that the variations between even and odd
nuclei are not shown. The shape of the curves is determined by the
ratio of the electromagnetic contribution to the Fermi energy at
nuclear densities, where ${\cal R}\approx\alpha m/\mu$ (see text).
The curves correspond to the values ${\cal R}/{\cal R}_0$ = 1/4
(blue), 1/2 (green), 1 (black), 2 (magenta), 4 (cyan), 8 (red), and 16
(yellow).  The possible range of allowed nuclei decreases with
increasing $\alpha$ (or ${\cal R}$). The energy scale for the SEMF is 
determined by the value of the Fermi energy (equation [\ref{semftwo}]), 
which is set by the strength of the strong force. For this figure,
$\varepsilon_F$ corresponds to the value in our universe; in other
universes, the binding energy scales linearly with $\varepsilon_F$. }
\label{fig:nukebind} 
\end{figure}

One can find the atomic number $Z_{max}$ of the most bound nucleus 
for a given value of the atomic weight. The function $Z_{max}(A)$ 
is determined by the condition 
\be
{\partial \over \partial Z} M(A,Z) = 0 \quad \Rightarrow \quad 
Z_{max} = \left( {A\over2} \right) 
{1+C/4B A^{1/3}+(m_{\rm n}-\mpro)A/4B \over 1+CA^{2/3}/4B} \,.
\ee
Using this value for $Z_{max}(A)$, we can find the maximum nuclear
binding energy as a function of atomic number $Z$. The result,
expressed in terms of the binding energy per nucleon 
$[E_{\rm b}(A)/A]_{Z_{max}}$ is shown in Figure \ref{fig:nukebind} 
for varying values of the strength of the Coulomb term. The curves
show how the shape of the binding energy curve changes with the value
of the ratio ${\cal R}$ of the Coulomb term to the Fermi energy. 
As ${\cal R}$, equivalently $\alpha$, increases, the binding energy
curve becomes more peaked and nuclei with large atomic numbers become
increasingly unbound. However, the range of bound nuclei remains 
relatively large, well beyond the atomic number of iron, as long as 
the Coulomb term does not increase by more than about a factor of
$\sim4$ (corresponding to the cyan curve in Figure \ref{fig:nukebind}). 
The peak of the curve (where the energy per particle is maximum)
specifies the atomic number of the most bound nucleus and varies by a
factor of $\sim2$ as the ratio ${\cal R}$ varies by a factor of 16.

We can determine another benchmark value of the ratio ${\cal R}$ by
requiring the existence of nuclei for the most abundant elements found
in terrestrial life forms. These elements include carbon, hydrogen,
nitrogen, oxygen, phosphorus, and sulfur (CHNOPS) and make up 98\% of
known biomolecules. The largest of these nuclei, sulfur, has atomic
number $Z=16$ and mass number $A=32$. By requiring the stability of
nuclei up to this size, the ratio ${\cal R}$ cannot be more than
$\sim6$ times larger than the value in our universe. As a result, 
the fine structure constant is bounded from above such that  
$\alpha\simless6\alpha_0$. 

Note that the constraint that a universe must support large stable
nuclei is necessary but not sufficient. Even if bound nuclear states
exist, they can still be unstable to radioactive decay. If the
half-lives for either $\alpha$-decay or $\beta$-decay are too short,
then the nuclei would no longer be useful for habitability. A full
assessment of the possible radioactivity for nuclei in other universes
has not been carried out, but will place further constraints on this 
scenario.

In this formulation of the problem, the Fermi energy $\varepsilon_F$
sets the overall scale of the binding energy curve. The entire binding
energy thus scales up and down with changes in $\varepsilon_F$, which
in turn depends on both the nucleon mass $m$ and the scale $\mu$ that
determines nuclear densities. The typical nuclear binding energies are
5 -- 10 MeV per particle for the value of $\varepsilon_F$ (about 36 MeV)
in our universe. This value is bounded from above by the requirement
that the nucleons remain non-relativistic. The value is bounded from 
below by the requirement that nuclear energy levels are much larger 
than atomic energy levels, which drive chemical reactions and hence 
biological function. The parameters of the problem must thus 
obey the ordering
\be
\alpha^2 \emass \ll {3 \mu^2 \over m} \ll m \,.
\label{nukeorder} 
\ee
Our universe displays a well defined ordering of these scales, with
atomic energy levels $\sim$eV, nuclear binding energies $\sim$MeV, and
nucleon masses $\sim$GeV. As outlined above, the ratio ${\cal R}$
cannot vary by more than a factor of $\sim4$ without the universe
losing too much of its periodic table. This constraint, in turn,
implies that the ratio of the first two terms in equation
(\ref{nukeorder}) cannot vary by more than a factor of $\sim16$, so
that the hierarchy between atomic and nuclear energy levels will be
maintained within any viable universe.

Although the Semi-Empirical Mass Formula provides a useful framework
to consider possible variations in nuclear structure across the
multiverse, it is certainly too simple to capture the entire nuclear
landscape.  In this model, the nucleons are considered to be
self-contained entities so that the interactions can be attributed to
the exchange of a single pion. A more complete treatment, including
the complexities of QCD, are now being carried out for specific
nuclei \cite{eckstrom,epelbaum2013,lahde,meissner}.  Although a full
exploration of the nuclear parameter space remains to be carried out,
such calculations will undoubtedly add structure and nuance to the
simplistic picture presented here (see also 
\cite{jaffe,schellekens2008}).

\vskip1.0truecm 

\bigskip 
\noindent

\end{document}